\documentclass[12pt,oneside,a4paper]{article}
\usepackage{authblk,color,hyperref}
\usepackage{ulem,lipsum}
\usepackage{units,eurosym,morefloats}
\usepackage[amssymb]{SIunits}
\usepackage{amsmath}
\usepackage{amsfonts}
\usepackage{amssymb}
\usepackage{graphicx}
\usepackage{wrapfig}
\usepackage{caption}
\usepackage{subcaption}

\definecolor{blus}{rgb}{0.1,0.1,0.8}
\definecolor{GreenYellow}{cmyk}{0.15,0,0.69,0}
\definecolor{Yellow}{cmyk}{0,0,1,0}
\definecolor{Goldenrod}{cmyk}{0,0.10,0.84,0}
\definecolor{Dandelion}{cmyk}{0,0.29,0.84,0}
\definecolor{Apricot}{cmyk}{0,0.32,0.52,0}
\definecolor{Peach}{cmyk}{0,0.50,0.70,0}
\definecolor{Melon}{cmyk}{0,0.46,0.50,0}
\definecolor{YellowOrange}{cmyk}{0,0.42,1,0}
\definecolor{Orange}{cmyk}{0,0.61,0.87,0}
\definecolor{BurntOrange}{cmyk}{0,0.51,1,0}
\definecolor{Bittersweet}{cmyk}{0,0.75,1,0.24}
\definecolor{RedOrange}{cmyk}{0,0.77,0.87,0}
\definecolor{Mahogany}{cmyk}{0,0.85,0.87,0.35}
\definecolor{Maroon}{cmyk}{0,0.87,0.68,0.32}
\definecolor{BrickRed}{cmyk}{0,0.89,0.94,0.28}
\definecolor{Red}{cmyk}{0,1,1,0}
\definecolor{OrangeRed}{cmyk}{0,1,0.50,0}
\definecolor{RubineRed}{cmyk}{0,1,0.13,0}
\definecolor{WildStrawberry}{cmyk}{0,0.96,0.39,0}
\definecolor{Salmon}{cmyk}{0,0.53,0.38,0}
\definecolor{CarnationPink}{cmyk}{0,0.63,0,0}
\definecolor{Magenta}{cmyk}{0,1,0,0}
\definecolor{VioletRed}{cmyk}{0,0.81,0,0}
\definecolor{Rhodamine}{cmyk}{0,0.82,0,0}
\definecolor{Mulberry}{cmyk}{0.34,0.90,0,0.02}
\definecolor{RedViolet}{cmyk}{0.07,0.90,0,0.34}
\definecolor{Fuchsia}{cmyk}{0.47,0.91,0,0.08}
\definecolor{Lavender}{cmyk}{0,0.48,0,0}
\definecolor{Thistle}{cmyk}{0.12,0.59,0,0}
\definecolor{Orchid}{cmyk}{0.32,0.64,0,0}
\definecolor{DarkOrchid}{cmyk}{0.40,0.80,0.20,0}
\definecolor{Purple}{cmyk}{0.45,0.86,0,0}
\definecolor{Plum}{cmyk}{0.50,1,0,0}
\definecolor{Violet}{cmyk}{0.79,0.88,0,0}
\definecolor{RoyalPurple}{cmyk}{0.75,0.90,0,0}
\definecolor{BlueViolet}{cmyk}{0.86,0.91,0,0.04}
\definecolor{Periwinkle}{cmyk}{0.57,0.55,0,0}
\definecolor{CadetBlue}{cmyk}{0.62,0.57,0.23,0}
\definecolor{CornflowerBlue}{cmyk}{0.65,0.13,0,0}
\definecolor{MidnightBlue}{cmyk}{0.98,0.13,0,0.43}
\definecolor{NavyBlue}{cmyk}{0.94,0.54,0,0}
\definecolor{RoyalBlue}{cmyk}{1,0.50,0,0}
\definecolor{Blue}{cmyk}{1,1,0,0}
\definecolor{Cerulean}{cmyk}{0.94,0.11,0,0}
\definecolor{Cyan}{cmyk}{1,0,0,0}
\definecolor{ProcessBlue}{cmyk}{0.96,0,0,0}
\definecolor{SkyBlue}{cmyk}{0.62,0,0.12,0}
\definecolor{Turquoise}{cmyk}{0.85,0,0.20,0}
\definecolor{TealBlue}{cmyk}{0.86,0,0.34,0.02}
\definecolor{Aquamarine}{cmyk}{0.82,0,0.30,0}
\definecolor{BlueGreen}{cmyk}{0.85,0,0.33,0}
\definecolor{Emerald}{cmyk}{1,0,0.50,0}
\definecolor{JungleGreen}{cmyk}{0.99,0,0.52,0}
\definecolor{SeaGreen}{cmyk}{0.69,0,0.50,0}
\definecolor{Green}{cmyk}{1,0,1,0}
\definecolor{ForestGreen}{cmyk}{0.91,0,0.88,0.12}
\definecolor{PineGreen}{cmyk}{0.92,0,0.59,0.25}
\definecolor{LimeGreen}{cmyk}{0.50,0,1,0}
\definecolor{YellowGreen}{cmyk}{0.44,0,0.74,0}
\definecolor{SpringGreen}{cmyk}{0.26,0,0.76,0}
\definecolor{OliveGreen}{cmyk}{0.64,0,0.95,0.40}
\definecolor{RawSienna}{cmyk}{0,0.72,1,0.45}
\definecolor{Sepia}{cmyk}{0,0.83,1,0.70}
\definecolor{Brown}{cmyk}{0,0.81,1,0.60}
\definecolor{Tan}{cmyk}{0.14,0.42,0.56,0}
\definecolor{Gray}{cmyk}{0,0,0,0.50}
\definecolor{Black}{cmyk}{0,0,0,1}
\definecolor{White}{cmyk}{0,0,0,0}

\newcommand{\B}{{\bf \color{PineGreen}Lorenzo:}\ }
\newcommand{\Bb}{{\it \color{PineGreen}Lorenzo}}
\newcommand{\A}{{\bf \color{PineGreen}Maike:}\ }
\newcommand{\Aa}{{\it \color{PineGreen}Maike}}
\newcommand{\C}{{\bf Caretaker:}\ }

\newcommand{\R}{{\bf \color{blue} Rutherf:}\ }
\newcommand{\G}{{\bf \color{BlueViolet} Geiger:}\ }
\newcommand{\SL}{{\bf \color{blue} Sau Lan:}\ }
\newcommand{\SLw}{{\it \color{blue} Sau Lan Wu}\ }
\newcommand{\Hh}{{\bf \color{blue} Haimo:}\ }
\newcommand{\Hhh}{{\it \color{blue} Haimo}\ }
\newcommand{\Lw}{{\bf \color{blue} Lawrence:}\ }
\newcommand{\Lwl}{{\it \color{blue} Lawrence}\ }

\newcommand{\EXP}[1]{{\bf\color{magenta}{Exp: #1}}}
\newcommand{\hop}{\medskip}
\newcommand{\gs}{{\bf \color{blue} Good Scient.:}\ }
\newcommand{\bs}{{\bf \color{BlueGreen} Bad Scient.}\ }

\newcommand{\bss}{{\it \color{BlueGreen} Bad Scientist}\ } 

\newcommand{\As}{{\bf \color{blue} Jane}\ }

\begin{document}
\title{\bf  ``What's (the) Matter?", \\ {\large A Show on Elementary Particle Physics with 28
Demonstration Experiments}}

\author[1,2]{Herbi K. Dreiner\thanks{For electronic correspondence please contact H.\,K.\,Dreiner at  dreiner@uni-bonn.de.}}
\author[3]{Max Becker}
\author[4]{Mikolaj Borzyszkowski}
\author[1]{Maxim Braun}
\author[1]{Alexander Fa{\ss}bender}
\author[1]{Julia Hampel}
\author[1]{Maike Hansen}
\author[1,5]{Dustin Hebecker}
\author[6]{Timo Heepenstrick}
\author[1]{Sascha Heinz}
\author[1]{Katharina Hortmanns}
\author[3]{Christian Jost}
\author[1]{Michael Kortmann}
\author[4]{Matthias U. Kruckow}
\author[1]{Till Leuteritz}
\author[3]{Claudia L\"utz}
\author[3]{Philip Mahlberg}
\author[3]{Johannes M\"ullers}
\author[1,2]{Toby Opferkuch}
\author[1]{Ewald Paul}
\author[3]{Peter Pauli}
\author[3]{Merlin Rossbach}
\author[1]{Steffen Schaepe}
\author[1]{Tobias Schiffer}
\author[1]{Jan F. Schmidt}
\author[4]{Jana Sch\"uller-Ruhl}
\author[1]{Christoph Sch\"urmann}
\author[1,7]{Lorenzo Ubaldi}
\author[8]{Sebastian Wagner-Carena}
\affil[1]{\small Physikalisches Institut, Universit\"at Bonn, Nussallee 12, 53115 Bonn, Germany}
\affil[2]{Bethe Center for Theoretical Physics, Universit\"at Bonn, Nussallee 12, 53115 Bonn, Germany}
\affil[3]{HISKP, Nussallee 14-16, 53115 Bonn, Germany}
\affil[4]{AIfA, Auf dem H\"ugel 71, 53121 Bonn, Germany}
\affil[5]{Humboldt-Universit\"at zu Berlin, Institut f\"ur Physik, Newtonstrasse 15, 12489 Berlin, Germany}
\affil[6]{Physikalische Chemie, Wegeler Strasse, 53115, Bonn, Germany}
\affil[7]{Raymond and Beverly Sackler School of Physics and Astronomy,  Tel-Aviv University, Tel-Aviv 69978, Israel}
\affil[8]{Jefferson Physical Laboratory, Harvard University, Cambridge, MA 02138, USA}

\date{}
\maketitle

\begin{abstract}
We present the screenplay of a physics show on elementary particle physics which has been developed by the Physikshow 
of Bonn University.  The show is addressed at non-physicists aged 14 to 99 and is intended to communicate some basic 
concepts of elementary particle physics including the discovery of the Higgs boson in an entertaining fashion. It is also
intended to demonstrate a successful outreach activity which heavily relies on the university physics students. This paper is 
addressed at anybody interested in particle physics, hopefully the combination of the screenplay and the detailed descriptions 
of the experiments will be illuminating. This paper is also addressed at fellow physicists working in outreach, maybe the 
experiments and our choice of simple explanation will be helpful. Furthermore, we are very interested in related activities 
elsewhere, in particular also demonstration experiments relevant to particle physics, since as far as we are aware, little of this 
work is published.

Our show involves 28 live demonstration experiments. These are presented in an extensive appendix, including photos and 
technical details. The show is set up as a quest, where two students from Bonn with the aid of a caretaker travel back in time to 
understand the fundamental nature of matter. They visit Rutherford and Geiger in Manchester around 1911, who recount their 
famous experiment on the nucleus and show how particle detectors work. Then they travel forward in time to meet E.~Lawrence 
at Berkeley around 1950, teaching them about the how and why of accelerators. Next, they visit Sau~Lan~Wu at DESY, Hamburg, 
around 1980, who explains the strong force. The final stop is in the LHC tunnel at CERN, Geneva, Switzerland in 2012. Two 
experimentalists tell them about colliders and our heroes watch live as the Higgs boson is produced and decays. The show has 
been presented in English at Oxford University and University College London (3/2014), as well as Padua University and ICTP 
Trieste (3/2015). It was first performed in German at the Deutsche Museum, Bonn (5/2014). The show has eleven speaking parts 
and involves in total 20 people. 
\end{abstract}

\tableofcontents

\section{Introduction}
\label{sec:intro}

With the rise of contemporary science it has become more pertinent to find a means of communicating groundbreaking
and essential scientific results to the general public. Galileo Galilei, arguably the founder of modern science, took a 
bold step in 1630 when he published his {\it Dialogo} \cite{galilei} in Italian, over the traditional Latin, making his 
thoughts and results accessible to a much broader audience. When the Royal Institution \cite{Royal-Inst} was founded 
in London in March 1799 ``with the aim of introducing new technologies and teaching science to the general public" the 
bridge between the academic and public domain strengthened. One of its early presidents, Michael Faraday, was a 
strong proponent of presenting physics to the public, including the use of live experiments. During his tenure, he 
created the famous Christmas lectures, an annual series of entertaining lectures on a single scientific topic which the 
museum still holds to this day. For example, on May 27th, 2014, at the Royal Institution, one of us (H.K.D.) was able to 
attend Jon Butterworth's lecture launching his popular book ``Smashing Physics" on the LHC and the Higgs boson 
\cite{Butterworth:2014eka}, where he also showed two live experiments. In Germany the Deutsches Museum of 
Masterpieces in Science and Technology \cite{dm-url} was founded in 1925 with the intention of displaying major 
scientific results through engaging demonstrations. Today it boasts hundreds of unique and interactive exhibits that help 
``[bridge] the gap between research and education.'' We have visited the Deutsches Museum M\"unchen three times
with the Bonn physics show with various programs, and have also performed five times in the branch of the Museum, in 
Bonn, Bad Godesberg. The Deutsches Museum technique quickly caught on and inspired several similar museums, for 
example the Science Museum, London \cite{science-london-url} and the Museum of Science and Industry in the city of 
Chicago \cite{science-chicago-url}. A museum founded by a former particle physicist, Frank Oppenheimer in 1969, and 
famous for its hands-on experiments is the Exploratorium in San Francisco \cite{science-sf-url}.

The Bonn physics show was launched in Dec. 2001 as an opportunity for the Bonn University physics students to be 
creative in developing new forms of presenting physics to a broad audience, in particular for school kids aged
10 to 18. From the beginning the shows typically filled with demonstrations in the context of classical physics, were 
sometimes in the form of actual plays and always filled with music and jokes, very different from a traditional lecture.

With the launch of the LHC in September 2008, and the discovery of the Higgs boson in 2012 
\cite{Atlas:2012gk,Chatrchyan:2012xdj}, there has been a renewed public interest in elementary particle physics with a
focus on the Higgs boson. In response there has been a wide range of books written for a broader public  
\cite{Butterworth:2014eka,Carroll:2012ewa,Randall:2012loa,Lederman:2013caa} and numerous public lectures. In contrast,
the principle behind our show ``What's (the) Matter?", which we present here, is to use our extensive experience in physics 
shows as well as the rich tradition of engaging the public in physics through {\it live}  shows with many {\it live} demonstration 
experiments to the latest developments in {\it modern} physics and in particular the recent discovery of the Higgs boson. The 
challenge was clearly to find experiments which are relevant, understandable, exciting {\it and} can be performed live. We show 
as many direct particle effects as possible, for example using radioactive probes, but naturally often revert to analogies. 

To this effect we have developed  a show based on a simple story line. A previous show was devised more as a lecture
(see the description in Sect.~\ref{show-precursors}). But we have found through our other shows which mainly involve
classical physics (not the subject of this paper), that a story line better holds the attention of the audience.

The show is set up as a quest, where two proponents from Bonn travel back in time to understand what is matter. They 
first visit Ernest Rutherford and his assistant Hans Geiger in Manchester around 1911. He tells them about the atomic 
nucleus, radioactivity and how to detect it, as well as the idea behind scattering experiments. They then travel forward 
in time to Ernest Lawrence at the University of California, Berkeley, around 1950, where they learn about the how and 
why of accelerators. On the next stop they visit Sau Lan Wu at the PETRA accelerator  at DESY, Hamburg, around 
1980, who explains quantized gauge interactions and the strong force. The final stop is in the LHC tunnel at CERN, 
Geneva, Switzerland in 2012. Two experimentalists explain colliders and our heroes watch live as the Higgs boson is 
produced and decays.

For us it has always been important to connect physics concepts with live physics experiments. Thus the play includes
27 experiments, which we perform live onstage, with the obvious risk that some demonstrations might night work. 
However this is actually the heart of our show, and we have thus included an extensive discussion of all the
experiments involved.

The purpose of this paper is to present our current show in detail, such that it might benefit other people presenting 
particle physics to the public. Since very little of outreach activity is published, we strongly encourage other groups to 
contact us, to let us know what they are working on, so that we can exchange ideas and demonstration experiments.

The outline of this paper is as follows. In the next section we describe our own previous shows, which have led up
to this production. We also give highlights of the history of other related activities, as far as we are aware of. In 
Sect.~\ref{sec:logistics}, we  discuss the logistics required to perform this show, and in particular also to take it on the 
road. Sect.~\ref{sec:the-play} contains the screenplay of the show, as performed in Padua, Italy, March 2015. The 
experiments are explained as given by the script of the play. In Sect.~\ref{sec:conclusion} we conclude. In App.~A, 
we describe the experiments in detail, including the technical specifications on how to build and also how we find it best 
to perform them. This might be the most useful section for many readers.

\section{Precursors to our Show and Related Activities}
\label{show-precursors}
Physics experiments have been demonstrated publicly for a very long time. In Germany, for example,
Otto von Guericke, the mayor of Magdeburg, demonstrated his vacuum pump and the Magdeburg
hemispheres\footnote{He also invented the vacuum cannon, which we employ in the play, and discuss
in App.~\ref{app:vacuum-cannon}.}
 at the Reichstag in Regensburg in 1654 and several times to the general public afterwards 
\cite{guericke1}. In Europe and the USA there were public demonstrations of electrical effects in upper 
class salons and at fairs in the 18th century \cite{early-electricity}. As mentioned in Sect.~\ref{sec:intro}, in 
Great Britain, Faraday regularly gave public physics lectures including demonstrations in the early 19th century. 
Unfortunately, it is beyond the scope of this paper to give a comprehensive overview of the history of
public science demonstrations. We thus focus on our own history and related present day activities.

\begin{figure}[t!]
  \includegraphics[width=\textwidth]{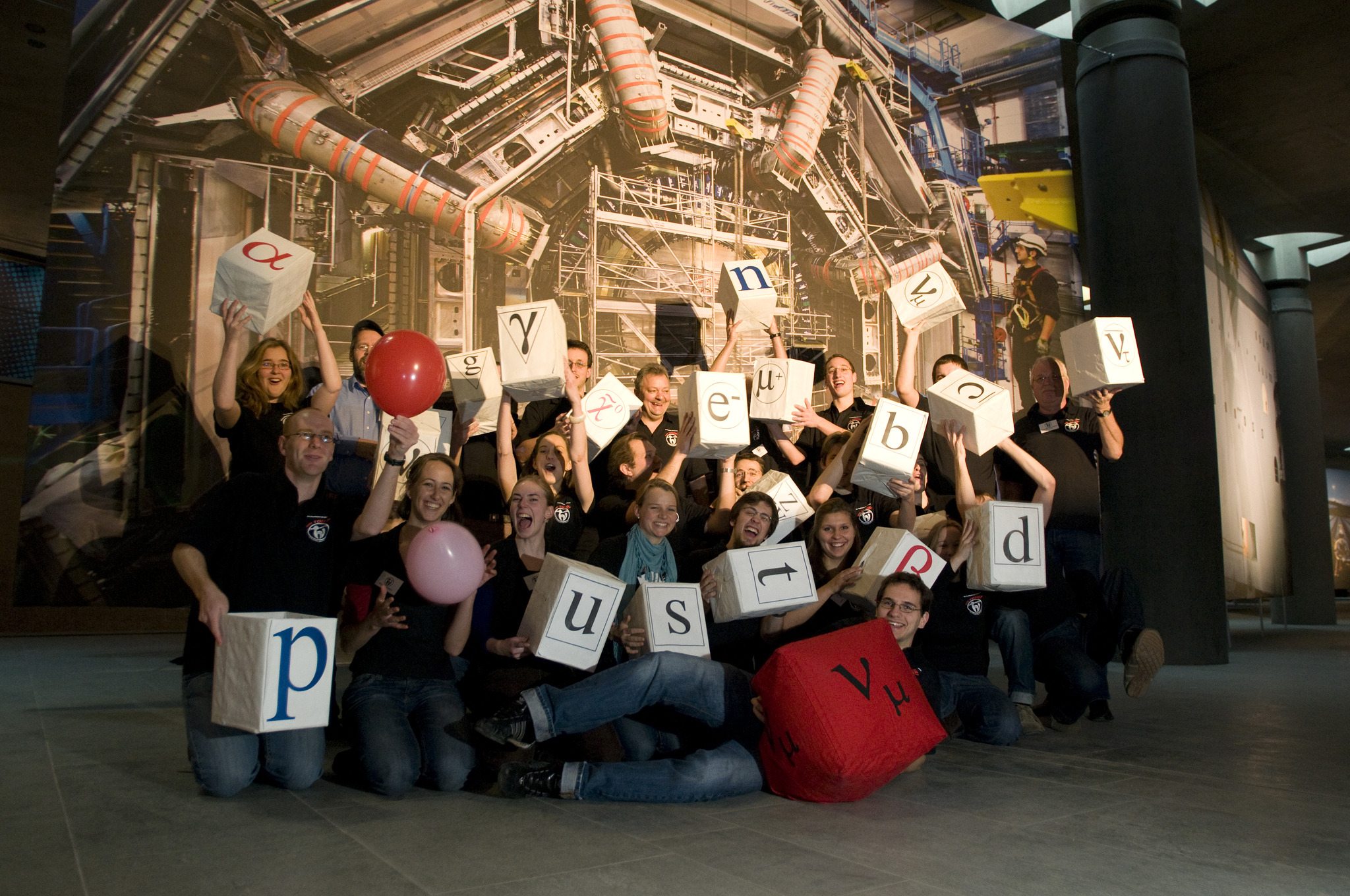}
	\caption{Group photo at the Weltmaschine exhibition in the subway station Bundestag in Berlin in 
	front of an oversize photo of the ATLAS detector, Nov. 16th, 2008. Photo by Christian Mrotzek. }
	\label{berlin-group}
\end{figure}

Around 1970 Prof.~Bassam Z. Shakhashiri started a program of public chemistry lectures and shows 
\cite{shakhashiri-bio} entitled `Science is Fun -- in the Lab of Shakhashiri' in the chemistry department at
the University of Wisconsin, Madison \cite{shakhashiri-web}. In 1984, this inspired Prof.~Clinton Sprott in
the physics department at the University of Wisconsin to start a physics show entitled `The Wonders of 
Physics'  \cite{Sprott,sprott-book}.  For recent activity see their webpage \cite{sprott-url}. One of us 
(HKD) attended the University of Wisconsin, Madison (1984-1989). The Bonn Physikshow was 
started in Dec. 2001 by two of us (HKD, MKo), with a first show in Nov.~2002 
\cite{Dreiner:2007cn,cern-courier,science-school,guardian-I,guardian-II}. A distinguishing feature of the 
Bonn Physikshow is that the students develop and present the shows largely on their own. We (MKo, 
SH, HKD) offer technical assistance. The guidance is mainly provided by more senior students. In addition,
we have developed a new show almost every year since 2002, giving each class of new physics students 
the opportunity to develop their own show. Between 15 and 25 students participate each year. It takes 
roughly 6 months to develop a new show. During the show itself we originally had two people moderating the
show and explaining the experiments, with the others performing them. Starting in 2004 we began 
incorporating simple storylines into the show, and in the process giving more people speaking parts. The 
storyline often puts experiments we had used in previous shows in a completely different context; it also helps 
to hold the attention particularly of the younger audience members, and can highlight the extraordinary features
of even very simple experiments. The storyline has become a central feature of all of our shows and something 
we highly recommend. We would be particularly interested in hearing of related activity elsewhere employing 
storylines or similar. However we should mention, that we are not professional actors or stage crew. Instead we 
are physicists and physics technicians. The emphasis is always on the experiments and the correct presentation
of the physics, although we do enjoy a good joke!

In Bonn, the show initially focussed on classical physics with typical experiments for example in  mechanics, 
electromagnetism, and acoustics. We also employed high $T_c$ superconductors and occasionally 
radioactivity, \textit{e.g.}~uranium glass. In 2004, CERN turned 50 years old and to celebrate we created for the
first time a show on particle physics. It was developed together with Prof. Michael Kobel (now TU Dresden) and 
Dr.~R.~Meyer-Fennekohl, as well as several Ph.D. students and postdocs in particle physics at Bonn University. 
Since then EP has been a full member of the team. This show was devised more as a lecture with many 
demonstration experiments, including the experiments discussed in \ref{app:oil-drop}, \ref{app:helmholtz-colis}, 
\ref{app:jacobs-ladder}, \ref{app:cloud-chamber}, \ref{hebe-beschleuniger}, \ref{app:vacuum-cannon}, 
\ref{app:airtable}, an extended version of \ref{app:medicineballs}, and \ref{app:teslacoil}, as well as a water 
\^Cerenkov experiment in a thermos (Kamiokanne) \cite{Kamiokanne}. The show was entitled `From Quarks to 
Quasars' and included the topics `Atoms and Nuclei', `Particles and Forces', `Symmetries' and `Astrophysics and 
Cosmology'. There was no storyline.

\begin{figure}[t!]
\center
  \includegraphics[width=0.8\textwidth]{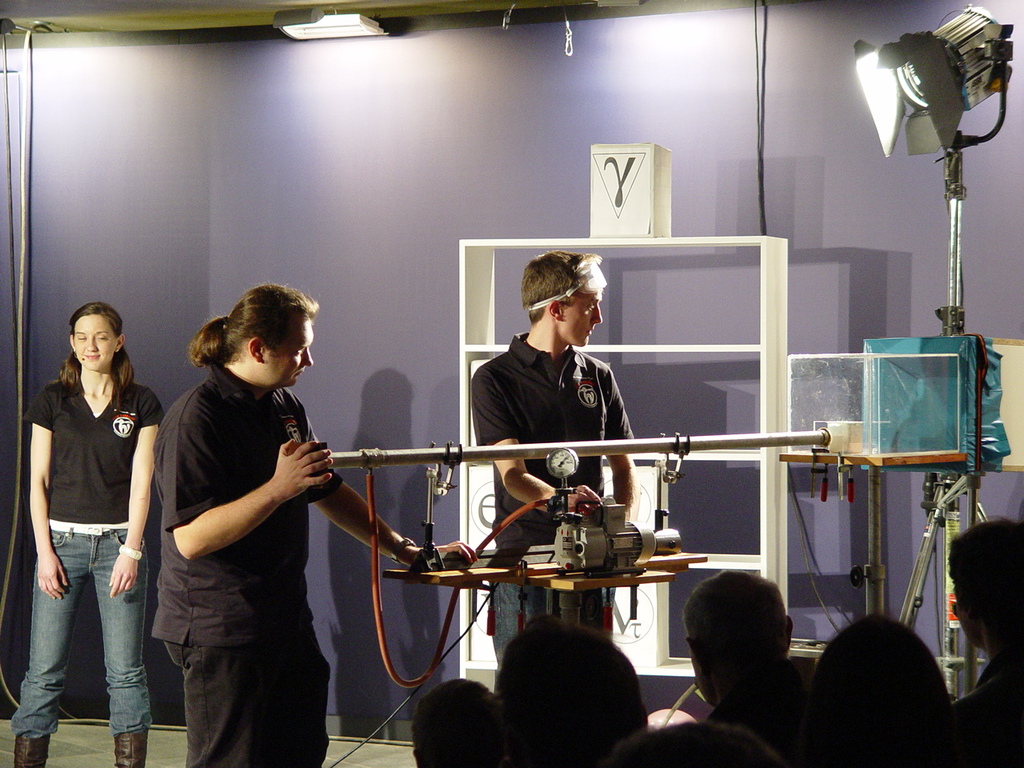}
	\caption{Vacuum cannon, the experiment described in \ref{app:vacuum-cannon}, onstage in Berlin, with 
	Cornelia Monzel, SSch and Nicki Bornhauser (from left to right). Inside the clear plastic box is the 
	quark target. For a detailed view see Fig.~\ref{accel-2}. Behind, the white box covered by the blue plastic bag 
	is the beam dump, with toilet paper rolls inside. In the background is the particle shelf. }
	\label{accel-1}
\end{figure}

\begin{figure}[h!]
\center
  \includegraphics[width=0.8\textwidth]{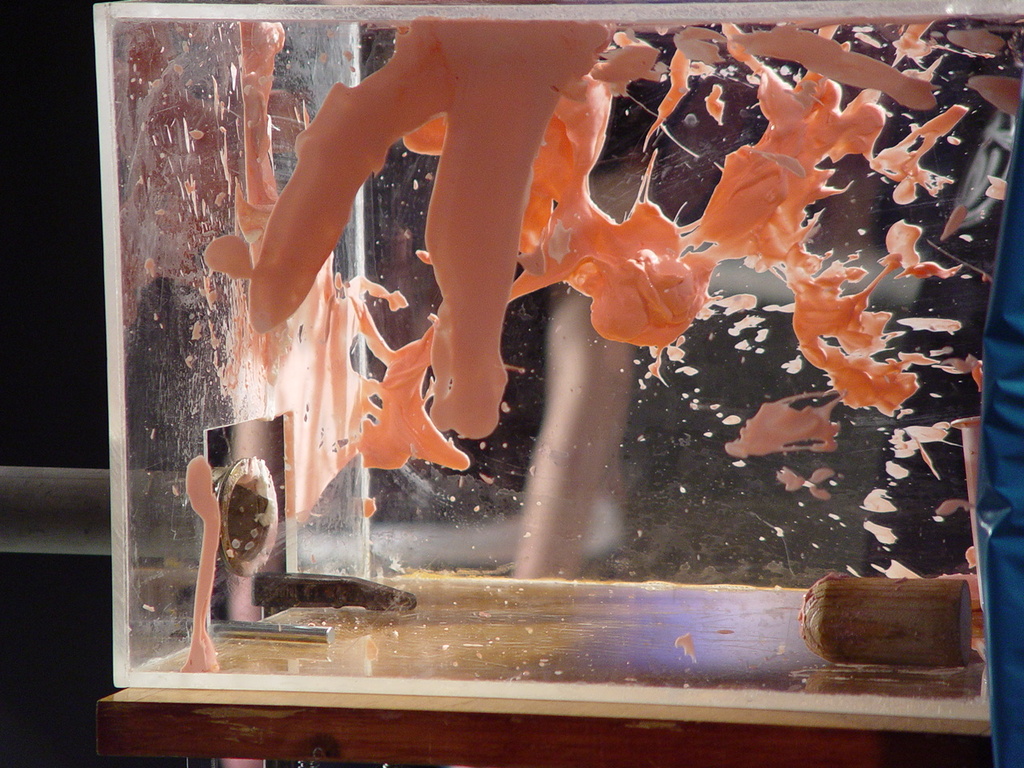}
	\caption{Collision event of the vacuum cannon as displayed in our plastic detector. The wooden projectile can be 
	seen on the lower right.}
	\label{accel-2}
\end{figure}

In the Autumn of 2008, with the initial launch of the LHC, the German funding agency (ministry) for large 
experiments (Gro{\ss}ger\"ateforschung),  BMBF (Bundesministerium f\"ur Bildung und Forschung), organized an 
exhibition on the LHC \cite{BMBF}. This  involved active research particle physicists as tour guides and was located 
in a new subway station in the center of Berlin: `Bundestag', see Fig.~\ref{berlin-group}. The exhibition had the 
somewhat grandiose title `Weltmaschine'  (World Machine). In Nov. 2008, we were invited to perform a 90~min particle 
physics show in Berlin, at the exhibition. We modified the 2004 show, including a new presentation 
modus, with always just one or 2 experiments on stage, due to the limited space. Experiments were rapidly exchanged 
through wheeled carts and many stage hands. As a guiding point, we had a bookshelf  on wheels, where a clown added
boxes labelled by the particle symbols as the particles were introduced in the show, see the background in 
Fig.~\ref{accel-1}. A highlight was the vacuum cannon, see Figs.~\ref{accel-1}, \ref{accel-2}, as well as 
App.~\ref{app:vacuum-cannon}.
This show was  very successful and we were invited to repeat it in the Ehrensaal of the Deutsches Museum, M\"unchen  
(March, 2009), at the German national particle physics laboratory DESY, Hamburg (Sept. 2009), and at the 
University of Heidelberg (Dec. 2009). In September 2010, we performed the show in French in the Globe at CERN, 
Geneva, Switzerland. This was our first trip abroad.

In 2012, in the framework of the DFG (Deutsche Forschungsgemeinschaft, a national German funding agency, similar 
to NSF in the USA) research grant CRC 110  `Symmetries and the Emergence of Structure in QCD' \cite{crc-110}, the
Bonn Physikshow was awarded a large grant to produce a new particle physics show and to travel with this show.
The result of this is presented in this paper. As can be seen, we developed a completely new show now based on a 
storyline, and with many new experiments. The show was written in English and first performed in the Martin Wood 
Lecture Theatre at the physics department of Oxford University.

We recently became aware of a very nice show entitled `Accelerate' performed in Oxford in Dec. 2011, before the 
discovery of the Higgs boson. A full length video of the show can be seen at \cite{Oxford-Accelerate}.  An extensive 
write-up of ten experiments is given at \cite{Oxford-Accelerate-2}. Compared to our show, their show is more like a 
lecture, \textit{i.e.} there is no story line. It has a female and male moderator, who regularly involve the audience, in 
the film these are school children.  It focuses on how accelerators like the LHC work and illustrate the science involved.
To illustrate electrostatic charges they use a van der Graaf. They demonstrate wake-field acceleration with beach balls
together with the audience. They also use the cathode ray tube, App.~\ref{app:helmholtz-colis} and a smaller version
of the cloud chamber App.~\ref{app:cloud-chamber}.

\section{Logistics}
\label{sec:logistics}

\begin{figure}[h!]
\center
  \includegraphics[width=0.8\textwidth]{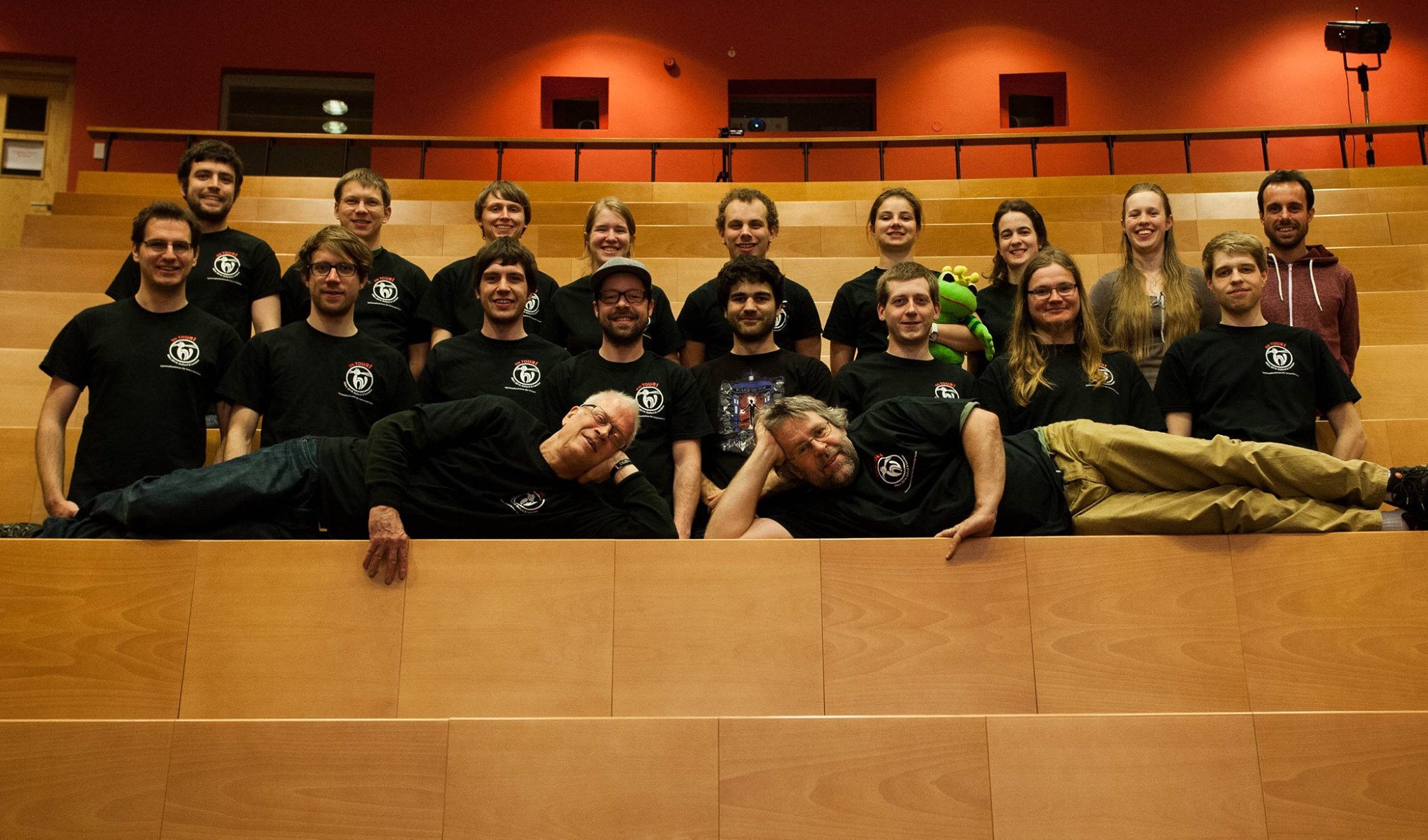}
	\caption{Group photo in the Martin Wood lecture theatre in the Physics department of Oxford University.
	Back row left to right MR, CJ, JSchm, JH, DH, CL, JSch-R, MH, LU. Front row left to right MBe, PM, TH, SH, JM,
	MKr, TSch, and MBo. Lying in front EP (left) and HKD (right).}
		\label{oxford-group}
\end{figure}

The show we describe here was developed for a first set of four performances in the Martin Wood Lecture Theatre in 
the University of Oxford Department of Physics, in March 2014, see also Fig.~\ref{oxford-group}. It was immediately 
followed up by two shows in the Cruciform Lecture Theatre at University College London.  Thus from the beginning it 
was clear that we would be traveling with the show. A significant effort is involved in transporting all of the required 
equipment, as well as the about 20 people. Here we give a few details which might be helpful, if somebody wishes to 
organize and perform something similar.

In the past we have travelled extensively with the physics show, starting in 2005 to the Deutsches Museum, Bonn, 
Bad Godesberg. We have returned: 2013, 2014 and 2015 (2x). Further trips have included the Deutsches Museum 
of Masterpieces in Science and Technology in M\"unchen (2006, 2009, 2013, 2017 planned), Berlin (2008), 
G\"ottingen University (2008), Wallraff-Richartz-Museum K\"oln (2009), DESY (2009, 2014), Heidelberg University 
(2009), CERN (2010), Sternwarte Solingen (2010),  BayKomm Leverkusen, the outreach division of the company 
Bayer (2012, 2015), the LVR Museum Oberhausen (2015), the LVR Museum Solingen (2016), and the Physikzentrum 
Bad Honnef (2016). With the show presented here, we have also travelled to Padua and Trieste, Italy (March 2015) 
and plan to travel to Copenhagen and Odense (August, 2016). In March 2016 we travelled with 9 people to Beijing, 
China, and for a week taught 15 physics students at Peiking University and Chinese academy of Science, Institute of 
Theoretical Physics (CAS, ITP) how to perform a physics show using their local equipment, culminating in a show
in Chinese for local highschool students. We have thus  gained a significant amount of experience with many logistical 
aspects.

\subsection{Equipment}
We have a collection of 28 demonstration experiments for this show, which we perform live onstage. Some are 
quite large, as described in the appendix, App.~\ref{sec:appendix}. For example the fire tornado is 2\,m tall and 0.63\,m 
in diameter, see App.~\ref{app:firetornado}. For all of these experiments, we bring everything that is needed to run them, 
including the power cords and other required cables. If we are abroad, we bring a large number of adapters, as 
unfortunately the sockets in Europe (especially in Britain) are not universal.

Depending on the location, we set up a control table either upfront, on the side of the stage or at the back of the lecture 
hall. This table has two laptops as well as a 6-channel mixer. During the show, from this table, two students control the 
projection, the lighting, as well as play the music and sound effects. We run all control cables for the lighting and 
cameras (which we all bring) to this one table. We have to-date always made use of the local sound system, which we 
hook-up to our control instruments at the table. We usually make use of the local microphones, although we also bring 
some of our own, as we need headsets for all people who both perform experiments onstage, as well as speak.

In addition to the physics experiments we have two 1000\,W halogen lamps (Hedler, type H25s). They are controlled 
via a dimmer pack (Elation Unipak II) using a DMX (digital multiplex) interface. We furthermore use two LED lights 
with 36 x 3\,W each (Eurolite LED Theatre 36 x 3\,W CW/WW). We also used one 1000\,W spotlight (Moonlight 1000, 
Model MN1000, GX9,5 220/240V/1000W). The flood lights are essential to get a proper stage lighting, which is usually 
not available in physics lecture halls. The spot light is important for various parts of the story, where we darken the stage 
and just light up one small part, \textit{e.g.}  the person dressed as a Blues Brother on the balcony in Padua, or the 
Caretaker at the end of the show. We also have two cameras which we use to project smaller experiments onto the screen. 

We employ an overhead projector for the Rayleigh oil drop experiment, see App.\,\ref{app:oil-drop}.

To-date we have always brought our own radioactive materials. We must transport these in a special metal box, 
complying with the latest European Union transport rules. We furthermore bring various materials needed for
the experiments, such as the Quark for the vacuum cannon App.~\ref{app:vacuum-cannon}, or the plastic balls
discussed in App.~\ref{app:ballcollider}. Several of the actors wear costumes, which we also bring along. For most
shows we have developed new Physikshow T-shirts.\footnote{We thank Karina Kortmann for designing several
shirts.}

In total, with all the experiments, this fills two vans. We use VW Transporters provided by Bonn University, 
each corresponding to a maximum load volume of about 6\,m$^3$, which we make almost complete use of. 
The maximum added weight load is 900kg. We drive the vans ourselves.


\subsection{People}
In the physics show {\it `What's (the) Matter?'} two students (MH and LU) lead through the show. They are 
accompanied by a caretaker (JSchm), who also runs the time machine. In each ``time zone" they meet
two scientists, who explain new physics. Altogether there are thus eleven speaking parts. In principle it would be 
possible to have the actors of the first two sections (Manchester, Berkeley) doubling up for the second 
two sections (DESY, CERN), reducing the number of speaking r\^{o}les to seven. However, typically we are 
not experienced actors and one r\^{o}le per person is enough. In addition, outside of their part, the actors 
help on lighting (spot) and stage management, taking experiments on- and offstage.

As mentioned above, we have two people at the technical table for sound, lighting, projection and music.
The spot is handled in turn by someone not involved onstage. 

We must move a significant amount of equipment on- and offstage, during the show, to avoid cluttering the 
stage. (In Oxford the stage was also fairly small.) In order for this to proceed smoothly, we have two stage 
managers to organize the sorting, setup and dismantling of all the experiments, before, during and after the 
show. These are important full-time jobs, which ensure the show runs smoothly. Between parts two (Berkeley) 
and three (DESY) we exchange most of the experiments onstage. This requires a short break.

We have one person who is responsible for the radioactive materials, full--time. This has mainly been EP, and 
should be a senior experienced scientist or technician. This requires formal safety training with 
radioactive materials.

Almost all experiments are projected via cameras onto a screen, which is best well above the stage. This
enables everybody to see important details of the experiments. For this we have two people full-time
on the cameras onstage. The camera positions must also be rehearsed, as well as where the actors
stand with respect to the experiment and camera. The oil drop experiment \ref{app:oil-drop}, is projected onto
a possibly separate screen. The water wave experiment \ref{app:water-wave} is projected directly onto
a wall.

In Oxford and London we had one technician, SH, which is the absolute minimum. In Padua and 
Trieste MKo also joined, and doubled as a camera man. We thus had two very experienced 
people who could help out at any moment, if any experiment failed or there were problems with the lighting, 
the sound or the projection. This was very helpful, as something always goes wrong\footnote{In Trieste the 
vacuum cannon failed, because the end piece wasn't greased properly. SH was able to fix this in real time 
and by the time the cannon reappeared in the CERN section it was ready for use. Thus, although not foreseen, 
we instead performed the experiment properly in the last section.}\!\!.

Thus we need at least 19 people for this show, with experienced actors 15 should suffice. In Oxford we were 19, 
in Padua 20. See Fig.~\ref{oxford-group} for the Oxford group photo.

\subsection{Rehearsals}
Since it is essentially a play we are performing, we require significant rehearsal time in Bonn. For a new show at 
least 8 days all day. The students taking part are not always the same as they have many other commitments. 
Thus we also needed a week of rehearsals before our trip to Padua. Furthermore each lecture hall is different. We
try to learn as much about the local stage beforehand as possible, sometimes marking the stage size by for the 
rehearsals in Bonn. In order to get all the equipment hooked up properly we often have to place experiments 
differently on different stages. We thus need at least one full day in a new location to set up all the experiments 
and technical equipment and rehearse who is standing where, who brings in what experiment between the acts 
\textit{etc}. On a two-stop trip such as the UK (Oxford, London) and Italy (Padua, Trieste) the team is well organized 
by the second stop and we need much less local rehearsal time.

\subsection{Music}
We make extensive use of music during our shows, which we play from a laptop. In Germany we therefore pay 
{\it GEMA} fees for all shows. For comedic purposes, we also make use of sound effects, \textit{e.g.} a plop sound
when the oil drop is applied in the Lord Rayleigh experiment, see App.~\ref{app:oil-drop}.

\subsection{Costs}

\subsubsection{Lodging and Board:} The main costs of the show are board and lodging for typically 19 or 20 people 
for several days. We have travelled for one-place-only to M\"unchen and Hamburg. Allowing for the one full day of 
local rehearsal and then three to four shows spread over two days, we need at least three overnights. This 
corresponds to sixty overnights. For two-stop-trips the team is well in-sync by the second stop and things proceed 
faster. We stayed in Oxford for four nights, but only two nights in London. We stayed four nights in Padua and also only
two nights in Trieste.

\subsubsection{Transport:} The trip Bonn--Padua--Trieste--Bonn is 2,252\,km, according to Google maps. With 
two vans and a cost of  \EUR 0.43/km, this corresponds to about \EUR 1,936.72. Our university also has a 
17-seater bus,  which costs
\EUR 0.63/km plus expenses for the driver (hotel and per diem). This is another \EUR 1,418.76 (plus hotel and per
diem).

\subsection{Audience}
When we started our physics shows in 2002, we devised the (general) show for children aged 10 or 11 years and older.
There was an article in the local newspaper beforehand, announcing the show, and about 750 people of all age groups 
showed up, packing our 550 seat auditorium. (We have since introduced a booking system.) It turned out that also 
children as young as 5 years already very much enjoyed the shows. Now we typically perform twice on the weekend 
for a general audience and have an extra show on a Friday morning for school classes. For the latter we specifically 
advertise the show for students in 4$^{\mathrm{th}}$ grade through 7$^{\mathrm{th}}$ grade, \textit{i.e.} for ages 9
through 12.

The particle physics show as presented here we advertise for teenagers aged 15 and older. However also here we have 
found that there is enough action in the show that younger kids take great pleasure in attending. They might not 
understand all explanations, but will still hopefully get an appreciation of the physics at the LHC.

In all cases grown-ups are also very welcome. However, our impression is that adults with little or no background in physics 
feel more welcome to a show which is advertised for kids 10 years and older, than 15 years and older. In the latter case they
often suspect that advanced secondary school physics might be a prerequisite and they don't come.


\section{The Play}
\label{sec:the-play}

The following is the screenplay as performed in Padua, March 2015. The show was just under 2 hours
long, including an about eight minute intermission after the Berkeley part, required for resetting the
stage with the new experiments.

\subsection{Prologue: Bonn, Germany}
\label{sec:prologue}

\begin{figure}[t!]
    \centering
        \includegraphics[width=0.5\textwidth]{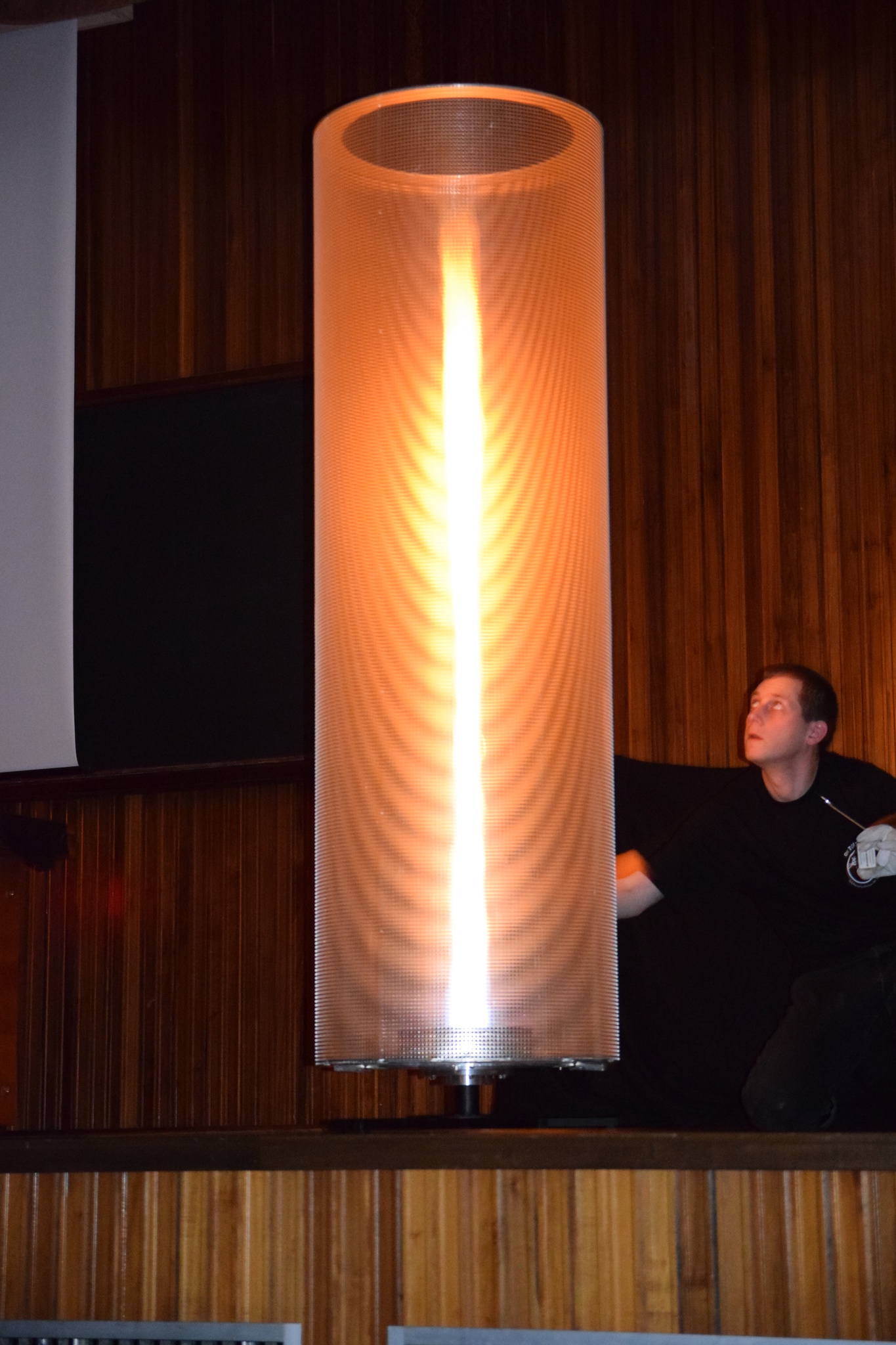}
    \caption{Fire tornado onstage at Trieste. The cage is about 2\,m high. On the right is MKr.}
    \label{fig:fire-tornado1}
\end{figure}

{\it \noindent (Dark stage.)}

\medskip

\EXP{Fire tornado (see Fig.~\ref{fig:fire-tornado1} and App.\,\ref{app:firetornado})}. 

\medskip

\noindent {\it (Two students on towels, in a spotlight, hanging out
  on the beach along the river Rhine in Bonn, Germany. Behind them the fire tornado, Fig.\,\ref{fig:fire-tornado1}.
  Contemplative silence.  Passing a large bag of HARIBO back and forth. {\Bb} occasionally tosses gummi bears 
  out, see Fig.~\ref{fig:prolog-haribo}. To the side a caretaker is sweeping. Next to him is a `Lost \& Found' box.)}

\vspace{0.5cm}

\begin{itemize}

\item[\A] {\it (Gazing up at the stars)} Lorenzo?

\medskip

\item[\B] {\it (Absent minded, eating HARIBO)} Huh? 

\medskip

\item[\A] Did you ever wonder where it all
came from?

\medskip

\item[\B] Maike, do you mean HARIBO? I'm pretty sure it's from Bonn.

\begin{figure}[h!]
    \centering
        \includegraphics[width=0.95\textwidth]{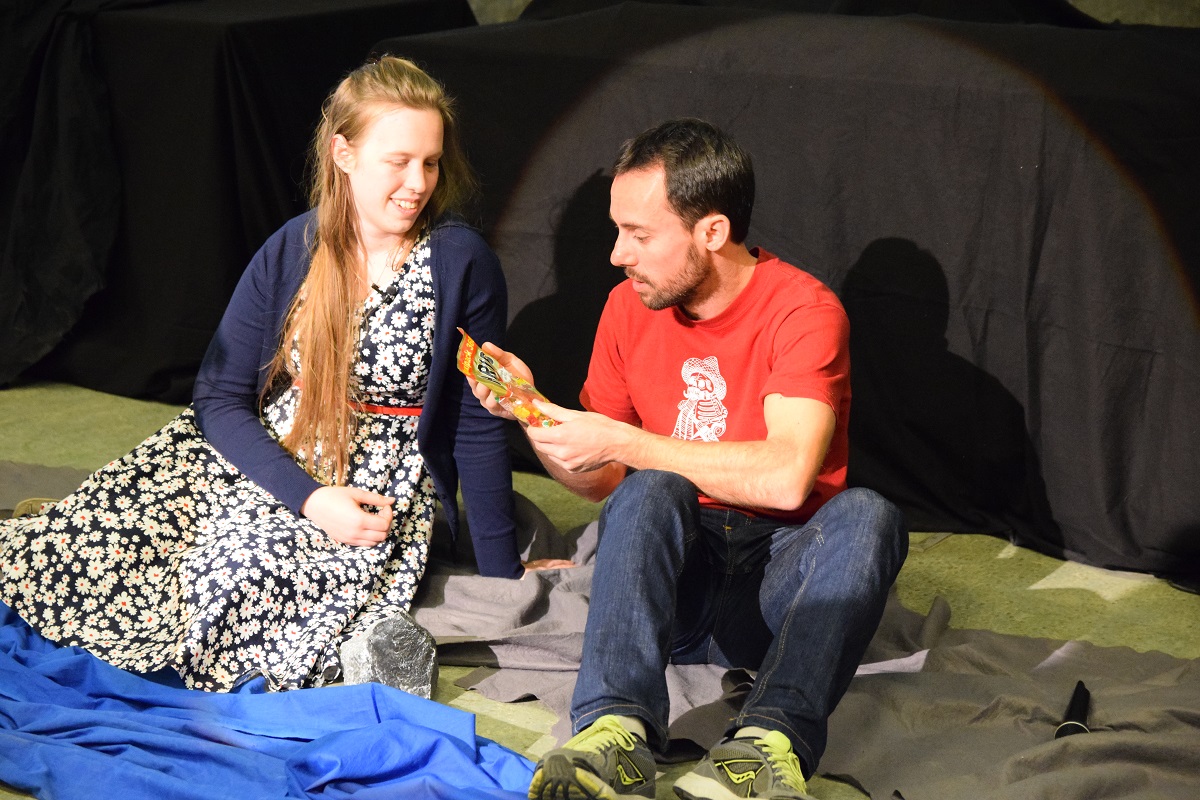}
    \caption{Maike (MH) and Lorenzo (LU) contemplating the universe and the yellow 
    gummi bears.}
    \label{fig:prolog-haribo}
\end{figure}

\medskip

\item[\A] Really?

\medskip

\item[\B] Yeah, sure. Founded in Bonn in 1920 by {\bf HA}ns..{\bf RI}egel..{\bf BO}nn. HA -- RI -- BO.

\medskip

\item[\A] Anyway, no, I mean ...  everything ... you know: the stars, the planets ...

\medskip

\item[\B] Oh, that's what you mean, how the universe began, how it will end. Or simply: what is ``matter"?  ... 
Paah, beats me. {\it (Continues eating gummy bears.)}

\medskip

\item[\A] Yes, exactly. ({\it Holds up a stone.}) Where does this rock come from? And what
  is it exactly that it is made of? Where does MASS come from?

\medskip

\item[\B] I know that my mass mainly comes from Haribo. I definitely eat too much. But it seems
impossible to stop. Sometimes I eat so much, I start sweating. Now, if I just smell them,
I can feel the perspiration coming.

\medskip

\item[\A] Argh! Too much detail. How about this: where does the mass of the HARIBO bears
  come from? I mean they don't eat other gummy bears, do they?

\medskip

\item[\B] Aha! Another yellow one!  I have always wondered why there are so many yellow gummy bears. I
don't like the yellow ones. Don't they remind you of coagulated wee?

\medskip

\item[\A] Lorenzo! Come on, didn't you also hear about the LHC and the Higgs boson. It's
  supposed to be really important. What do you think?  ... Or do you think?

\medskip

\item[\B] ({\it sighs}) Okay, okay, you are like a dog with a bone! Just a second, let me
  think. {\it(Poses in deep thought, briefly.)} ..... Ok,I give up. I have no idea. If you
  really want to know ... why don't you ask the caretaker over there? {\it (Points to a
    strange figure to the side who has a large broom and is sweeping.)}

\medskip

\item[\A] Huh? What is he doing here? And why is he sweeping here in the park, by the Rhine?

\medskip

\item[\B] I don't know, I just saw him \ldots well go ahead, you might as well ask.

\medskip

\item[\A] Okay: Excuse me, Mr. ..., {\it(Holds up a rock.)} What is matter? What is it made of? And most of all:
  where does mass come from?

\medskip

\item[\C] { Oh, hello Maike. Hello Lorenzo. Nice to meet you.  {\it(Turning towards the
      audience,)} I've always wondered when this will happen. {\it (Back to Maike and Lorenzo)}
    Okay, here is your answer: mass comes from red gummy bears!}

\medskip

\item[\B] What???

\medskip

\item[\C] Worried Lorenzo?  Haha, just kidding.  Anyway, these are very big questions. You are
  shooting for the moon, so to speak ... excuse the pun.  You must go on a personal journey, a
  long and arduous quest, during which you Lorenzo must grow, so that both of you can fully
  understand the answers you will find. Are you prepared? Your travels will take you through
  space AND time ....

\medskip

\item[\A] Oh, he's just kidding again...

\medskip

\item[\C] No, no, no! You must get ready. Grab your towels, your gummy bears and, most
  important, take an umbrella.

\medskip

\item[\B] An umbrella?

\medskip

\item[\C] Wait a moment! {\it(Rummages around in his lost and found box, choses finally an
    umbrella.)} Ah, there it is! This one has been in here for a long, long time. Spin the
  umbrella and off you go.

\hop

{\it(\Aa\  and \Bb\  hold the umbrella above their heads and open it)}

\hop

\item[\C] No, that is the British way. You must hold it forward. {\it(Towards the spotlight which
    is shining from the back of the auditorium.)}

\medskip

\item[\A] Will we see the LHC? Or even the Higgs boson? 

\hop

\item[\B] Will I lose any weight?

\medskip

\item[\C]  You must start at the beginning. ... Maybe then you can see the answers yourself...

\medskip
{\it (Time travel starts, see Fig.\,\ref{fig:prolog-haribo2}, accompanied by music and a film.)}

\medskip

\begin{figure}[h]
    \centering
        \includegraphics[width=0.95\textwidth]{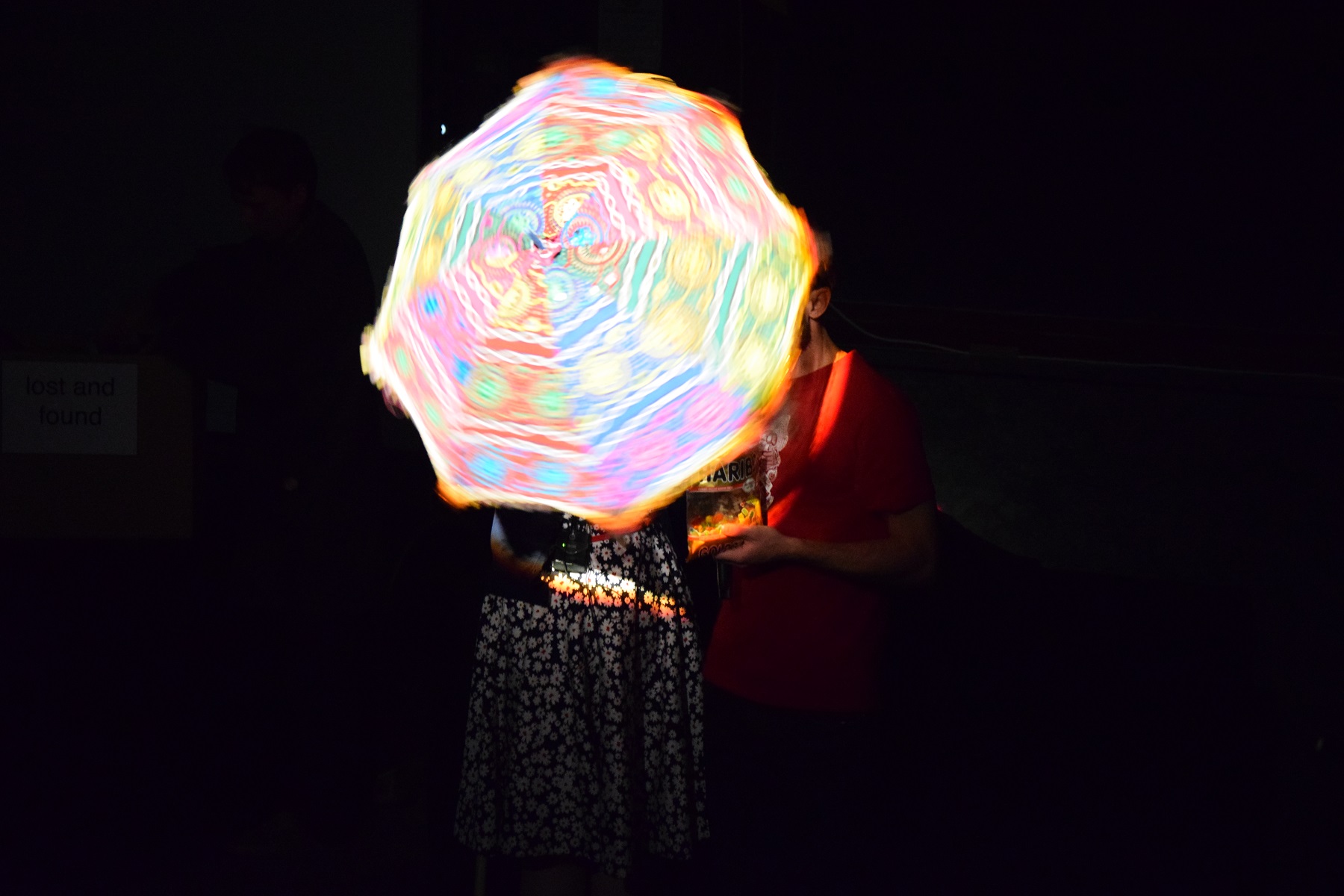}
    \caption{Maike (MH) and Lorenzo (LU) partially hidden behind the rotating umbrella, which represents the time machine.}
    \label{fig:prolog-haribo2}
\end{figure}

\end{itemize}

\subsection{Manchester, UK}

\begin{itemize}

\item[\A] Wow! That was some journey. Lorenzo, are you ok? {\it(Lorenzo nods.)}  Where are we
  here?  {\it(Lifts an old object.)} And also ... when?

\hop

{\it (Caretaker directs \Aa \ and \Bb \ into new space highlighted by
  spot. In the dark, Rutherford and Geiger are frozen at their lab tables.)}

\hop

\item[\C] Welcome to Manchester ... the best football town ... in England. {\it(Looks at his watch.)} It is 1911. This 
is Ernest Rutherford and his German assistant Hans Geiger. They should be able to help you with your questions. 
Have a good time!  {\it (The Caretaker claps his hands twice, Rutherford and Geiger come to life.)}

{\it (Caretaker starts cleaning blackboard/screen for oil drop experiment.)}

\medskip

\item[\R] Welcome to my laboratory here in Manchester. What can I do for you?
 
\hop

\item[\A] We are on a journey through space and time and would like to know {\it(Holds up
    the rock.)} what matter is made of. And, why do objects have mass?

\medskip
 
\item[\R] You have come to the right place! {\it (Pause, looks up, throws out his chest.)} Let
  me make a bold statement: everything, all matter, is made of small bits, which we call atoms.

\hop

\item[\A] You mean, like small Lego blocks. {\it (Holds up some Lego blocks.)}

\medskip

\item[\R] Yes that is a good analogy.

\medskip

\item[\B] How big are these building blocks? {\it(Holds up his Haribo bag.)} And what are
  their colours?

\medskip

\item[\R] For that, let ME show you an experiment, which my assistant will conduct 
  over here.  This is \textit{Herr Geiger} from Germany.

\medskip

\item[\G] {\it (Clicks his heels.)} {\it Guten Tag.}  

\medskip

\item[{{\bf \color{PineGreen}Maike,}}\B] {\it Guten Tag.}

\medskip

\item[\G] Let me show you an experiment that demonstrates how
  ridiculously small these smallest particles, the atoms, actually are.

\medskip

\EXP{ Oil drop experiment (see App.\,\ref{app:oil-drop})}

{\it Caretaker stops cleaning the screen, when the projector is switched on and continues with other
  furniture/blackboard/\dots}

\medskip

\item[\R]  We have here a small dish filled with water. It is placed on an overhead projector so that you can see it 
on the screen. Herr Geiger will add some fine spores. They are floating on the water --- you can see them as small 
black dots.

\medskip

 Then he puts a drop of oil on the surface.

\medskip

 {(\it Geiger let's oil drop on surface, ``PLOP"-sound comes over
   loudspeakers.)}

\medskip

It pushes away the spores. At the end, the layer of oil has a thickness of one oil molecule
only.  Beforehand, we can measure the volume of the oil drop and the area of the surface covered
by the oil drop. With these numbers we can calculate the size of an oil molecule using
the formula shown on the slide. 

\medskip

({\it Formula projected with beamer.})

\medskip

Here you see the drop volume. We then divide by the area of the oil drop, as measured in our experiment. 
The calculation yields that the size of an oil molecule is equal to one part in ten billionth of a meter, or written 
in a mathematical way: 10 to the -10 meters. {\it (This number is also on the slide.)}

\medskip

\item[\A] Wow, so they are really, really small!

\medskip

\item[\B] {\it (Thinking ... counts on his fingers.)} That means that even this super sharp knife {\it(Shows a 
sharp kitchen knife.)} here, at the smallest part of the blade, is about 1 million atoms wide? That is amazing!

\medskip

\item[\R] Yes, that is about right. Very good. And moreover, most atoms are very stable and
  always the same. They are composed of even smaller particles.
  
\medskip

\item[\B] How can these smaller bits come together to always make identical atoms??

\medskip

\item[\R] You have come to the right man. In fact, I, Ernest Rutherford from New Zealand(!), am
  the world expert on the substructure of the atom.

\medskip

\item[\B] Oohh, really?

\medskip

\item[\R] Yes, I am from New Zealand.

\hop

\item[\B] No, I meant about the atom.

\hop

\item[\R] Ah yes, sorry, ... {\it(swells chest)} indeed, I was the first person, to look deep
  inside the atom. ... Well my assistants helped me\dots \textit{(shrinking)}

\medskip

\item[\B] Aaah!

\medskip

\item[\R] Well they actually did all the dirty work.\dots \textit{(shrinking further)} \dots
  Endless waiting for little light blips. That was a task for Geiger -- he is good at counting!
  
\medskip

\item[\B] So was it Geiger who actually did all the work??

\medskip

\item[\R] Ehm, ok, no matter.  Come over here and have a look with us. We have prepared a
  further experiment.

\hop

\EXP{Scattering experiment (see App.\,\ref{app:woodenscattering}, and Fig.~\ref{multiple-scattering}.)}

\hop

\item[\G] Let me show you how to look into the small world of the atom. Commonly this is done
  with scattering experiments.  That means you scatter projectiles, particles, off of an
  arbitrary target and then learn something about it.\dots This setup demonstrates how
  scattering experiments are done.

\hop

Lets have a look at the screen over there. We have some steel balls which we use as
projectiles. (\textit{Shows one steel ball to the audience.}) After rolling down this ramp, they
hit the target placed in the center and get deflected. The scattered projectiles are collected 
in these pockets along the side.

\hop

\textit{(3 particle bunches on circular target.)}

\hop

Lorenzo, what do you think?

\hop

\item[\B] Apart from the very forward direction, it looks like the projectiles are more or less
  uniformly distributed.

\item[\G] Very good, well observed.  The many particles in the forward direction did not hit our
  target in the center, passed it on the left or right and thus went straight through. The other
  particles hit the circle and were deflected almost uniformly.

 {\it (Caretaker collects the previous steel balls with a magnet, puts them in his overcoat pocket
    and exits stage.)}

  And now I replace the circular target with a triangular one. Spot the difference\dots

\textit{(3 particle bunches on triangular target.)}

\medskip

\item[\A] Ohh, that looks quite different.  The scattering pattern has changed. \dots The
  projectiles are no longer distributed uniformly, but now seem to show three populated spots.

\textit{(\Aa \ points at the three spots.)}
\hop

\item[\G] Exactly! We again have one accumulation where the particles missed our target straight
  ahead.\dots Here and here we observe a higher density, where the particles hit the flat side
  of the triangle and are deflected to these pockets. \dots We have repeated the experiment with
  both targets many times and taken some photos.
	
\textit{(Show comparison photographs on the screen. Seen here in Fig.~\ref{multiple-scattering}.)}

\begin{figure}[t!]
  \includegraphics[width=0.5\textwidth]{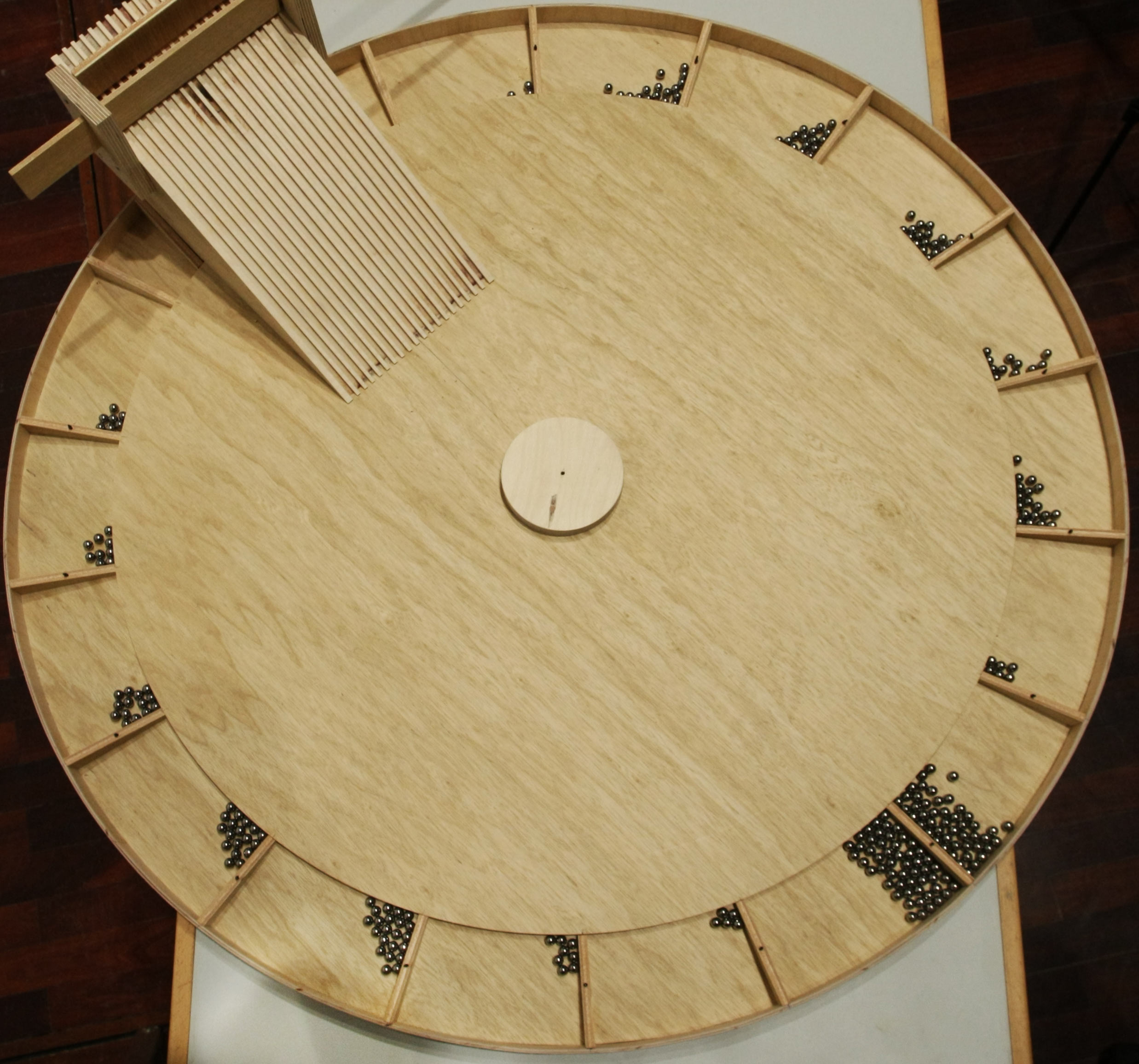}
  \includegraphics[width=0.4905\textwidth]{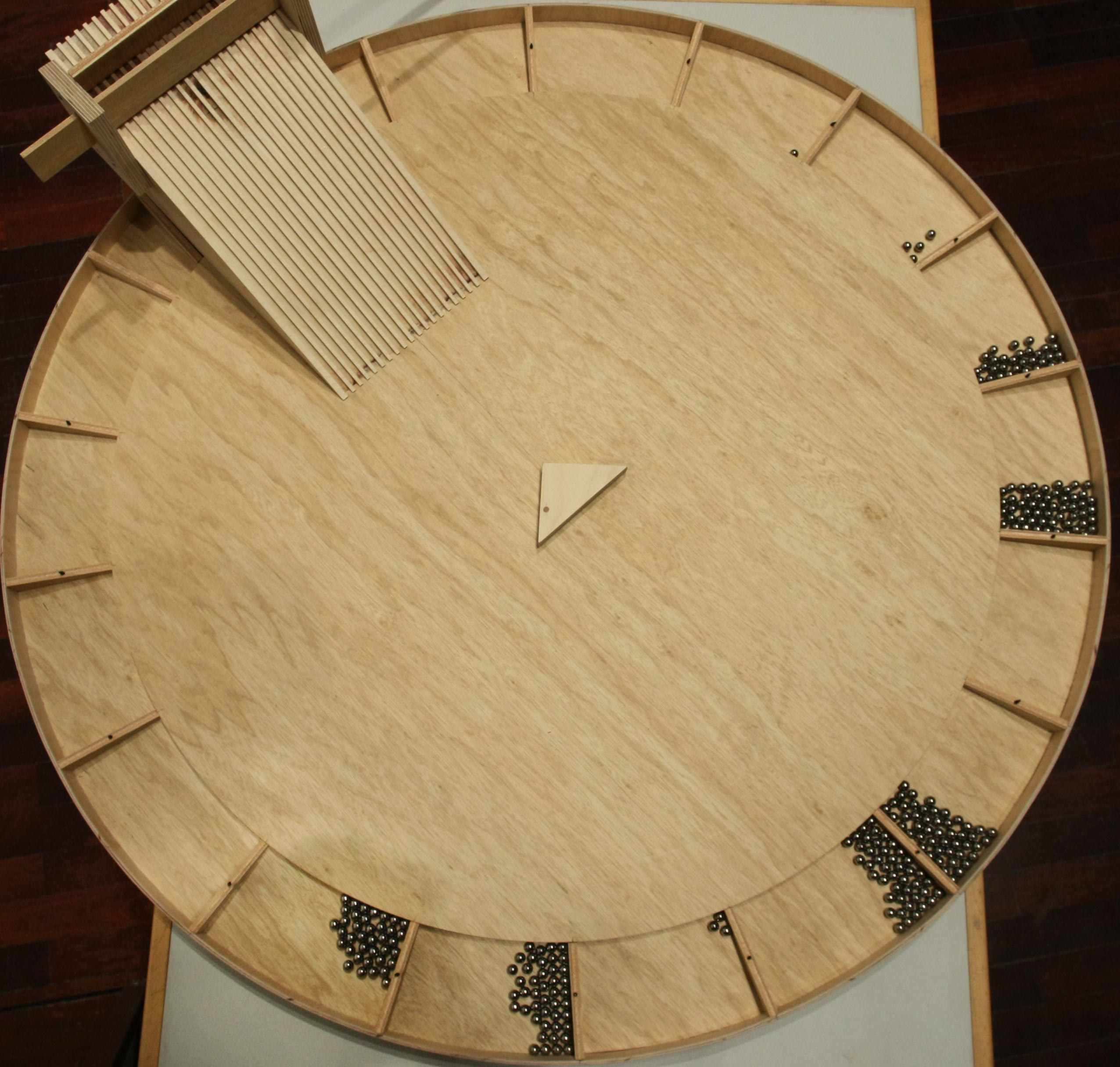}
	\caption{Result of scattering multiple bunches on the left off of a circular target, and on the right off of a 
	triangular target. On the left, except for the forward direction, the projectiles are roughly evenly distributed.
	On the right with the triangular target, the projectiles accumulate in two bins on either side, respectively.}
	\label{multiple-scattering}
\end{figure}

\hop

So that is the general concept of a scattering experiment. You shoot a bunch of particles at an
unknown target and you learn something about its structure from the way the particles are
scattered. This works even if the target is so small that you can not see it with the naked eye.

\medskip

\item[\R]  Fine\dots fine\dots {\it (Film starts. Rutherford explains the film.)}\\

\medskip

\EXP{Rutherford scattering -- film}

\medskip

In my original experiment I used an apparatus like it is presented to you on the screen.  Alpha
particles emitted from a radioactive Radium source were used as projectiles.  The thin gold foil, here in
the center, formed the target.

I observed ---and you can follow my example now--- that only a few of the alpha particles
passing the gold atoms are scattered. But by far the most pass the layer of atoms without
scattering.  Just like in our wooden model. From this real experiment, we can deduce the
structure of the gold atoms.

You see, that the mass of the atom is concentrated in a very small volume, which I will call the
nucleus. It is the small darker circle placed in a large, lighter, almost empty sphere, which does
not deflect the alpha particles.

So, we get an idea of the atom's structure. In the center of each atom is a nucleus, which is
100,000 times smaller than the atom itself. It is made of protons which are positive and
neutrons which are neutral. Outside it are small electrons, negatively charged, on planet-like
orbits {\it (Looks up towards heaven.) } -- Bohr excuse me. And it is these orbits which make
the size of the atom.

\hop

\item[\G] In a further experiment, I can even show you these little electrons {\bf directly}
  over here. 

\hop
  
\EXP{Helmholtz coils (see App.\,\ref{app:helmholtz-colis})}

\hop

\item[\G] Lets have a look on the screen: you can see a vacuum tube which is filled
  with a very small amount of air. In this vacuum tube we have an electron gun that is shooting
  electrons upwards at high speeds. You can see them as this blue line.

\textit{(Showing with a laser pointer the straight line of electrons in the dark.)}

\item[\G] These accelerated electrons collide with the remnant air and make it glow. \dots

\medskip

\item[\B] So we see the excited atoms that the electrons have bumped into along the way?

\medskip

\item[\G] Exactly. Now let's see what I can do with the electrons \ldots

\textit{(Mission Impossible music starts playing, bends electrons to cirlces of different radii.)}

\medskip

\item[\A] How do you do that?

\medskip

\item[\G] Let the camera zoom out. Outside the tube you can see we have two coils. These are simply 
copper wires, wrapped up many times. When an electric current flows through them, they generate a 
homogeneous magnetic field.  Since the electrons are electrically charged and moving, the magnetic 
field bends them onto curved paths. You saw, if I changed the field I could make the circle bigger or smaller.

\medskip

\item[\B] Okay, so now we understand the structure of the atom and we know that we are all made of them.

\medskip 

\item[\A] But where does the mass of the atoms come from?

\medskip

\item[\R] Ah yes, your original question. Well {\bf your} mass is simply the sum of the masses
  of your atoms. There are very very very many atoms inside you. But they each have an incredibly 
  tiny mass. Inside the atoms the mass comes almost solely from the protons and neutrons. From 
  the nucleus. The electrons, which we saw here, make up less than 1 part in a thousand and are 
  just along for the ride.

\medskip

\item[\B] Have we made progress here? Where does the mass of the electrons, protons and
  neutrons come from?

\medskip
 
\item[\R] {\it (shrugs)} I am the Master of the atom, but {\bf even I} don't have a clue!
  Anyway -- let me show you something funny. These atoms don't always just sit there and do
  nothing. There are several which are special. Like the Radium I showed you in my 
  scattering experiment. They spontaneously send out energetic particles. They are
  radioactive. 

\medskip

{\it (They walk over to radioactivity experiment.)}

\medskip

\item[\R] The radioactive material we use here is very safe. Don't be afraid, their
  range in air is only about a few centimeters.  {\it(To the audience.)} There is nothing to
  worry about. {\it(Back to visitors.)} But I will let my assistant perform this experiment.

\medskip

\EXP{Radioactive decay and Jakob's ladder  (see App.\,\ref{app:jacobs-ladder})}

{\it (Caretaker re-enters stage and carefully observes the experiment.)}

\medskip

\item[\R] What we have here are two bent wires with a high voltage between them. Look, what
  happens when we light a match at their closest point.

({\it The match is lit and we see the extended spark rising through the wires.})

\item[\G] The heat of the match separates the electrons from the atomic nucleus. This
makes the air electrically conducting, so that a flash between the wires is triggered. 

\medskip

\item[\R] And now, let's see what happens with this radioactive source Geiger is holding in his hands \ldots

({\it Spark is triggered by radioactive source})

\item[\G] The little dot at the tip of this tube is emitting nuclei of Helium, so--called alpha particles, at a very
high energy. They can also separate the electrons from the atoms in the air.
  The voltage then causes a lightning flash. Since the air inside gets very hot, it rises
  upwards.

\medskip

\item[\R] Ah, he is a good one our Hans!  This is a way to detect little particles that you
  can otherwise not see with the naked eye. Geiger, you should work on such a detector.

\medskip

\item[\G] {(\it Turns away from Rutherford, mutters in a lower voice.)} Yes, but I will call it Geiger--Counter...

\item[\B] Ernest, thank you so much for your help. I guess we must find someone who knows
  the origin of the mass of all these protons and neutrons.

\medskip

\item[\C] {\it(Claps his hands, Rutherford and Geiger freeze.)} Well, I guess this will
  take some t- i- m- e \dots

\medskip

\item[\A] Oh it's you!

{\it (Caretaker walks towards \Aa\  and \Bb, notices the radioactive sources and stops.)}

\item[\C] {\it (To the audience.)} If I may say something as well, Enrico Fermi, in Rome at the time got the
Nobel Prize in 1938 for his work on radioactive materials! ({\it Taps on Geigers shoulder.})

{\it (Show slide with picture of Enrico Fermi.)}

\medskip

\item[\C] ({\it To \Aa\ and \Bb.}) Unless you want to wait for 40 years -- let's see if I have another shortcut for
  you \dots   {\it (Finds another umbrella in his box.)}

\item[\B] Just a second. Here, Geiger, have some gummy bears from Bonn. ({\it Puts some
    gummy bears in Geiger's pocket.}) They are from the future. And they are delicious.
  Good byyyyyyyeeeee....

{\it (\Aa \ \& \Bb \ spin the umbrella...)}

\end{itemize}

\subsection{Berkeley, California}
\label{sec:berkeley}

{\it (\Aa, \Bb \ and the Caretaker step into the new space, highlighted by the spotlight. Ernest
  Lawrence and his female assistant, frozen in their lab.)}

\vspace{0.5cm}

\begin{itemize}

\item[{\bf Caretaker:}] Here you are in Berkeley, California. The time {\it(Checks his funny
    watch.)} is 1952. Before you, you see Prof. Ernest Lawrence. He is the all time--expert on
  particle accelerators. This here, hmm, is one of these assistants you have never heard of. Here
  is the next step on your quest. {\it(Briefly claps hands twice.)}

\item[\A] Hello, this is Lorenzo and I am Maike. We are from Bonn.

\hop

\item[\Lw] Oh, Bonn, where Wolfgang Paul just became the director of the Physics Institute!

\hop

\item[\B] Indeed! Can you maybe help us understand matter and find out where mass comes from?

\hop

\item[\Lw] Yes, welcome, you have certainly come to the right place here in my laboratory, in
  Berkeley California. Before we come to your question, let us look at some new particles you
  might not have heard about.

\hop

{\it(They walk to the cloud chamber.)}

\hop

\item[\Lw] Here, I have an interesting detector. It is called a cloud chamber. It can ``see'' some of the 
smallest particles we know. We have mounted a camera, so up on the screen you see the paths of the 
particles live as they come in.


\hop

\EXP{Cloud chamber (see App.\,\ref{app:cloud-chamber})}

\medskip

{\it (While \Lwl explains cloud chamber, \Aa \ and \Bb \ go oooh and aaah.) }

\hop

\item[\Lw] Those fat long straight lines are protons, you have maybe heard about. The really fat
  short lines are helium nuclei, also called alpha particles. That is what Rutherford used, to do his 
  famous experiment. The longer squiggly lines you also see here are electrons (some of them are 
  also positrons, the anti-particle of the electron). And all these particles come from cosmic rays, 
  from outer space. We are not producing them here. And they fly through everything. 

\hop

{\it (Caretaker claps twice, Lawrence and assistant freeze. Man in Black Suit, white shirt, dress hat, and sun
  glasses walks in from audience. He addresses the audience. See Fig.~\ref{fig:blues-brother}.)}

\hop

\begin{figure}[h!]
    \centering 
        \includegraphics[width=0.90\textwidth]{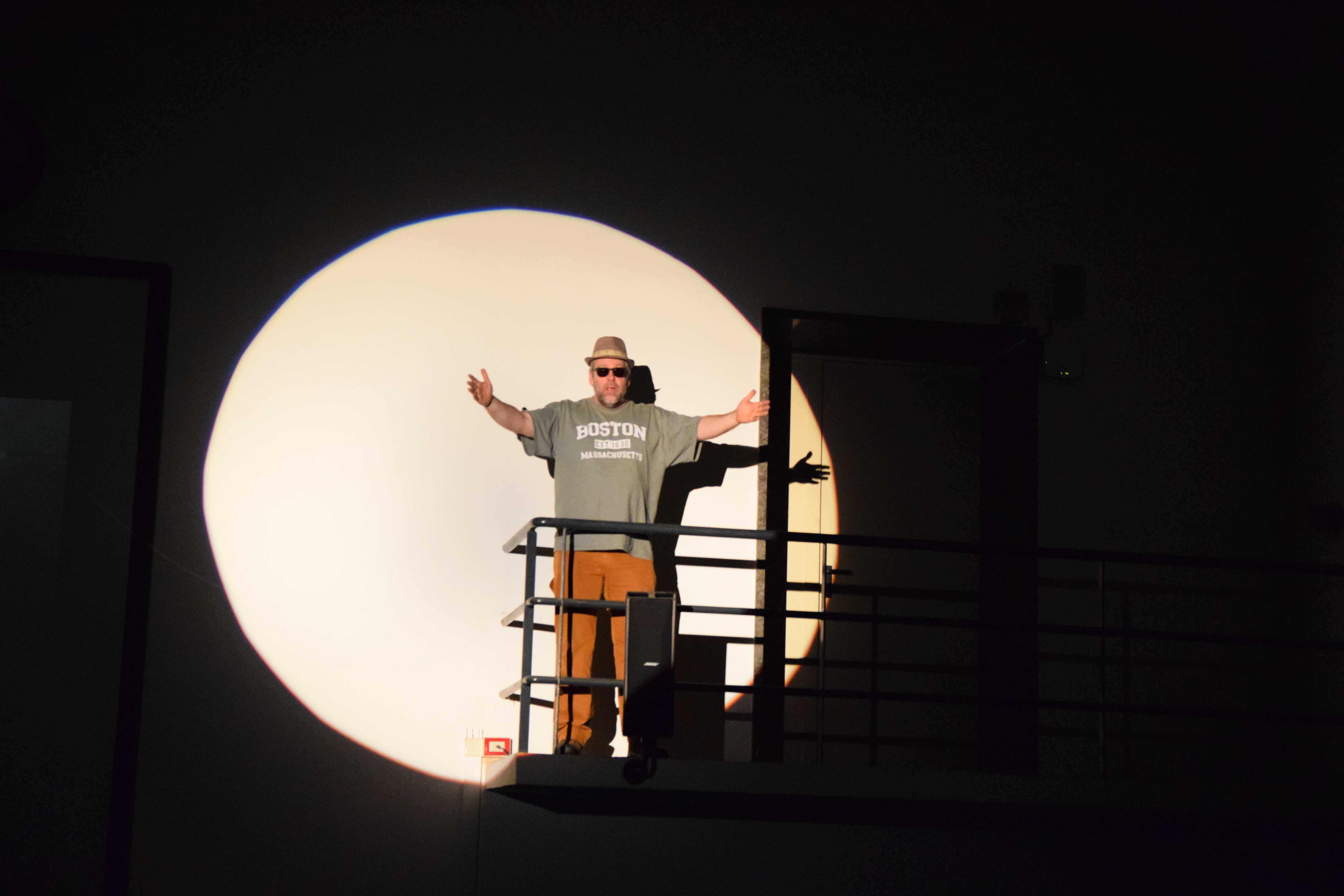}
           \caption{Blues Brother (HKD) on the balcony in the Padua lecture hall.}
    \label{fig:blues-brother}
\end{figure}

\item[{\bf Blu Bro:}] You mean these cosmic rays come from the universe and they fly through everything?
They fly through {\bf Y O U!}  {\it(points at Lawrence)}, {\bf M E!} {\it(points at himself)}, 
{\bf T H E M!} {\it(points at the audience)}, \\
{\bf E V E R Y B O D Y!!} {\bf
    E~V~E~R~Y~B~O~D~Y!!} 
    
    {\it (Music from Blues Brothers: Everybody Needs Somebody, karaoke version. Man, Maike and Lorenzo start singing.\footnote{ {\it 
 Everybody is bombarded //  By the cosmic rays // They're from outer space //
 Did you know? (Man at Maike) // I didn't know // What about you?  (Man at Lorenzo) //
  Neither did I //
  What about you? (at the audience) //
  They go through you, you, you! (3 times) //
In the airplane //
you get even more! //
Sometimes I feel //
I feel a little muon inside //
when it flies through my body //
I never ever ever have a place to hide. //
(music) //
Sometimes I feel //
I feel a little muon inside //
when it flies through my body //
I never ever ever have a place to hide. // They go through you, you, you! (3 or 5 times)
 }} Caretaker claps again, scene comes back to life.)}

\hop

\item[\A] You mean we are constantly bombarded by all this stuff.

\hop

\item[\Lw] Yes.

\hop

\item[\B] And what are
those long straight thin lines? They seem to happen quite often.

\hop

\item[\Lw] Yes, very good. Those straight lines are \ {\bf m~u~o~n~s}. They are a totally new kind
  of particle. They live only very briefly, but you can see they are still out there. They were
  discovered here in California.  This muon is 200 times heavier than the electron, but otherwise
  basically the same. 

\hop

\item[\A] Aha, so the muon is just a big brother of the electron, but what is it actually for?

\hop

\item[\Lw] The simple answer is: we have no idea. You could say: it is matter but it doesn't
  matter.
  
\hop

{\it (\Bb \  laughs briefly but hysterically at the joke.)}

\hop 
  
\item[\Lw]   It is too short lived to be of relevance for stable matter, like us. Then again, maybe
  it is an important piece in the matter puzzle.

\hop

{\it (Caretaker takes folded newspaper out of his overcoat pocket and starts reading.)}

\item[\A] Ah, that's interesting, are there more of these unusual particles?

\hop

\item[\Lw] Yes, come look at this experiment, it shows that it is {\it even more} complicated:
  anti-matter! Every particle has an anti-particle, it is equal in mass but opposite in charge.
These are safe sources, but my assistant Jacqueline will perform the experiment.

\hop

\item[\As] The name is Jane.

\hop

\begin{figure}[h!]
    \centering 
        \includegraphics[width=0.90\textwidth]{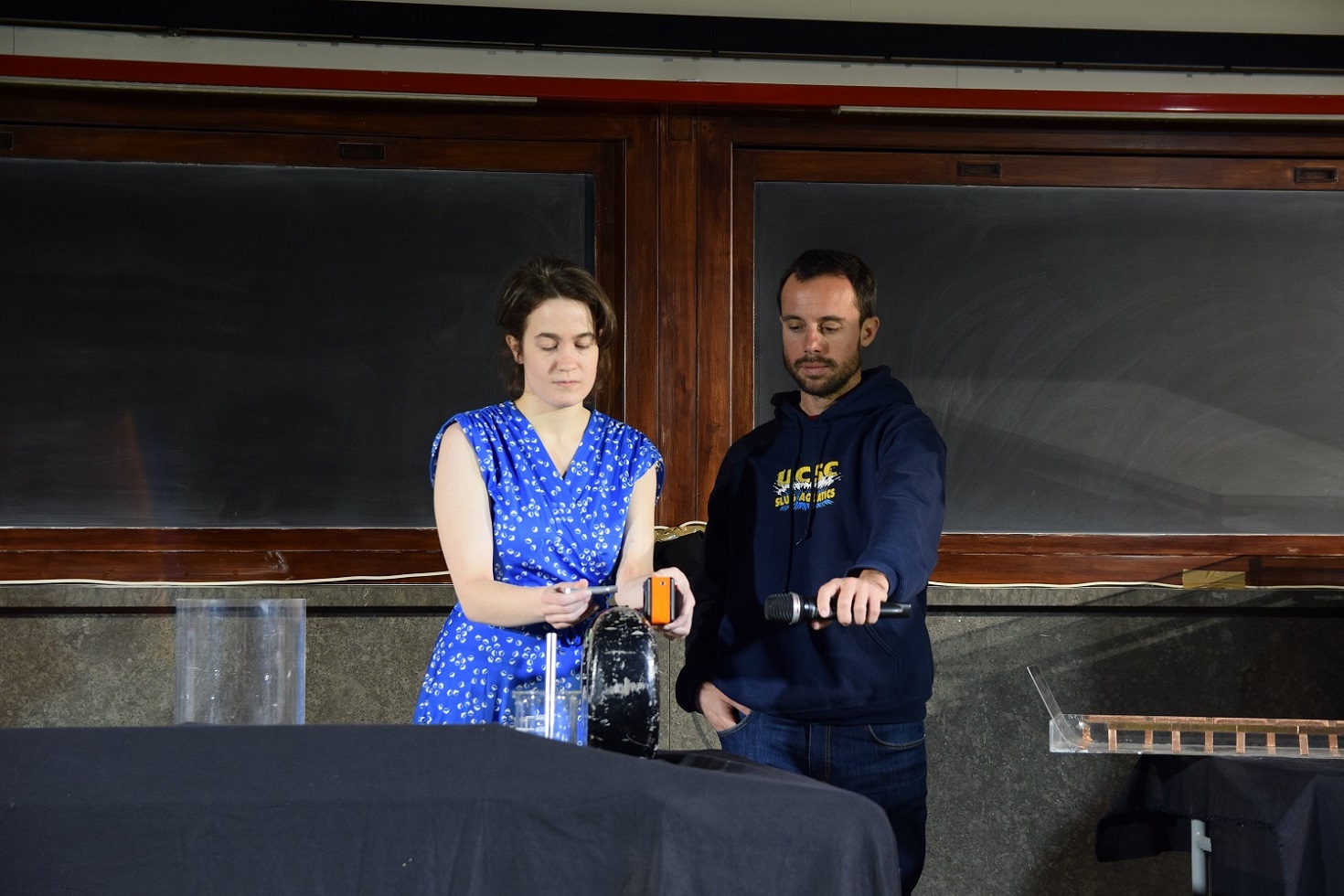}
           \caption{The beta-decay matter and anti-matter experiment, with JSch-R (Jane) and LU. The orange box
    is the Geiger counter. The black object underneath and slightly to the left is the magnet. One source 
    is the silver pen in JSch-R's right hand. The other source is in the glass.}
    \label{fig:jana-lorenzo-beta}
\end{figure}

\item[\Lw] Yes!

\hop
 
 \EXP{Experiment with Sr-90 ($\beta^-$) and Ge-68 ($\beta^+$) in  magnetic field (See 
 App.\,\ref{app:betaplusminus} and Fig.~\ref{fig:jana-lorenzo-beta}.)}
 
 {\it (Assistant performs the experiment while \Lwl explains.)}
 
\hop 
 
\item[\Lw] The first radioactive source emits fast electrons that are detected with the Geiger counter.

\hop

({\it Spotlight on Geiger who briefly appears from the side door with a big smile.)}

\hop

\item[\Lw] If we now insert the source in a magnetic field we hear that the clicking decreases. Where did the electrons go?
We can try to detect them above the magnet. {\it (Assistant moves the Geiger counter above the magnet.)} Nothing. Let's try below the magnet. {\it (Assistant moves the Geiger counter below the magnet.)} Now we hear the clicks again! The magnet bends the electrons downwards.

  \hop
  
\item[\A] {\it(To Lorenzo.)} Just as we bent the blue electron ring in the cathode ray tube?

\hop
  
\item[\Lw] Indeed. The second source emits positrons, the anti-particles of electrons. Again, we can detect them with the Geiger counter.
When we put the source in the magnetic field the clicking decreases. Let's see if we find them where we found the electrons previously. {\it (Assistant moves the Geiger counter below the magnet.)} No, they are not there. Let's try above the magnet.  {\it (Assistant moves the Geiger counter above the magnet.)} Here they are! They have positive charge, so the magnet bends them upwards!

\hop

\item[\B] Anti-matter... Awesome! {\it (To Maike.)} Maybe I can finally build my own warp drive!

\hop

\item[\A] Forget it Lorenzo! There is a much better use for that. Antimatter is used in medical imaging, 
they call it positron emission tomography (PET).  But coming back to our original question, we wanted 
to learn about the proton ...

\hop

\item[\Lw] Sorry, I have seemingly digressed with all these particles. Right, you wanted to look
  inside the proton. For that we need accelerators.

\hop

\item[\A] Accelerators? Why can't we just take a microscope?

\hop

{\it (Caretaker yawns, lays down with the hat over his face and falls asleep.)}

\item[\Lw] Well let me first explain how a microscope works.  {\it(Points to slide with image.)} It sends light waves
  onto a sample. The light then goes through these lenses and reaches our eye. We can only see the object, if the
  wavelength of the light is {\it smaller} than the size of the sample. Here, I have an
  experiment with water waves. As I am allergic to manual labor my assistant Simona will demonstrate it to you.

\hop

\item[\As] The name is Jane.
  
\hop

\item[\Lw] Yes yes.

\hop

\EXP{Water waves (see App.\,\ref{app:water-wave})}

\hop

\item[\As] Here I have a water bath with a light underneath. Through the mirror I am able to
project the reflection of the water onto the screen. When I switch on the wave generator, you can see the waves
moving across the surface, projected onto the screen. Let's now see what happens when I put
this large object, {\it (Holds it up into the spotlight.)} larger than the wavelength, in the path of the waves.

\hop

{\it (Music: Surfin'  USA. \Aa \ and \Bb \ dance.)}

\hop

\item[\A] Ahhh, I see how the object blocks the path of the waves. So if you sit behind it you can
  see where it is, from the wave pattern behind it. ({\it Points on the screen with laser pointer.})

\hop 

\item[\As] Exactly! But now see what happens if I put a smaller object in there. An object, which
  is smaller than the wavelength.

\hop

{\it (Music: Surfin'  USA. \Aa \ and \Bb \ dance.)}

\hop

\item[\A] Mhh, the waves reunite and there is no distortion in the pattern further back.

\hop

\item[\As] Yes, very good, that is our problem. As soon as the object is smaller than the wavelength we can
  not see it anymore. The waves just go around it. But now in our wave bath we can increase the frequency, and thus
decrease the wavelength. {\it (Cranks up the wave generator.)}

\hop

Now, you can see the distortions in the back again!

\hop

\item[\A] So why don't we just increase the frequency of the light in our microscope, in order to
  see the smallest particles?

\hop

\item[\As] For technical reasons it is not possible to do that with light waves. Instead we use the method Rutherford also 
used in his famous experiment.

\hop

({\it Spotlight on Rutherford who briefly appears from the side door with a big smile and triumphantly raising his hands.)}

\hop

We use particles instead of light. When one looks very close at a particle it becomes wavelike. So we can decrease the 
wavelength of the particle to get a better resolution. To get a smaller wavelength we have to increase the energy of the 
particle. So we want to accelerate particles to high energies.  

\hop

{\it (Caretaker stands up slowly, has a stretch, walks to {\it \color{blue} Lawrence}, takes a cup out of his overcoat
  pocket and drinks.)}

\begin{figure}[h!]
    \centering
    \includegraphics[width=0.90\textwidth]{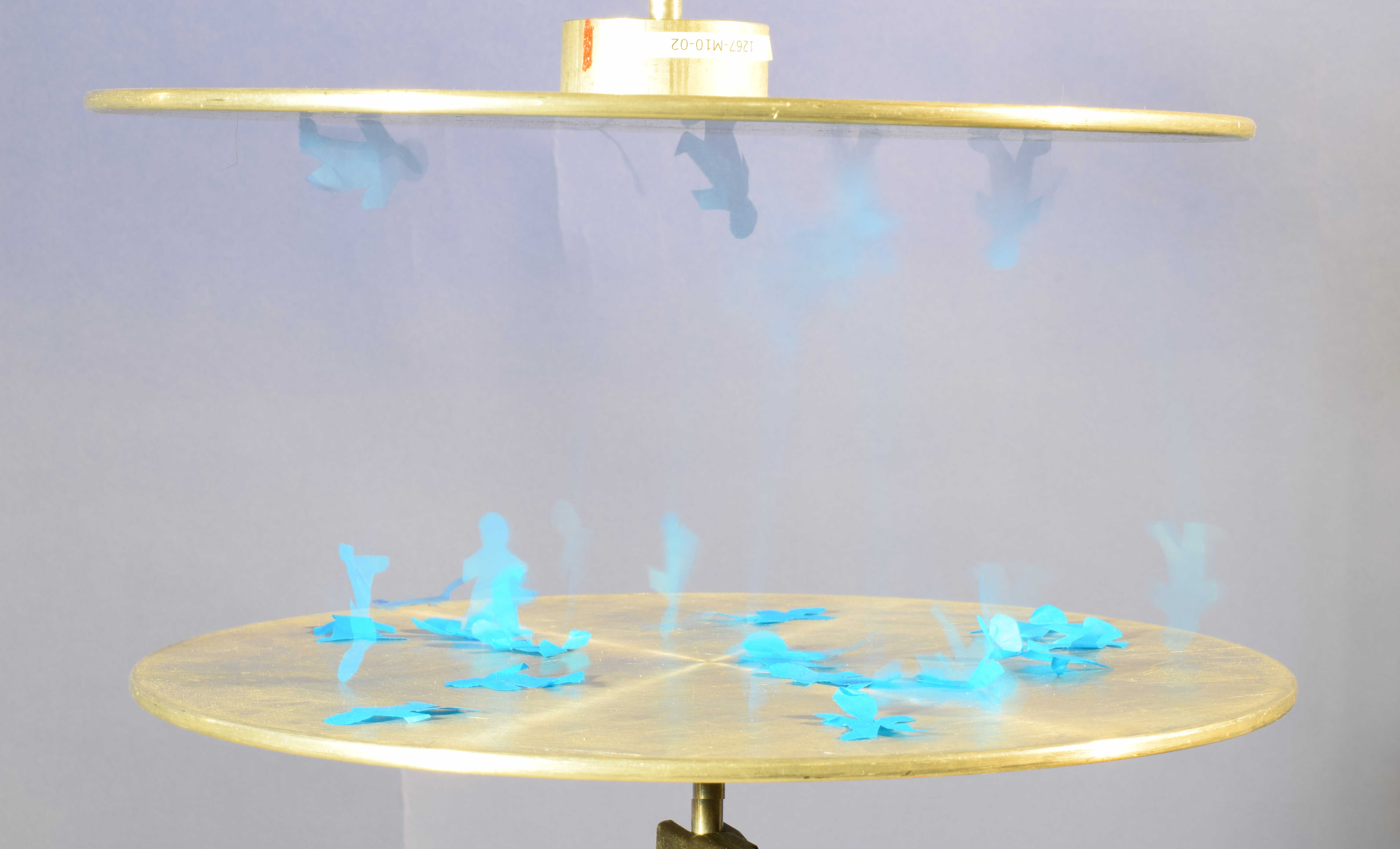}
   \caption{Dancing paper men experiment. The picture is taken with a high exposure time to capture the motion 
    of the dancing paper men. See also the setup in Fig.\,\ref{fig:dancing-men-1}.}
        \label{fig:dancing-men}
\end{figure}

\item[\A] Now I get it, you accelerate the particles to higher and higher energies to see smaller
  and smaller particles. ... But ... how do you accelerate them?

\hop

\item[\As] That is easy, we just use an electric field. I can show you that with another experiment.

\hop

\EXP{Plate  Capacitor (see App.\,\ref{app:dancingpaperwomen} and Fig.~\ref{fig:dancing-men}.)}

\hop

\item[\As] Here I have a large plate capacitor. The lower plate is connected to Earth.
I shall put some Italian soccer players on it.

\hop

\item[\A] Soccer players?

\hop

\item[\As] Well, a paper version of the Italian soccer team. That is why they are blue. Now I use this cat fur to 
rub electrons onto this plastic stick. From there I put them on the top plate. ... 

\hop

{\it(Paper people start dancing to music. \Lwl and Caretaker move slightly up and down with the beat.)}

\hop

\item[\B] Wow ..., but why did they keep falling down?

\hop

\item[\As] Have you ever seen the Italian soccer team play? ... No, with the plastic stick the top plate got charged
  negatively. The bottom plate and the paper people then get charged positively. Due to the
  difference in charge the paper people get attracted and rise to the upper plate. There, by
  contact they get negatively charged and are attracted downwards. Then the same process starts
  again.

{\it (Caretaker offers a part of his newspaper to \Lwl and starts reading again.)}

\hop

\item[\B] Amazing! But they just go up and down. How can you achieve higher energies?

\hop

\item[\As] Well, for that we simply use several capacitors in a row.  Here I have this linear
  accelerator.

\hop

\EXP{Linear electric Accelerator (see App.~\ref{app:linear-accelerator})}

\hop

\item[\As] It has these copper strips which I will charge, alternating positive and negative. To
  power the whole thing, instead of killing more cats, I use this modern machine {\it(See 
  Fig.~\ref{fig:wimshurst}.)}, which I hook up here. Inside the accelerator I place these 
  conducting balls. {\it(Runs the machine to music.)}

\hop

\item[\B] Interesting, that is pretty good. But to be honest, that doesn't seem very fast.

\hop

\item[\As] If you want to reach really high energies, you would have to build a much longer
  accelerator. To get the required energy it would have to be several kilometers long ... that doesn't
  seem feasible.

\hop 

\item[\Lw] Yes yes yes, thank you Sue, but to solve this problem, I Ernest Orlando Lawrence have 
invented another kind of accelerator. {\it(Raises circular accelerator.)} Here the
  particles are accelerated in a circle, so I called it a circular accelerator.

\hop

\item[\A] To me it rather looks like a huge salad bowl.

\hop

\EXP{Salad bowl Circular Accelerator (see App.~\ref{app:salad-bowl})}

\hop

\item[\Lw] Well, it's an accelerator! It is very similar to the linear one with alternately
  charged copper strips. However, instead of using that old fashioned machine, I use something
  modern, which comes out of this box. It is called {\it electricity}. Furthermore, I also need
  a conducting ball to place inside. {\it(Takes a ball from the linear accelerator and runs the 
  circular accelerator to music.)}  And you see the ball travels on a circular path.

\hop

\item[\B] Ahh that is cool, so you can accelerate the particle again and again, and you don't lose
  any. That is very clever. But, I have a question, in this experiment, you change the charge of
  the ball each time it crosses a copper strip, right?  
  
\hop

\item[\Lw] That is correct.

\hop  
  
\item[\B]  Is that also possible with real particles?

\hop

\item[\Lw] No, unfortunately not. An electron is always negatively charged, and a proton always
  positively. In real particle accelerators we have to change instead the electric fields, just as the particle passes. 
  We also have an experiment to show this. It is a mechanical analogue, using the gravitational field. My assistant 
  Francesca
  can explain this.
  
\hop

\item[\As] The name IS Jane.

\hop

\item[\Lw] Yes, yes...

{\it(Two people carry in the mechanical synchrotron model to music. The Caretaker assists.)}

\begin{figure}[h!]
    \centering
        \includegraphics[width=0.95\textwidth]{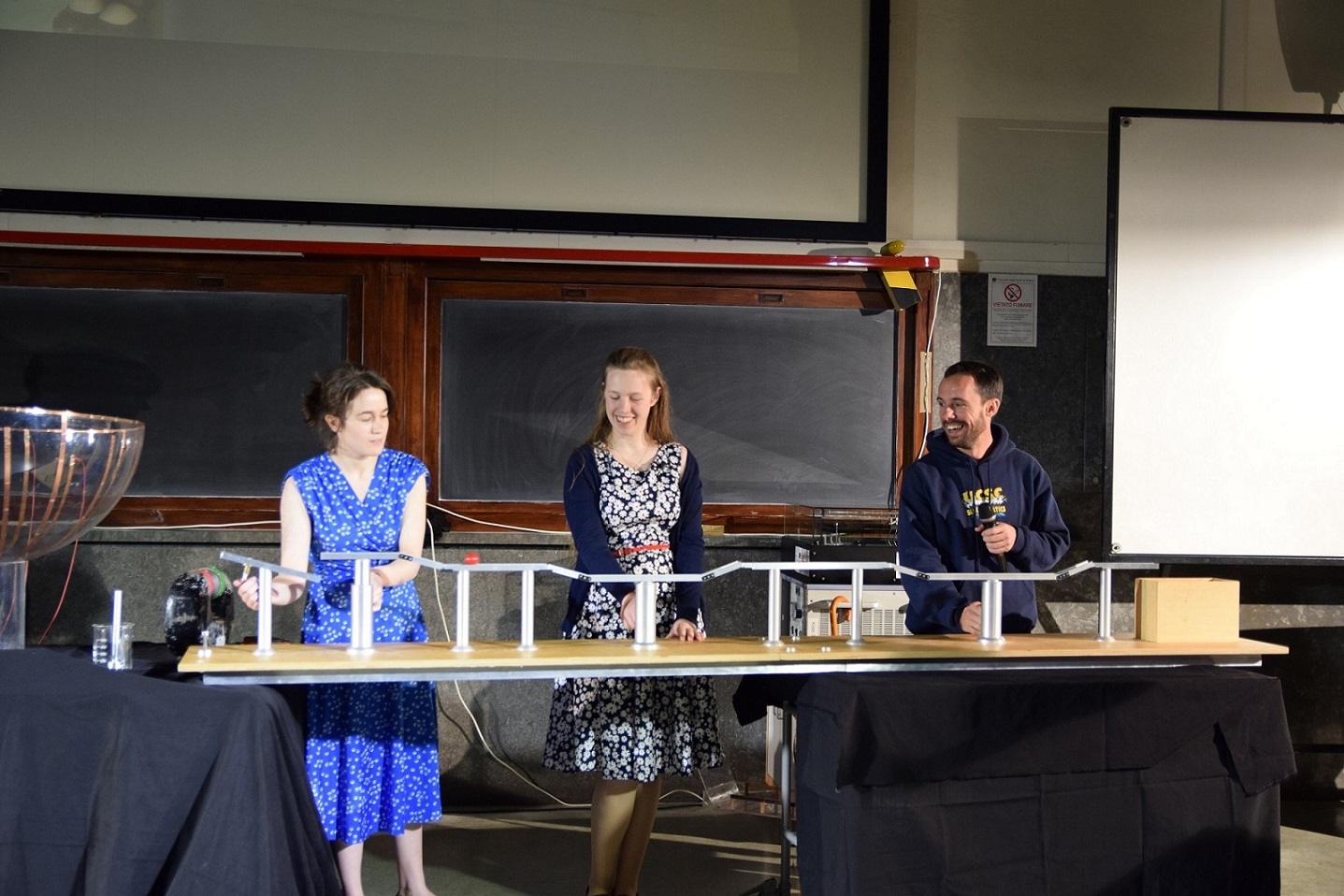}
    \caption{Mechanical synchrotron. The balls are released on the left and then raised at the right time
    where the slightly wider aluminum supports are. From left to right we have JSch-R, MH, LU.}
    \label{fig:mechanical-synchrotron}
\end{figure}

\hop

\EXP{Mechanical synchrotron model (see App.\,\ref{hebe-beschleuniger} and Fig.~\ref{fig:mechanical-synchrotron}.)}

\hop

\item[\As] This is a mechanical accelerator. I can let a ball roll down from this one end. Then there are three points 
where the rail can be raised manually to accelerate the ball. I could use your help, would you mind?

\hop

{\it (Maike and Lorenzo assist with the accelerator. Lorenzo gets it wrong the first time and looks clueless.) }

\item[\B] Hhmm, did I do something wrong?

\hop

\item[\As] Yes Lorenzo. You have to raise the track as the ball rolls by! It's not that difficult. Should we try again?

\hop

\item[\B] I'm sorry. Yes, let's do it again.

\hop

{\it (The second time Lorenzo does it right. They repeat it a third time too.)}

\hop

\item[\As] Here we have accelerated the ball using gravity. In a proper accelerator I would have to switch the
  electric field as the particle goes by. I would have to synchronise it with the particle's flight. That is why such a machine
  is called a synchrotron.

\hop

\item[\B] Oh, now I understand! So, in a real accelerator you have to switch the electric field just in the instant when the electron or proton is flying by? Oh, and by doing it in a circle you just sit there and wait every time it
  comes by. And you also don't lose any! Clever! 

\hop

\begin{figure}[h!]
    \centering
    \includegraphics[width=0.81\textwidth]{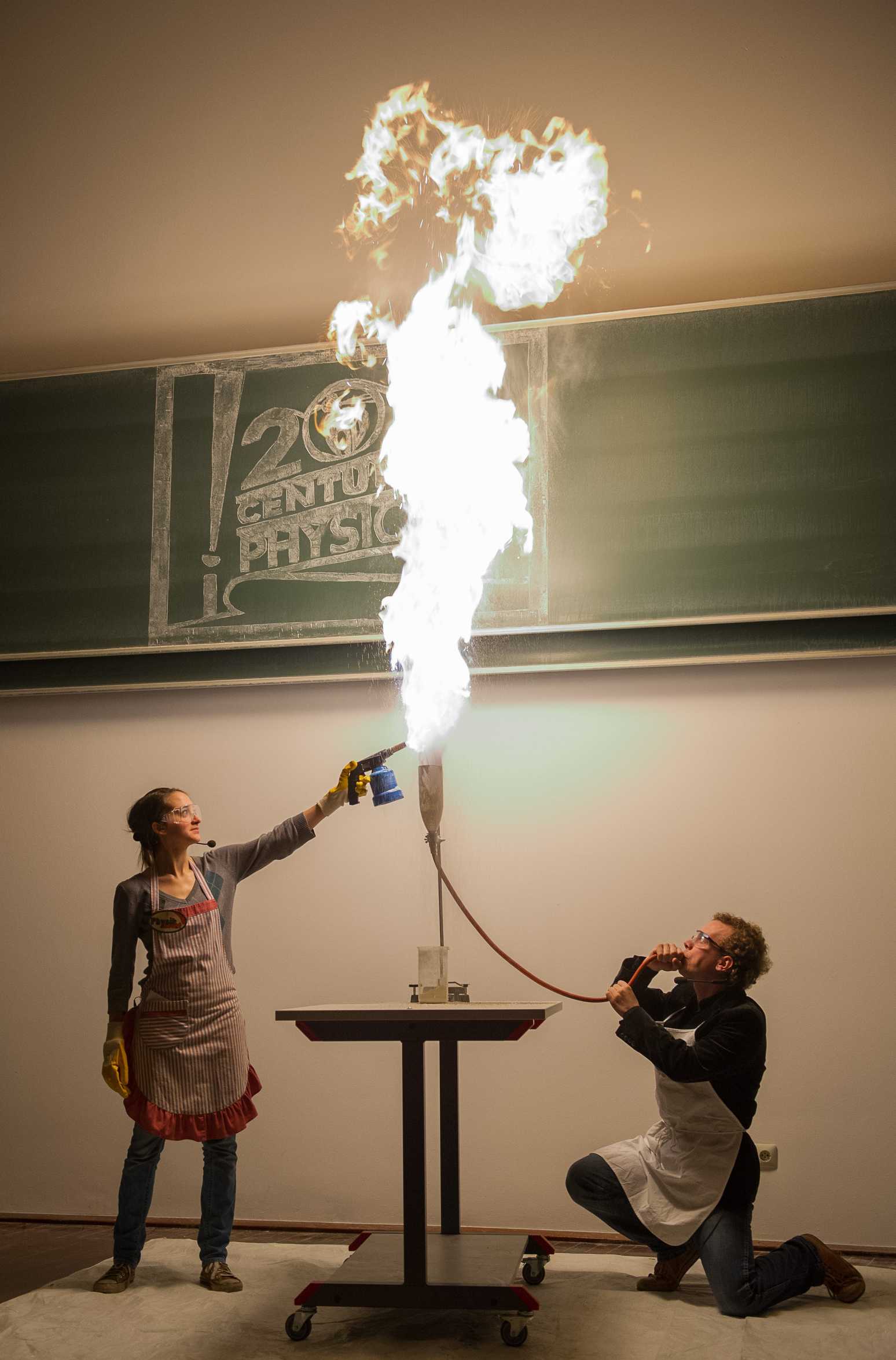}
    \caption{Burning flour cloud with Sina K\"urtz on the left and DH on the right. The cut-off PET bottle 
    is filled with flour. The latter is ejected upwards, by blowing through the red rubber hose. The floor is covered by
    a fire-proof tarp. This is from a different Bonn physics show. Photo by Volker Lannert.}
    \label{fig:sina-dustin}
\end{figure}

\item[\As] Exactly! Let me show you what a real accelerator looks like.

\hop

{\it (Project on to the screen a photo of the ELSA accelerator in Bonn.)}

\hop 

\item[\A] Isn't that the ELSA accelerator in Bonn?
It must be a very intricate machine! And how fast are these particles going?

\hop

\item[\As] These particles rapidly reach speeds close to the speed of light. 
Speaking of light, let me show you an experiment, just for fun. Maybe one of you could assist me.

\hop

\item[\B] Me, me, me!

\medskip

\EXP{Flour explosion (see App.~\ref{app:flour-power}, and Fig~\ref{fig:sina-dustin}.)}

\hop

\item[\As] We call it flour power.

\hop

{\it (Caretaker claps twice, scientists freeze. Caretaker brings new Umbrella.)}

\hop

\item[\C] I think it is time to move on. You've just learned how to construct a microscope
  to look at the proton. Now you should go see one in real life. {\it(Maike and Lorenzo start to
    spin the umbrella.)} This will be a more complicated intercontinental, space-time travel, and
  will take some \  {\tt t i m e}. Maybe 5 minutes. We will take a short break, but please remain seated.

\end{itemize}

\subsection{DESY, Hamburg, Germany}
\label{sec:desy}

{\it (\SLw and \Hhh frozen in their lab. Caretaker, \Aa \ and \Bb \ arrive. Maike and Lorenzo have their clothes 
swapped, i.e. Lorenzo is wearing a dress. Maike and Lorenzo look at the other and then themselves.)}

\vspace{0.5cm}

\begin{itemize}

\item[\A] What happened? Oh no, look at you!

\hop

\item[\B] And you!

\hop

\item[\C] Hmm, it seems something has gone wrong.

\hop

\item[\B] Very wrong!!

\hop

\item[\C] Did you maybe spin the umbrella the wrong way? Or maybe you have crossed a worm hole on the way...

\hop

{\it (Maike and Lorenzo shrug helplessly.)}

\hop

\item[\C] Anyway, the quest must continue! Lorenzo, be a man ... or whatever. {\it (Pause)} So, welcome to Hamburg, Germany!  
Here we are in the international research center DESY. The time is 1980.  These are Sau Lan Wu and her assistant Haimo 
Zobernig, they will be able to help you to answer your questions about the proton.
{\it (Caretaker claps twice; Sau Lan and Haimo come to life. Caretaker steps to the back
  of the stage and stands motionless like a guard.)}

\hop

\item[\A] Hello Ms.\,Wu, Hello Haimo!  

\hop

\item[\SL] Huh? Hello! Who are you?? 

\hop

\item[\A] This is Lorenzo, I am Maike, we came from Bonn on a journey through space and time.

\hop

\item[\SL] Oh, Bonn, isn't that where Wolfgang Paul built the first strong focusing electron
  accelerator in Europe?

\hop
\item[\A] Yes!

\hop 
\item[\SL] {\it (Looks at Lorenzo.)} Is that trendy in Bonn?

\hop

\item[\B] {\it (Embarrassed)} Ehm, this is German fashion from the future... {\it (Pauses)} 
Can you help us with our quest? We want to know what the proton is made of and where
  mass comes from.  A simple answer would be much appreciated... {\it{(Sighs, looks at audience
      for approval.)}}

\hop

\item[\SL] Well, to study the  proton, or to look deep into matter, you need large accelerators.
  They are like our microscopes.

\hop

\item[\A] Oh, Mr.\,Lawrence in Berkeley told us about those... but what happens when you actually want to look inside?
	How can an accelerator be like a microscope?

\hop

\item[\SL] My assistant Haimo will show you a simple accelerator we have built!

\begin{figure}[h!]
    \centering
           \includegraphics[width=0.9\textwidth]{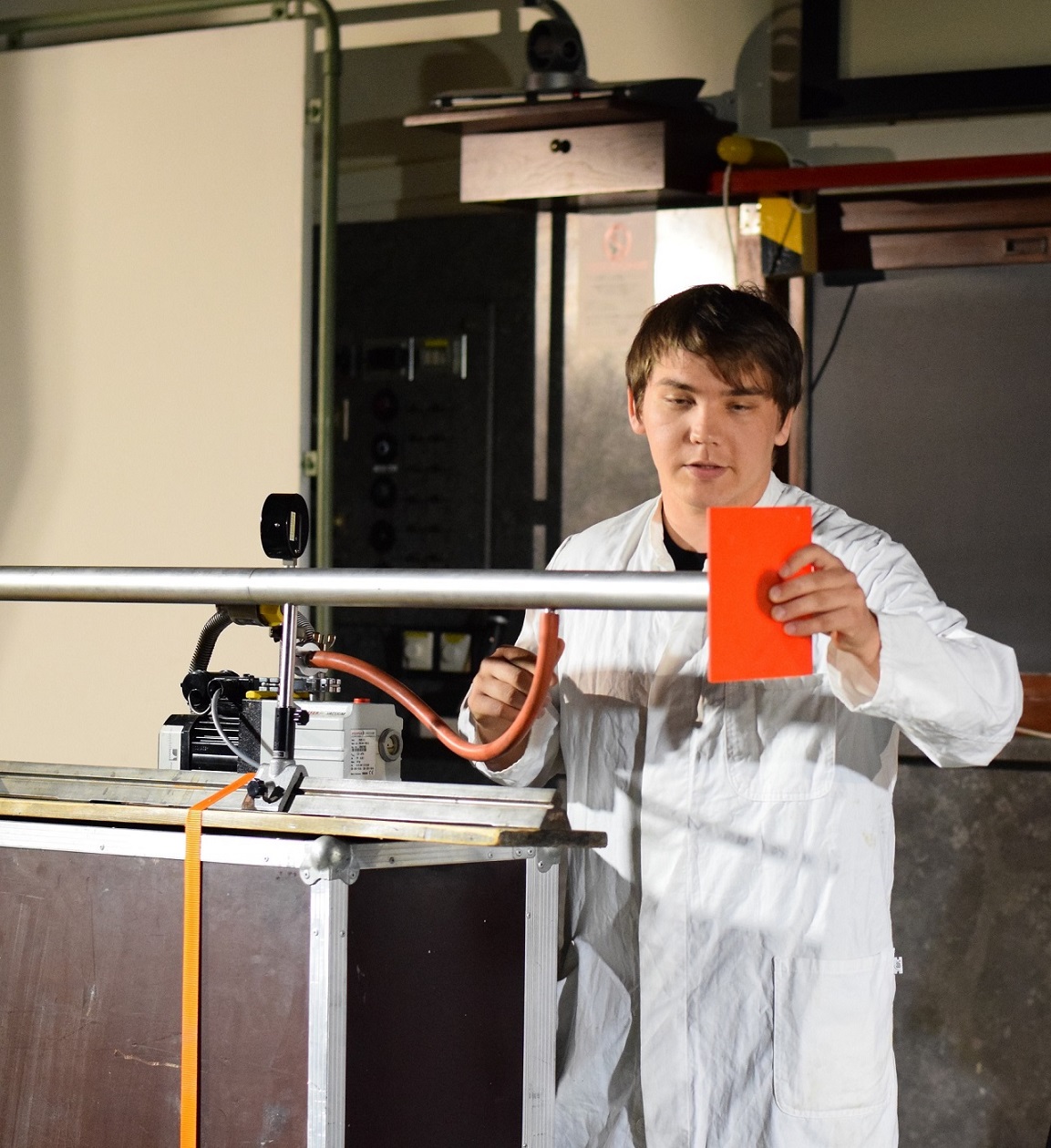}
   \caption{Haimo Zobernig (MBr), ready to fire the vacuum cannon.}
    \label{fig:vacuum-cannon-maxim}
\end{figure}

\hop

\EXP{Vacuum cannon (see App.~\ref{app:vacuum-cannon}, and Fig.~\ref{fig:vacuum-cannon-maxim}.)}

{\it (Caretaker is interested in the experiment.)}

\hop

\item[\Hh] This experiment consists of three parts: the accelerator, the detector, and this ...
  box.  The accelerator in this case is just this hollow metal pipe.  Connected to it is this
  vacuum pump.  On this side, we put in our projectile, this wooden ball!  And then ... and then ... hmmm?

\hop

\item[\SL] ...Haimo, we need a target that we want to investigate!

\hop

\item[\Hh] ..Oh, yes, right, uhm....

\item[\C] Maybe I have something you can use. {\it (Takes Quark out of his ``Lost \& Found" box.)}  I just bought it
  in a local supermarket. It is full of something called ``Quark''.  {\it(Gives it to Haimo, who holds it in front of
    the camera, to prove it is Quark. Then puts it inside the detector.)}

\hop

\item[\Hh] Yes, yes, this is our target and it will do just fine as a model of a proton!  We put
  it right here in the detector.  Okay, now we seal the pipe on both ends and start the vacuum
  pump.  When the pressure is low enough, I will remove this cap and we will see what happens.

\hop

{\it (Big bang, when quark explodes.)}

\hop

\item[\C] {\it(Frowns upon realizing he has to clean up everything. To himself:)} I have made a huge mistake....

\hop

\item[\SL] Lorenzo, would you like to read out the detector?

\hop

\item[\B] Can you eat that?

\hop

\item[\SL] Sure. {\it (Tries some of the quark.)} But it tastes a bit {\bf strange} to me.
 
\hop

\item[\A] Hmm, no, I think it is {\bf TOP}!  {\it(Shows two raised thumbs.)}
 
\hop

\item[\B] Delicious, this is such a {\bf beautiful} experiment!
 
\hop

\item[\SL] You are so {\bf charming}.

\hop

{\it (Flash a slide showing the six quarks of the SM.)}

\hop

\item[\C] {\it (Pulls an endless sheet of paper towel out of his ``Lost \& Found" box.)} Now I
  have to clean it {\bf UP}! Then I'll take it {\bf DOWN}, to the basement.
 
{\it (Caretaker starts to clean the detector with paper towels from his box and a large rubbish bag.)}

\hop

\item[\SL] What we can actually see with this experiment is that protons are made of quarks.  This
  was first seen with a huge electron accelerator in Stanford, California in the 1960s.  Real
  accelerators are far more complicated since they don't use vacuum technology for the
  acceleration.
 
\hop

\item[\B] So what do these ``quarks'' look like? Like that cottage cheese? {\it(Points to remnants of the 
quark explosion, that the Caretaker is still cleaning up\ldots)}
 
\hop

\item[\SL] No, that was just an analogy.  But there is something unusual about these quarks,
  because you actually never see them separated, but only in pairs or triplets.  Here, let me show
  you this little testicle model! {\it(Seductively plays with the balls.)}

\hop

\EXP{Testicle Model (see App.~\ref{app:testiclemodel})} 
 
\hop

\item[\Hh] The balls are enclosed with this elastic band.  You see you can stretch it, but you
  can't actually pull out an individual ball. 
	
\hop

\item[\B] Ouch! ({\it Touches his balls.})

\hop

\item[\A] You baby!  {\it(To Sau Lan.)} As you stretch the rubber skin it seems you must use more and more
  force. It gets harder and harder to pull them apart?
	
\hop

{\it (Caretaker finishes cleaning the detector, moves it to the back of the stage and puts away the
  rubbish bag. Now he wipes Quark off the floor/tables circuitously.)}

\item[\SL] Yes, exactly, and that is why even by pouring in more energy you can't see a free ball.
  This is the same for quarks.  The whole model represents a proton with 3 quarks inside.  These
  are held together by the strong force that is symbolized by the elastic band.
	
\hop

\item[\Hh] It is different from the electromagnetic force.  There you easily ionize atoms --
  separate  electrons from the nucleus  -- to make the path of the electron visible in a cloud
  chamber!
 
\hop

\item[\A] So if it is impossible to pull out an individual quark, how do you know they are inside?

\hop

\item[\SL] Excellent question.  Let me show you another nice experiment.  Here we have this green
  balloon, which represents the original conception of a proton.  Also we have this red balloon,
  which represents today's conception of a proton.  We are now going to throw these balloons at
  each other to see how they behave.  We could maybe use some help from our audience here!  Who
  wants to come on stage?  It's a very easy task!

\hop

\item[\B] {\it (To kid from audience.)} What's your name?

\hop 

\EXP{Tossing Balloons (see App.~\ref{app:balloontossing})}

\hop

\item[\SL] Ok, you stay right here.  So first, I'm going to throw the green, balloon. Could you please catch it 
and just throw it back? {\it (Throws the green balloon.)} Thank you, that was easy. It is just homogeneously
filled with air. And now the red balloon... are you ready?  {\it (Throws the red balloon.)} Alright, thank you!  
Big applause for N.N.! {\it (Pop the red balloon to reveal what's inside.)}

\hop

\item[\Hh] As you could see, the balloon with solid balls in it behaved differently from the one only filled with 
air. In a similar way we figured out, that the proton must also consist of smaller objects!  Of course at 
Stanford it was a bit more complicated, but the same idea.

\hop

\item[\B] {\it(Plays with the testicle model, stretches the balls.)} Ok, I get the idea with these quarks inside the proton, but what about this force that holds them together?

\hop

\item[\SL] I have another experiment! Haimo, would you be so kind to explain it?
  
\hop

\EXP{Air table and strong force (see App.~\ref{app:airtable})} 

\hop

\item[\Hh] On this table, we have three pucks, representing three protons.  When we now turn on the
  air table, compressed air blows through many small holes in the table.  Therefore the pucks
  glide frictionless over the table.  The protons both carry a positive electric charge.  We
  expect that they push off each other.  This is realized by small magnets inside the pucks.

\hop

\item[\B] Wait... but in a nucleus, there are many protons with equal charge, very close together!
	Now, how do you explain THAT? 

\hop

{\it (Caretaker unscrews the tip from his hat cleans it elaborately with a paper towel.)}
	
\item[\Hh] This can be explained by the strong force, which is symbolized by the velcro I now put on the
  pucks.  It holds the protons together.  At short distances it is very strong.  But it drops off
  rapidly afterwards.  If the protons are far away, the electromagnetic force dominates, so that
  the two protons are pushed off.  But at very small distances, the strong force makes them stick
  together, just as the velcro.  This way, protons inside a nucleus stay together.

\hop

\item[\A] Ok, so the strong force doesn't reach very far, but when it kicks in, it is VERY strong. 
	But coming back to your rubber model. 
	How does this rubber skin look in the real world?
	What is it, that holds the quarks together?

\hop

\item[\SL] That is a very good question. We are now looking at particles that are very very small.
	Remember a human hair is already as wide as 40 trillion protons next to each other. And the 
	quarks live inside such a proton. In this world the laws of nature are given by quantum theory.

\hop

\item[\Hh]	And in the quantum world,  energy always comes in little packets, that have a fixed size.
	Let me show you this little experiment!

\hop

\EXP{Photon Clicker (see App.~\ref{app:photonclicker})}

\hop
	
\item[\Hh]The red laser light, that you can see here, consists of millions of millions of photons per second.  
Before this detector, I have some filters that block out the largest amount of those photons.  The detector 
is hooked up to these loudspeakers. {\it(Turns them on.)} As you can hear, the loudspeakers crackle from 
all the photons. {\it (Then turns laser intensity way down.)} Now you hear individual clicks. 

\hop

\item[\B] Wow, so we can actually hear the individual photons arriving here in this little detector?

\hop

\item[\Hh] Yes, that is correct.  And this packet, the photon, which we only know as light so far, is also a
  force carrier of the electromagnetic force!  We now know that every force has a force carrier.
  
\hop

{\it (Caretaker finishes cleaning the tip and screws it on the hat. Then he stands motionless again.)}
  
\item[\SL]   For the strong force, this particle is called the gluon and was discovered by me here at DESY!

\hop

\item[\B] Really?

\hop

\begin{figure}[ht]
  \center{\includegraphics[width=0.85\textwidth]{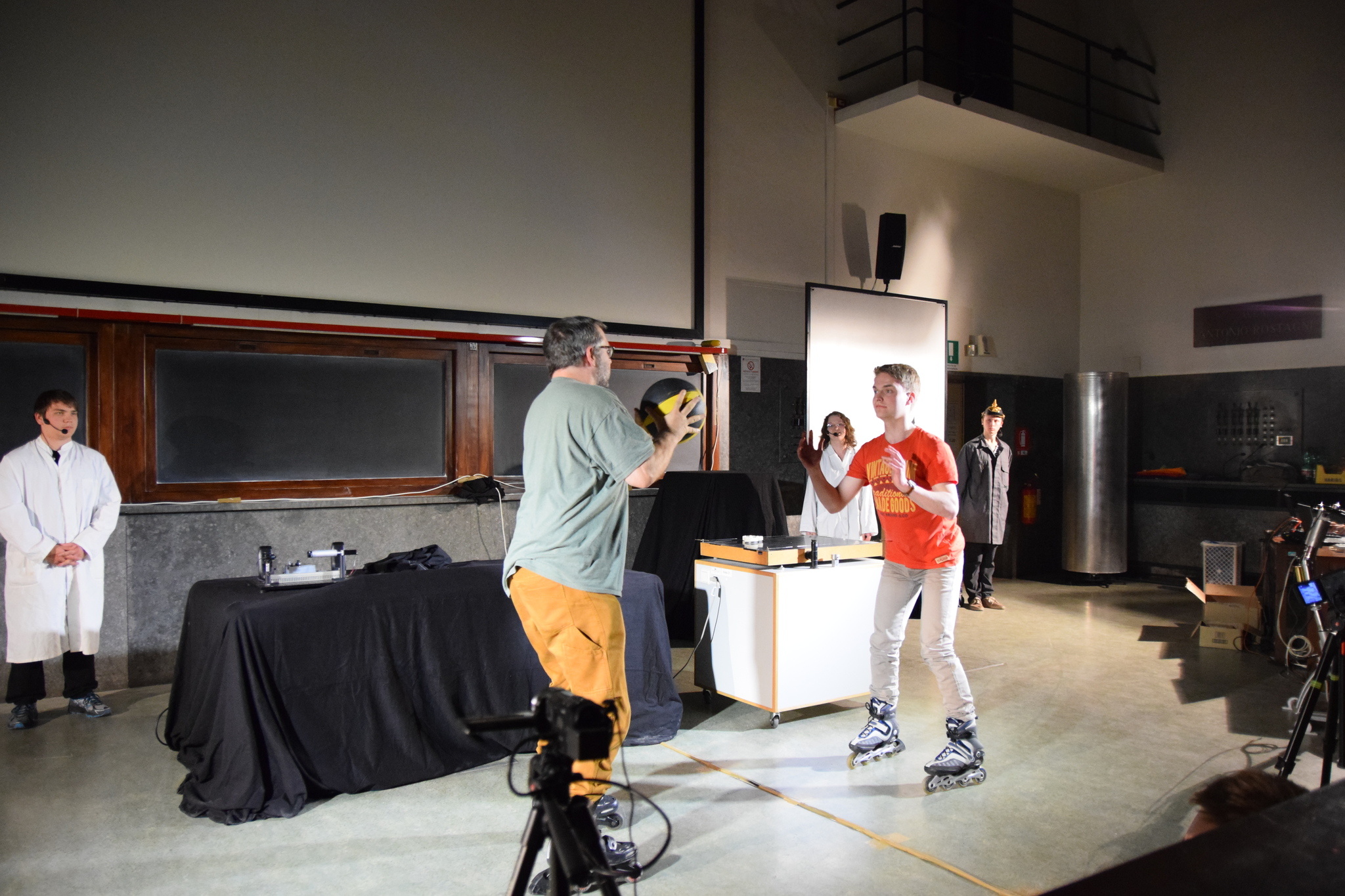}}
  \caption{AF (right) and HKD (left) exchanging a medicine ball during the
  show in Padua. On the left in the back is MB as Haimo Zobernig and on the right is KH as Sau Lan Wu.
  Just to the left of AF you can see the air table of experiment \ref{app:airtable}. In the background on the
  right is the Caretaker (JSchm) with a German spiked helmet.}
\label{stage-medicine-balls}
\end{figure}

\item[\SL] Yes! ... Together with Haimo, and the others at DESY. 
Let me show you how these forces work!  For this experiment we shall need the help of
Herbi and Philip.

\hop	

{\it (HKD and AF come in on their inline skates, one of them carrying a 5\,kg medicine ball.)}

\hop 

\item[\SL] As you can see they are wearing inline skates. {\it(They lift their feet for all to
    see.)} Furthermore, Herbi has a 5\,kg medicine ball. This symbolizes the exchange particle,
  like the photon, or the gluon. They will now show us how they can exert a force on each other,
without touching  and instead using the medicine ball.

\hop

\EXP{Tossing medicine balls on inline skates. (See App.~\ref{app:medicineballs}, and Fig\,\ref{stage-medicine-balls}.)}

\medskip

\item[\SL] As we have just seen, the medicine ball exchanges a force between those two, which
  pushes off these two physicists, just like the two positive charges repel each other!  The
  gluon inside a proton does the opposite, it makes the quarks stick together.  The strong force
  is essential to hold the quarks inside the proton and the neutron together. So they don't fall apart. It is also
  essential to hold all the protons and neutrons inside the nucleus together.  Unfortunately
  in our analogy, we can not do an attractive force.

\hop

\item[\Hh] Surprisingly it is this interaction of the quarks and the gluons which is responsible for the main part of
  the proton mass!  That is about all we can teach you about this....  We still do not
  have a full understanding of how this all works. It is one for the theorists, so it could take
  forever!  But... since you are here... have you ever been inside a big collider?

\hop

\item[\B] No, can we go there? That would be really cool.

\hop

\item[\Hh] Well, no, but I can show you what it feels like to be accelerated to a velocity close
  to the speed of light, just like the particles inside the accelerator. Lorenzo, sit down on this bicycle.

\hop

\item[\B] Bicycle?  I mean I am pretty fit {\it(Flexes his non-existant muscles.)}, but a bicycle
  at the speed of light?  Do I need to swallow some drugs first?

\hop

\item[\Hh] No, no worries, a few espressi are enough. Just go on! 
	At the bottom of the screen you will see, at what speed you are driving!

\hop

\EXP{Relativistic bicycle (see App.\,\ref{app:rel-bicycle}, and Fig.\,\ref{pic:relativistic-bicycle}.)}


\begin{figure}[h]
  \center{\includegraphics[width=0.55\textwidth]{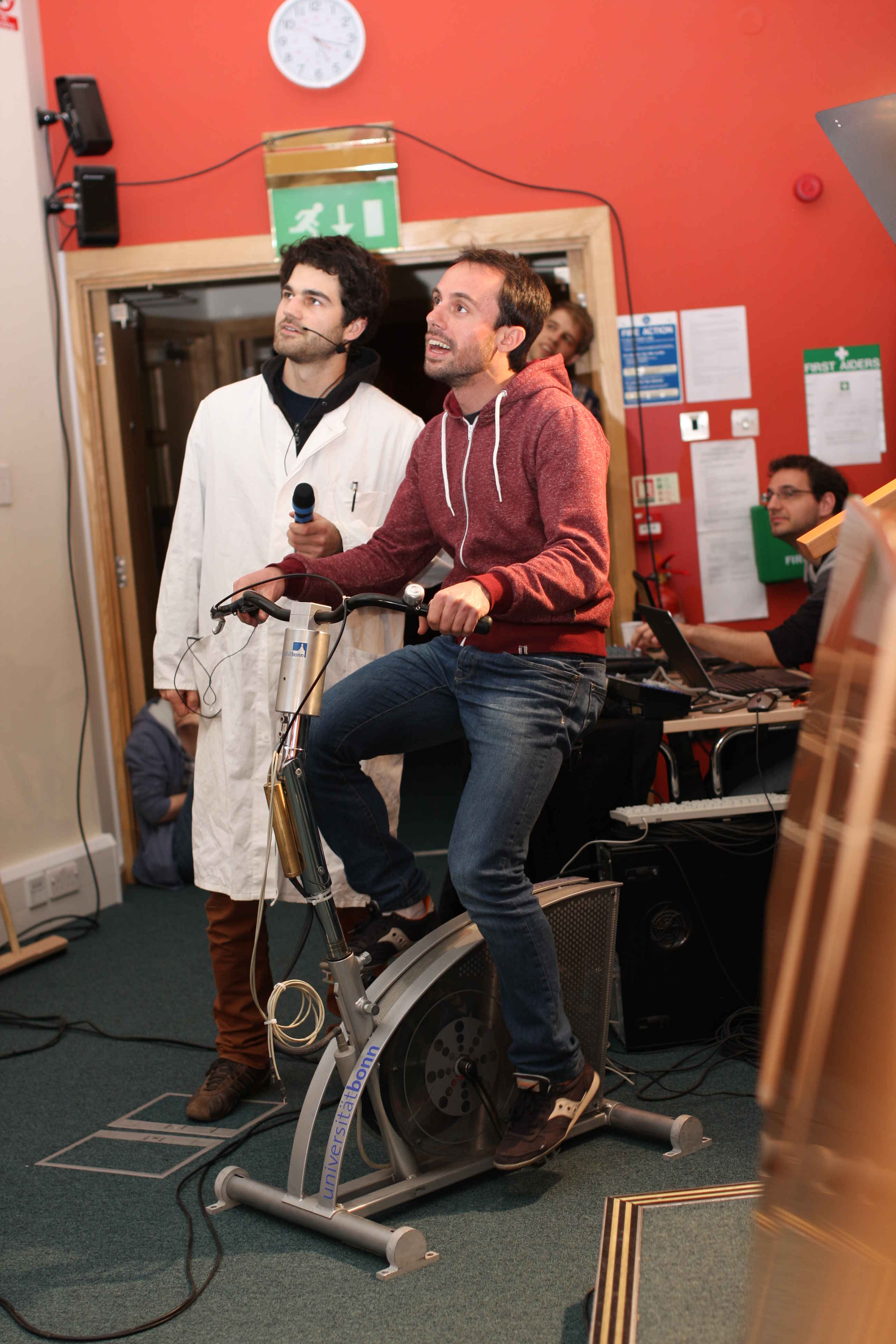}}
  \caption{LU on the relativistic bicycle, for details see \ref{app:rel-bicycle}, during rehearsals at Oxford University. Next to 
  him is JM. In the back is MBe at the table where the lighting, sound, music and the projection are controlled.}
  \label{pic:relativistic-bicycle}
\end{figure}

\item[\B] I have a fixie at home {\it(Swells his small breast.)} ... {\it (hushed voice)} it's actually illegal, but kind of fun. {\it(Starts pedaling.)} Whoa! Kind of psychedelic! Help!

\hop

\item[\Hh] That's what Einstein told us about how the world works close to the speed of light. 
	The faster you go, the more the world stretches around you!

\hop

\item[\B] Wow.. that was quite something! We have learned a lot about the strong force here, and now we know where most of the proton mass comes from....
	But we still don't know where the quarks themselves get their mass from!

\hop

{\it (Caretaker claps his hands. Sau Lan Wu and Haimo Zobernig freeze.)}

\item[\C] At this stage of knowledge and in order to find the answer to your initial question, you need to return to the present.

{\it(Maike and Lorenzo do the umbrella thing.)}

\end{itemize}

\subsection{CERN, Geneva, Switzerland}

{\it (A big dark hall, with a large photo of the Atlas detector on the back wall. The spot is on the Caretaker, Maike and 
Lorenzo.)}

\vspace{0.5cm}

\begin{itemize}

\item[\C] Here we are at CERN, the center of the particle universe. This is where Carlo Rubbia,
  from Pisa, discovered the $W$--boson, for which he got the Nobel Prize in 1984.  The time is
  2012. We are in the LHC tunnel, the tunnel of the largest machine ever built.  I hope, these two
  ``scientists'' can help you with your questions about the quark mass. Good luck! {\it(The Caretaker 
  claps his hands twice, the stage is now fully lit, the scientists come to life. The Caretaker takes a folding 
  rule out of his overcoat pocket and measures something.)}

\hop

\item[\gs]  Huh? Did you see that? I have no clue how {\it they} {\it(Points at them.)}  got down here but since they are  
here \dots

\hop

\item[\bs] \dots we can use their bodies to conduct some very important studies on the
  influence of deadly radiation!

\hop

\item[\gs] No! We can't test radiation on them.  We would {\bf never} do that.. at CERN. In fact, it is
  impossible to power up the accelerator if anybody is down here.

\hop

\item[\bs] ({\it To Good Sceintist, asking.}) Feed them to my pet black holes?

\hop

\item[\gs] No! No the courts have decided, we do not  make black holes at CERN!!

\hop

\item[\bs] ({\it To Good Scientist, pleadingly.}) But then can we at least test the deathly laser turrets on them?

\hop

\item[\gs] ({\it to \bss}) No!!! Cut it out! You are embarrassing us. There aren't any laser
  turrets either. 
 
\medskip

\item[\B] Hello! Excuse me. Helloo! Sorry, this is Maike and I am Lorenzo. We are from Bonn.

\medskip

\item[\gs] Oh from Bonn, the home of the physics Nobel Laureate Wolfgang Paul.  Welcome. Can I help you with any
  questions you may have about CERN and the LHC?

\hop

\item[\A] Wow! That would be great!
  
\item[\bs] CERN, the center of the universe. Where we created the biggest man-made {\it(Looking at Maike.)} or even
woman-made!, bang! 

\item[\gs] ({\it Rolls eyes}) Right. So, here we are in the LHC tunnel. The LHC was built by more
  than 10,000 scientists and engineers from over 100 countries. (Important large contributions
  were made here in Oxford/Padua.) The LHC tunnel is a circle with 27\,km circumference. It is about
  100\,meters under ground.  Let me show you a film.

\hop

\EXP{LHC Film}

\hop

{\it (Caretaker notices the glass of wine and examines it in detail.)}

\item[\B] And what is so special about this LHC accelerator?

\hop

\item[\bs] In the LHC we accelerate protons to very very very very very very very 
high energies, actually the highest particle energies ever achieved by man or {\it (looks at Maike)} woman!

\hop

\item[\A] And how much is that very very very very\dots?

\hop

\item[\bs] very very very. \\

Consider this lovely experiment. This is a special kind of transformer that has 
 many many many many many coil windings on the upper side
  but only a few on the lower side. Let's see what it can do:

\hop

\EXP{Tesla Transformer (see App.~\ref{app:teslacoil}, and Figs.\,\ref{img:tesla-1}, \ref{img:tesla-4}.)}

\begin{figure}[ht]
  \center{\includegraphics[width=0.85\textwidth]{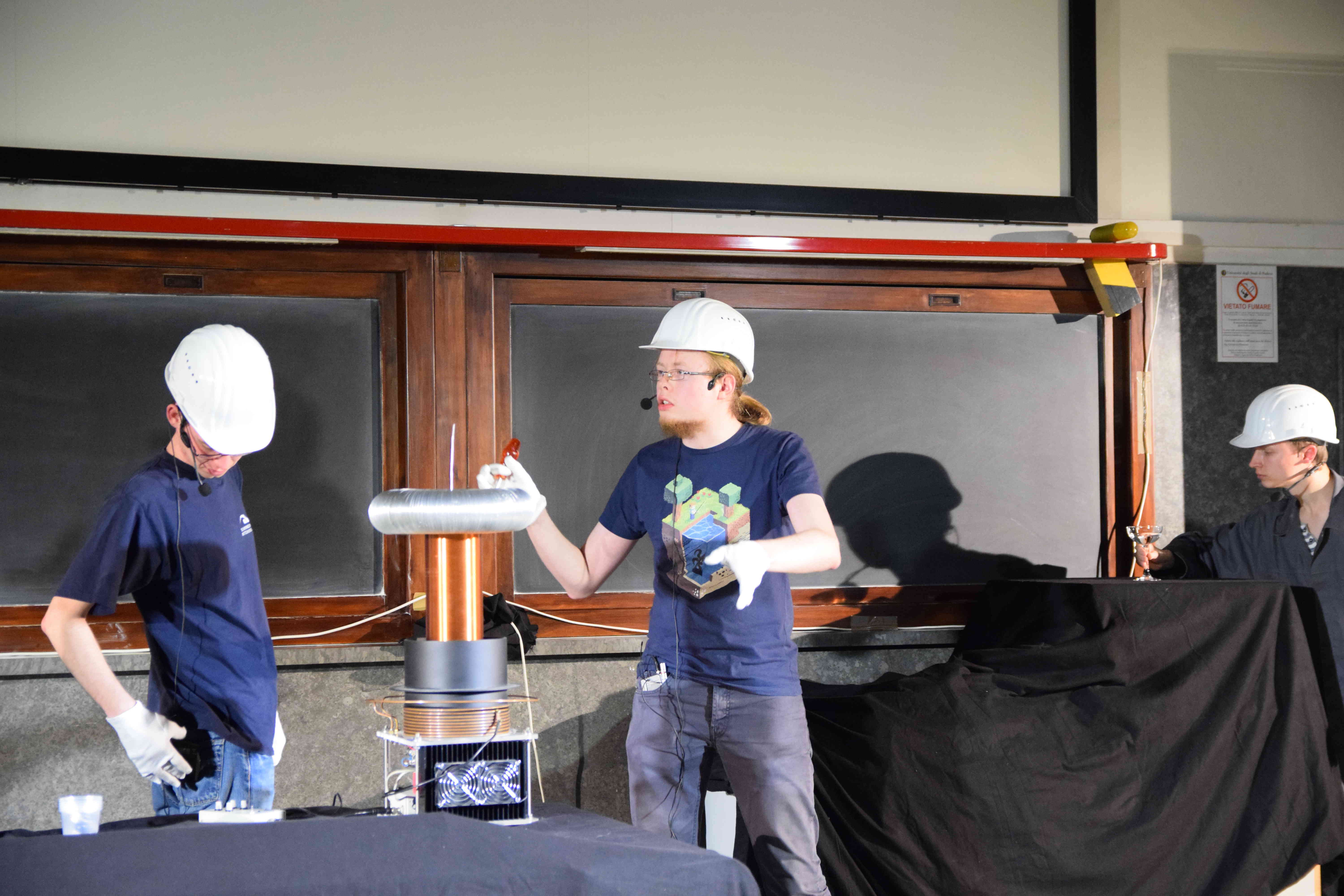}}
  \caption{TL and CSch as the two CERN scientists in the Padua show. Between them
  our Tesla coil. To the right is the Caretaker (JSchm) placing the glass of water which is needed later.}
  \label{img:tesla-1}
\end{figure}

\hop

\item[\gs] Those lightning bolts were discharges of about 200,000 volts. At the LHC we have
  energies which correspond to a potential of 4 trillion or 4 million million volts!

\hop

\begin{figure}[htb]
  \center{\includegraphics[width=0.55\textwidth]{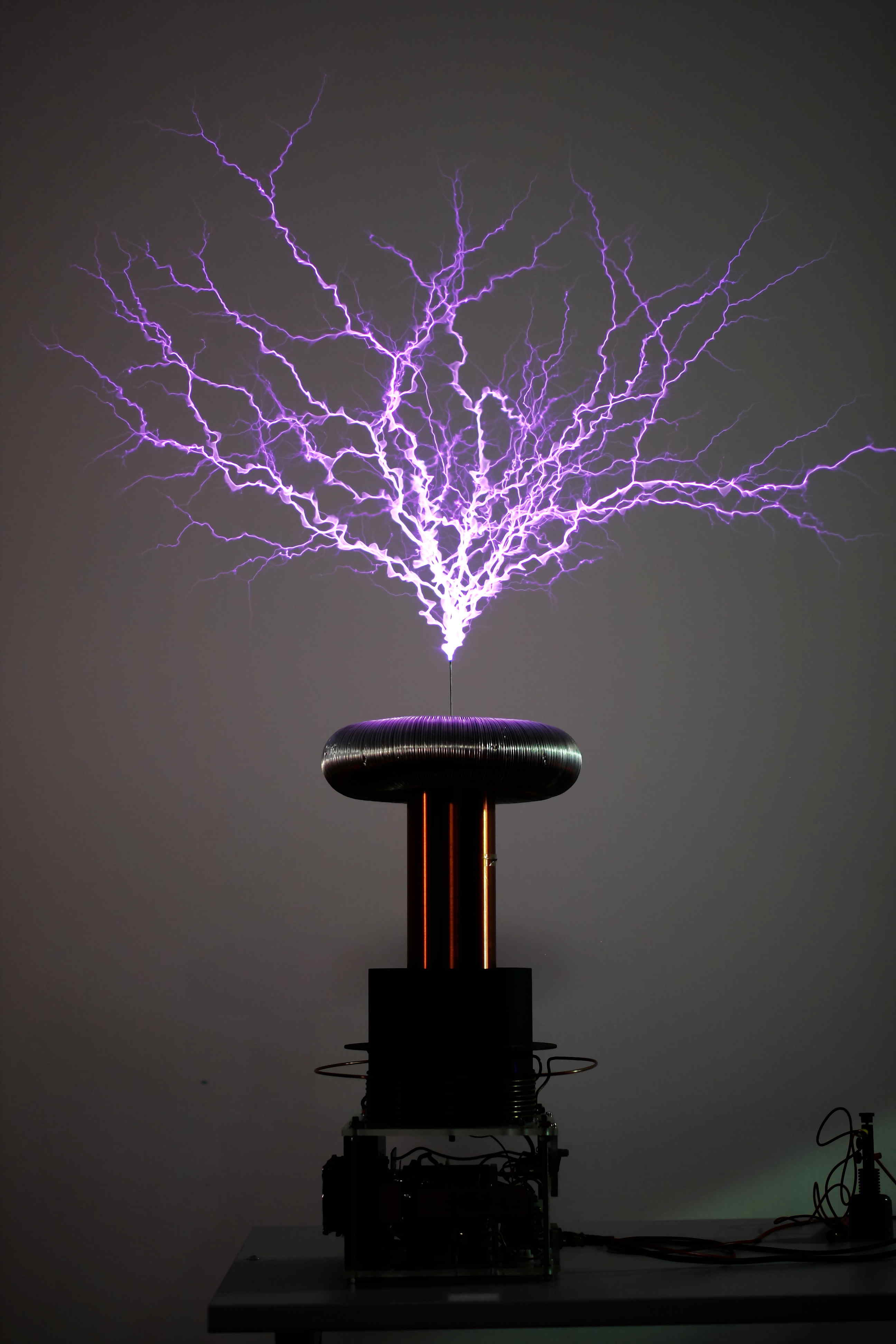} }
  \caption{Tesla coil during discharge.}
\label{img:tesla-4}
\end{figure}

\item[\bs] And as you saw in the film, we accelerate protons to these energies going around
  clockwise and also another set of protons going around counter-clockwise.

\hop

\item[\B] This seems pretty silly, who cares which way they go around?

\hop

\item[\gs] Well let me show you this wonderful historic experiment we borrowed from our colleagues
  at DESY, in Germany.

\hop

\EXP{Vacuum Cannon exhibited}

\hop

{\it (Caretaker starts playing with two balls, throwing/catching/juggling.)}

\item[\B] Ahh, we already know that. 

\hop

\item[\bs] Oh damn! I was so hoping we could do this again\dots

\hop

\item[\gs] Anyway, let me use it for the argument I wish to make. Here you accelerate the
  particle and shoot it at this target. We call it a fixed target experiment, because the target,
  the quark, does not move. The target is big and massive and you are sure to hit something, and
  you are also very likely to have an interesting reaction, as you saw.

\hop

\item[\B]  That makes sense and seems very practical.

\hop

\item[\bs] But as you maybe noticed, almost all the Quark flew forward.  A large part of the
  energy of the accelerated particle can not be used to ``{\bf bang}"
  ... part of the energy is required to maintain the forward motion.

\hop

\item[\A] Hmm, so what if you bang two moving things together? {\it(Bangs her fists
    together.)} 

\hop

\item[\bs] Yes, excellent! That is exactly what we do here at CERN. This way there is no net motion. I can
  convert the full energy into something new, like a big fat owl.

\begin{figure}[h!]
	\center{\includegraphics[width=0.85\textwidth]{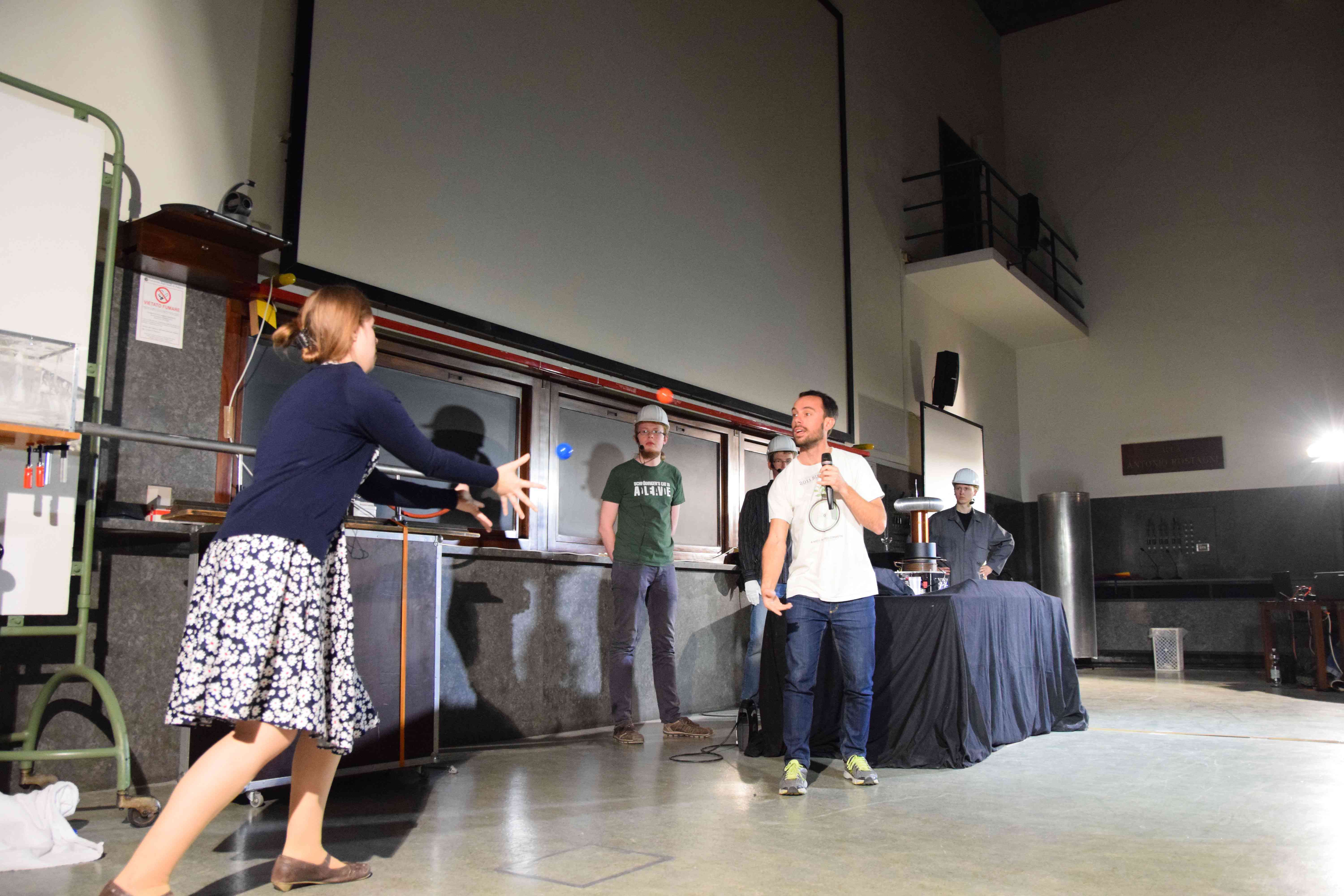}}
	\caption{First step: Hit single balls (protons) onto each other. On the left MH, on the right LU. Between 
	them further back CSch further back the Caretaker (JSchm) with  a hard hat.}
	\label{Fig:TwoProtons}
\end{figure}

\hop

\item[\B] Ok, that's really cool. So if we take these two balls {\it(Points at the Caretaker,
    who brings him the balls.)} and bring them up to super duper speeds and then ... Can
  we maybe try it out ... right here?
      
\hop

\item[\gs] Sure.

\hop

\item[\bs] Let them bang!

\hop

\EXP{Balls I (see App.~\ref{app:ballcollider}, and Fig.\,\ref{Fig:TwoProtons}.)} 

\hop

{\it(\Aa\ and \Bb\ throw a ball each. 
Try it twice.})

\hop

\item[\B] Hmm, that didn't work so well. How do you make sure that you have collisions? Do
  you keep shooting until they hopefully actually hit each other?

\hop

\item[\gs] Well we are a bit smarter here at CERN than you. We don't send individual protons
  around the ring. We use bunches.

 {\it(The Caretaker pulls out two large boxes full of colored plastic balls, gives one to \Aa\ and one
    to \Bb.)}
    
\hop

\EXP{Balls II (See App.~\ref{app:ballcollider}, and Fig.\,\ref{Fig:ManyProtons}.)} 

\hop

{\it ({\bf
    ``splat!"} {\bf ``kaboom!"}. Both as
    comic book pop art on screen.)} 

\begin{figure}[h!]
	\center{\includegraphics[width=0.85\textwidth]{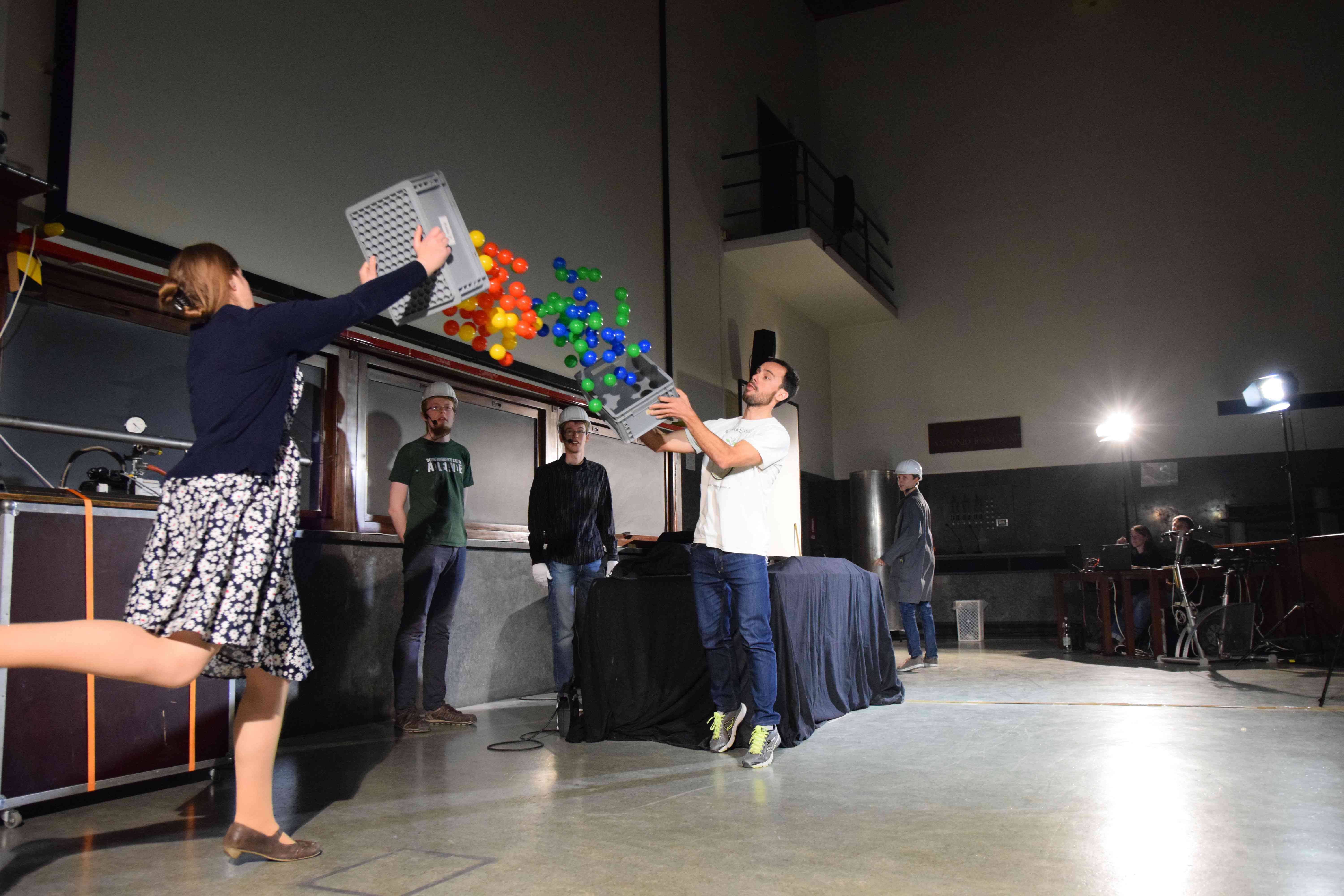}}
	\caption{Second step: Hit two boxes of balls (bunches of protons) onto each other.
	To the left MH, to the right LU. Between them the two CERN scientists (CSch, TL).
	In the far background the Caretaker (JSchm).}
	\label{Fig:ManyProtons}
\end{figure}

\hop

{\it (The Caretaker sweeps the balls away.)}

\item[\A] That worked much better. I even saw some collisions! {\it(Stops, turns to the
    audience, whispers.)} did you?

\hop

\item[\B] Ok, so what do you use the LHC for?

\hop

\item[\bs] Haven't you heard: we have made {\bf T H E\ \ H I G G S\ \ B O S O N}. We
  discovered it in July, 2012, here at CERN, with the LHC!

\hop

\item[\B] What? You have discovered the Higgs boson?

\hop 

\item[\bs] Exactly.

\hop

\item[\B] Wow, the Higgs Boson, Maike, the Higgs boson. That is amazing, Ye-haw! {\it(Music! dances a jig with \Aa . 
The caretaker and the CERN scientists join in.)}    The Higgs boson! \\
{\it(Music suddenly stops. Everybody freezes. \Bb\  calms
    down, stops, thoughtful, and slowly.)} \\ But what actually {\it is} the Higgs boson?

\hop

\item[\bs] I was hoping you might ask that. {\it(Triumphantly:)} So the Higgs boson is a
  quantum manifestation of the Higgs field!

{\it (The Caretaker continues sweeping the balls.)}

\hop

\item[\B] A quantum what?

\hop

\item[\gs] Yes, do you remember the photon?

\hop

\item[\B] Oh yes, we ``heard'' the photons click, back at DESY.

\hop

\item[\gs] The photon is the quantum, the smallest packet of the electromagnetic
  field, which is a vector field.

\hop

\item[\B] Oooookaaay.

\hop

\item[\gs] Similarly there is a Higgs field and the Higgs boson, which we observed, is the
  quantum, the smallest packet of the Higgs field. The Higgs however is a scalar field.

\hop

\item[\B] Quantum? Scalar?? Bosons??? Oh boy! I need some more gummy bears...

\hop

\begin{figure}[h!]
	\center{\includegraphics[width=0.6\textwidth]{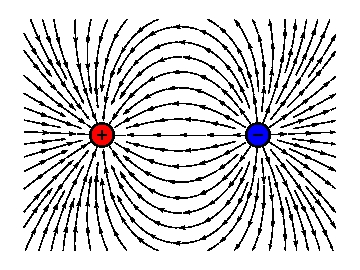}}
	\caption{Graphical depiction of the electric field between two opposite charges. Image made using Mathematica.}
		\label{Fig:paper-electric-field-5}
\end{figure}

{\it (The Caretaker pulls a Haribo bag from his pocket and gives it to \Bb, who starts eating.)}

\hop

\item[\gs] Let me begin by explaining what a field is. Let's start with an electric field. Which is a vector field. On the
  screen you see two electric charges, one positive, one negative, see Fig.\,\ref{Fig:paper-electric-field-5}.
  Between them an electric field forms, represented by
  the lines with the arrows. \\ {\it(To \Bb)} Sorry, would you mind to stop eating 
  and actually listen? Thank you. \\ So, if I put something that is charged inside the field, it gets a
  property which has not only a value, but also a direction. Hmm, why don't we demonstrate that here with you?
   Please stand at these two sides of the stage.
  
\begin{figure}[h!]
	\center{\includegraphics[width=0.7\textwidth]{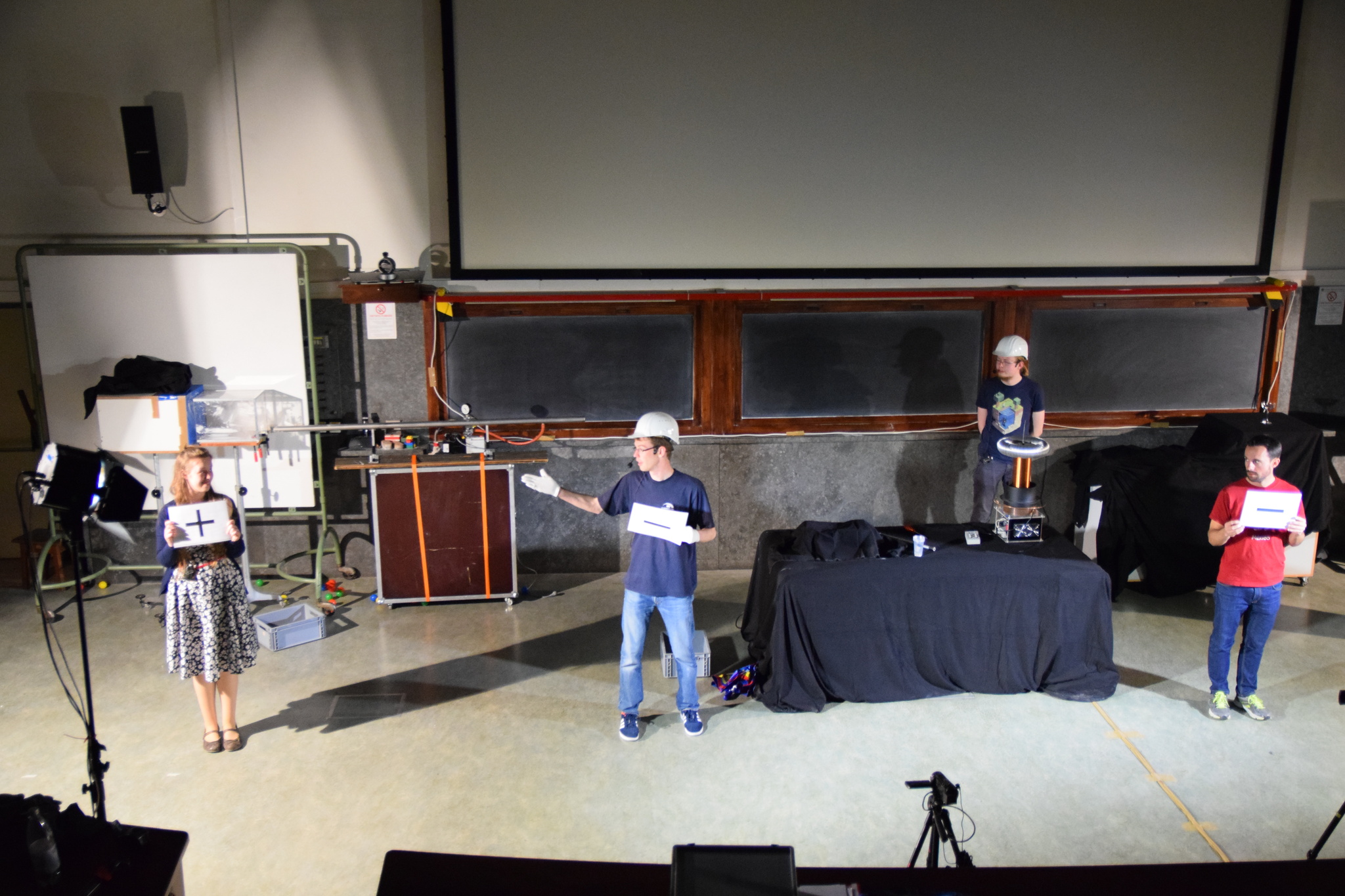}}
	\caption{Setup for the electric field demonstration. On the left MH with the minus sign. On the right LU
	with the plus sign. In the middle TL with a minus sign. At the back, at the blackboard, CSch.}
		\label{Fig:paper-electric-field-1}
\end{figure}

  \textit{(The Caretaker and the \bs hand charges to \Aa\  and \Bb.)}

\begin{figure}[h!]
  \center{\includegraphics[width=0.42\textwidth]{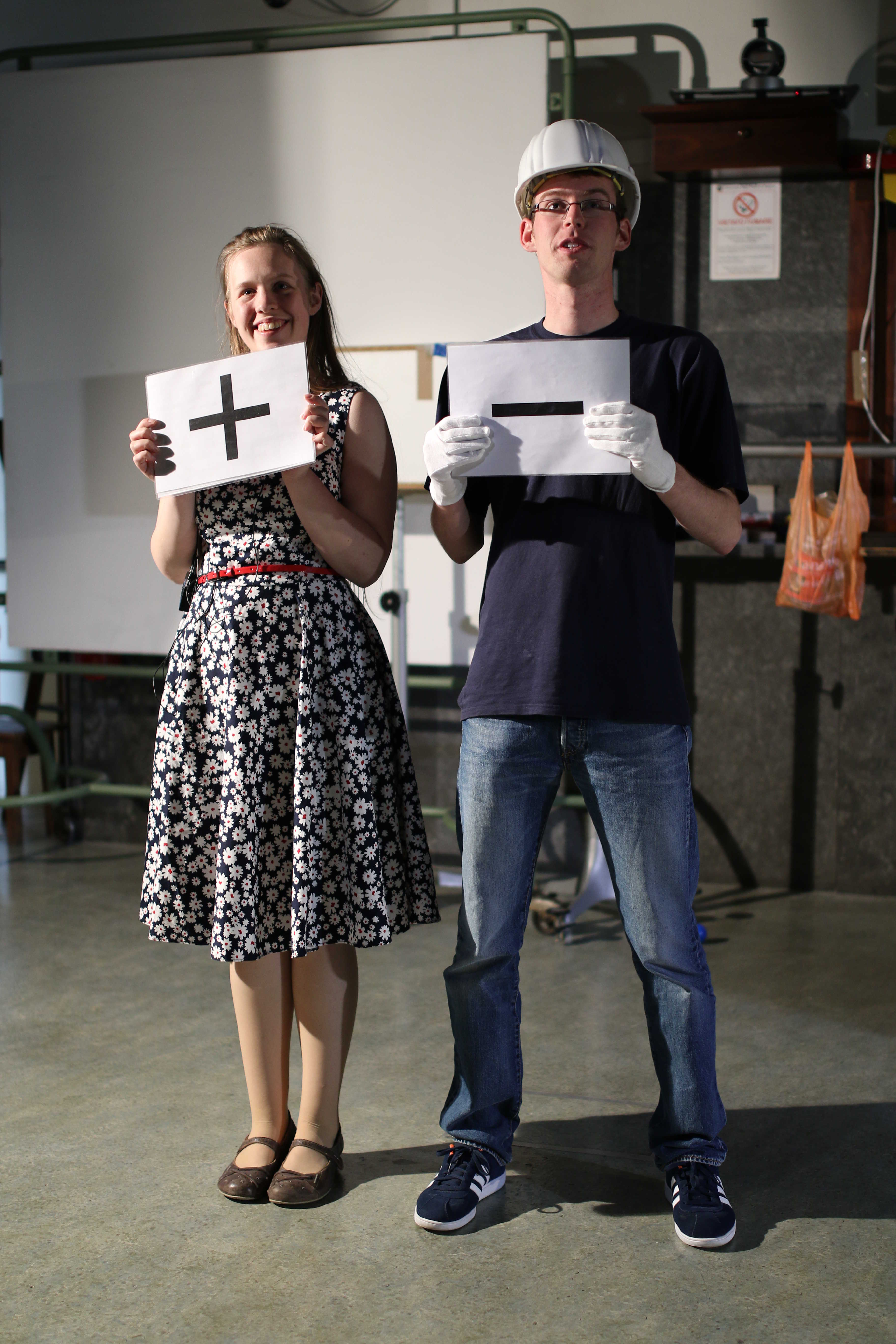}\; \; \; \; \includegraphics[width=0.42\textwidth]{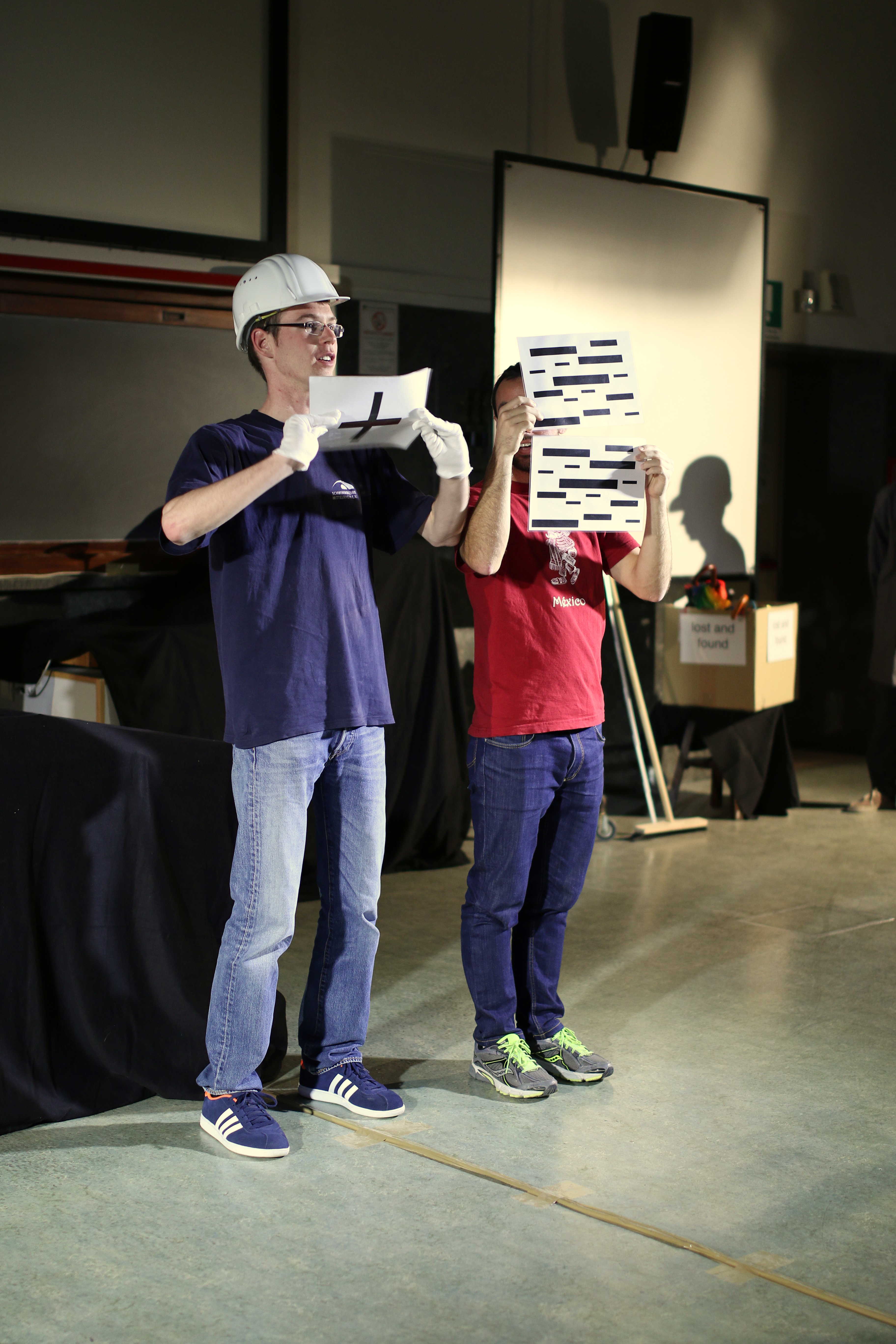}}
  \caption{In the left photo MH with a single positive charge. Unseen to the right is LU with the corresponding 
  single positive charge. TL as one of the CERN scientist places himself with a probe negative charge inside the resulting electric field. In the right photo Lorenzo has dramatically increased his negative charge
  and thus the resulting electric field.
    }
  \label{pic:paper-electric-field-2}
\end{figure}

\hop

\EXP{Paper electric field and acceleration of one charge (see App.~\ref{app:paperelectric}, and Figs.\,\ref{Fig:paper-electric-field-1}, \ref{pic:paper-electric-field-2}.)}

\hop

\item[\gs] So now I am charged myself. I sit in the field you generate and feel attracted to
  \Aa. So I start moving in her direction. But since you are only very weakly charged the energy I
  get is low, so I move slowly.  Now, if I just flip my charge 
   I am attracted towards \Bb\ and I move that way, but my speed stays the same as before.

{\it(The Caretaker takes the gummi bears from \Bb\  and gives some to camera / technicians / audience / \dots)}

\item[\gs] Now if you double the charges you are holding, the field gets stronger. So my energy (or your
  attraction) is larger and I am moving faster. {\it(Runs once back and forth between \Aa\ and \Bb.)}

\item[\gs] And finally if you create a very strong field {\it(\Aa\ and \Bb\ hold up many many charges.)} I 
will be very fast! ({\it To audience.})  You'll have to watch carefully! One!--Two!--Three! {\it(On the count 
of Three flips charges back and forth once but does not move.)}

\hop

\item[\A] You are indeed very fast! So we have understood, that ... if a particle interacts with a vector
field it gets a property which has not only a value ... but also a direction, right?

\hop

\EXP{Scalar field analogy, weather map of Italy. (See App.\,\ref{app:scalar-field}, and Fig.\,\ref{img:italy-map}.)}

\begin{figure}[ht]
  \center{\includegraphics[width=0.45\textwidth]{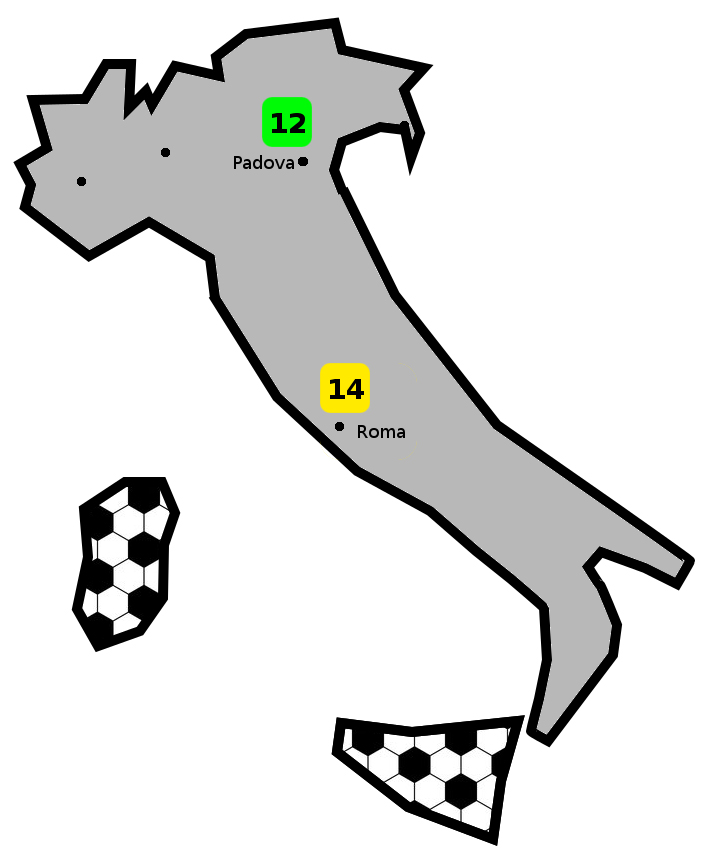}}
  \caption{A temperature map of Italy.}
  \label{img:italy-map}
\end{figure}

\hop

\item[\bs] Yes, exactly, but for a scalar field this is different. Let us consider the temperature, as a field. This is 
shown here on the screen with a map of Italy (Fig.\,\ref{img:italy-map}).  The temperature is a scalar field. If I 
now take this glass of wine ... {\it (Looks more closely at the glass.)} Oh, ... it is only water 
{\it (Looks up at the audience.)} has anybody seen Jesus?

\hop

No matter. If I put this glass in Padua, the water will take on the local temperature of 12 degrees. 
A property, but no direction! \\
If I put the glass instead in Rome, it will take on the local temperature of 14 degrees, again a property, but no direction.\\

Salute. {\it(Drinks from the glass.)}

\hop

\item[\B] Ok, ok, ok, this is all a bit much for me. What does this have to do with the Higgs boson?

\hop

{\it (The Caretaker tries to compare the height of \Aa\ and \Bb\ with the folding rule.)}

\item[\bs] Let me take this blue cloth. It shall symbolize the Higgs field. I shall also need the help of two assistants.

{\it (Helpers get on stage.)}

 \hop
 
\item[\bs] Nature has decided that the Higgs field should {\bf always be everywhere}. So imagine
  this cloth all through space and {\bf eternally} billowing and fluctuating.

  \hop

\EXP{Higgs Field and black cloth (see App.~\ref{app:higgscloth}, and Fig.\,\ref{fig:LightParticle}.)}

\begin{figure}[h]
	\center{\includegraphics[width=0.85\textwidth]{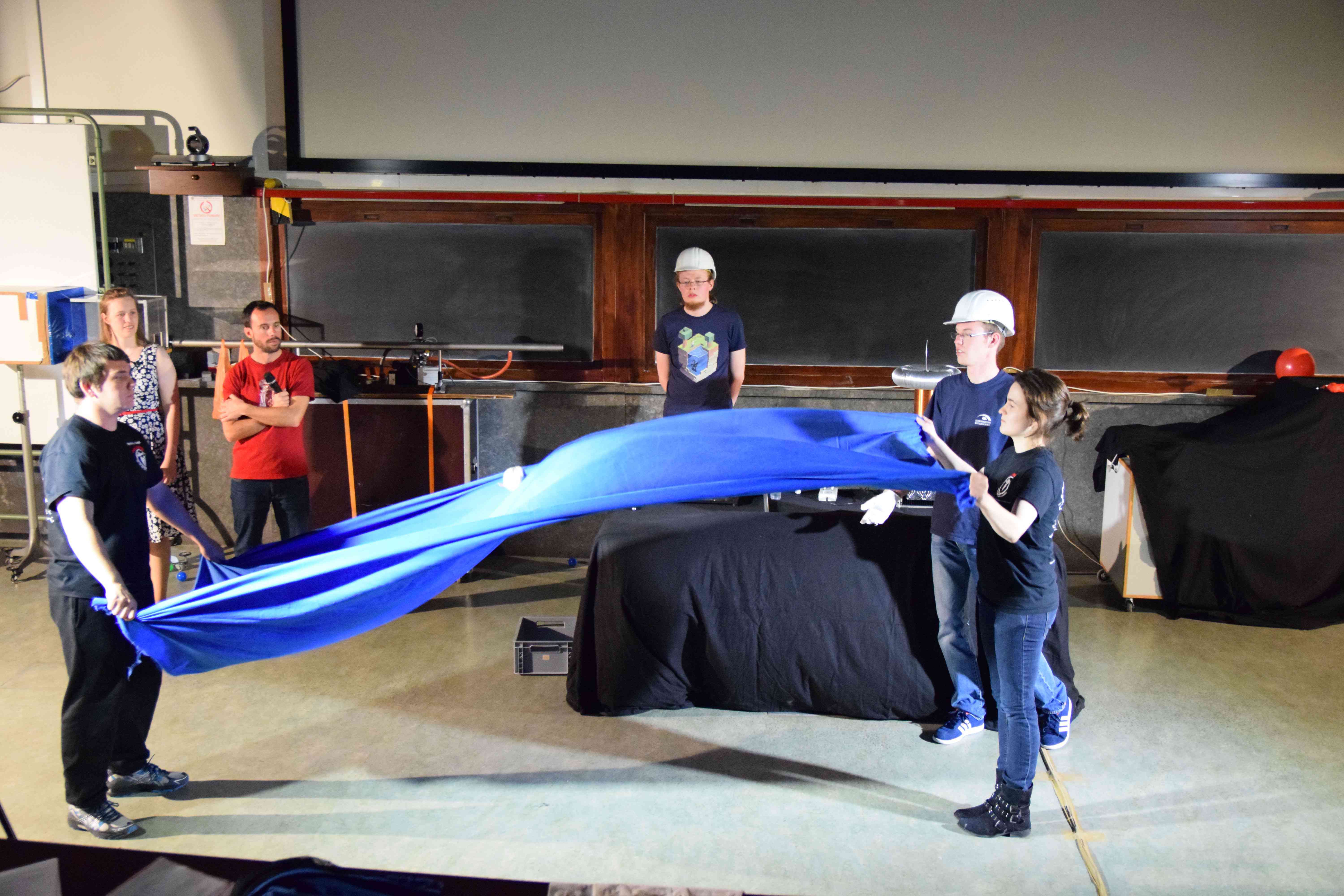}}
	\caption{Light particle (small plush sheep) moving through the Higgs field.}
	\label{fig:LightParticle}
\end{figure}

\hop

\item[\bs] As particles move through this background scalar(!) Higgs field they also get a property,
and that property is {\it (Wait for it, thumps his belly)} MASS!

\hop

\item[\B] Mass? We have been wondering about mass on our whole journey. So it is related
  to the Higgs field and not to red gummy bears! {\it(Looks at caretaker menacingly.)}

\hop

\item[\gs] Yes, different particles interact differently with the Higgs field. And the property
  they get from interacting with the Higgs field is universal: everywhere and eternal. So it is
  its own! {\it(The two people are still making waves with the blue cloth.)} Here, this small sheep
  represents an electron, it has a small mass, it interacts weakly with the Higgs field. ... This
  stuffed owl, however, represents a massive top quark this interacts much much much much 
   stronger with the Higgs field and is thus much heavier!

\hop

\item[\B] So the particles get their different masses by how they interact with the Higgs
  field?

\hop

\item[\gs] Yes, exactly! Just like here:

\hop

{\it (The Caretaker lays a path on the floor with the folding rule and follows it with large steps.)}

\EXP{Eddy current experiment as a Higgs field analogy  (see App.~\ref{app:ewaldeddy})}

\hop

\item[\bs] I have here a simple metal bolt. I also have a small magnet, which you can see, holds up this screwdriver.
Finally I have a hollow aluminum tube ... which is NOT magnetic\\  {\it(Holds magnet against tube, it doesn't stick.)} \\
This aluminum tube shall represent the Higgs field.\\
{\it (First drops metal bolt through the tube, into a plastic cup below.)}\\
 So the soulless metal bolt flies through uninhibited: this represents the massless
  photon. It does not interact with the aluminum tube, the Higgs field. \\
  {\it (Next drops the magnet through the aluminum tube, it falls very very slowly.)}\\
  The magnet however interacts with the metal tube, creating a current in the
  tube, this causes the magnet to fall slowly, this is then a massive particle.

\hop

\item[\A] OK, but now what about the Higgs particle?

\hop

\item[\bs] Recall: the Higgs particle is the excitation of the Higgs field. So we must excite the Higgs field, and if we 
do this strongly enough, well then we should produce a Higgs boson.

\hop

\item[\B] Can you do that here?

\hop

\item[\bs] Yes, I shall excite the field ... \\ 
{\it(Pulls out two heavy hammers.)} \\
with these hammers. But I will also need a lot
of energy. So I will need all of your help.

\hop

\item[\B] Maybe we should count together!

\hop

\item[\bs] {\it(With \Bb\ and the audience.)} One!--Two!!--Three!!!

\hop

\EXP{Hammer to cloth, produce a green frog, as the Higgs boson.}

\hop

\item[\A] Wow, so we have {\bf finally} understood the mystery of mass and the Higgs boson. Yeah! Time to head
  home, no?

\hop

\item[\B] Yes! No, Wait!  One last question. So how did you do {\bf that} here at the LHC?

\hop

\item[\gs] Well, you saw! With this giant collider we have smashed together protons, right here,
  in this spot. On several very very very very rare occasions we then produced a nice fat Higgs
  boson. The Higgs boson lives for a short bit and then  {\bf ``POOF"}
  decays. And we can see the decay particles....{\it(Rotating yellow lights go
    on, a warning horn sounds.)} Oh no, get out! Get out of the tunnel.  The collider has been
  turned on! 

\hop

\item[\bs] The protons are coming!! Everyone get out. Here they come...

\hop

\item[\gs] {\it(To audience, as he is about to run out.)} Sorry, but {\bf you} have to stay...

\hop

\EXP{Collider with people and LEDs (see App.~\ref{app:lightsuits})}

\hop

{\it (The room goes dark. Two people in white LED suits come from either side of the lecture hall. They
  pass in the middle and at either end disappear again. On the next pass they collide in the
  middle. The white lights extinguish. Another person in a green LED suit rises and moves slowly
  before also his light extinguishes and two people in blue LED suits run up the two aisles of the
  auditorium.)}

\hop

\item[\C] {\it (Spot light on Caretaker.)} Wow, that was amazing. I would like to see that again. How 'bout you? Maybe we can
  rewind that. \\ 
  
  {\it (Hear the sound of a tape rewinding. Caretaker ``rewinds`` with his hand. The blue LED people (photons) return back
    down the aisles. Their lights extinguish as they meet and the green LED person lights up,
    slowly moving backwards. (You can hear the beeping sound of a van backing up.) Then he extinguishes
    and the two white LED people reverse to the sides of the stage. )}

\hop

\item[\C] Okay, but now a bit slower. \\
    
    {\it (The Caretaker claps his hands and
    slowly the collision again proceeds forwards, this time the Caretaker explains what is happening.)}
    
\hop

\item[\C] Here come our two protons in shining white hurtling down the LHC tunnel. In the center, in the middle
of the detector they collide. On this rare occasion, they produce a big fat green Higgs boson. The Higgs lives 
only a very short time and then decays to two blue photons. And these fly through our detector and are thus observed.
This is how the Higgs boson was discovered at the LHC.

\hop

\item[\C] So, our two heros have found the Higgs boson and have completed their quest. We hope you
  have enjoyed this journey ... I have. Now I have finished MY work, and I guess it is time to
  call it a day and head home. {\it (Picks up a black umbrella. Turns around, opens the umbrella
    and puts it over his shoulder. In gold writing on umbrella ``The End'', see Fig.\,\ref{fig:TheEnd}. Slowly spins 
    the umbrella
    as the spot light shrinks down to just the umbrella ... then extinguishes.}

\begin{figure}[h!]
	\center{\includegraphics[width=0.85\textwidth]{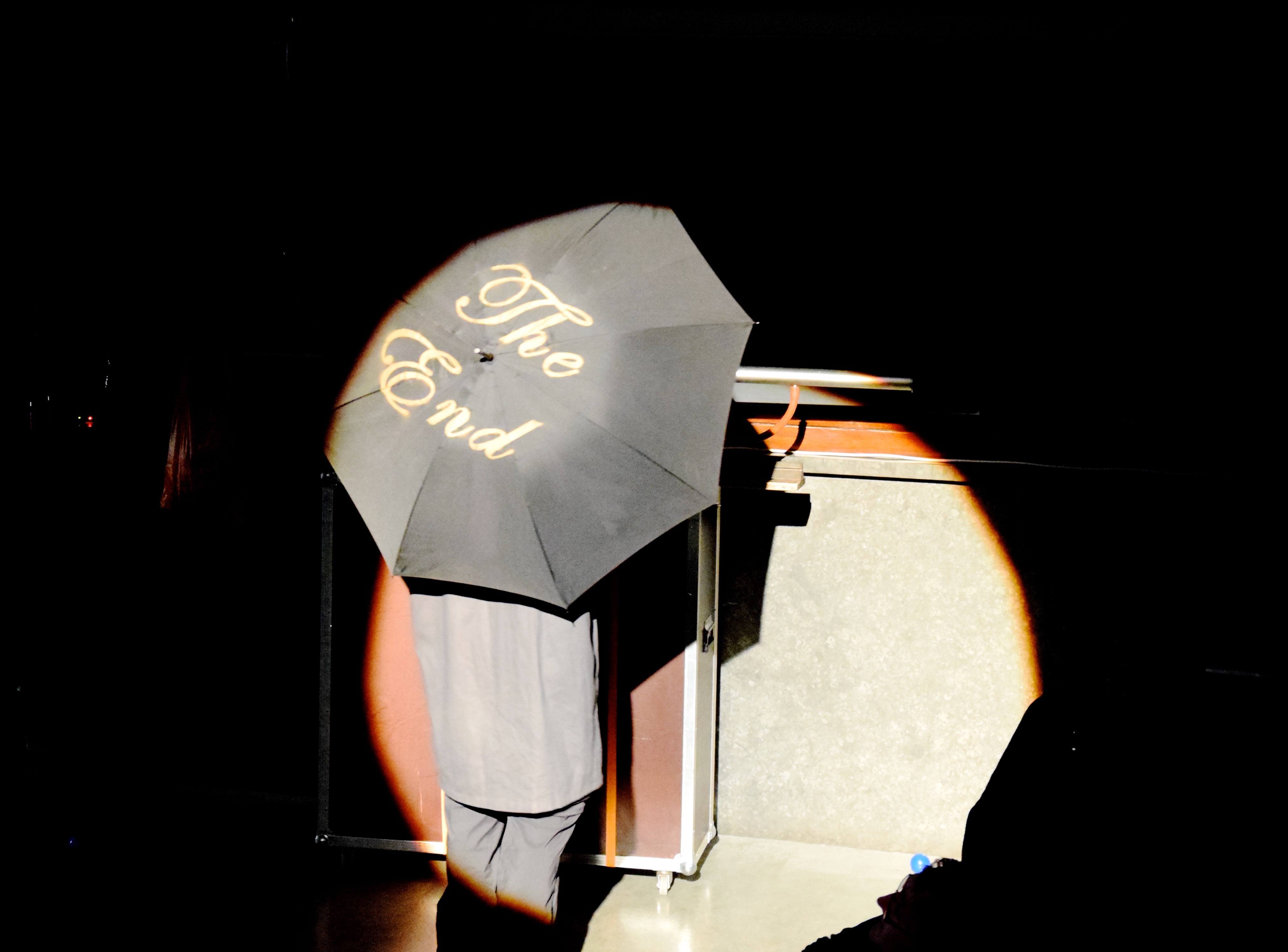}}
	\caption{Closing scene with the Caretaker.}
	\label{fig:TheEnd}
\end{figure}

\end{itemize}

\vspace{1cm}

\centerline{\bf \large THE END}

\vspace{1cm}


\section{Conclusion}
\label{sec:conclusion}
We have presented the details of a physics show we have developed on modern particle physics, including
the Higgs boson. The show includes 28 live experiments performed and explained onstage.

\section*{Acknowledgements} 

We thank the SFB-TR CRC 110 \textit{Symmetries and the Emergence of Structure in QCD} for extensive 
financial support, which made the trips to Oxford and London (2014) as well as to Padua and Trieste (2015) 
possible. The SFB furthermore supported a Physikshow trip to DESY, Hamburg in Spring, 2014. In the 
last fourteen years we have also received extensive financial and moral support from the Physikalisches 
Institut der Universit\"at Bonn, including many work hours from Michael Kortmann and Sascha Heinz, as well 
as the use of the department workshops. We have also received over many years financial support from the 
other institutes in the Fachgruppe Physik und Astronomie der Universit\"at Bonn: the HISKP, the IAP and the 
AIfA.

We would like to thank Katinka Ballmann, Doug Cowen, and Sidney Perkowitz for going over the entire manuscript.

From the beginning and over many years Patricia Z\"undorf has provided logistical support on all levels as
well as never ending encouragement to the Bonn Physikshow. Until recently she was supported by Dagmar
Fa{\ss}bender, who is now retired. For several years now, and in particular for this show and our trips to 
Great Britain and Italy we have been extensively supported by Petra Wei{\ss}, who also handled most of the 
financial issues.

This show is based on two earlier Bonn particle physics shows, the first performed in 2004 and the second performed
from 2008 to 2010. We have benefitted greatly from the experience gained in those shows and have retained quite a 
few experiments. We would like to thank the members of those shows for the enjoyable collaboration and the invaluable 
input they have given: Timo Altfelde, Katinka Ballmann, Markus Bernhardt, Nicki Bornhauser, Sebastian Fuss, Mathieu 
Gentile, Stefan G\"ors, Peter Henseler, Marc Hofmann, Walter Honerbach, Markus J\"ungst, Alexander Karim, David Keitel, 
Michael Kobel, Claudia Kob\"oke, Anna-Lisa Kofahl, Peter Kofahl, Karsten Koop, Rebecca Koop, Na\"emi Leo, Christoph 
Luhn, Jessica Mende, Remer Meyer-Fennekohl, Nico M\"oser, Cornelia Monzel, Martin Niestroj, Tim Odenthal, Stefan Patzelt, Ludmila 
Piters-Hofmann, Marc Prinz, Jana Puschra, Christoph Rosenbaum, Iris Rottl\"ander, Melanie Schmitz, Jan Schumacher, Markus 
Schumacher, Duc Bao Ta, Sofia Terhalle, Tobias Troost, Andreas Valder,  Judith Wild, Karina Williams, and Daniela Wuttke.

In the 14 years of the ``standard" (non-particle physics) Physikshow over two hundred students have participated. It is 
not possible to mention all of them here.  I would just like to mention a few more beyond those listed above, who have 
been decisive for the development of the show: Jana B\"urgers, Robert Geisselbrecht, Dominick Klaes, Branislav 
Poletanovic, Andrea Raccanelli, Tobias Strehlau, and Alice Yeo.

For this show, Dagmar Fa{\ss}bender (Bonn University) sewed the LED-light suits. The wooden scattering 
experiment was designed by one of us (MKo) and built by W.\,Lenz and him. 

We thank Karina Kortmann for designing several of the Physikshow on-tour T-shirts.

We would like to thank Bonn University for their support over the years. Andreas Archut and the Presse-Abteilung 
have been extremely helpful in promoting the show. We thank the University for trusting us and getting us the 
best gig ever, at the Deutschlandfest on the Bonner M\"unsterplatz in front of over 10,000 people in 2011. And we 
are very grateful for receiving the Alumni Prize of Bonn University in 2006.

We would like to thank Andrea Niehaus and the crew at the Deutsche Museum, Bonn, Bad Godesberg, for 
initially believing in us and giving us our first opportunity outside of Bonn University. They have been 
supporting us over the last 10 years. We also thank them for hosting the premier of this show in German and in 
Germany. We thank Rainer M\"ahlmann from the Deutsche Museum, M\"unchen for many years of support and our
first invitation outside of Bonn. Thank you also for trusting us despite a chaotic last rehearsal and a sleepless
night.

We thank John Wheater for inviting us to Oxford University and Sian Owen (Oxford University) for her superb 
support and organization in the many months leading up to and then also during our stay in Oxford, in March, 
2014. We thank Daniela Treveri-Gennari, as well as Marc, Luca and Giulio Taggert for their culinary and 
leisure--time (Freizeit) support during our stay in Oxford.  We thank Jon Butterworth and Oli Usher (UCL London) 
for their extensive support leading up to and during our visit at UCL in London. In difficult circumstances they made 
things work. Thank you also to Anna and Jay Watson for their hospitality during our visit in London.

We thank Francesca Soramel and Fabio Zwirner (University of Padua) for the invitation to Padua and the 
support for our trip. We thank them for their wonderful hospitality and the great atmosphere in their lecture
hall. It has to be said: for us, this was our best show ever! We thank Giampaolo Mistura for all his local 
organizational efforts. We thank Sara Magrin and Roberto Temporin for all their technical support in the 
Padua lecture hall and their very warm hospitality. We thank the Baccolinis for regional support. Scarpone!

We thank Fernando Quevedo (ICTP Trieste) for inviting us to Trieste, the support for our trip, the warm 
hospitality and the housing right on the beach! We thank Joe Niemela and the local organization for their
support and for filling the large lecture hall!

We thank Volker Lannert for all the great photos he has taken at Physikshow events over the years and for
always making them available to us.

We thank Daniel Class and Don Lincoln for allowing us to use the image in Fig.\,\ref{exchange-force-boat}.

We thank Nathalie Burwick for composing the music, which accompanies the time machine during the shows.

\begin{appendix}

\section{Technical Details of Demonstration Experiments in the Show}
\label{sec:appendix}

In this appendix we summarize the details of the various experiments presented in the show. They
are given in the order they appear there.  Where possible, we include a brief history of the experiment, and 
its relevance to (particle) physics. We then describe the materials employed and give the technical 
specifications needed to build the respective apparatus. Next we describe the typical presentation in the 
 show. The context in the story line and the actual wording of the explanation during the show 
is given in the text in Section \ref{sec:the-play}. We conclude the description of each experiment with a 
discussion of relevant safety issues. 

Before discussing the individual experiments we give some more 
general pointers which are pertinent to all the experiments.

\subsection{Presentation technique}
\label{app:ring-schleuder}

In general when we present an experiment in the Bonn physics show, we first describe the setup in detail. It 
takes the audience some time to understand what they are seeing, and in most cases they can not appreciate
the results, unless they have some familiarity with the ``initial conditions"\!\!\!. Therefore, before performing the 
experiment, we guide them through the various parts of the apparatus and explain how they are connected. 
In the process we make sure not to give away the main effect, the punch line of the experiment, so-to-speak. 

\begin{figure}[t!]
 \includegraphics[width=0.50\textwidth, height=6.75cm]{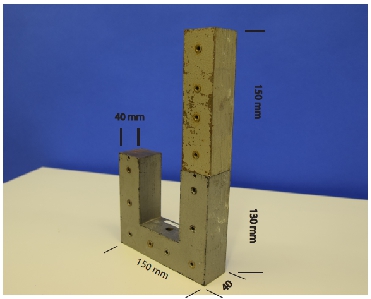}\hspace{0.3cm}\includegraphics[width=0.5\textwidth]{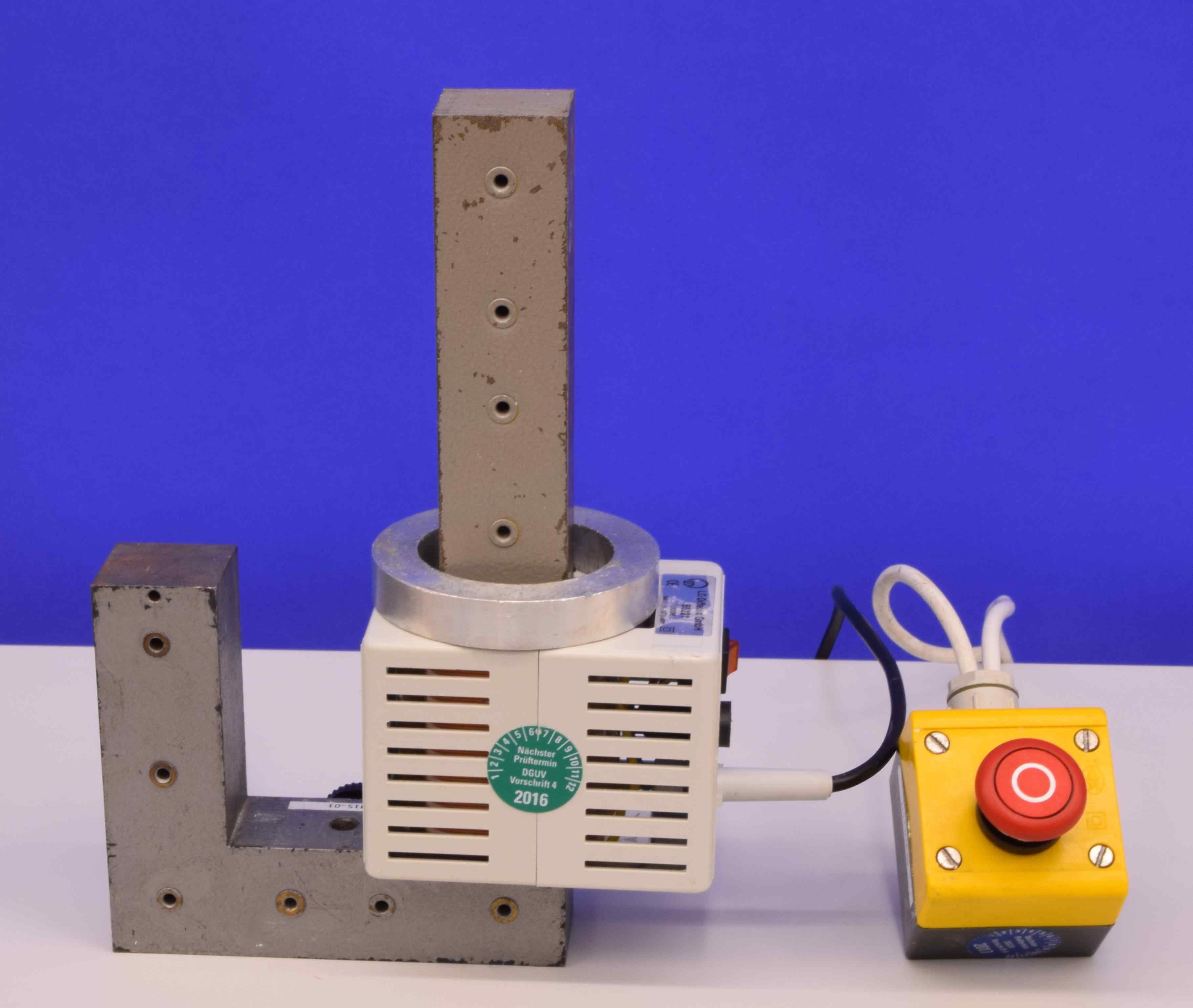}
	\caption{The jumping ring experiment, which is not part of this show. It is discussed for illustrative purposes.
	As can be seen on the left, we have a symmetric U-shaped iron yoke. 
	On one arm we place a 150\,mm extension. On this we place a coil with 500 windings. The coil is 
	connected to the mains power (240 V, 50 Hz) via the red button switch. We often have an audience 
	member hit the switch.
			}
	\label{fig:ring-schleuder}
\end{figure}

In order to make this a bit clearer, we consider an example-experiment, which we often perform. This is on Lenz's law, 
also known as the jumping ring. The setup is shown in Fig.\,\ref{fig:ring-schleuder}. Instead of just hitting the red 
button and watching the ring fly up, we first draw the audience's attention to the iron yoke, and then the coil, which 
is placed on the extension of the yoke.  It consists of 500 windings of copper wire; the maximum load is 2.5\,A. The 
coil is hooked up via the red button to the main voltage supply, 240\,V, 50 Hz, in Germany. Next, we place an 
aluminum ring (outer diameter: 90mm, inner diameter 60mm, 16mm thick) on the extended leg of the iron yoke. The 
dimensions of the iron yoke can be seen on the left in Fig.\,\ref{fig:ring-schleuder}. It consists of lamination steel, 
which suppresses eddy currents.

When explaining the setup, we do not mention that the ring could fly upwards, many people do not know this. 
The surprise is a main effect of this experiment. We then have an audience member, typically a child, come forward 
and on the count of three, the child hits the red button launching the ring upwards. Here we often play a sound effect 
of disappointment, as the ring does not fly very high, maybe about 1\,m. We then repeat the experiment with a ring 
which has been cooled for a long time in liquid nitrogen. Depending on hitting the right phase, this ring can fly 3 
meters into the air, significantly higher than for example in \cite{ring-schleuder-youtube}. 

If there is additional time, it is also instructive to use a ring which has a small slit in it, which one could even hide from 
the audience. Since now no circular current can flow in the ring, it does not jump and you can just hear the coil vibrate. 
To make it clear that there is a slit, we insert a white sheet of paper. A further variation is to use a closed iron ring of the 
same geometric dimensions. This ring also does not jump, as the coil forms an electromagnet which attracts the iron 
ring stronger then the repulsion from Lenz's law. To demonstrate that it is indeed iron we use a permanent magnet.

The essential point being: each experiment is different, but it is important to give the audience time to absorb what 
they are seeing, and to guide them through the setup {\it prior} to performing the experiment. And whatever astonishing 
feature the experiment may hold, should not be revealed before hand.


\subsection{Fire Tornado}
\label{app:firetornado}
In our show, the fire tornado is symbolic of a camp fire, and is presented in the prologue, Sect.\,\ref{sec:prologue}.  
It has nothing to do with particle physics, and is thus not explained there, although usually we do explain it.

The fire tornado is a fairly well known experiment, with many example videos publicly available, \textit{e.g.} on 
YouTube \cite{firetornado}. It can be used for entertaining purposes, as we do in this show, as well as for 
demonstrating angular momentum conservation, temperature dependence of the density of air, the stack effect, 
and the different colours emitted by alkali metals when burning. Fire tornadoes require a fire, in our case 
provided by igniting safety paste in a large metal dish. The latter is placed in the center of a rotating platform. 
The hot air from the fire rises and draws in fresh air from the side. If this air is set in rotation, conservation of 
angular momentum leads to a higher angular velocity, as the air approaches the fire, at smaller radii, leading to 
the full tornado. This is similar to a figure skater drawing her/his arms in as she/he rotates. 

Fire tornados can occur naturally in intense wild fires. An example has been filmed in Australia 
\cite{firetornado-III}. Fire tornados can also be created, a nice version is where the air is set in rotation by
several box fans at fixed distance around a camp fire, but blowing at an angle to the radial direction 
\cite{firetornado-II}. This is not practical for an indoor show. We discuss here a safer and easier to repeat 
setup, using a vertical rotating  cylinder, as can be seen for example in \cite{firetornado}.   

\begin{figure}[t!]
\center \includegraphics[width=0.50\textwidth]{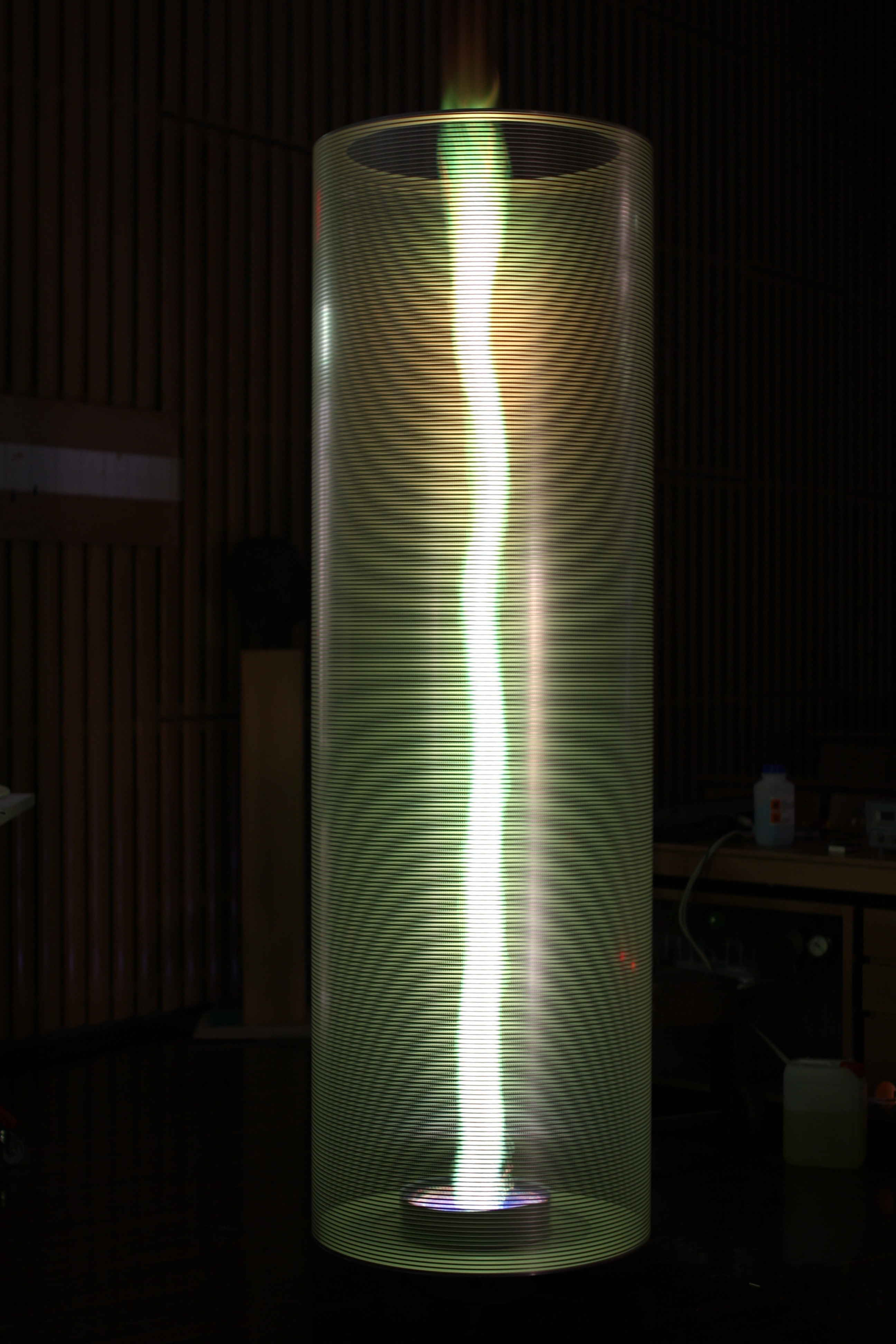}
 	\caption{Fire tornado in our 2\,m high rotating cage.}
		\label{fig:fire-tornado-2}
\end{figure}

\subsubsection{History}
We do not know who devised the first fire tornado for demonstration purposes, but there is a variety of
instructions to build mostly smaller devices on the internet.

\subsubsection{Materials and Technical Details}
The experiment is shown in Fig.\,\ref{fig:fire-tornado-2}. It requires a thin sheet of metal which is rolled into a 
cylinder. We use a 2\,m high cylinder with a 63\,cm diameter. The metal sheet is 2\,mm thick. The metal sheet 
must be perforated with holes, to allow air flow, and also for visibility of the fire tornado inside the cylinder. Our 
metal sheet has 5\,mm x 5\,mm size holes. The cylinder is placed on a well-oiled metallic turntable, and fixed 
with braces, so it can not tip over. Inside the turntable we have a fire-proof, pan shaped vessel. We place safety 
burning gel (or paste) inside the vessel. There is an additional larger hole at the bottom of the side of the cylinder, 
through which we can ignite the gel with a long match.

\subsubsection{Presentation}
The paste is set on fire and the rather small flame is shown to the audience. Some appropriate music possibly with a 
spinning and/or fire context is played, while the turntable is spun. Even with low spinning velocities the air starts 
rotating and a tornado builds up quickly within the cylinder. This is one of the few experiments where we do not
explain the setup beforehand, as it is pretty much self-explanatory. We usually explain the effect afterwards, 
however not in the particle physics show. Here it was just supposed to represent a camp fire by the Rhine in Bonn.
It is often nice to dim the lights during this experiment.

One can add alkali metals to the paste which give the flame nice colors. In the particle physics show we did not do
this.

\subsubsection{Safety}
If handled properly the fire tornado is a safe device, however there are a few things to take into account. First,
only safety gel/paste should be used as a burning material. The fire can then not spread easily if the device falls over 
by accident. The floor below the device and within at least one meter radius should be fireproof. The amount of burning 
gel/paste should be chosen to allow the fire to burn only as long as needed for the demonstration. When filling the vessel, you should have in mind that while spinning, the burning gel (paste) will accumulate at the edge. Many lecture 
halls require the completion of a safety form when performing any experiment with an open flame, such as the fire 
tornado. Thus also when traveling to a new location, make sure all the local safety requirements are fulfilled. You might
also need to turn off the smoke alarm during this experiment. This should also be done in accordance with your
local safety code.


\subsection{Lord Rayleigh's Oil Drop Experiment}
\label{app:oil-drop}

\begin{figure*}[h!]
	    \centering
           \includegraphics[width=0.46\textwidth]{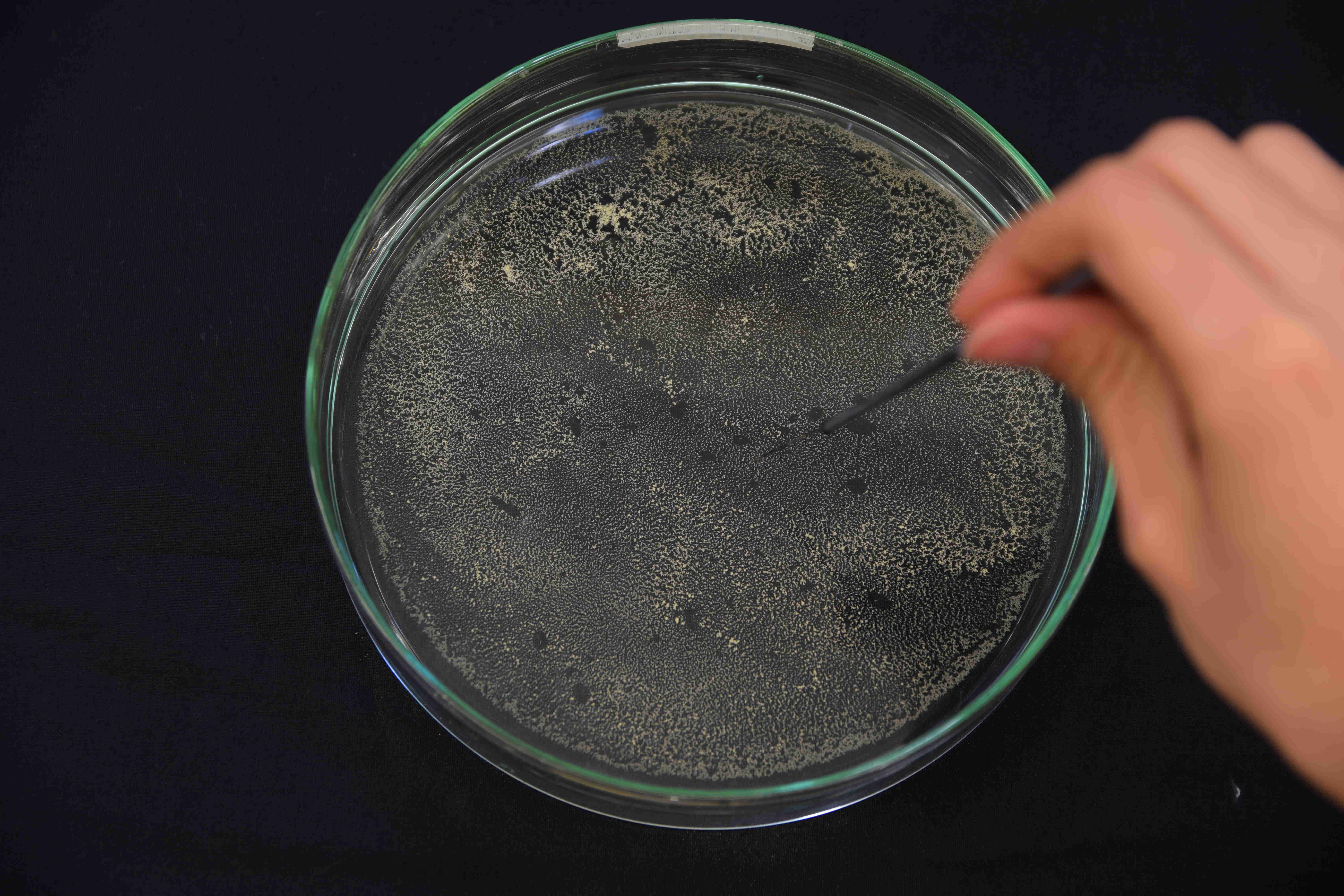}\; \includegraphics[width=0.46\textwidth]{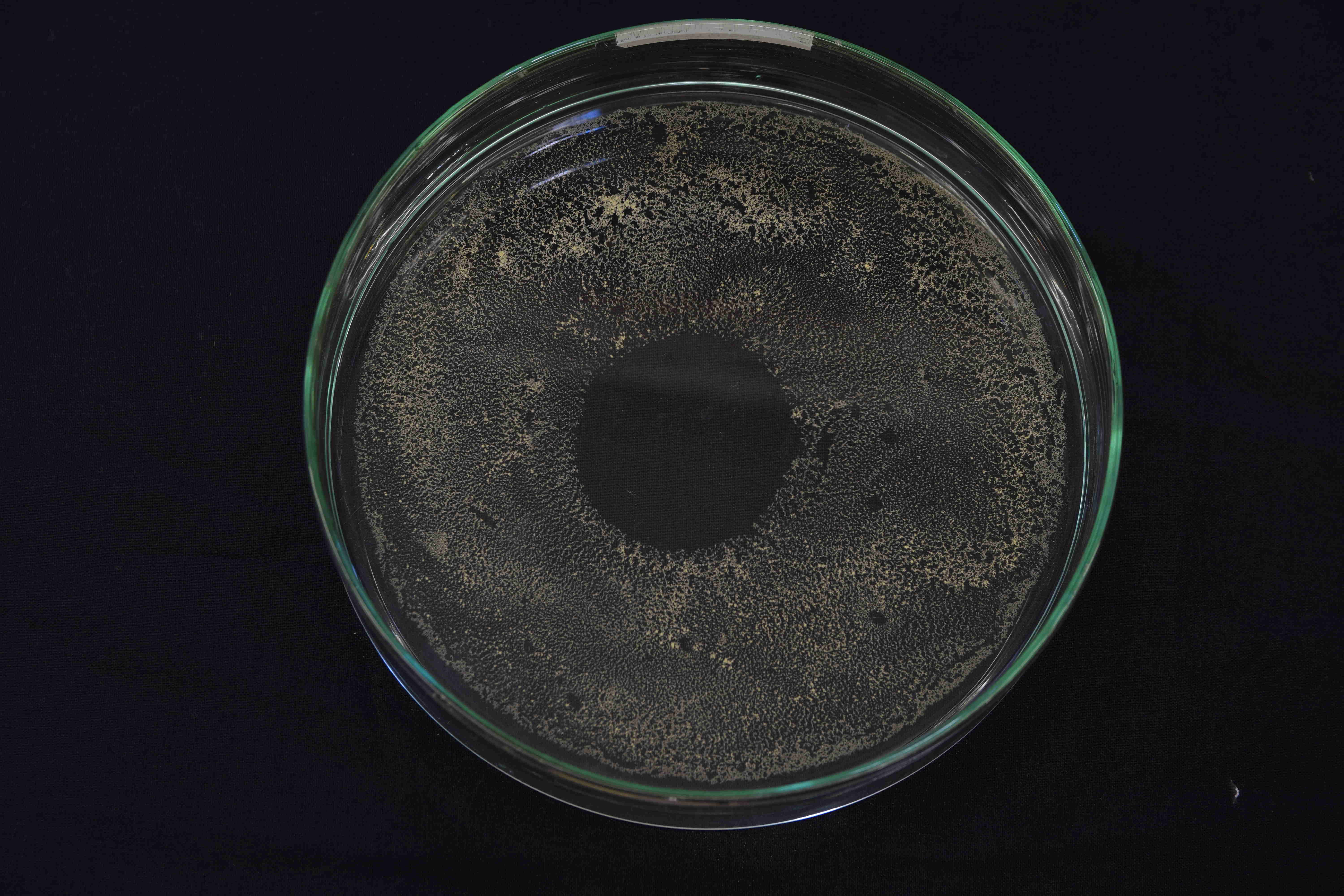} 
           \\ (a) \hspace{6cm} (b)
       \caption{Lord Rayleigh's Oil Drop Experiment. (a) Petri dish with the lycopodium powder spread on the surface. The needle is just above the surface of the water. (b)~After the drop of a mixture of oil and ethanol is placed in the center, it displaces the spores and a  circular oil film is clearly visible.}
    \label{oil-drop}
\end{figure*}

The oil drop experiment, shown in Fig.\,\ref{oil-drop}, is simple to execute but returns a surprisingly accurate 
measure for the thickness of an oil molecule and therefore the scale of the size of the atom.

\subsubsection{History} 
The first usage of this experiment to measure the thickness of oil was performed by Lord Rayleigh in 1890, 
but the experiment itself dates back to 1774 with Benjamin Franklin's attempts to describe the  spontaneous 
spreading of oil molecules \cite{Franklin}. In his experiments Lord Rayleigh found the thickness to be $16 
\times 10^{-8}\,$cm \cite{OIL}, although by implementing a cleansing technique on the surface another 
scientist, Agnes Pockels, obtained an improved value of $13 \times 10^{-8}\,$cm only a few years later 
\cite{Pockels,OIL}. This latter measurement is accurate to two significant figures according to more modern 
measurements.

\subsubsection{Materials} 
The experiment, as shown in Fig.\,\ref{oil-drop}, requires water, a Petri dish, some oil, which is in fact a mixture 
of ethanol and oil acid in the proportion 1:2000. Furthermore we use a fine tipped needle or extremely precise 
graduated pipette, lycopodium powder\footnote{This can be purchased online, see for example: 
\url{www.zooscape.com/cgi-bin/maitred/GreenCanyon/questp511741/r11}.}\!\!, and some form of dispenser for 
the powder. Ours consists of a beaker covered by a taught cloth with a few small holes which is essentially a very 
fine shaker. The presentation of the experiment works best when placed on an overhead projector.

\subsubsection{Presentation and Technical Details} 
For the experiment, we fill the Petri dish about half way with water. We shake out enough lycopodium powder to 
cover the entire water surface, see Fig.\,\ref{oil-drop}(a). With the needle we extract a small drop of the oil and 
ethanol mixture from the bottle and carefully apply it to the center of the Petri dish. Here a highly accurate 
graduated pipette can also be used. Touch the needle to the surface of the water as close to the center as possible. 
The powder will be pushed aside by the oil droplet, allowing us to measure the area of the thin surface formed. It's 
important that the oil drop used is small enough that the Petri dish surface is only partially covered by the oil film, as 
shown in Fig.~\ref{oil-drop} (b).

In the show, we place the Petri dish on an overhead projector so that it can be seen by the entire audience.
We also assume the volume of the oil drop is known, and we do not actually measure the area of the oil
surface because it would take too long. We instead just present the relevant equation on a slide: 
\begin{equation}
\mathrm{oil\,\,drop\,\,thickness}=\frac{\mathrm{oil\,\, drop\,\, volume}}{\mathrm{measured\,\, oil\,\, film\,\,
surface\,\, area}}\;,\label{oil-eqn-slide}
\end{equation}
and give the value of the resulting thickness, which we interpret as being the size of an oil molecule. This is 
the only mathematical equation we present in the show. 

\begin{wrapfigure}{r}{0.50\textwidth} 
    \centering
    \includegraphics[width=0.45\textwidth]{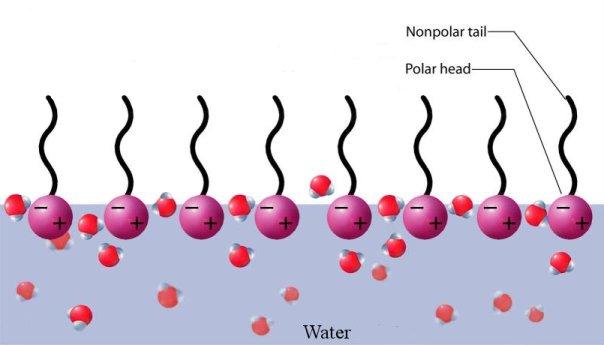}
    \caption{A visual depiction of the alignment of the molecules \cite{oil-wiki}.}
    \label{fig:polar-heads}
\end{wrapfigure}

While we don't address this in the show, the fact that the film is one molecule thick requires only some basic 
chemistry to explain. Oil molecules have a polar head and non polar tail which align so that the polar head 
points towards the water and the non-polar tail points away. This means that the molecules will orient themselves 
so that their entire length runs perpendicular to the surface of the water, see Fig.\,\ref{fig:polar-heads}. Since we 
determine the thickness of the film we thus measure here the size of the entire molecule: polar head plus tail.

\subsubsection{Extensions}
\begin{figure*}[h!]
     \centering
     \begin{subfigure}[t]{0.46\textwidth}
        \centering
        \includegraphics[width=\textwidth]{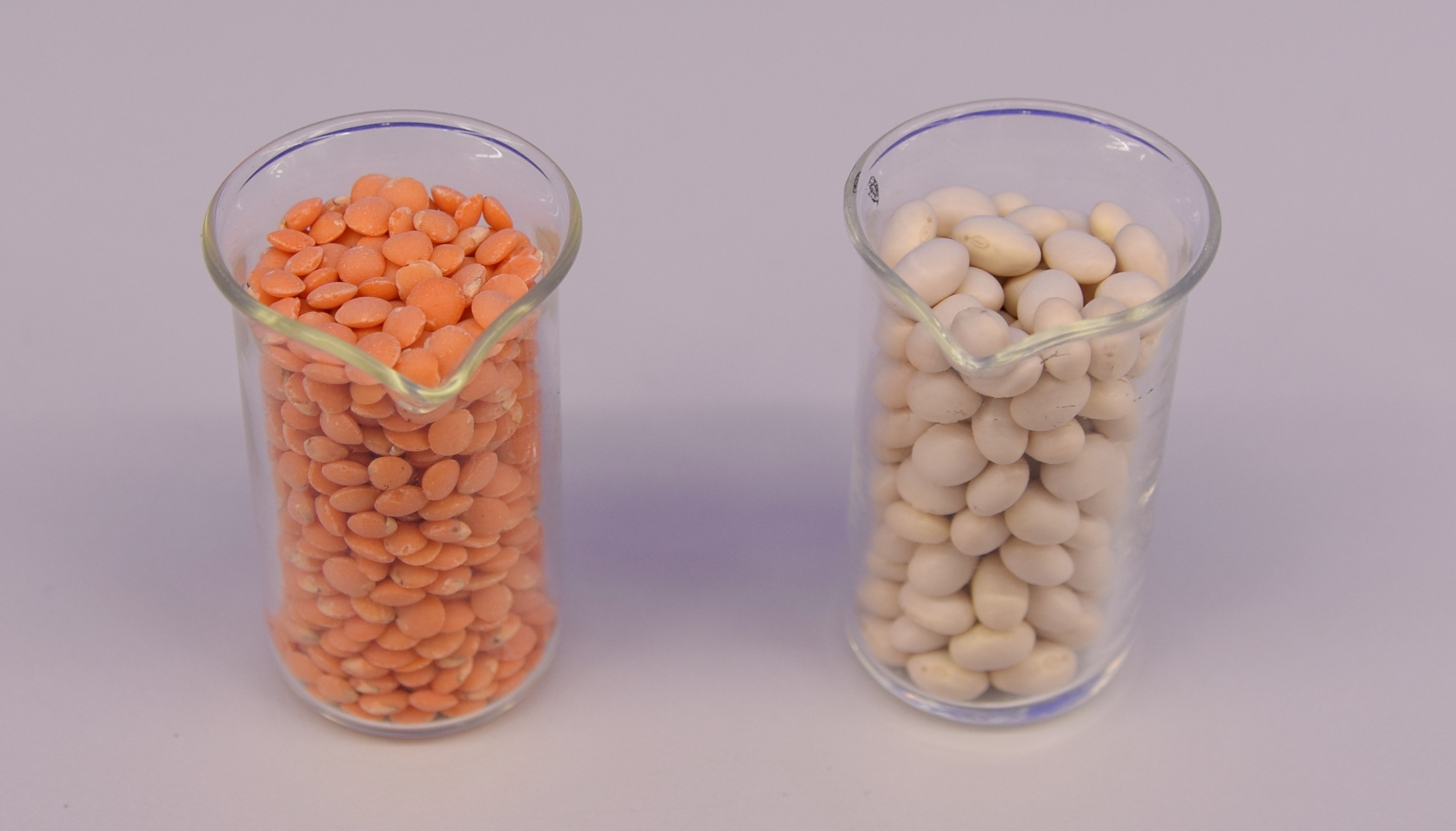}
        \caption{Lentils and beans in equal sized beakers.}
    \end{subfigure}%
    ~ 
    \begin{subfigure}[t]{0.46\textwidth}
        \centering
        \includegraphics[width=\textwidth,height=0.568\textwidth]{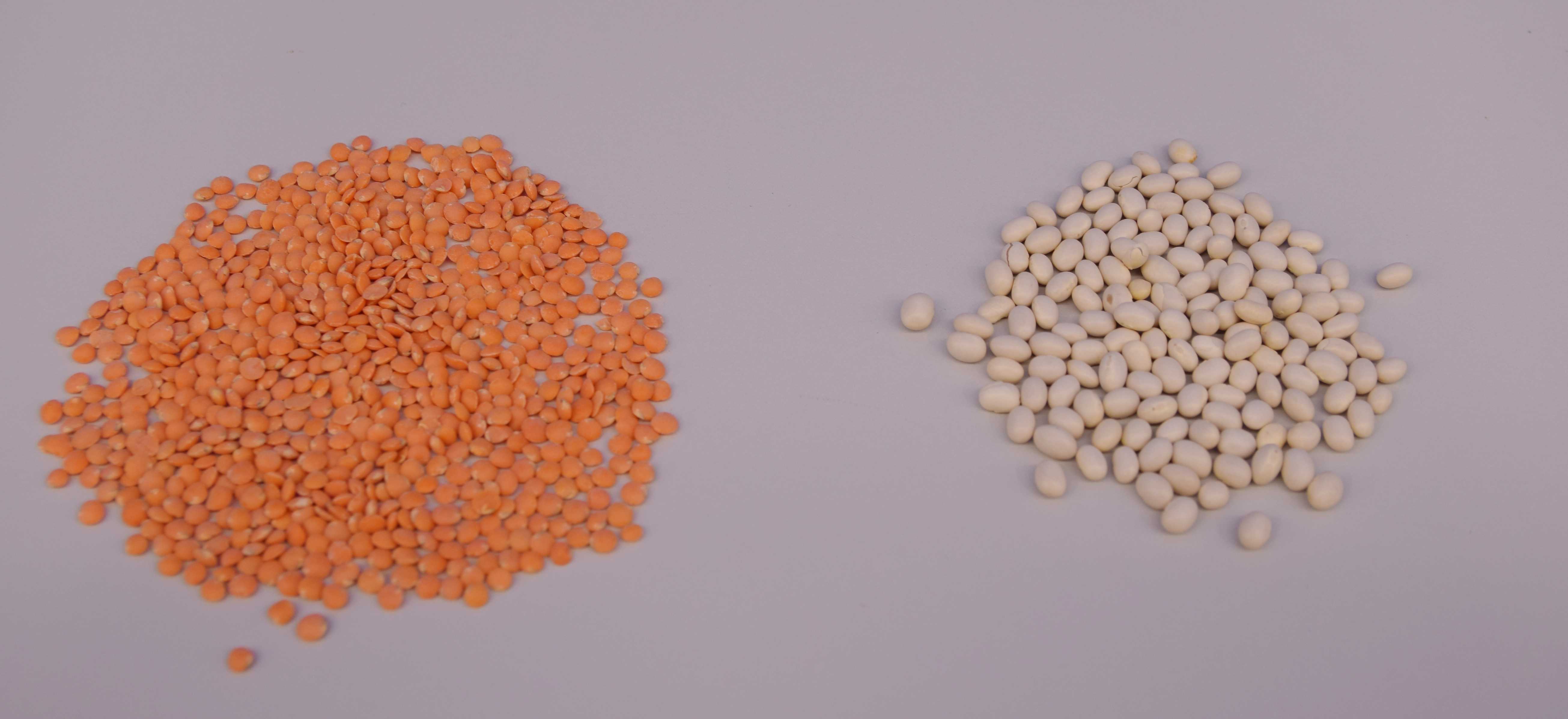}
        \caption{Lentils and beans spilt on table.}
    \end{subfigure}    
    \caption{Simple analogy experiment with lentils and white beans.}
    \label{fig:lentils-and-beans}
\end{figure*}

Given more time it is instructive to add a further experiment, as an analogy. One of us (HKD) often uses this
in public lectures on atoms. We fill one small glass with lentils and one identical glass with beans, so there is 
an equal volume of both. Use for example red lentils, and white beans, see Fig.\,\ref{fig:lentils-and-beans} (a).
We pour both of them out to form two separate circles in a single layer of lentils/beans, see 
Fig.\,\ref{fig:lentils-and-beans} (b). Because lentils have a much smaller diameter they form a much larger 
surface area than the beans. This quick demonstration gives an intuitive confirmation of  Eq.~(\ref{oil-eqn-slide}). 

\subsubsection{Safety}
The needle and the oil-ethanol mixture should be kept safely away from any children in the audience.


\subsection{Wooden Scattering Experiment}
\label{app:woodenscattering}

\begin{figure}[t!]
  \includegraphics[width=\textwidth]{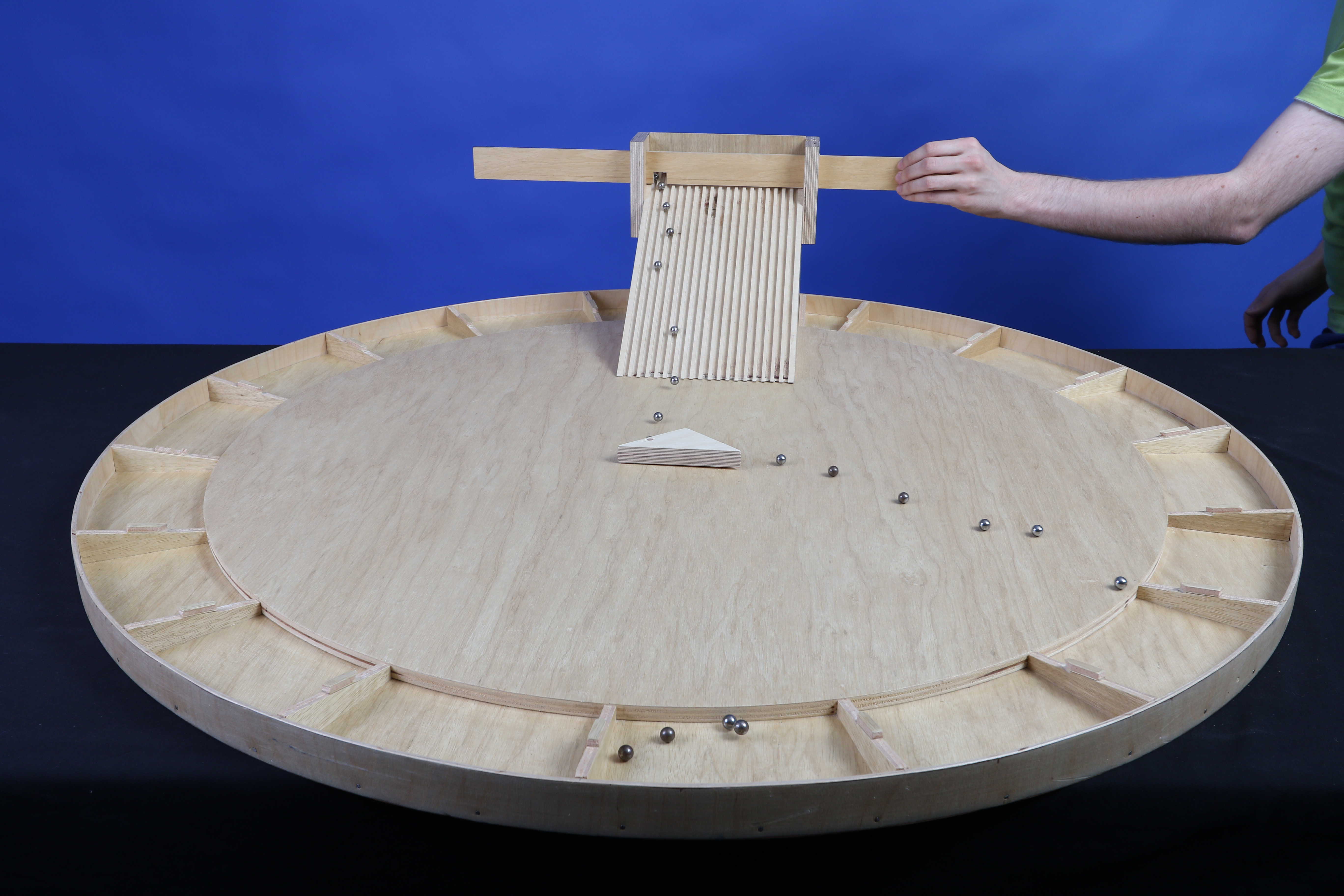}
	\caption{Wooden scattering experiment. The steel balls roll down the incline and hit the central target, 
	here a triangular piece  of wood. The target can be replaced by other shapes. The outside ring is 
	lowered and divided into 18 equal-sized pockets, of scattering angles
	$\Delta\phi=20^{o}$, where the steel balls are collected and counted.}
	\label{wooden-scattering}
\end{figure}

Our wooden scattering experiment is shown in Fig.~\ref{wooden-scattering}. It is a simple demonstration
of scattering steel balls off of a central target. The shape of the target can be varied.

\begin{figure}[h!]
\center
  \includegraphics[width=0.7\textwidth]{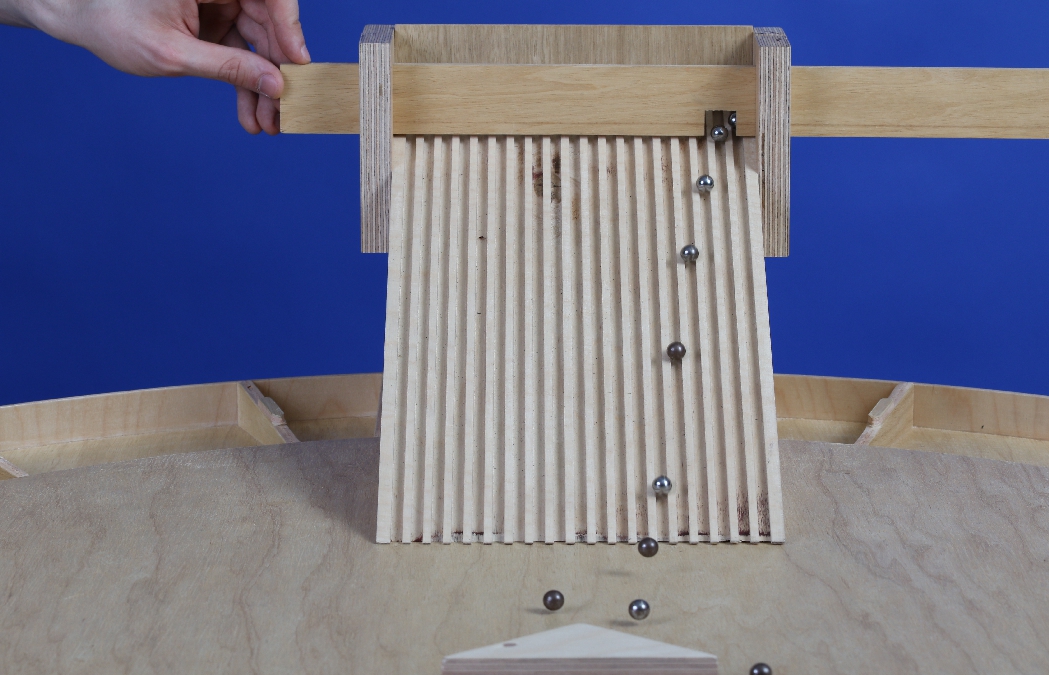}
	\caption{The mechanism for releasing the steel balls down the ramp. This ensures that the balls do
	not scatter off of each other.  In the back we have a loading mechanism, so that while we release one
	set of steel balls, we can refill a second set, just behind.}
		\label{wooden-scattering-2}
\end{figure}

\subsubsection{History}
Certainly over the years many similar demonstration experiments have been built. We were inspired by a 
similar design at DESY, Hamburg, seen by one of us (EP), some years ago \cite{Bechtle}. It shows that iron 
balls (as beam particles) scattered off of targets of different shapes yield different distributions. A second improved 
experimental setup was built at DESY \cite{Zoufal}  for  the exhibitions ``Particle Zoo'' and  the touring 
exhibition on the LHC, ``Weltmaschine" \cite{BMBF}. The DESY experiment was rectangular. We modified the design
to be circular, as shown in Fig.~\ref{wooden-scattering}. It was built by one of us (MKo), together with 
W.\,Lenz, the cabinet maker in the Physikalische Institut, Bonn.

\subsubsection{Materials and Technical Details}
The experiment is built in 3 parts:

1. The main part of the experiment consists of a circular board with an outer diameter of 110\,cm equipped with 
18 lowered pockets for collecting the balls after they have scattered (or not) off of the central target. The pocket 
size corresponds to a scattering angle range of $\Delta\phi=20^o$. The upper board has a diameter of 89\,cm. 
The pockets are lowered by 12\,mm and separated by 25\,mm high dividers. The dividers should be thin, to have 
well defined pockets. The lower circular board is bounded by a 25\,mm high wooden band.

\begin{figure}[h!]
\center
  \includegraphics[width=0.7\textwidth]{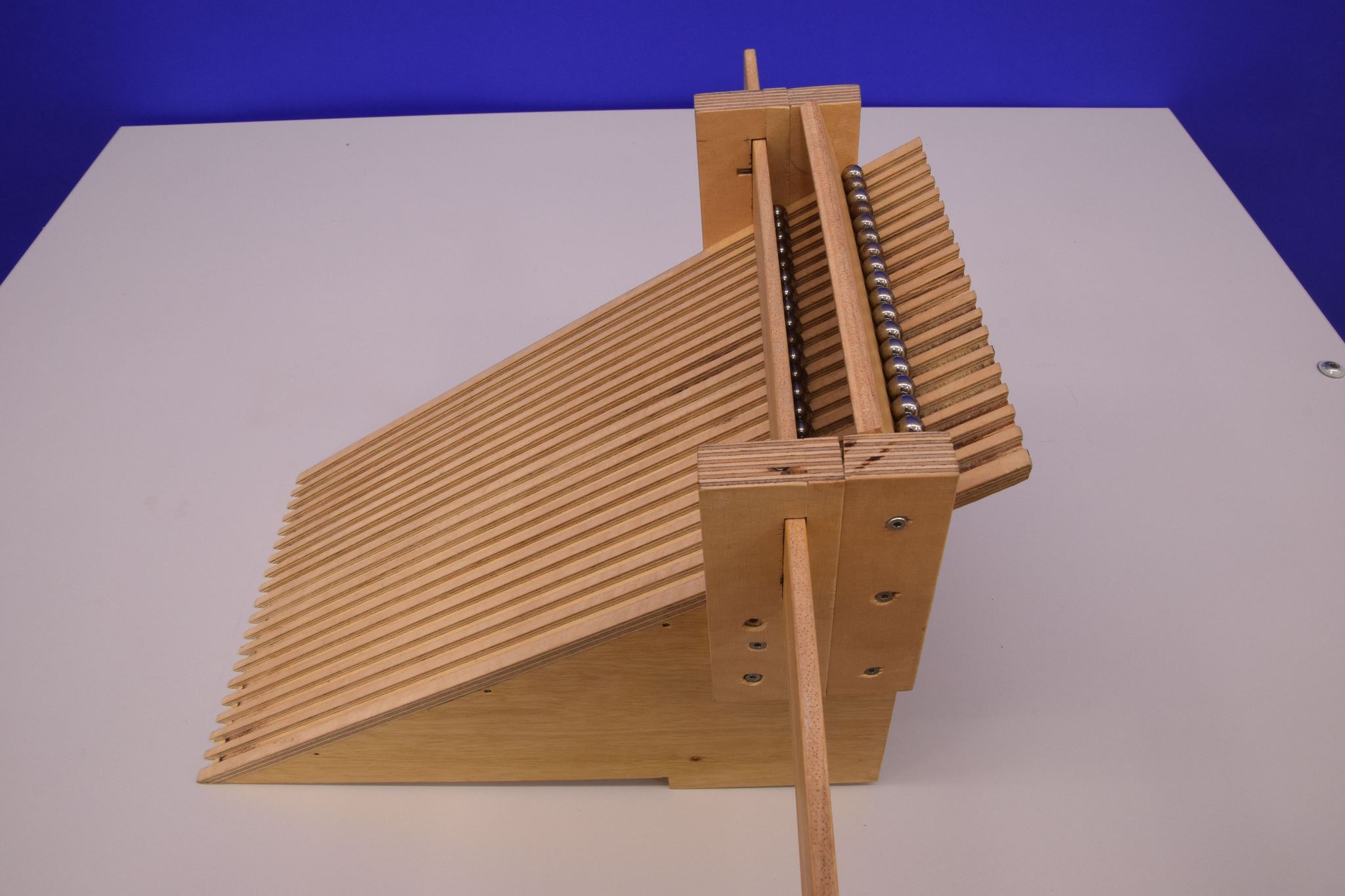}
	\caption{The reloading mechanism. The board holding the upper row of steel balls is easily rotated,
	releasing all balls to the lower level.}
		\label{fig:wooden-scattering-3}
\end{figure}

2. On one side of the circular experiment we have a ramp with 19 small grooves for guiding 19 balls in straight 
lines down the ramp. The grooves have been cut with a circular (buzz) saw into a multiplex wooden board. The 
ramp is 20\,cm wide and about 20\,cm high. The ramp has a sliding board with a small hole, to allow the balls to 
descend the ramp individually, as shown in Fig~\ref{wooden-scattering-2}. This avoids scattering between the 
steel balls. Behind the sliding board, we have a hinged board, as shown in Fig.~\ref{fig:wooden-scattering-3}. 
Behind this 
hinged board we can store 19 further balls. After the first 19 have been released down the ramp, we can swing 
the hinged board to reload the balls for scattering. While the second set is released, we can reload the hinged 
board, making it possible to do a multiple set of scatterings rapidly during a show. To speed up the refill, we 
prepare several containers with the exact number of balls beforehand. The balls are made of chromium-plated 
steel and have a diameter of 10 mm.

3. In the center of the circular board we fix targets of various shapes and sizes. Pegs in the targets slot into 
holes in the top board. In the show we use a circular target with a diameter of 12\,cm and a triangular shaped 
target with long edge perpendicular to the beam direction of 12.8\,cm. The geometric height of the triangle is
6\,cm. The targets have a thickness of 11\,mm.

\subsubsection{Presentation}
\begin{figure}[h!]
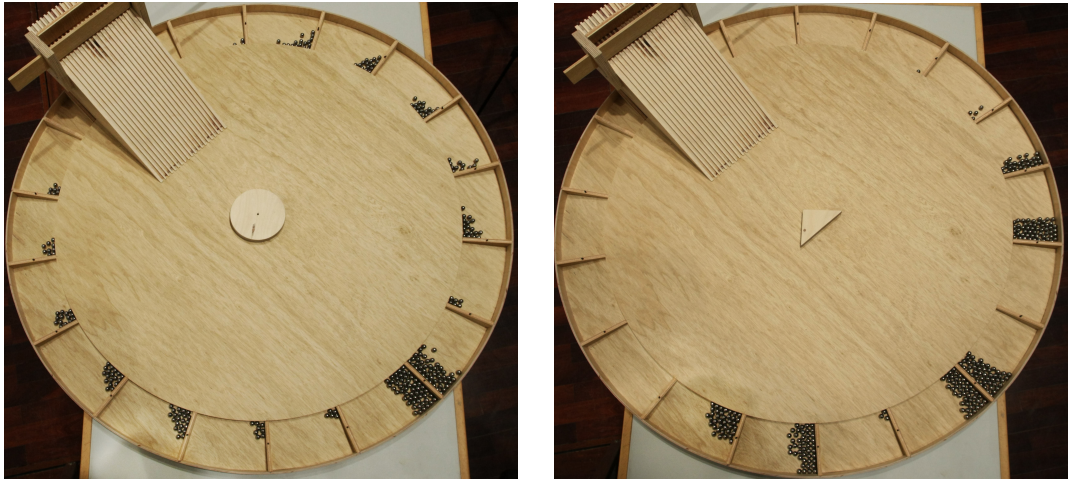

\center
\includegraphics[width=0.43\textwidth]{scattering_circular_target.jpg}\;\;\;
\includegraphics[width=0.43\textwidth,height=6.3cm]{scattering_triangular_target.jpg}
	\caption{The result of several runs of our scattering experiment, on the left with a circular target, and
	on the right with a triangular target.}
		\label{fig:many-runs}
\end{figure}

Before starting the experiment, it is important that it is setup exactly horizontally. In the show we do three runs 
of 19 balls for the circular target, and then again for the triangular target. As the 
target is not as wide as the ramp quite a few balls pass the target unscattered and enter the two forward 
direction pockets. For the circular target the remaining balls are distributed roughly evenly over the other 
pockets. For the triangular shaped target there is an accumulation in two directions. This can be seen after the
three respective runs. However, in order to emphasize the point, we also show two photos of the results for 
multiple runs on the circular and triangular target, respectively, see Fig.\,\ref{fig:many-runs}. Between the runs on 
the two targets, the Caretaker appears on the scene and collects the scattered steel balls with a magnet, which 
then vanish in his overcoat pocket.

\subsubsection{Safety}
There is no obvious safety risk in running the experiment. Small children in the audience should not be allowed to get the
steel balls, for example after the show, as they might swallow them.

\medskip

\subsection{Cathode Ray and Helmholtz Coils Experiment}
\label{app:helmholtz-colis}

The Helmholtz coils are an extension of the cathode ray tube experiment regularly presented at the high school 
level. In German high schools the entire setup, as presented here, is often employed.

\subsubsection{History} 
We briefly recount some aspects of the history of the cathode ray tube, especially since some of the
important developments occurred in Bonn \cite{CRT,Pais:1986nu}. The earliest relative of the cathode ray
tube was the gas discharge tube, invented in 1838 by Michael Faraday. It was similar in construction, but
lacked sufficiently low pressure and more modern electron emission techniques. Thus early experiments
gave a glowing cloud instead of a collimated beam. In 1858 in Bonn, Johann Heinrich Wilhelm Geissler, a 
glass blower, invented both the mercury pump and the metal--glass seal, creating a superior vacuum. 
Employing these advances, he developed the Geissler tubes. These were the first examples of what is 
traditionally used in the classroom today as a cathode ray tube. The Geissler tubes are evacuated and filled 
with small quantities of neon or hydrogen gas to form a colorful light beam in the electron's path. The first 
major investigations with the tubes were performed by Julius Pl\"ucker in Bonn \cite{Pluecker}. In the 1870's 
William Crookes developed the Crookes Tube, but his technology was focused more on image projection, not 
visualization of the electron beam.

The cathode ray tube is set in the middle of the Helmholtz coils. The latter consists of two coils, separated a 
distance $d=R$, where $R$ is the radius of the coil. This generates a nearly homogeneous magnetic field 
inside the  cathode ray tube.

We have not been able to determine, whether Helmholtz actually invented the Helmholtz coils. Incidentally,
Helmholtz was a professor of physiology in Bonn from 1851 to 1858.

\begin{figure*}[h]
        \includegraphics[width=0.95\textwidth]{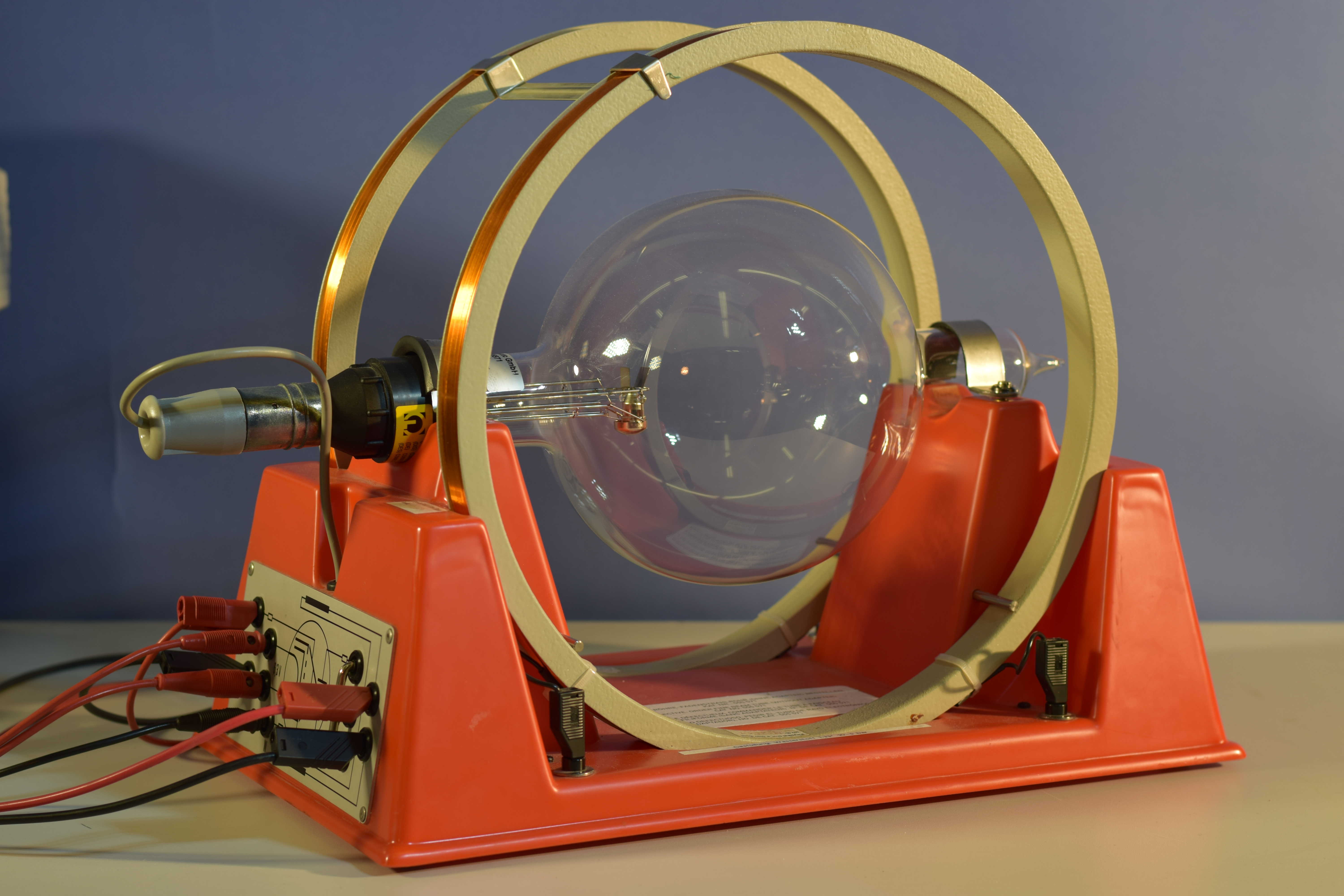}
        \caption{The Helmholtz coils with the glass tube and electron beam emitter.}
        \label{fig:Helmholtz1}
\end{figure*}

\subsubsection{Materials} 
The Helmholtz coils together with the cathode ray tube used in the show are manufactured by 
Leybold Didactic for educational purposes \cite{Leybold-Helmholtz}. Our setup is shown in Fig.\,\ref{fig:Helmholtz1}. 
The central part consists of a glass tube with a large spherical mid--section. The glass tube is evacuated and then 
filled with a small amount of hydrogen gas. The final pressure in the tube is $1.33\times10^{-5}$\,bars. The electron 
gun in the central sphere shoots the electrons vertically upward. Around the glass vessel are two large coils in the 
same plane as the electron beam. They produce an approximately homogeneous magnetic field in the glass sphere.
If the magnetic field is not perpendicular to the electron beam, the electron forms a spiral path.

\begin{figure}[h]
        \includegraphics[width=\textwidth]{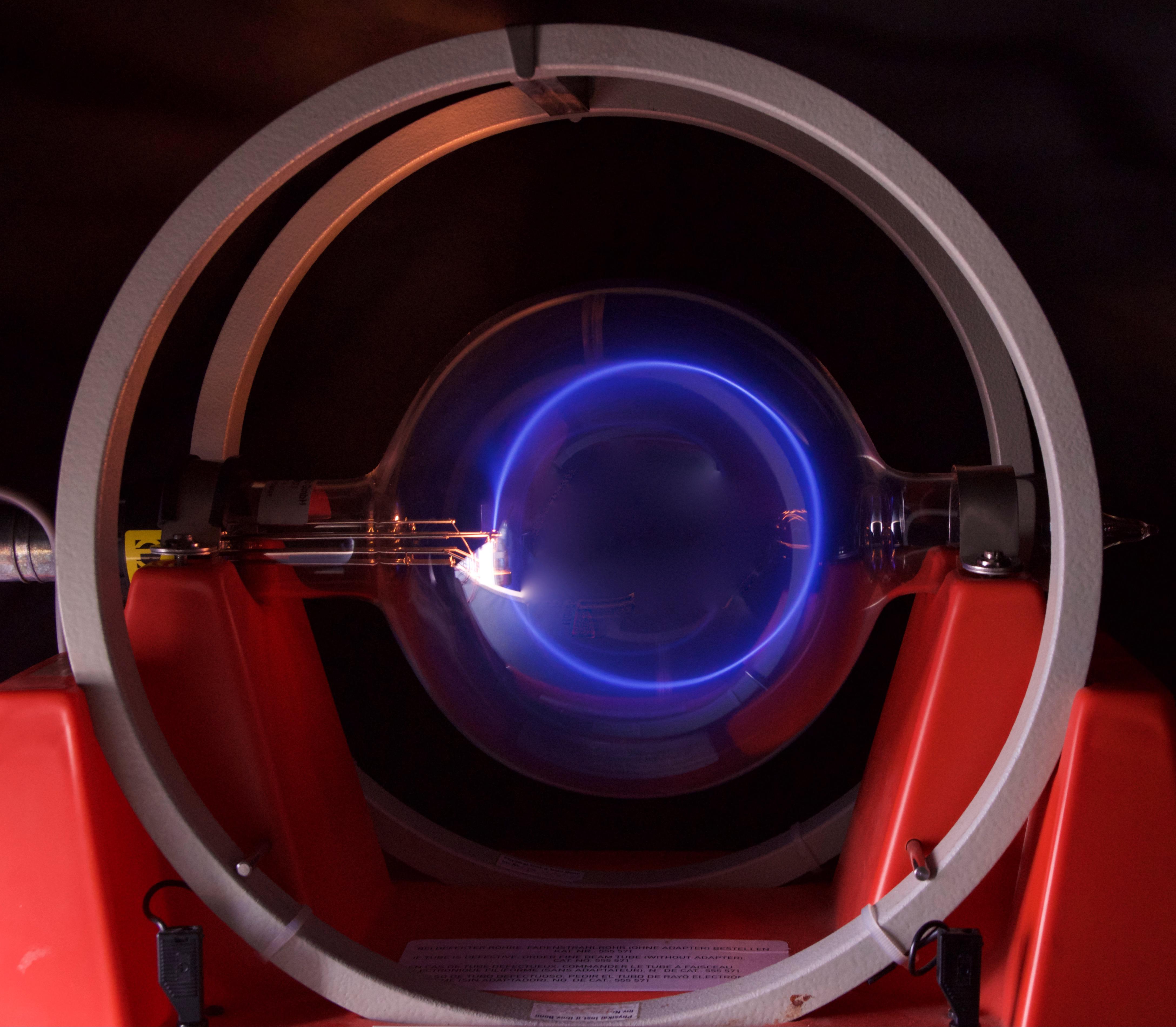}
        \caption{The Helmholtz coils: the circular blue glow shows the path of the electrons interacting with the 
        hydrogen gas.}
        \label{fig:electron-circle}
\end{figure} 

\subsubsection{Technical Details} 
We typically use an accelerating voltage for the electron gun of 200\,V. The current for the Helmholtz coils is around 
1\,A. The resulting magnetic field is about 0.68\,mT, which gives an electron beam orbit radius of about 7\,cm.

While this is not done during the show, the apparatus can be used to measure the electric charge to mass 
ratio of the electron. Setting the centripetal force equal to the Lorentz force one has
\begin{equation}
\frac{m_ev^2_e}{r}=q_e v_e B,
\end{equation}
where $m_e,\,v_e,\,q_e$ are the mass, the speed and the charge of the electron, respectively, $B$ is the 
magnetic field and $r$ is the electron orbit radius. Using the fact that the kinetic energy of the electron $
\frac{1}{2} m_e v^2_e = q_e U$, where $U$ is the applied accelerating voltage, we can solve for the electron 
charge over mass ratio
\begin{align}
\frac{q_e}{m_e} &= \frac{2 U}{B^2 r^2}\,.
\end{align}
Inserting our values, we obtain $\displaystyle{\frac{q_e}{m_e}}= 1.8 \times 10^{11} \,\displaystyle{\frac{\mathrm{C}}
{\mathrm{kg}}}$ 
which is close to the known value.

\subsubsection{Presentation}
In the show, we first explain the details of the setup with the camera image projected onto the screen. At this stage the 
magnetic field is zero, thus when we dim the lights, the blue glow of the electron beam points vertically upwards. We 
then play suspenseful music, for example the theme from the film {\it Mission Impossible}, as we adjust the magnetic 
field. The electron beam forms a circle, which varies in size, see Fig.\,\ref{fig:electron-circle}. The essential point is that 
the audience here gets to ``see" an elementary particle for the first time, via the interaction with the gas.

\subsubsection{Safety}
Follow the safety instructions of the electrical equipment. As there is very low pressure in the glass tube, one should be careful
not to break it, \textit{e.g.} during transport.


\subsection{A Simple Charged Particle Detector: Radioactive Decay and Jacob's Ladder}
\label{app:jacobs-ladder}

This is a nice and simple experiment to show how ionizing radiation, or otherwise invisible particles, can 
be detected in a way directly observable by a live audience. The discharges are visually spectacular close 
up, but the small size of the experiment requires a projector for it to be appreciated in a large lecture hall.
care must be taken, to only use radioactive sources conforming with all safety regulations.

\begin{figure}[h!]
  \centering
  \includegraphics[width=0.95\textwidth]{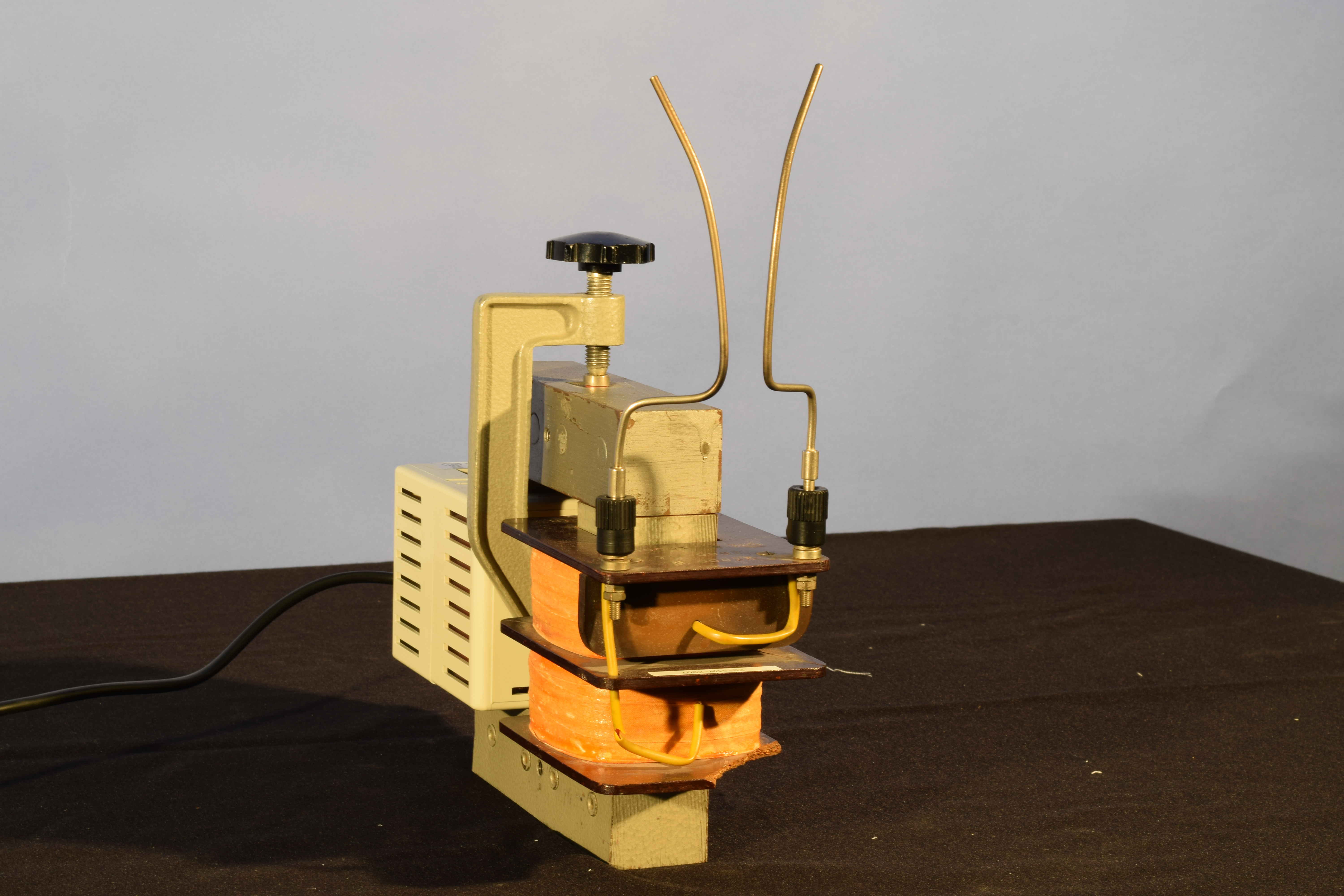}
	\caption{Our Jacob's Ladder}
	\label{jacobs-ladder-setup}
\end{figure}

\subsubsection{History} 
The exact history of the Jacob's ladder experiment is difficult to trace. In principle, the version we use is a 
redesigned Geiger counter which allows for a more readily observable reaction to the ionizing particles. The 
first Geiger counter was invented in 1911 by  Hans Geiger as a means of counting radioactive alpha particles. 
In 1925 Geiger and Walther M\"uller, enhanced the device, so that it could detect many forms of ionizing 
radiation \cite{Geiger}. If an individual has been accredited with the first implementation of the Jacob's ladder 
specifically for the detection of ionizing radiation, it is not well documented.

\begin{figure}[ht]
       \centering
        \includegraphics[width=0.95\textwidth]{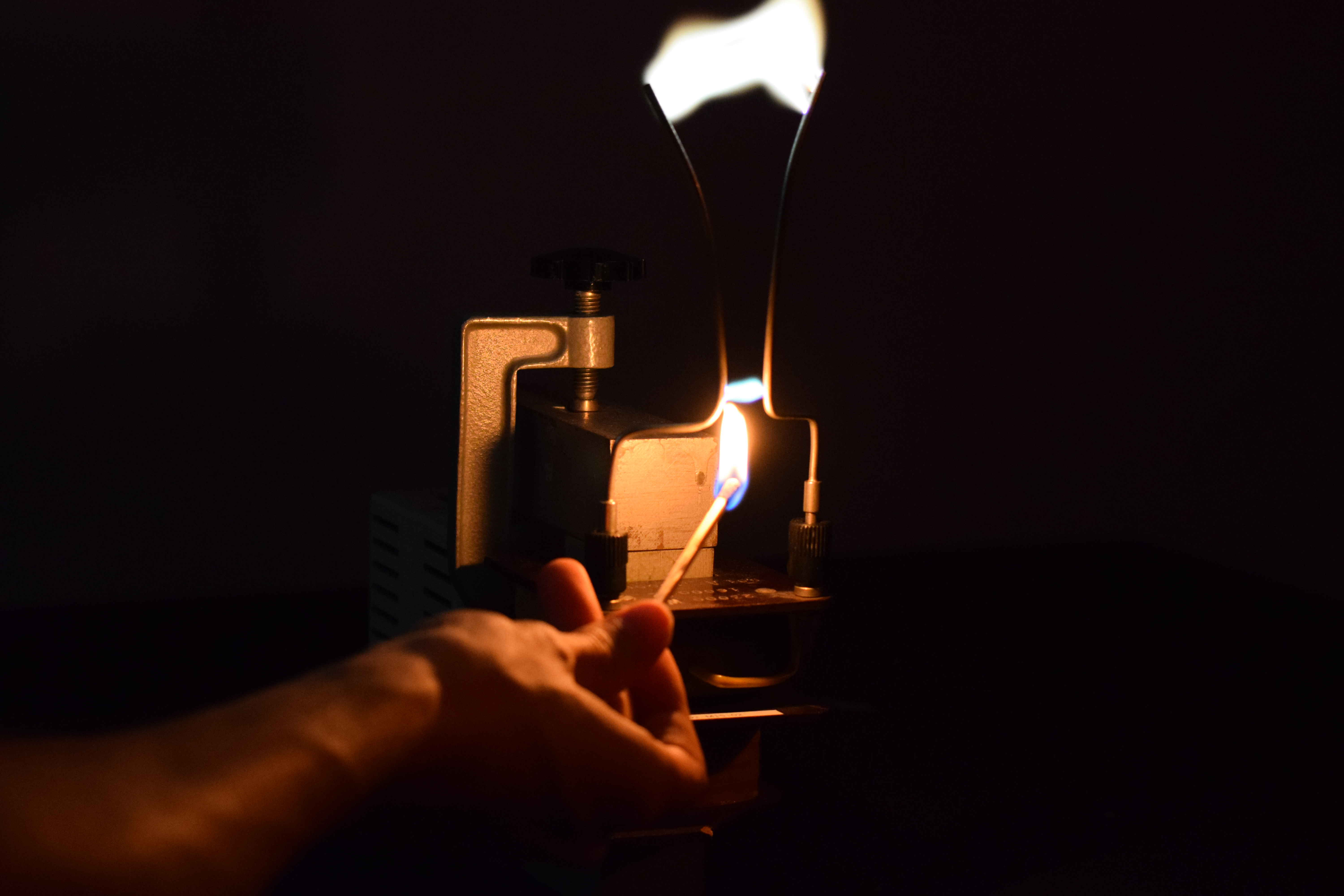}
        \caption{The climbing spark effect with a match as the initial ionizing source. Here the input potential was around 
        100\,V, corresponding to about 4,600\,V at the gap, and the two bars were about $1.5$\,cm apart.}
\label{jacobs-ladder-match}
\end{figure}

\subsubsection{Materials} 
The setup consists of a primary coil with 500 windings and a secondary coil of 23,000 windings. They are 
connected by a closed rectangular iron yoke made of lamination steel, as in App.\,\ref{app:ring-schleuder}.
This yields an amplification factor of  46. The primary coil is connected to the main voltage of 220\,V via a 
variable transformer. Thus the output AC voltage of the primary coil can be varied between 0 and 220 V. The 
secondary coil is connected to two bent wires forming the ``Jacob's Ladder" arrangement, as shown in 
Fig.~\ref{jacobs-ladder-setup}. The wires are 3\,mm thick. At the closest point at the bottom, we set the gap 
to about 7\,mm, when detecting the ionizing radiation. The gap can be larger if one uses a match to light the 
spark. In the show we fix it to the narrower value as we perform both of these experiments immediately in 
sequence.

It is essential to be able to vary the gap voltage continuously. If it is too low, no spark discharge will occur. If it 
is too high there are serial discharges unrelated to the ionization provided by the agent. We set the voltage as 
high as possible without getting natural discharges. 

We use a long match to first trigger the discharge. We then used an Am$^{241}$ source, which is an 
$\alpha$-emitter, with a radioactivity of 340\,kBq. Unfortunately, this source cannot be transported according
to current law. We thus show a film in the show.

\subsubsection{Presentation} 
We first turn on the voltage to a level below where natural discharge occurs. Then we light a match and hold 
it just below the gap at the bottom between the two bars, see Fig.~\ref{jacobs-ladder-match}. The gas in the 
flame of the match is ionized, meaning it forms a plasma. This has a much higher conductivity than the air and 
we get a spark discharge. Because the gap is smaller at the bottom the spark starts there. The spark itself is now 
a plasma and is hot. The plasma rises, and with it so does the path of least resistance for the electrons. This is
the observed rising arc. The wires get wider  towards the top and at some point the arc can not be maintained. In 
Fig.~\ref{jacobs-ladder-match} one can see the lower spark ignited after the upper one.

\begin{figure}
    \centering
    \includegraphics[width=0.9\textwidth]{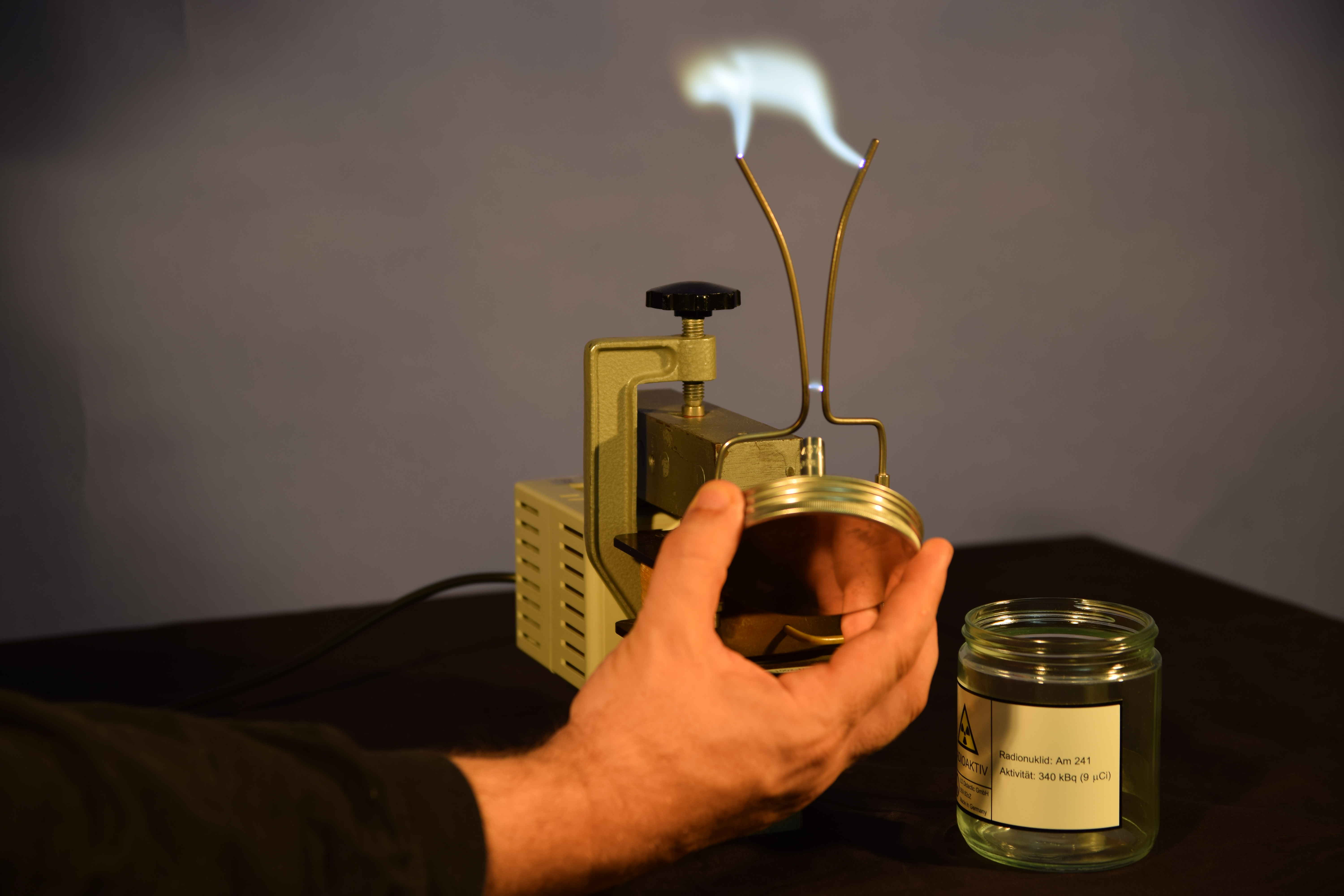}
    \caption{The Jacob's Ladder activated here by a radioactive 
    Am$^{241}$ source. With a radioactivity 
    of $340$\,kBq the source required a 200\,V potential and 5\,mm of separation between the bars. We are
    not allowed to transport such a strong source. }
    \label{jacobs-ladder-alpha}
\end{figure}

The match is a good way of opening the experiment since it provides a visible cause to the ionization of the air, 
and the audience is familiar with a match\footnote{There are several experiments one can perform to show that 
the flame forms a plasma. For example putting a lit candle between two capacitor plates. When the plates are 
charged the flame tilts to one side. Unfortunately, there is no time for this in our show.}\!. 

Now we repeat the same process with our radioactive source to demonstrate the ionization of the air by the otherwise 
invisible particles emitted in radioactive decay. The voltage must be increased to 200\,V. Our recommendation is that 
beforehand the presenter should play with the variable transformer to get an idea for what voltage is appropriate for 
the two stages of the experiment. This will depend on the humidity, so the best approach is to test the values the day 
of the experiment. Note that during tests the metal rods heat up, lowering the required natural discharge voltage.

The ionizing agents are now the highly energetic particles emitted by our source. In the photo, see 
Fig.~\ref{jacobs-ladder-alpha}, we use an Am$^{241}$ source. This triggers the discharge correlated with the approach 
of the source to the gap. We thus have a simple charged particle detector, for everybody to observe live.

The radioactivity safety laws allow us to use this source in the lecture hall onstage. However, the European 
transportation laws for radioactivity are stricter, meaning that the source is not legal for us to transport. In Oxford and 
Padua we therefore used a weaker source, which is legal to transport, but for which it is somewhat trickier to trigger 
the sparks.

In the remainder of the show we use a Geiger counter connected to a loudspeaker. It is more abstract, but still
quite immediate, and the audience now hopefully understands the underlying principle.

\subsubsection{Safety}

This experiment requires caution on several levels. The main concern is that we are applying thousands of 
volts across the gap which is enough to cause physical harm. All audience members should be kept at a 
distance from the experiment, and the presenter must take care not to touch the metal bars while he uses 
both of the agents. Furthermore, after using the apparatus, the power supply to the Jacob's Ladder should 
be turned off and disconnected. It's common for audience members to approach the stage during an 
intermission or after the show, and we want to assure that the device presents no danger.

The use of radioactive sources requires caution and the strict adherence to the safety rules. Three of us, HKD, 
MKo and EP, have participated in the required training program. EP and HKD have also participated in the training 
program for transport of radioactive material. At any moment during the show or on our travels, one of the above is 
always responsible for the radioactive materials. It is also his responsibility to lock the sources away as soon as 
they are finished being used in the show.


\subsection{Cloud Chamber}
\label{app:cloud-chamber}

\begin{figure}[h]
        \centerline{\includegraphics[width=0.6\textwidth]{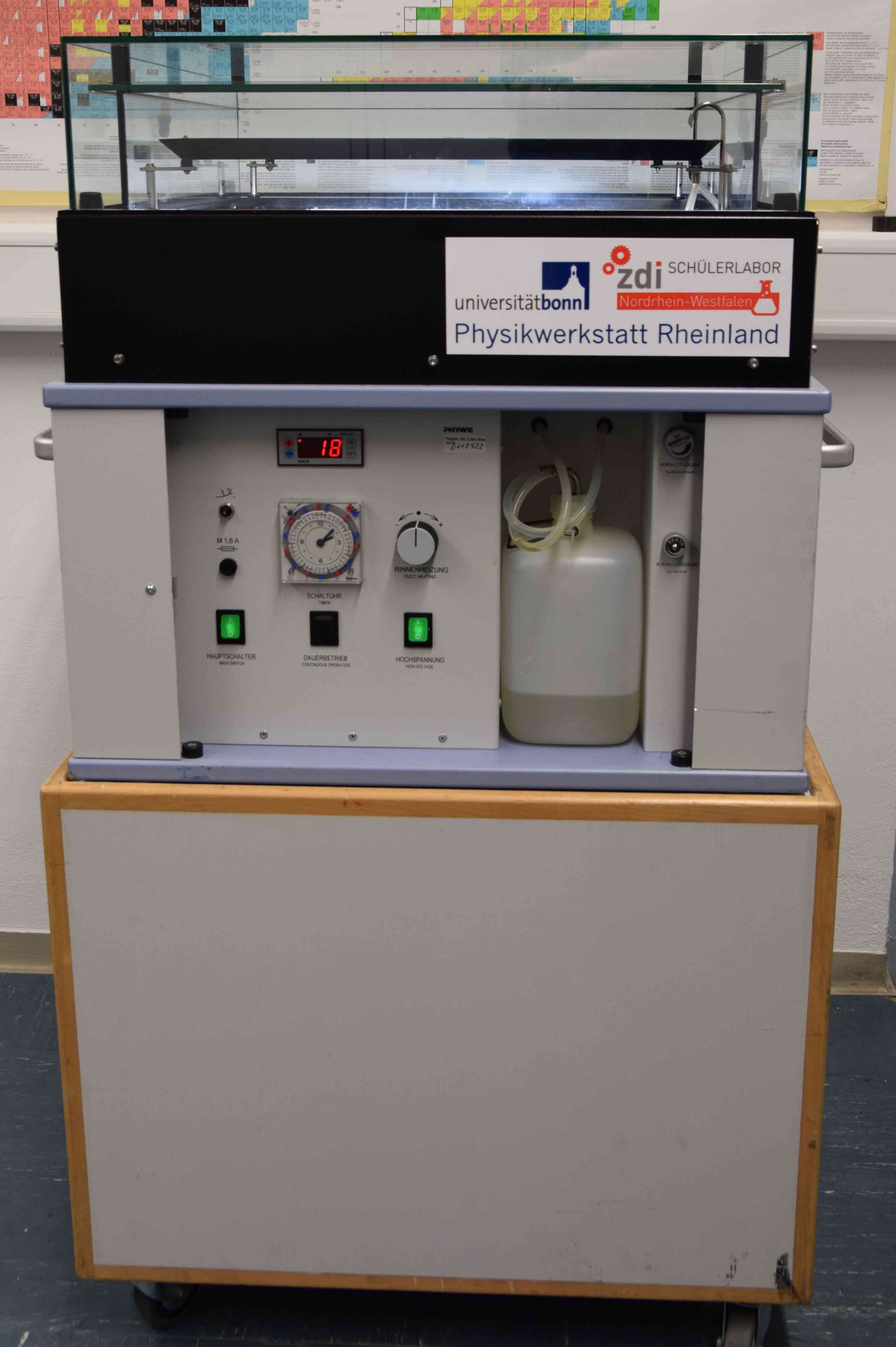}}
        \caption{The {\tt PHYWE} cloud chamber. The upper black part with the glass casing contains
        the actual chamber. Underneath is the electronics and the refrigeration unit. The plastic container on the
        right holds the 2-propanol. The wooden box is on wheels for easy transport. It also raises the chamber
        to a height easily visible for adults.}
        \label{fig:cloud-chamber}
\end{figure}

This experiment we use is built and sold by {\tt PHYWE}. It is shown in Fig.\,\ref{fig:cloud-chamber}.
For more information on this specific chamber see the website \cite{phywe}.

\subsubsection{History}
The cloud chamber was invented by the Scottish physicist Charles Thomson Rees Wilson (1869-1959), in 1911
\cite{Wilson:1912bva}, therefore it  is also called the Wilson chamber. For this work, Wilson received the Nobel 
prize in physics in 1927, together with Arthur Compton \cite{wilson-nobel}. A cloud chamber contains a 
supersaturated vapor. When high energy charged particles pass through the cloud chamber they ionize molecules
in the gas along their path. In the supersaturated layer, these ions then act as seeds, where the vapor condenses,
forming ``contrail" like features, which are easily visible with proper lighting. In the original cloud chamber the 
supersaturation was achieved by expanding the chamber. Our apparatus is a diffusion cloud chamber, as invented 
by Langsdorf in 1936-37 \cite{langsdorf}, which allows for continuous operation \cite{review-diffusion}. Here the 
bottom side of the chamber is cooled well below freezing, in our case about $-32\,^\degree$C. A small distance 
above the cold lower edge a supersaturated region develops.

Historically the cloud chamber was very important in particle physics. It was essential for the discovery of the 
positron by Anderson in 1933 \cite{Anderson:1933mb}, the muon by Neddermeyer and Anderson 
\cite{Neddermeyer:1937md}, and Street and Stevenson \cite{Street:1937me} in 1937, as well as the kaon, 
discovered by Leprince-Ringuet and L'Heritier \cite{LeprinceRinguet:1944mh}, and Rochester and Butler 
\cite{Rochester:1947mi} in the 1940s. This is thus a true particle physics experiment, which can be 
incorporated into a show.

\subsubsection{Materials and Technical Details}
We use a comercial  {\tt PHYWE} cloud chamber. To operate it only some 2-propanol is needed.  It is possible 
to devise a simple diffusion cloud chamber with dry ice, as noted already in 1950 \cite{dry-ice}. Dry ice has a 
temperature below $-78.5\,^\degree$C. Documentation for construction using dry ice is also widely available 
on the internet, see for example \cite{dry-ice-youtube}.

\begin{figure}[h]
        \centerline{\includegraphics[width=0.66\textwidth]{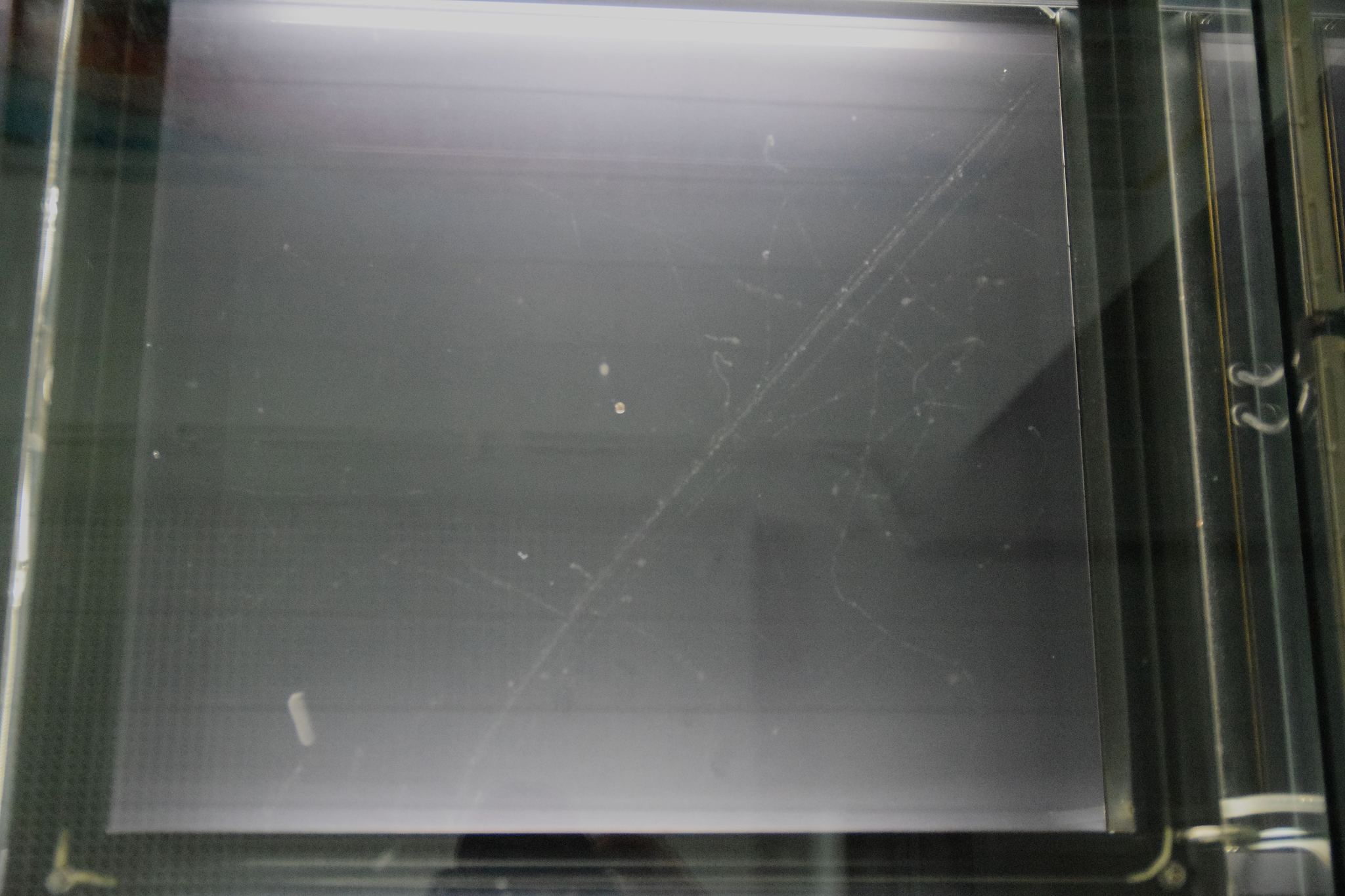} \;
        }
        \caption{A top view photo of the {\tt PHYWE} cloud chamber, while in operation. An unusually long track can be seen
        here diagonally across the chamber, most likely a muon. In the lower left corner is also a thick track
        from a hadron, corresponding most likely to an alpha particle.}
                \label{fig:cloud-chamber-tracks}
\end{figure}

The {\tt PHYWE} commercial cloud chamber we use has an enclosed volume of about 450\,mm x 450\,mm x 
200\,mm, covered by a glass construction. It weighs about 80\,kg. The lower box contains the refrigeration unit. 
The 2-propanol evaporates from trays at the outer frame of the upper part in the glass housing. The entire 
apparatus must therefore stand on level ground. The 2-propanol is heated by a wire carrying an electrical current. 
This current can be varied, as can the flow of 2-propanol. In the bottom center there is a black metal plate cooled 
by the refrigeration unit. The chamber is illuminated internally from the side. The 2-propanol vapor takes about 
30\,min to cool down sufficiently.

\subsubsection{Presentation}
The cooling unit of the cloud chamber is unfortunately fairly loud. We turn it on 40\,min before needed 
and turn it off, just after it has been demonstrated.

For the show we place a small camera directly above the glass plate. The camera is connected to the projector 
via the control table. The lecture hall is darkened during the observation of the particles. If it is not dark enough,
the cloud chamber and the camera should be covered with a black cloth.

The shape of the trails, as seen in Fig.\,\ref{fig:cloud-chamber-tracks}, quite clearly indicates the nature of the 
particle passing through. They also happen sufficiently often for this to be explained live during the show. Very 
short thick tracks correspond to $\alpha$ particles. They are produced within the chamber from radioactive radon 
decays.\footnote{Rn$^{222}$ has a half-life of $\tau=3.3\,$days and is an $\alpha$-emitter. Radon is a gas and 
occurs in daily life. It is the main source of the observed $\alpha$-tracks.}  Very thin tracks with an irregular path 
arise due to electrons or positrons, which mostly come from radioactive decays, as well. Apart from that we 
observe very thin long trails. These are due to muons ($\mu$). They are (mostly secondary) particles produced by 
cosmic rays interacting with the atmosphere.

When we explain the cloud chamber with people standing immediately around, we explain the setup in detail, 
including the trays with the 2-propanol, the cooling unit etc. As an example of condensation, we often mention a 
cold glass outside in the summer, where water condenses on the outside. In the show presented here there is no 
time for this. Furthermore, we typically explain the origin of the particles, emphasize that they are constantly flying 
through everything and everybody, while we realise nothing. The cloud chamber enables us to see them. We mention the
analogy of the particle tracks with airplane contrails and then discuss the observed particles in some detail. In 
previous shows, we also used photos of tracks in the cloud chamber to introduce new particles such as the
strange quark.

\subsubsection{Safety}
The cloud chamber weighs about 80\,kg. It has two sets of handles on the side and can be carried by two (strong) 
people. If at all possible, it should be wheeled around on the lower box and transported in an elevator between floors. 
Appropriate care must be taken, if a radioactive source is used with the cloud chamber, which we did not in this show
Otherwise this experiment is self enclosed and safe.


\subsection{Beta$-$, and Beta$+$ Decay, Antimatter Live Onstage}
\label{app:betaplusminus}
\begin{figure}[h]
    \centering
        \includegraphics[width=0.8\textwidth]{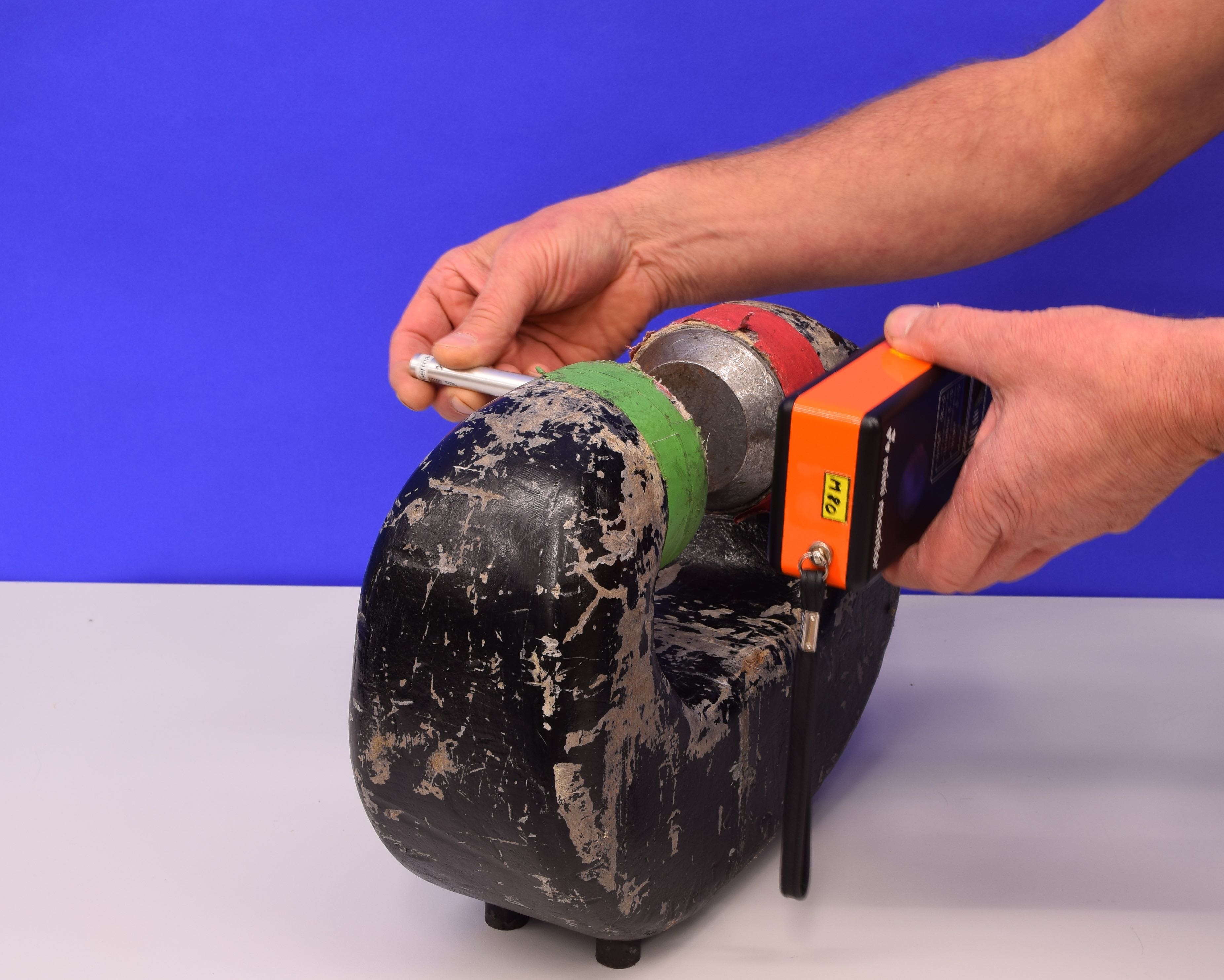}
         \caption{The horseshoe-shaped magnet. On the left, the silver, pen-like object is a beta emitter, on the right, 
         in orange and black, is a Geiger counter.}
    \label{fig:magnet}
\end{figure}

In this experiment we demonstrate anti-matter live onstage, using $\beta-$ and $\beta+$ emitters and a powerful magnet,
see Fig.\,\ref{fig:magnet}.

\subsubsection{History}
Radioactivity was discovered by Becquerel in 1896 \cite{becquerel}. He used a phosphorescent uranium salt
as a source. However, the detector, a photographic plate, was wrapped in thick black paper, which was most
likely thick enough to absorb all of the $\alpha$-rays. Thus the blackening of the photographic plate must have 
been due to the $\beta$-rays of the first daughter product in uranium decays: thorium-234. Becquerel could 
thus be considered the discoverer of $\beta$-rays. For a discussion of this history see for example \cite{Pais:1986nu}. 

In 1899, it was Rutherford who first showed that the ``uranium rays" had two distinct components with different 
penetration strengths; he denoted them as $\alpha$- and $\beta$-rays \cite{rutherford}. It took 10 years to 
establish that $\alpha$-rays are in fact helium nuclei. Bragg and Kleeman were able to show in 1904, that the 
emitted $\alpha$-rays had a fixed energy, a discrete spectrum \cite{bragg}. This was thus also believed to be 
the case for $\beta$-rays. There were many inaccurate experiments using photographic plates, which seemingly 
confirmed this conjecture. However the photographic plates have a highly non-linear response to the $\beta$-rays,
making a counting experiment difficult. It wasn't until 1914 that Chadwick was able to show in Hans Geiger's 
laboratory in Berlin, that the $\beta^-$-spectrum is continuous \cite{Chadwick:1914zz}. Remarkably the setup for 
his experiment is almost identical to our demonstration experiment, which we describe below. Chadwick had a 
$\beta$-emitter, containing lead (Pb-214) and bismuth (Bi-214), known respectively as radium B and C, at the time.
The $\beta$-rays were bent by a strong magnetic field and detected, this is the decisive point, by a Geiger counter.
In 1930, based on this experiment, Wolfgang Pauli \cite{Pauli} was able to postulate the neutrino, which was 
discovered in 1953 by Cowan and Reines \cite{Reines:1953pu}. Thus it took 18 years to establish the exact nature 
of $\beta$-rays. See \cite{Franklin:2000yt} for an extensive description of the history of beta rays and the discovery 
of the neutrino.

As discussed in App.~\ref{app:cloud-chamber}, the positron, the anti-particle of the electron, was discovered  by 
Anderson in cosmic rays in 1933 \cite{Anderson:1933mb} using a cloud chamber. Nuclear $\beta^+$-decay was 
first observed by Ir\`ene and Fr\'ed\'eric Joliot-Curie in 1934 after the $\alpha$-irradiation of Al$^{27}$ and B$^{10}$.
We note that already in 1928 Dirac predicted the anti-particle of the electron \cite{Dirac:1928hu}. It was the necessary 
consequence of combining the new quantum theory with special relativity.

\subsubsection{Materials and Technical Details}
We use two radioactive sources: Sr$^{90}$ (strontium) and Ge$^{68}$ (germanium). The Sr$^{90}$ (strontium) 
source is a $\beta^-$ emitter, releasing electrons at rate of 3.83 kBq. Sr$^{90}$ $\beta^-$ decays to Y$^{90}$ 
(yttrium) with a $\beta^-$ energy of about 550\,keV. Y$^{90}$ $\beta^-$ decays to Zr$^{90}$ with an energy of 
about $2.3\,$MeV. The Ge$^{68}$ (germanium) source is a $\beta+$ emitter, and has an activity of 1\,kBq. The 
Ge$^{68}$ undergoes electron capture to Ga$^{68}$ (gallium). The gallium is the source of positrons. It decays 
yielding $\beta^+$ with an energy of about 1.9\,MeV in the dominant 
decay mode. We use a strong horseshoe-shaped permanent magnet to bend the electrons ($\beta^-$) 
and positrons ($\beta^+$). The magnet has a permanent field strength of 210\,mT, measured between the 
pole shoes. Thus the radii of curvature for our $\beta^\pm$ are in the cm range. In the show, a hand-held 
Geiger counter with a loud speaker is used to detect the bent particles. The magnet is about 32\,cm wide,
26\,cm high and the gap is about 4\,cm. It weighs about 35\,kg.

\subsubsection{Presentation}
The setup is shown in Fig.~\ref{fig:magnet}. In presenting the experiment one person holds the $\beta-$ source 
and the Geiger counter. They are initially held in line well above the magnet. A loud clicking 
sound can be heard. Slowly the source is lowered into the gap of the magnet between the pole shoes, while 
the detector is lowered in parallel, but outside of the magnet. With the source at the point of strongest magnetic 
field the clicking vanishes. The polarization of the magnet should be set up such that the electrons are bent 
down towards the base of the magnet. This is to later avoid extra effects from pair annihilation of the positrons on the base of the magnet. The Sr$^{90}$ source is held 
fixed between the pole shoes, while the Geiger counter is first moved upwards and then downwards. The 
electrons are distinctly rediscovered having been bent downwards. Afterwards the experiment is repeated with the 
$\beta+$ scource, Ge$^{68}$, however now first ``looking" downwards, where the electrons were, and then discovering 
the positrons above the magnet. This demonstrates the opposite charge. 

\subsubsection{Safety}
Before using the sources onstage we inform the public of their nature and the required safety measures. The 
emission rate of both radioactive sources are below the limits for transport and public use.  We always take 
care that the probes are moved  to the setup very close to the time when the experiment  is carried out, and 
are moved back into the  radioactive safety box immediately after the experiment has been finished. It is very 
important not to drop the sources, and also to keep them away from the audience. Furthermore we take great 
care that the sources are never pointed at the audience.


\subsection{Ripple Tank: Resolving an Object with Water Waves}
\label{app:water-wave}
This is an experiment to demonstrate the resolving power of a wave depending on its wavelength.
We use a ripple tank illuminated through a glass bottom.

\subsubsection{History}
This is a standard experiment used in physics education to demonstrate the properties of waves,
such as reflection, refraction, and diffraction. We use it here to demonstrate the resolving power of
a water wave as a function of the size of the object and the wavelength of the water. This directly relates
to ideas in particle accelerators, as we discuss below. A film we have made of this experiment has also been 
used by Jon Butterworth in his public talks about his book `Smashing Physics' 
\cite{Butterworth:2014eka,private-JB}. 

\begin{figure}[t!]
\center
  \includegraphics[width=0.5\textwidth]{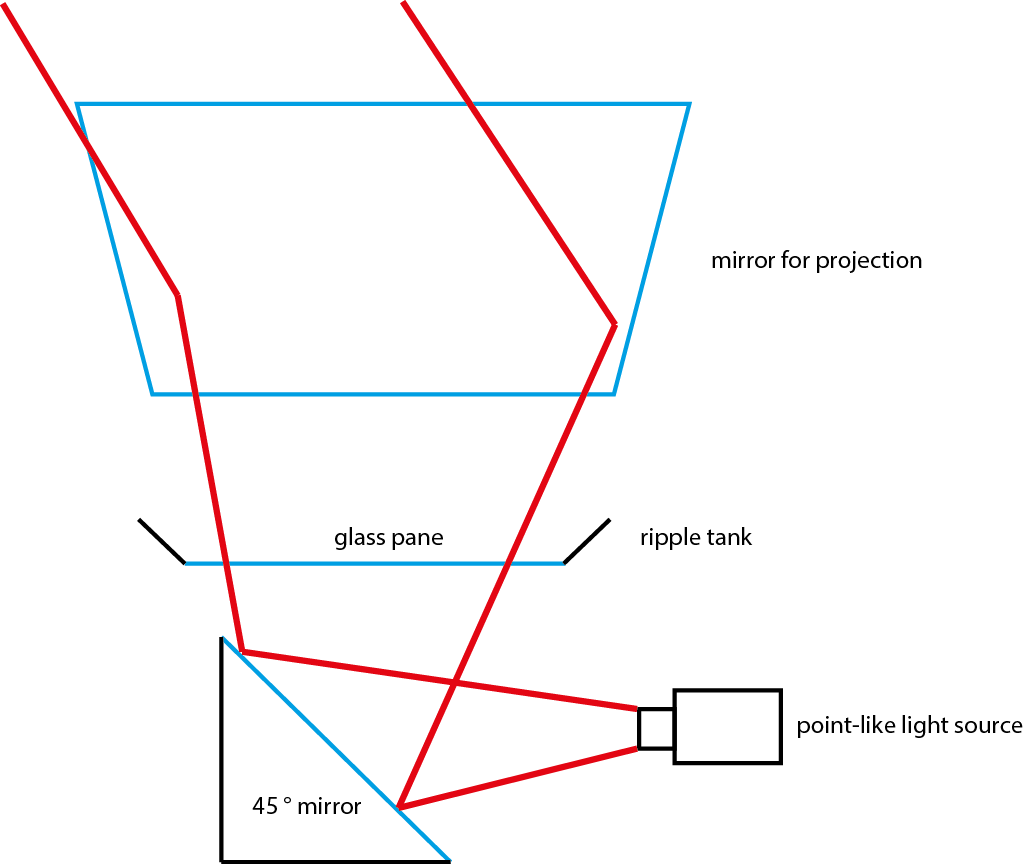}
	\caption{Schematic drawing of the setup for the ripple tank. On the right we have a point-like light source.
	A mirror deflects the light upwards to the ripple tank. The light enters the ripple tank through
	the glass plate bottom, is distorted by the water surface and then reaches a large mirror above
	the tank. This mirror redirects the light to a screen. 
		}
	\label{fig:drawing-waves}
\end{figure}

\begin{figure}[h!]
\center
  \includegraphics[width=0.7\textwidth]{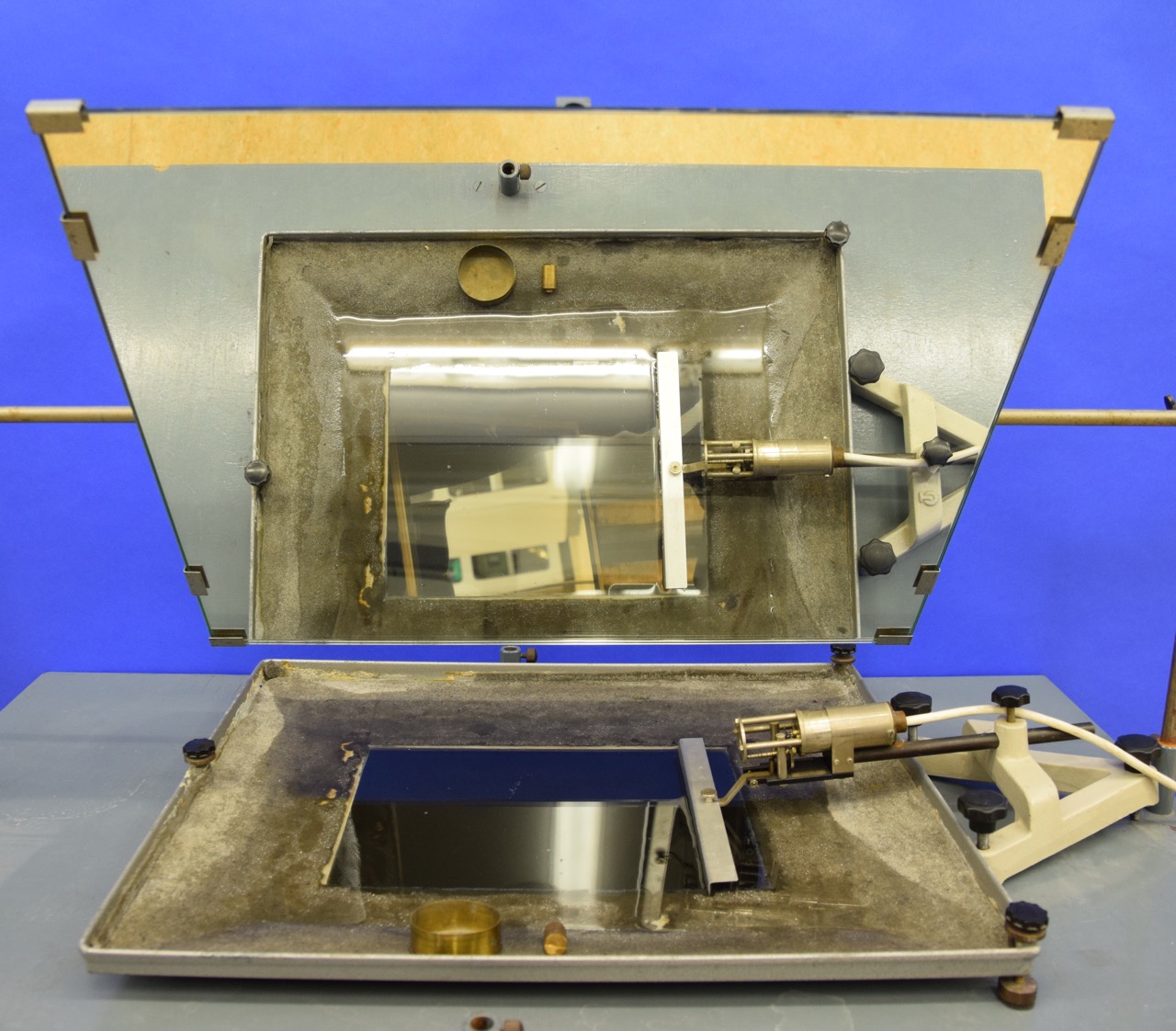}
	\caption{Our ripple tank with a square glass bottom. On the right, mounted on the tripod base, a cylindrical
	motor is connected to an aluminum bar. At the front, in the ripple tank, are the two
	cylindrical objects we place in the water. In the back hanging over the ripple tank is a large rectangular
	mirror. In the mirror you can see the ripple tank, and also look down through the glass bottom.}
	\label{fig:wave-dipper}
\end{figure}

\subsubsection{Materials and Technical Details}
In Fig.~\ref{fig:drawing-waves} we show a schematic drawing of our experiment, and in Fig.\,\ref{fig:wave-dipper} a photo of the 
experiment. On the right we have a strong point-like white light source. A mirror deflects the light upwards to the 
ripple tank. The ripple tank has a few centimeters of 
water in it. The light enters the ripple tank through the glass plate bottom, is distorted by the pattern on the 
water surface and then reaches a large mirror above the tank. This mirror is roughly at a 45$^\circ$\,angle and 
redirects the light to a screen\footnote{Thus a second screen is helpful during the show.}. Everything is mounted 
on a wheeled cart, and is thus easily wheeled on- and offstage. 

On one side of the ripple tank is a straight bar dipper, see Fig.~\ref{fig:wave-dipper}, which is connected to a motor, 
with adjustable frequency. When the motor is started plane waves travel across the full width of the tank and are 
clearly visible on the screen.

We also employ two cylindrical objects, with differing diameter: 2.5\,cm and 5.5\,cm. They are inserted vertically into
the water.

\subsubsection{Presentation} 
We briefly explain the setup. We then darken the auditorium and turn the wave generator on. The waves are clearly 
visible on the screen and the audience has an intuitive feel for what happens. We then insert the larger of the two
cylinders vertically into the water. It is visible as a dark circular shadow on the screen. At this point the size of the 
object and the frequency of the waves are chosen, such that the object leaves a clear distortion in the wave pattern
all the way to the back edge of the ripple tank. Next we insert the smaller cylinder; its diameter should be smaller than
the wavelength. Due to diffraction the waves bend around the cylinder, and are barely distorted behind it. By the time 
they reach the back of the tank they are plane waves again. We then increase the frequency of the wave generator, 
which reduces the wavelength. At some point, the distortion pattern clearly reappears behind the smaller cylinder all the way to 
the back of the tank. Thus at high frequency the smaller object can be resolved. 

In order to relate this to high energy physics, we note that a particle can be described as a wave. Smaller wavelengths correspond
to higher energies of the particle. We thus need powerful accelerators to look deep inside the protons.

In between, while observing the wave patterns, we play ``Surfin USA'' by The Beach Boys.

\subsubsection{Safety}
As water is used, it should be avoided to have open electrical circuitry close by.  The mirrors must be carefully wrapped
for safe transport.


\subsection{Dancing Paper Men and the Electric Field}
\label{app:dancingpaperwomen}

\begin{figure}[h!]
    \centering
    \includegraphics[width=0.85\textwidth]{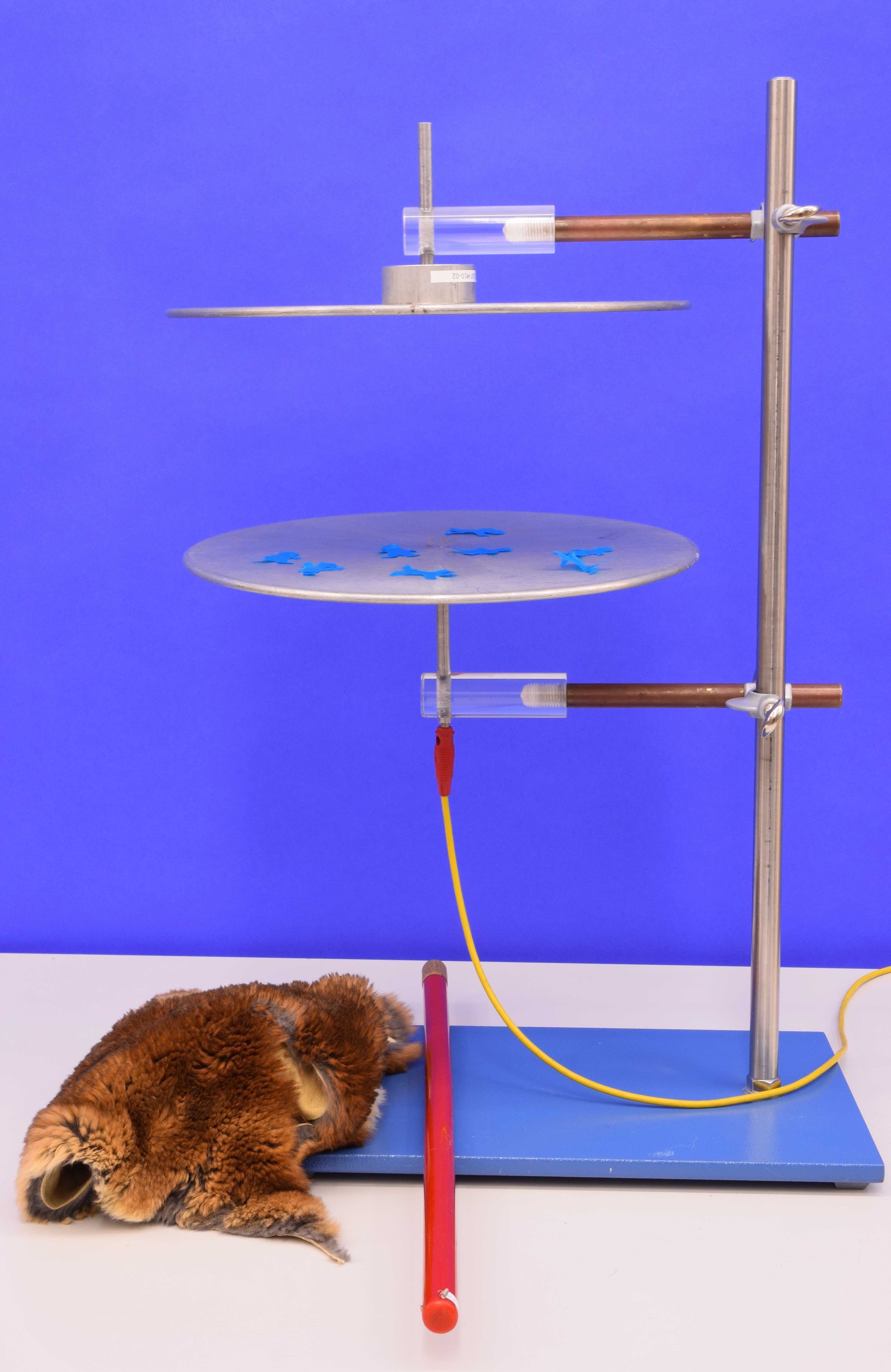}
    \caption{The complete experiment. On the left is the cat fur next to the red plastic rod. The two plates
    are electrically isolated from the support structure. The bottom plate is connected to Earth.}
    \label{fig:dancing-men-1}
\end{figure}

This is a very simple experiment which is also very entertaining. We often use it to introduce the concept of
the electric field, or  to simply demonstrate electrostatic effects.

\subsubsection{Materials}
The experiment consists of two metal plates, each $28$\,cm in diameter and $5$\,mm thick, which are placed parallel 
to the ground and about $20$ cm apart, see Fig.\,\ref{fig:dancing-men-1}. The plates are electrically insolated from 
each other and from the support structure. Additionally, the bottom plate is grounded, while the top plate remains 
disconnected. During the show we  use a plastic rod and an animal fur to create static charge from the triboelectric 
effect. This charge is applied to the top plate.

The experiment also requires some small tissue paper figures. While any shape will work, it is important to use
lightweight tissue paper (German: Seidenpapier). For our show in Italy we used small blue paper men in 
reference to the Italian national football team's history of constantly falling down without being fouled.

\subsubsection{Presentation}

To begin, we lay out the small paper men on the bottom plate of our large horizontal capacitor. The figures should be 
spread out so that they are not touching each other, otherwise they are more likely to clump up during the experiment,
and form a chain instead of bouncing. We then charge the top plate by first rubbing the plastic rod with the fur and
then drawing the plastic rod along the top plate. It takes several applications to sufficiently charge the upper plate.
The induced charge on the lower plate then charges the paper men and they are attracted to the top plate, see 
Fig.\,\ref{fig:dancing-men}. In the show, at this point, we typically play some rhythmic dance music. At the top plate, the 
paper men acquire the opposite charge and fall back down. We continue to apply some charge to the top plate, and the 
paper figures bounce back and forth, giving a dancing effect.

\subsubsection{Safety}
If the top plate is touched when charged one can get a small electrostatic shock, but there is no real physical danger.


\subsection{Linear Accelerator}
\label{app:linear-accelerator}

\begin{figure}[h!]
    \centering   
    \includegraphics[width=0.99\textwidth]{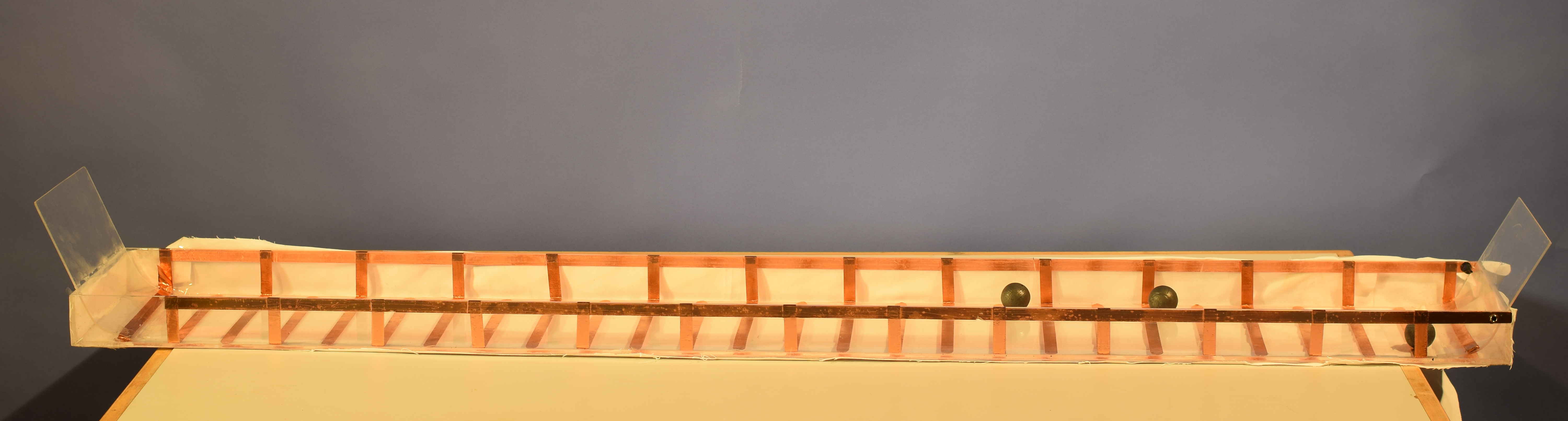}
 \caption{Our linear accelerator experiment. The two ramps can be seen at either end on the left and the right. 
 The box has been underlaid with white paper on the bottom and three sides for the photo.}
    \label{fig:lin-acc-1}
\end{figure}

The linear accelerator experiment helps illustrate some basic principles of accelerating particles. It's worth noting 
that here the electric charge of the small balls is changed, while in a true synchrotron the electric field itself would 
be flipped. The experiment is based on an idea by Jochen Dingfelder (Physikalisches Institut, Bonn) in light of the 
salad bowl accelerator below, App.~\ref{app:salad-bowl}, and was built by Sascha Heinz. The experimental setup 
is shown in Fig.\,\ref{fig:lin-acc-1}.

\subsubsection{Materials}
\begin{figure}[h!]
    \centering
      \includegraphics[width=0.99\textwidth]{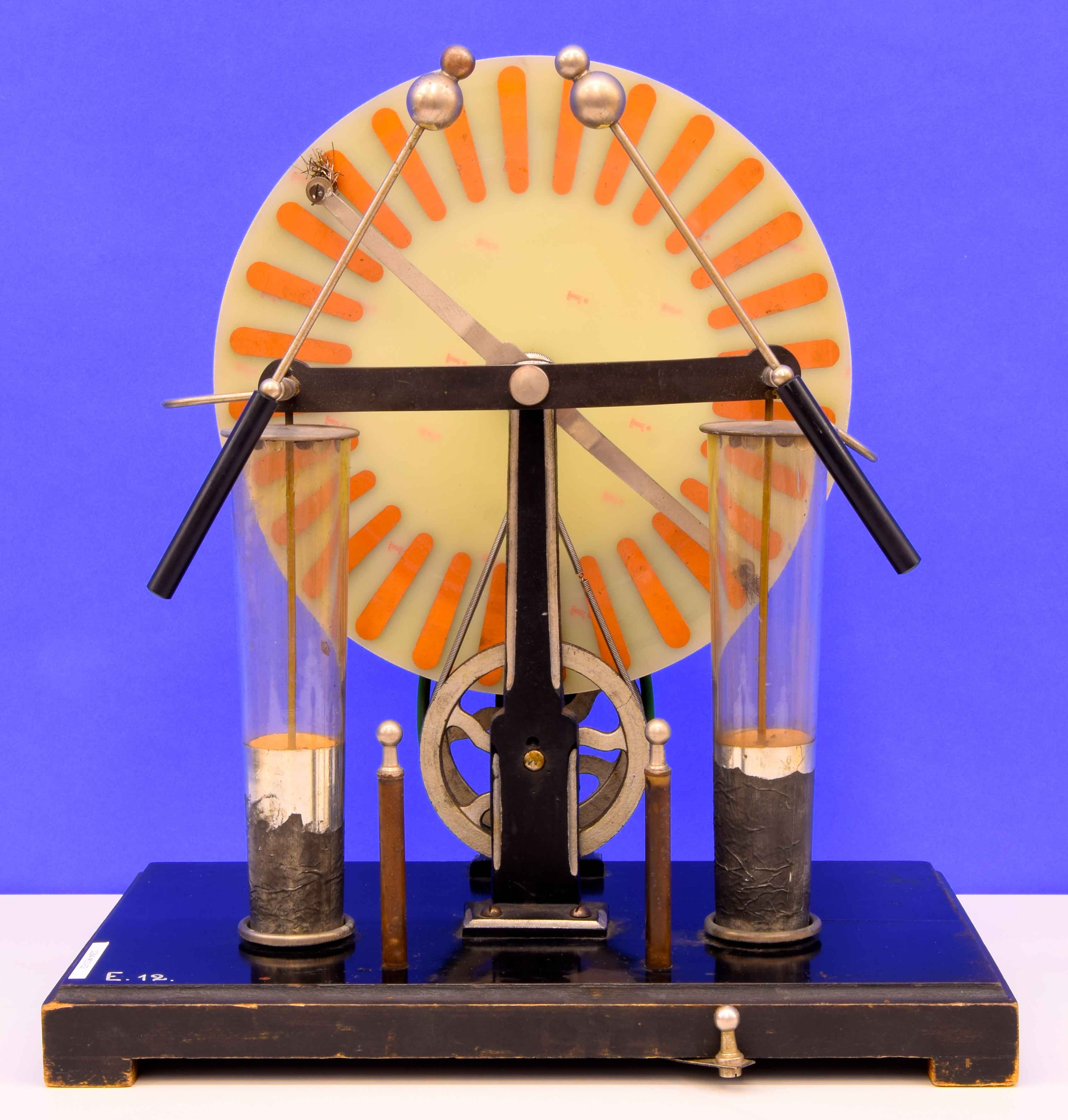}
  \caption{In the show we power the accelerator with this electrostatic generator out of our collection. It is a 
  Wimshurst influence machine, which produces a direct current (DC). We consider it less  abstract than a 
  modern power source. The cylindrical objects on the left and right are Leiden jars, which enable more charge 
  collection and thus higher voltages. This experiment is from around 1930. The rotating circular discs with the 
  copper contacts have been restored. }
  \label{fig:wimshurst}
\end{figure}

The body of the accelerator consists of a transparent plastic box of dimensions $140\,\mathrm{cm}\times
20\,\mathrm{cm}\times 6\,\mathrm{cm}$. At both ends there is a small ramp for the balls to roll up and back 
down again,  and avoid them bouncing out of the box, see Fig.\,\ref{fig:lin-acc-1}. Inside the box on the bottom, 
we have placed a series of $1\,$cm wide self adhesive conducting copper strips that run perpendicular to the 
longer side and are about $5\,$cm apart from one another. The strips are arranged so that every other strip is 
connected to a longer strip on the upper left of the long side, and the other strips are connected on the opposing 
side. On one end of the ramp we have affixed a small socket to each long strip, to easily connect the experiment 
to a power supply, see Fig.\,\ref{fig:wimshurst}. Demonstrating the effect requires at least one light conducting 
ball, but several balls are more fun. For our experiment we use Styrofoam balls coated in graphite.

\subsubsection{Technical Details}
For the experiment the two long strips are connected to an AC power source. The potential gap across each strip 
must be high; in our presentation we use around 25\,kV, created with the Wimshurst influence machine shown in 
Fig.\,\ref{fig:wimshurst}. However, the current is less than 0.5\,mA, so the experiment is safe to touch. The voltage 
can not be pushed much higher without creating discharges between the opposite metal strips. Once the 
potential is applied one or several balls can be rolled along the track.

When a ball touches one of the short metal strips on the bottom side
of the box, it acquires the charge of that strip. The ball  is then repelled by that strip, while being 
attracted to the two neighboring strips. The inertia of the ball guarantees that it crosses the strip and is continuously 
accelerated. In this way each metal strip serves as a gate. 

The analogy with a synchrotron has some limitations. In this experiment the ball changes charge at exactly the right 
moment. With a proton or an electron this is obviously impossible. In a synchrotron it is the electric field that is changed
just as the charged particle passes by. We demonstrate this in the mechanical analogy below: 
App.~\ref{hebe-beschleuniger}.

\begin{figure}[h!]
    \centering
      \includegraphics[width=0.99\textwidth]{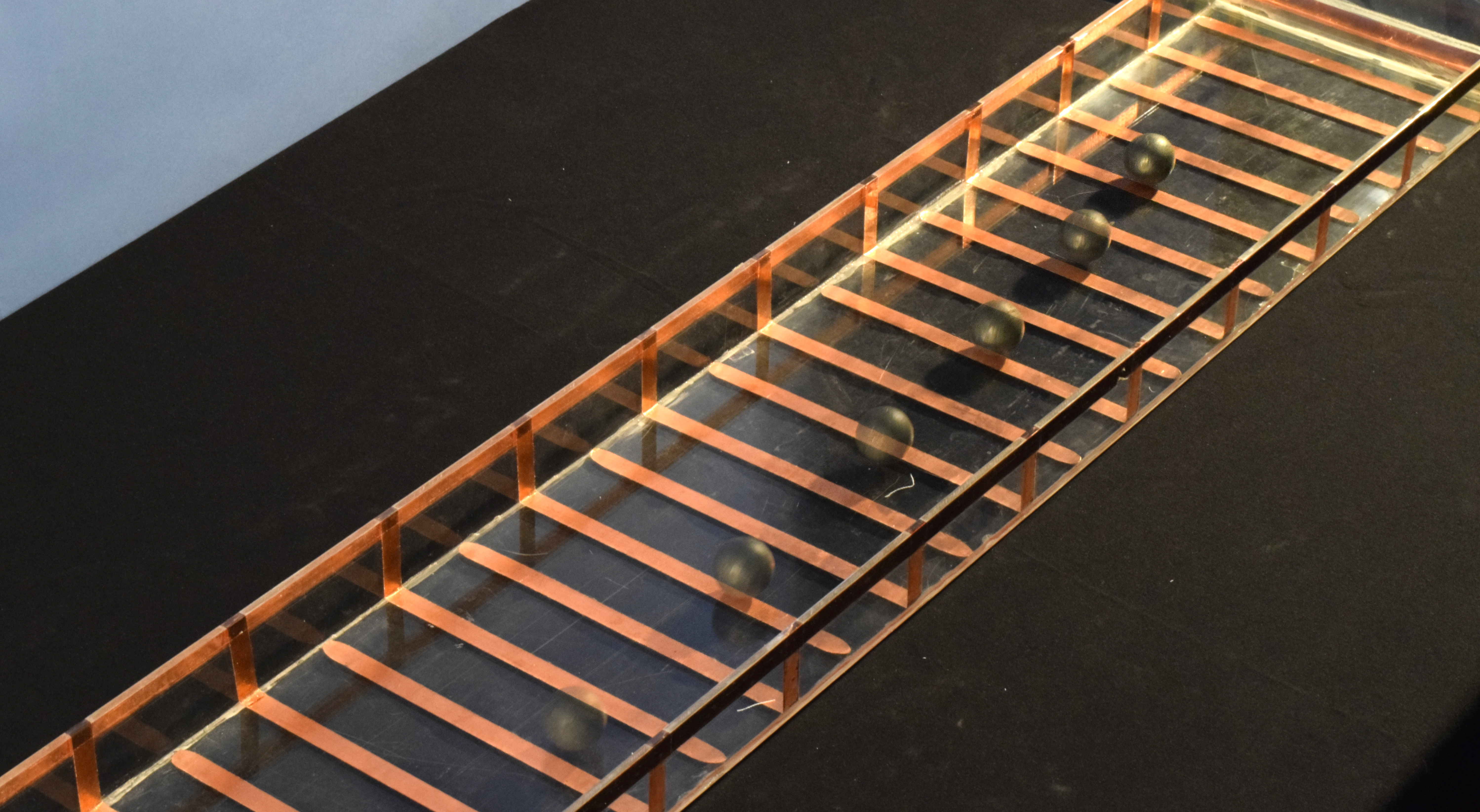}
        \caption{A multiple exposure picture of the linear accelerator in action.}
        \label{fig:lin-acc-3}
\end{figure}

\subsubsection{Presentation}
In the show we accompany the running of the experiment with music, typically something happy and upbeat.
Fig.\,\ref{fig:lin-acc-3} shows a multiple exposure picture of the linear accelerator in operation.


\subsection{Salad Bowl Circular Accelerator}
\label{app:salad-bowl}

The salad bowl circular accelerator experiment, shown in Fig.\,\ref{fig:salad-bowl-1}, employs the same technique 
as the linear accelerator above, App.~\ref{app:linear-accelerator}, but is circular.

\subsubsection{Materials}
\begin{figure}[h!]
    \centering
      \includegraphics[width=0.99\textwidth]{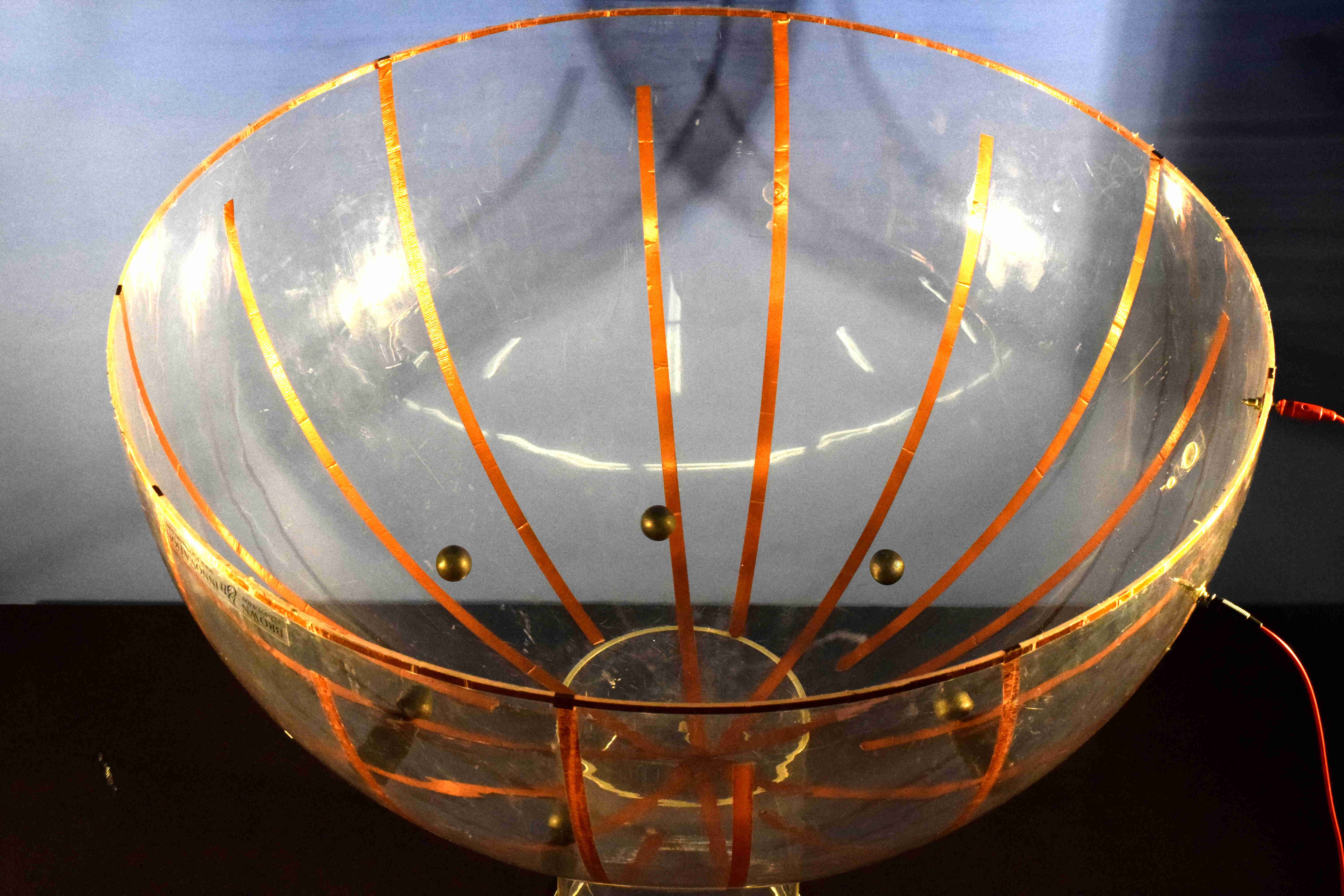}
        \caption{A multiple exposure picture of the Salad Bowl Experiment which highlights the path of the ball. On the 
        right the power hook-ups are visible.}
    \label{fig:salad-bowl-1}
\end{figure}

The body of the Salad Bowl consists of a plastic hemisphere of diameter 90\,cm, as can be seen in 
Fig.\,\ref{fig:salad-bowl-1}. Much like in the linear accelerator the inner surface of the plastic hemisphere is lined 
with 1 cm wide self adhesive conducting copper strips that trace out radii dividing the hemisphere into $16^{th}$. 
Every other copper strip is connected in the center. The other strips are connected to another copper strip that 
runs along the upper edge of the hemisphere. Thus the bowl alternates between one kind of strip and the other 
as the ball goes around. We use the same balls as for the linear accelerator in App.~\ref{app:linear-accelerator}.

\subsubsection{Technical Details}
The strips connected along the outside edge and those connected at the center need to be given a large potential 
difference. In our experience a voltage of 25\,kV and a current of 0.5\,mA is ideal. It allows the experiment to run well 
without posing any risk to the user. Higher voltages typically leads to discharges between oppositely charged strips. 
Once the potential is established, the ball should be started with an angular velocity, similar to starting a roulette 
wheel. The acceleration mechanism is identical to the linear accelerator, App.\,\ref{app:linear-accelerator}.

\subsubsection{Presentation}
As the ball rotates around the hemisphere, we play music, such as ``Get Around'' by the Beach Boys.\footnote{Be 
sure you have the legal rights to play the music in your shows. We pay the appropriate GEMA fees in Germany for each performance
of the show.}


\subsection{Mechanical Synchrotron}
\label{hebe-beschleuniger}

\begin{figure}[h!]
\center
  \includegraphics[width=0.85\textwidth]{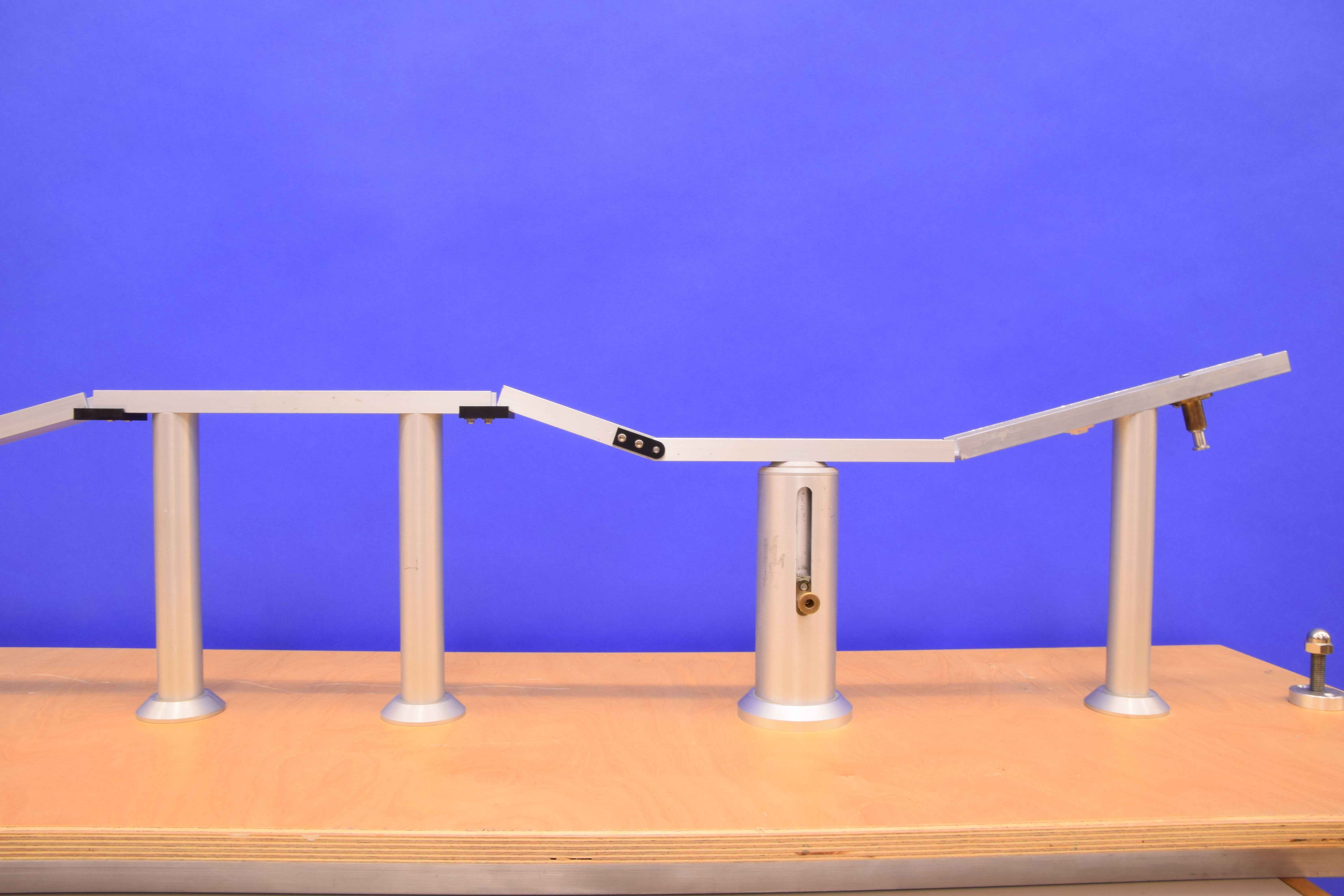}
	\caption{The first segment of the accelerator, with the release point on the right, and the first raising element.
	An even more detailed view is shown in Fig.\,\ref{fig:details-mechanical-sync-1}.}
\label{fig:details-mechanical-sync-2}
\end{figure}

\begin{figure}[h!]
\center
  \includegraphics[width=0.47\textwidth]{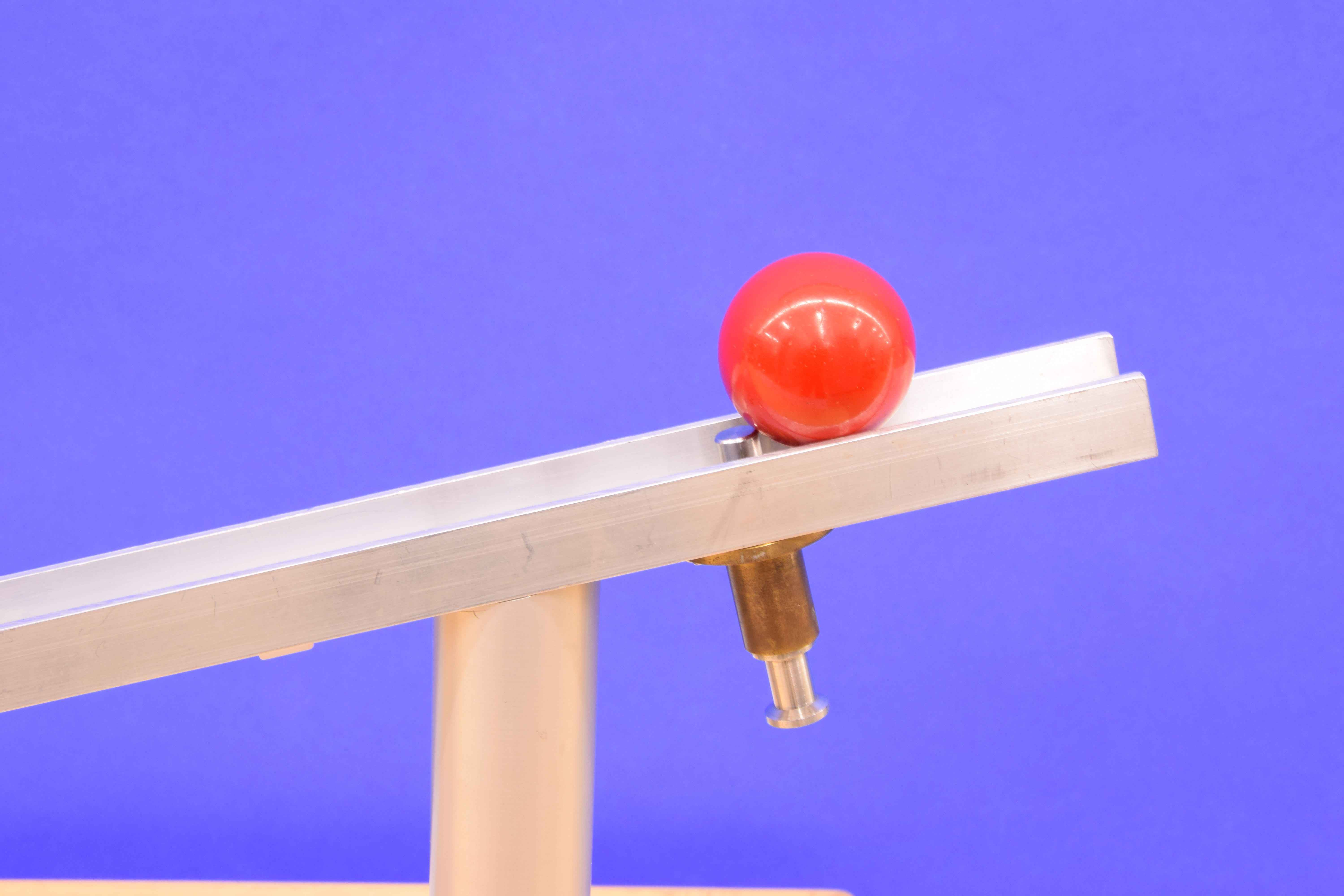}\; \includegraphics[width=0.47\textwidth]{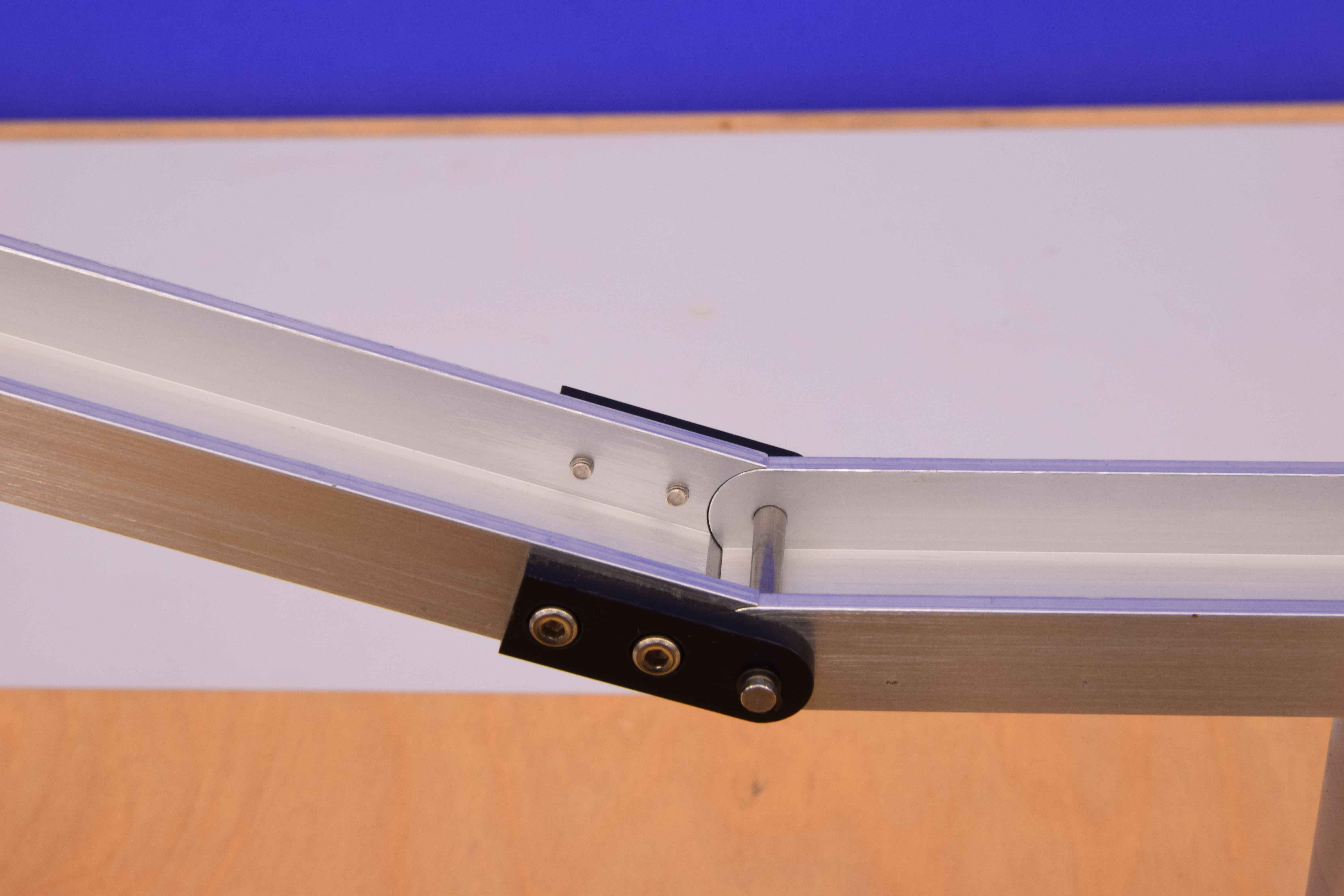} \\
  (a) \hspace{7cm} (b) \\
	\caption{(a) The release mechanism for the ball. (b) A hinge between two moving pieces.}
				\label{fig:details-mechanical-sync-1}
\end{figure}

This experiment is a mechanical model of a synchrotron accelerator. As presented here it is based on technical
drawings from DESY, Hamburg. It was developed in Bonn by Michael Kobel (TU Dresden). A full view is shown 
in Fig.\,\ref{fig:mechanical-synchrotron}. The Figs.\,\ref{fig:details-mechanical-sync-2} and
\ref{fig:details-mechanical-sync-1} show some details of the device.

\subsubsection{History}
Accelerators have played a crucial role in the history of particle physics, and the contruction of the LHC was 
essential for the discovery of the Higgs boson. This is why we have dedicated a complete section of the play to 
them, Sect.~\ref{sec:berkeley}.  Historically, accelerator developments and the directly connected discoveries 
were honored by several Nobel prizes: in 1939 (E. O. Lawrence), 1951 (J. D. Cockroft and E. T. S. Walten), 1959 
(E. Segr\`e and O. Chamberlain), 1961 (R. Hofstadter), 1984 (S. van der Meer) an 1990 (J. I. Friedman, 
H.~W.~Kendall and R. E. Taylor).

If an electrically charged particle is magnetically held on a circular path in a static electric field, its speed does
not increase. For an electron and a plate capacitor with two holes for the particle to move through, this may be 
intuitively clear. Between the plates the particle is accelerated, outside it is decelerated. A static electric field is
conservative. 

The idea of a synchrotron is to turn the electric field on when the charged particle enters the space between the capacitors, 
and off, just as it is about to leave. Thus it is accelerated at the right moment, but afterwards not decelerated. The electric 
field is no longer static, as it must be turned on and off, synchronized with the particle movement. Furthermore, as 
the particle gains energy, the magnetic field must be correspondingly increased. 

For the linear and circular accelerators in App.~\ref{app:linear-accelerator}, and App.~\ref{app:salad-bowl}, the 
electric field is held constant. However, the graphite coated balls, change their charge each time they pass over 
a brass strip. Our mechanical accelerator is thus a better analogy: we raise the potential, just as the ball passes 
by, \textit{i.e.} synchronized with its motion.

\subsubsection{Materials and Technical Details}
The accelerator consists of a linear aluminum rail track, resting on aluminum pillars. The rails have a width of 30\,mm 
and a height of 15\,mm. The thickness of the rails is 3\,mm. Three of the rail sections can be raised and lowered by 
levers, and the entire apparatus rests on a  wooden platform. The overall  length of the accelerator is 2.6\,m. The
wooden platform is 3.00\,m x 0.40\,m. It consists of two parts, which can be held together with clamps. The 
accelerated particle consists of a steel ball traveling on the rails. The release mechanism for the ball is shown on the 
left in Fig.\,\ref{fig:details-mechanical-sync-1}, a track hinge is shown on the right. We are happy to provide technical 
drawings upon request.

\subsubsection{Presentation}
In the show we have two people operate the accelerator, and one person, who releases the ball, see 
Fig.~\ref{fig:mechanical-synchrotron}. In the first run, the second person intentionally lifts his/her rail too early, thus 
decelerating the steel ball. Over the sound system, we play a sigh of grief: Ooooooh. In the second try it all works, 
and the ball is noticeably faster. To make it clear, we have a second successful run. It is also possible to measure 
the final speed of the ball, and display this to the audience.

Outside of a show this experiment is robust enough to use as a hands-on display. In that case students can run
a competition on who can produce the highest final speed.


\subsection{Flour Explosion}
\label{app:flour-power}
\begin{figure}[h!]
    \centering
    \includegraphics[width=0.81\textwidth]{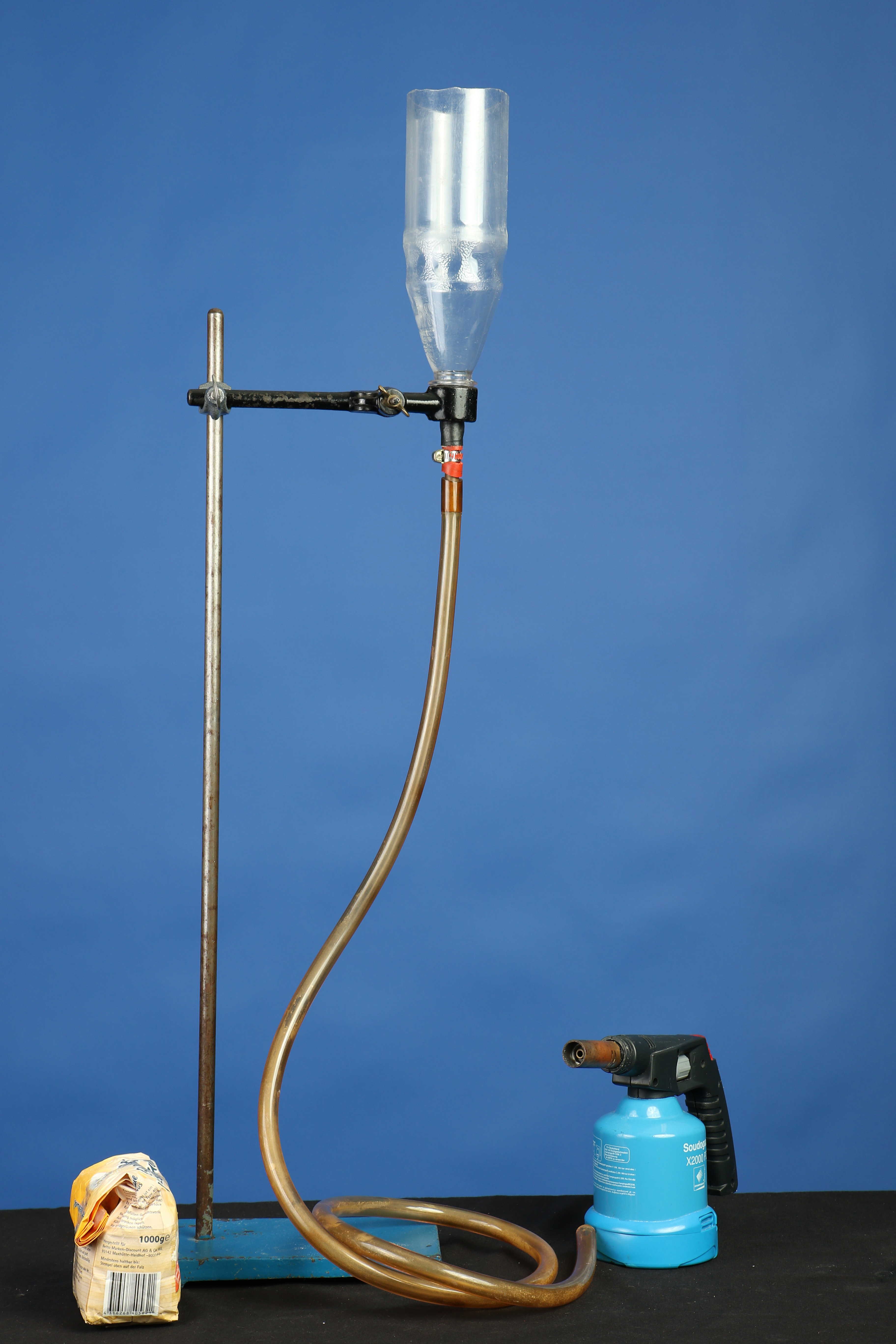}
    \caption{Close-up of our flour explosion experiment. On the left the conventional baking flour. On the right the gas
    blow-torch. the support structure holds the inverted and cut plastic bottle. A long tube is connected
    to blow into the bottom of the bottle from a safe distance.}
    \label{setup_flourexp}
\end{figure}

This experiment is added to the show for fun only and for a pun (``flower power") in the Berkeley part of the play, 
Sect.~\ref{sec:berkeley}. It has 
no connection to particle physics. Therefore it might be omitted if problems with safety regulations arise. It shows 
the flammability of powders. The experimental setup is shown in Fig.\,\ref{setup_flourexp}.

\subsubsection{History}
This experiment was developed for a previous physics show, on the physics of cooking. The 
experiment was used to have some extra entertainment while also to educate about the dangers of powder 
or dust explosions. A pile of flour on a plate does not burn. See also the discussion in \cite{dust-explosion}
on industrial and mining dust explosions.

We also considered a smaller version of this experiment within a glass box, but the effect was not as impressive 
and the ignition probability much lower. For the demonstration as discussed here, additional safety measures 
have to be taken, as described in App.\,\ref{firesavety}.

\subsubsection{Materials and Setup}
This experiment requires a rubber hose, about 1.5\,m long, with an inner diameter of about $5\,$mm, a blow torch, 
some duct tape, a sturdy plastic bottle and sufficient flour. The effect works better the smaller the grains of the flour. 
We fix the bottle with a structure, such as used for flask holders in chemistry labs. This in turn should be fixed to 
a table or cart to avoid uncontrolled fires.

In principle other powders can be used, such as pepper, paprika, toner powder, or lycopodium powder. Though 
these are usually quite expensive in larger quantities. Metallic powders can lead to stronger explosions.

We cut off the bottom of the plastic bottle. Compared to a standard funnel the bottle has the advantage of an 
elongated body which directs the flour upwards. The hose is inserted in the narrow end of the bottle and sealed 
well with the duct tape. A picture of this construction is shown in Fig.~\ref{setup_flourexp}. The bottle is then placed 
in the holding structure and filled with flour. About $100\,$g of flour is sufficient. Only the conical part of the bottle 
has to be filled with flour for the best effect.

\subsubsection{Presentation}
After motivating the experiment, one person briefly explains the setup while the other fills the flour in the 
bottle. Afterwards they both put on safety glasses and gloves. We also point out that this experiment should 
{\bf not} be performed at home. One protagonist holds the ignited blow torch with an elongated arm above the 
bottle opening. Some music is started, possibly hard-rock with some fire context (Rammstein, Bloodhound 
Gang).  As the other protagonist blows in the far end of the hose, both protagonists keep a safe distance 
from the bottle, see Fig.\,\ref{fig:sina-dustin}. After the first run there is usually enough flour left in the bottle 
that when shaken a second one can be done without refill.

\subsubsection{Safety}
\label{firesavety}
This experiment involves large flames. The distance to the audience should be at least 3\,m. The bottle needs to be 
securely fastened to an immobile structure to avoid it falling in any undesired direction. Above the bottle there should be 
more than 2.5\,m space for the flame. The material usually burns up in the air, nevertheless a non flammable floor covering 
should be used. This also allows for easy removal of the leftover material. Both presenters should wear safety glasses 
and gloves.


\subsection{Vacuum Cannon}
\label{app:vacuum-cannon}
\begin{figure}[h!]
    \centering
    \includegraphics[width=0.94\textwidth]{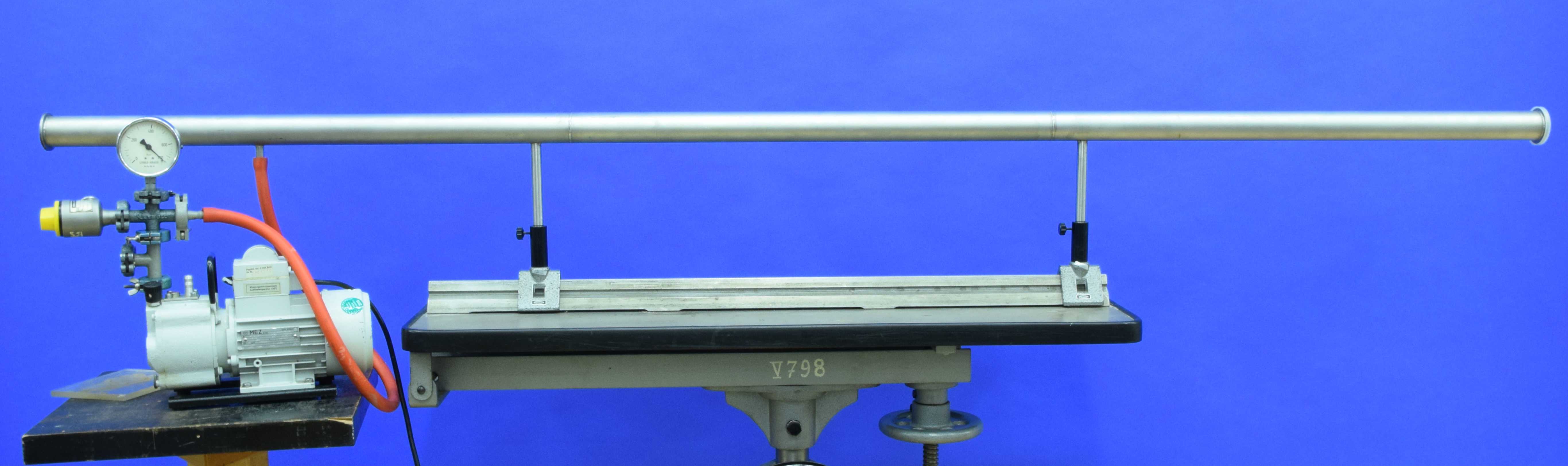}
    \caption{An overall view of our vacuum cannon, without the detector.}
    \label{fig:vacuum-cannon-1}
\end{figure}

The vacuum cannon, see Fig.\,\ref{fig:vacuum-cannon-1}, also known as vacuum bazooka, or ping-pong 
cannon is a well-known demonstration experiment. It is listed as PIRA 2B30.70 in the Physics Instructional 
Resource Association\footnote{We were not able to find it at the official site \cite{pira}, but found a 
corresponding description under this number at \cite{pira-vacuum-cannon}.} \cite{pira}. 

\subsubsection{History}
The first vacuum cannon was built by Otto von Guericke. It is first mentioned in 1672 in chapter 29 of the third 
book of \cite{guericke} and  called a ``Windb\"uchse". It was certainly built much earlier. Fig.\,16 in \cite{guericke} 
depicts a beautiful engraving of the historical experiment. Von Guericke also invented the first piston vacuum 
pump in 1654 \cite{guericke-vacuum-pump}, and performed for example his famous experiment with the 
Magdeburg hemispheres pulled by horses, see Fig.\,11 in the 23rd chapter of the third book of \cite{guericke} for 
a wonderful full page engraving with 16 horses. Other  Windb\"uchsen had been built earlier, possibly already in the 15th  
century in N\"urnberg, however these were based on overpressure \cite{zoellner}. 

As a fun modern demonstration experiment the vacuum cannon is mentioned in 2001 in 
Ref.~\cite{cannon-orig}. 
Modifications have been discussed in the literature \cite{vacuum-cannon-1,vacuum-cannon-2}, including also a 
theoretical analysis \cite{vacuum-cannon-th-1} resulting in an upper bound on the exit velocity as a function
of the length $L$ of the tube
\begin{equation}
v_{max}(L)=\sqrt{\frac{P_0}{\rho}} 
\left[ \frac{L}{L+\frac{m}{\rho A}} \sqrt{1+2\frac{m}{\rho A L}}\right] \;
\stackrel{L\,\gg\,\frac{m}{\rho A}}{\xrightarrow{\hspace*{1.6cm}}}\; \sqrt{\frac{P_0}{\rho}}\approx 290 \, \frac{m}{s}\,.
\end{equation}
Here $P_0$ is the atmospheric pressure, $m$ is the mass of the projectile, $\rho$ is the density of air, $A$ is 
the cross sectional area of the tube, $L$ its length, so that $\rho A L$ is the mass of the air in the tube after 
the projectile has been emitted. The asymptotic speed on the right is achieved for large tube lengths. The 
resulting numerical value assumes normal air pressure and density conditions ($P=$101\,kPa and 
$\rho=$1.25\,kg/m$^3$). A numerical analysis of the projectile acceleration, including fluid dynamics was 
performed in a Master's thesis in Ref.~\cite{vacuum-cannon-numerical}. See also the related work in 
\cite{windbuechse}.

The design widely discussed in the literature employed a 2 to 2.5\,m PVC or acrylic tube and a ping-pong ball as 
a projectile. The ends were sealed with tape. In order to launch the cannon the tape at one end was 
pierced with scissors or with a sharp knife.

In 2006, we designed and built our own cannon.\footnote{We heard of the vacuum cannon via Tilman Plehn, who told us 
of one built at Boston University.}  We have modified the setup, as we describe 
below, and have also changed the outlook or context of the experiment, to use it as an analogy for a particle physics, 
fixed-target experiment. Thus we also modify the exit setup, including a ``target" and a ``beam dump",
see Fig.\,\ref{fig:vacuum-cannon-3}.

\begin{figure}[h!]
    \centering
    \includegraphics[width=0.89\textwidth]{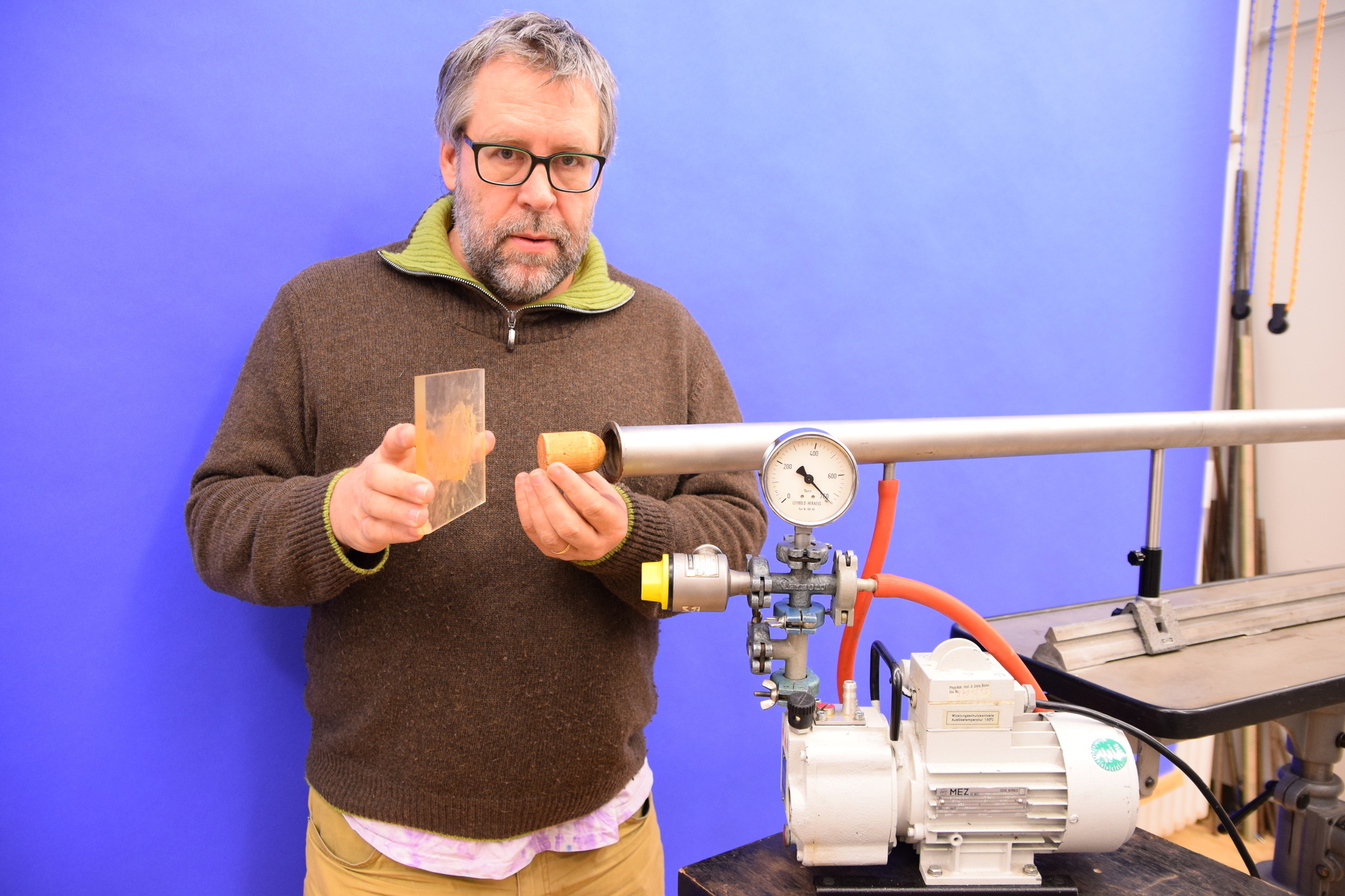}
    \caption{One of us (HKD) with the injection side of the vacuum cannon. The vacuum pump as well as the hook-up via
    the red rubber hose is shown. The wooden projectile is just at the entrance to the pipe, together with the greased
    plastic plate seal.}
    \label{fig:vacuum-cannon-2}
\end{figure}

\subsubsection{Materials and Technical Details}
An overview is shown in Figs.\,\ref{accel-1}, and \ref{fig:vacuum-cannon-1}. More details can be seen in 
Figs.\,\ref{accel-2}, and \ref{fig:vacuum-cannon-2}. The main cannon consists of a 2.2\,m long V2a steel (non rusting) 
pipe. It has an inner diameter of 40\,mm and a wall thickness of 2.25\,mm. At each end of the pipe there is a flange 
welded to the pipe, providing a larger area to seal the tube. In the middle, a connector has been welded to the pipe, 
where the vacuum pump is attached. Two further metal rods are welded to the long pipe, for mounting on an optical 
bench. See also Fig.~\ref{fig:vacuum-cannon-1}. At the exit end a simple plastic cap seals the pipe. At the entrance, 
we use a 20\,cm x 20\,cm plexiglas plate as a seal. At both cases we use lubrication grease to improve the seal. After 
evacuation, in order to fire the cannon, we pull down the plexiglas plate. Not much force is needed. We find this 
much better than sealing the pipe with tape and puncturing it with scissors or a sharp knife. It 
also allows us to reuse the cannon fairly quickly.\!\footnote{In the performance at ICTP Trieste, this was useful as 
the cannon misfired on the first attempt, a rare occurrence, and we could repair it by the DESY part, where it reappears. }

We use either a 130\,W, 2.5\,m$^3$/h or a 370\,W, 12\,m$^3$/h vacuum pump; both have a pressure gauge, which 
we display via camera onto the large screen in the shows. With the stronger pump the evacuation time is about 
10\,s, with the weaker pump about 1\,min. We usually prefer the weaker pump, as, when accompanied by 
appropriate music, it adds some drama in a live show, as the pressure gauge slowly drops. Furthermore the stronger 
pump requires three-phase alternating current, which is not available at most places we perform.

The cannon is pointed at a target, which is placed very close to the cannon exit, see Fig.\,\ref{accel-2}. We 
use ``quark", a dairy
product, which is readily available in supermarkets in Germany. It comes in 250\,gr or 500\,gr plastic containers;
we use the 500\,gr packages. It is similar to cottage cheese in the USA, or {\it frommage blanc} in France. We 
choose it, since the particles inside of a proton are also called quarks.\!\footnote{Howard Haber recalls to have 
heard the story, that James Joyce, heard the phrase ``{\it Drei Quark f\"ur eine Mark}" (transl.: ``three (packets) 
of quark, for one mark") from a market vendor. This then supposedly became ``Three quarks for Muster Mark!" in 
Finnegan's Wake. Joyce lived in Z\"urich for an extended period and is in fact buried there. However,  he must have
heard this while travelling in Germany, as Switzerland has had the {\it Franken} as its currency for about 200 years. 
This story is mentioned in Ref.~\cite{drei-quark}, as apocryphal. They also mention that supposedly there was a 
German commercial slogan to the above effect at the time. } Furthermore when struck by the projectile, the quark 
gives a good splatter. We often enhance this by adding food coloring to the quark.

\begin{figure}[h!]
    \centering
    \includegraphics[width=0.89\textwidth]{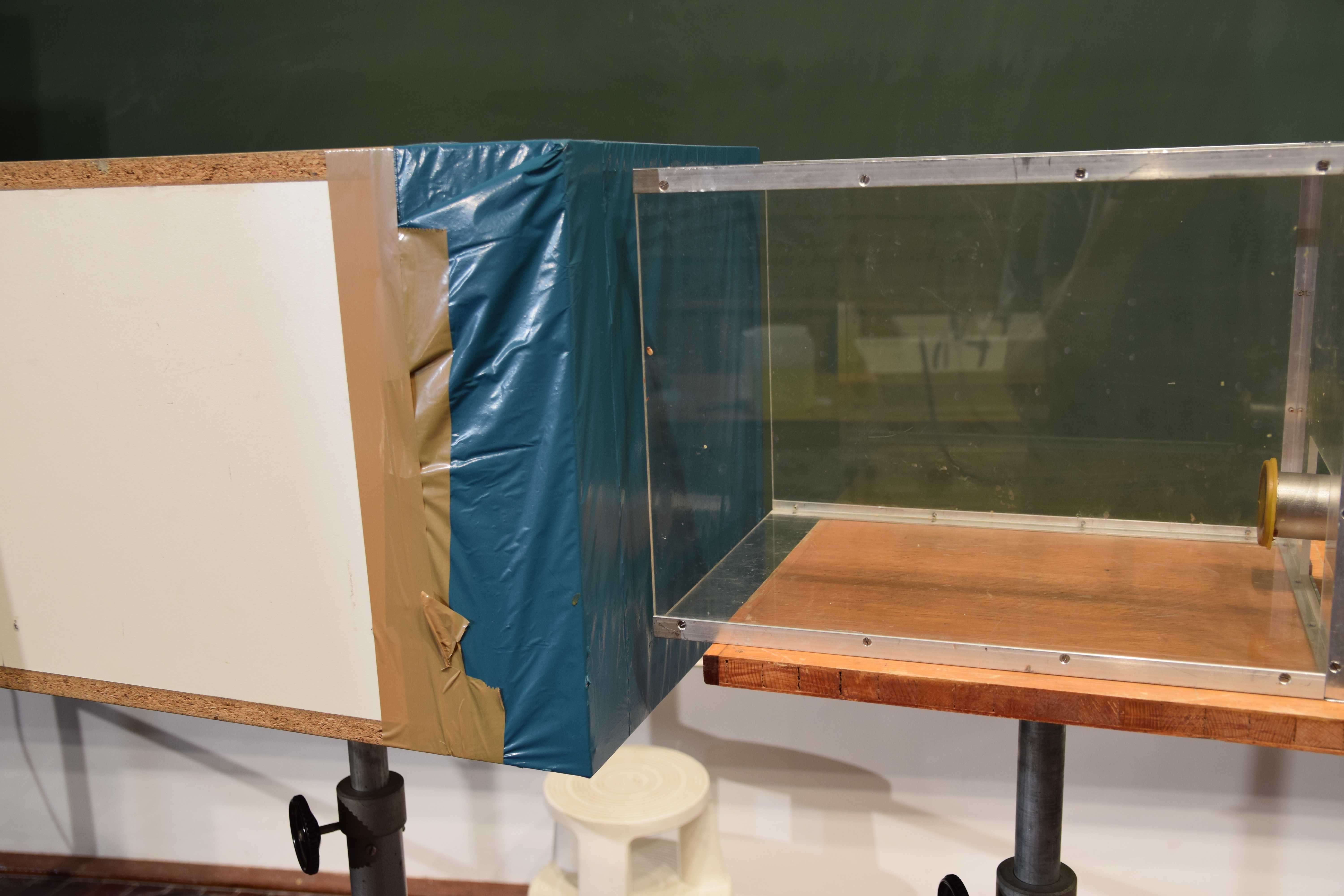}
    \caption{A detail of our vacuum cannon experiment. The end of the vacuum cannon is on the right, entering the plexiglas box. We place
    the Quark paket just in front of the pipe. On the left is the wooden box we use as a beam dump. In blue is the 
    garbage bag, taped to the box.}
    \label{fig:vacuum-cannon-3}
\end{figure}

The quark target is placed inside a transparent plexiglas box of dimensions 30\,cm x 30\,cm x 42\,cm, see 
Fig.\,\ref{fig:vacuum-cannon-3}.. The material is 6\,mm thick. On the upstream side, we leave an opening of 
dimensions 10\,cm x 11\,cm, where the projectile enters. We then place the end of the cannon inside the plexiglas 
box, directly against the quark container. The back side of the box, facing away from the cannon is completely open. 
This plexiglas box acts as our detector, recording the single event we create. It is fixed to the wooden platform it rests 
on with a screw clamp.

Behind the plexiglas box, on a separate stand, we have a wooden box of dimensions 40\,cm x 40\,cm x 60\,cm, made
of coated 19\,mm chipboard. This box is filled with toilet paper rolls, which absorb the shock of the projectile. The 
side of the wooden box facing the cannon is open. We cover it with a tough garbage bag, taped to the box. This also 
prevents the quark from entering the box and makes for easier cleaning. This box serves as our ``beam dump"\!,
see Fig.\,\ref{fig:vacuum-cannon-3}.

As a projectile we use a wooden ball with 40\,mm diameter, or a more bullet shaped one with the same diameter
and 68\,mm length. It can bee seen in Fig.\,\ref{accel-2} and Fig.\,\ref{fig:vacuum-cannon-2}. We have not measured 
exit velocities. 

After all seals have been applied and checked the tube is evacuated.  After the desired pressure is reached (about 
10\,mbar), as read on the pressure gauge, the pump is switched off to protect it from the shock of the inflowing gas. 
Depending on the type of the pump an intermediate valve might be needed for protection.

Finally, the plexiglas plate at the entrance end is removed in a fast downward movement. The inflowing air propels 
the projectile to the front. It pushes out the front cover and then hits the desired target. The quark container 
is ripped apart and the quark splashes all over the inside of the plastic ``detector", see Fig.~\ref{accel-2}.

Since quark is not readily available outside Germany (in Austria it is called ``\textit{Topfen}"), when we go on 
the road, we must buy the quark in Bonn and cool it during transport.\footnote{Apparently quark is now sold at 
Whole Foods in the USA \cite{whole-foods}.}

\subsubsection{Presentation}
The presentation of this experiment is usually enacted quite dramatically. Depending on the available space
on- and offstage the cannon is either onstage throughout the relevant sections of the play or wheeled in
on short notice. Similarly for the detector and the beam dump. If they are onstage beforehand for an extended 
period of time, they are covered with a black cloth. The presenter introduces it as a ``real particle accelerator'', 
which is revealed, by removing the cloth. Then the presenter explains the apparatus, in particular the 
vacuum pump, and how it is connected. If the apparatus is shown on the screen, the parts can be highlighted 
with a laser pointer.

Next a proton target is required. The quark is delivered by a crew member, or in this show by the caretaker. He 
holds it into the camera so the ``Quark" label can be easily read via the projection. It is placed inside the detector 
and the cannon is carefully lined up, with the exit just in front of the quark, inside the detector. We have several 
variations on how to present the accelerator, and also the quark. For example the latter can be brought in from 
offstage, held up high, almost like a religious relic. The presenters of the experiments wear lab coats and put on 
safety glasses before starting the vacuum pump. Then the projectile is inserted, the openings are sealed and the 
pump is started, giving the typical loud noise of a vacuum pump. Depending on the show and the 
speed of the pump, this time can be covered by more music or explanations of the host, \textit{i.e.} 
referring to the discovery of point-like particles in the atomic nuclei by fixed-target scattering experiment.

When the desired level of pressure is reached, the pump is switched off and the rear opening of the tube is uncovered. 
We usually stage that with some more action like adding personal protection, \textit{i.e.} goggles, and someone 
on stage yelling ``Fire!''.

The projectile hits the target, spilling quark in all directions, \textit{cf.} Fig~\ref{accel-2}. After the dust has settled one 
of the hosts acts as an analyzer, or does a read-out by tasting the spilled quark. Due to the German word ``quark'' 
this is an excellent opportunity for some cheesy\footnote{No pun intended.} puns, \textit{e.g.} ``Hmm, tastes strange!''.
See also the extensive dialogue in the play in Sect.~\ref{sec:desy}.

If one chooses the target appropriately the 
result can be spectacular. In the show we use quark, but we have also used watermelons, for example.

\subsubsection{Particle Physics Analogy}

In our show the vacuum cannon is used as a demonstration of a fixed-target scattering experiment. The cannon itself is the 
accelerator, the quark is the target, while the projectile represents the beam particle. The box in the back is the beam 
dump, see Fig.\,\ref{fig:vacuum-cannon-3}.

\subsubsection{Safety}

The projectile leaves the cannon with significant velocity. Care must be taken to ensure the cannon is only fired 
into the beam dump and never into open space. The spillage of the target material can often not be entirely 
contained within the target housing. Some backfire through the opening for the projectile can occasionally reach the 
presenters and more quark is spilled when disassembling the experiment. Hence, care should be taken to 
protect carpets or any other flooring that might be hard to clean or any nearby objects that cannot stand some 
drops of cottage cheese.

\subsection{Testicle Model}
\label{app:testiclemodel}
This is a simple model of three hard balls wrapped inside a balloon, as an analogy for the quarks
inside a nucleon.

\subsubsection{Materials}
\begin{figure*}[h!]
    \centering
            \includegraphics[width=0.47\textwidth,height=5.3cm]{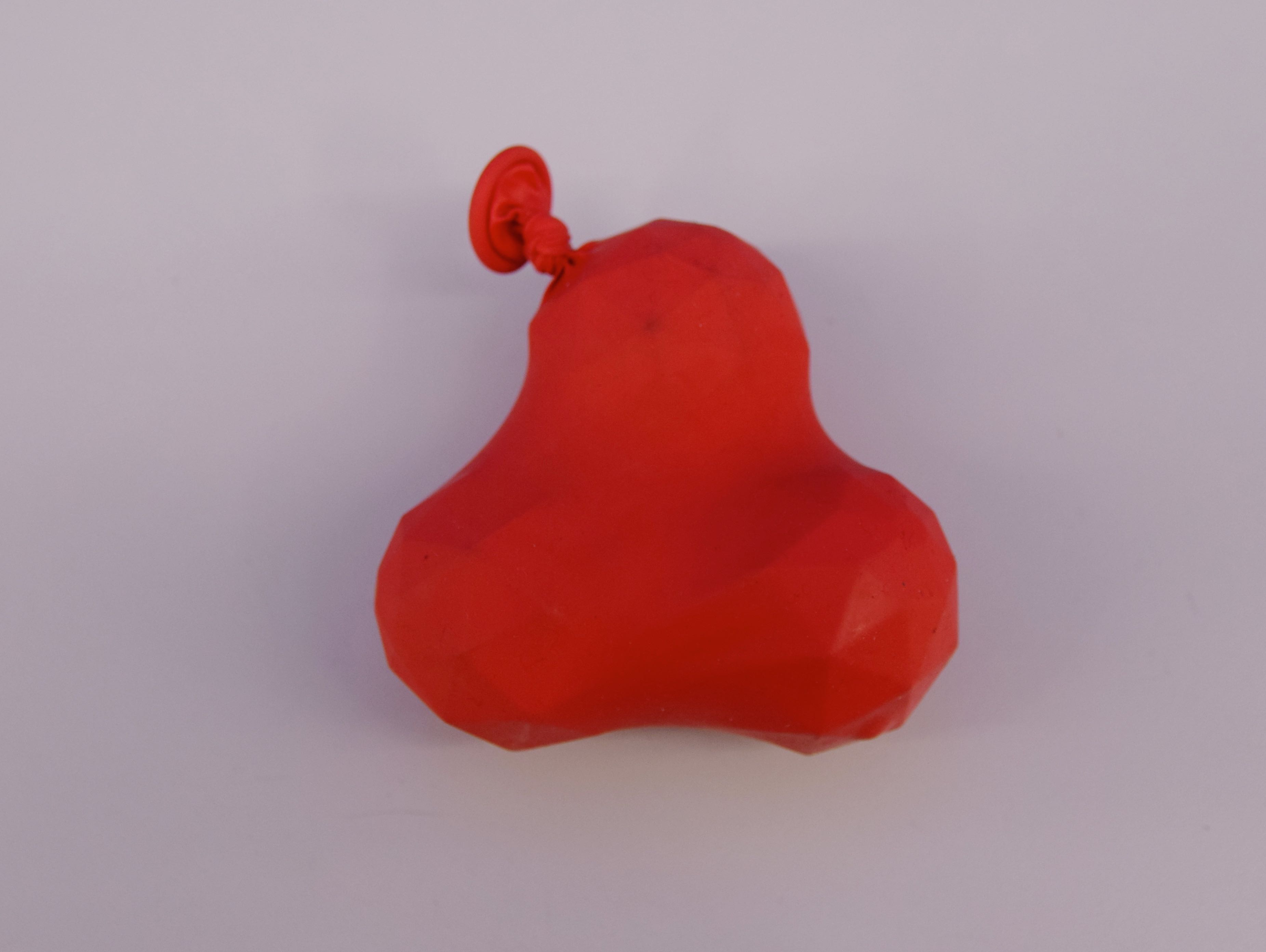} \,
        \includegraphics[width=0.47\textwidth,height=5.3cm]{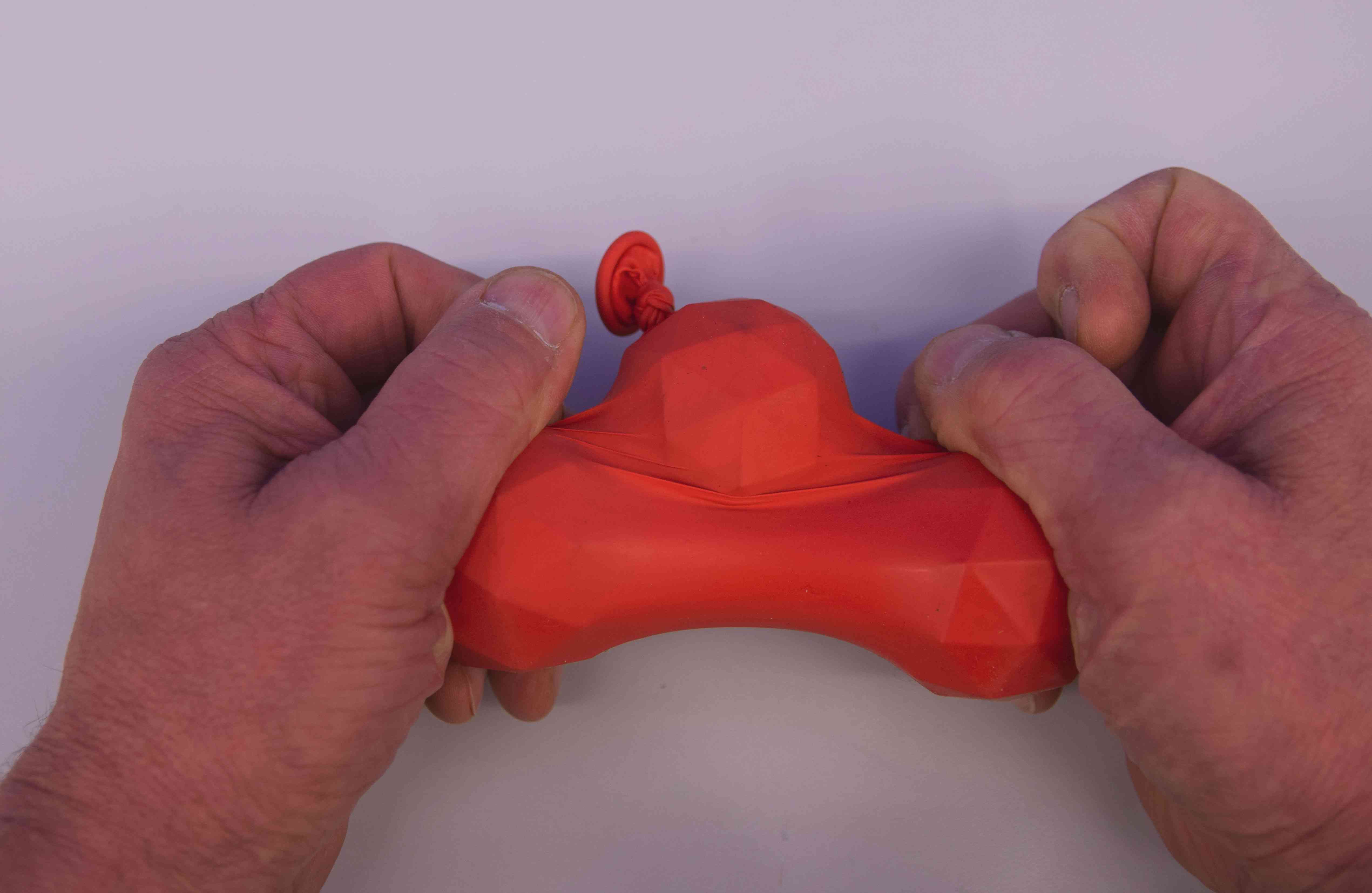} \\
(a) \hspace{6cm} (b)
    	\caption{(a) The testicle model (balloon bag model) for the proton. (b) Stretching the model.}
	\label{fig:testicles}
\end{figure*}

This is a simple self-made experiment: three wooden balls with a diameter of approximately 5\,cm are squeezed
inside an uninflated balloon and thus held together. It is shown in Fig.\,\ref{fig:testicles}. The surface of the balls 
should be smooth, so that the skin of the balloon is not ruptured.

\subsubsection{Technical Details}
The model represents the effect of the strong force in hadrons. The quarks, represented by the wooden balls, are 
almost free in their movements as long as they are close together. But if you try to remove one quark, the confining
force rises linearly with the separation, and you find it is not possible, see Fig.\,\ref{fig:testicles}.

\subsubsection{Presentation}
Since the model is relatively small, the stretching of the balls in the balloon is filmed with a camera and 
shown on screen. The experimenter pulls the balls away from each other and lets them snap back. By 
this, it is demonstrated that the nucleon consists of three parts, which can be detected, but not separated. 
In the show, as a joke, we sometimes call this the ``testicle model". We then play an ``ouch" noise with the 
fast recoil of the balls to imply pain. 

\subsubsection{Safety}
If you are careful with the snapping back of the balls inside the balloon, this experiment is harmless.

\subsection{Balloon Tossing}
\label{app:balloontossing}
This experiment is an analogy to show that a particle with substructure, such as the proton
behaves differently from a homogenous particle.

\subsubsection{Materials and Technical Details}
For this experiment, two medium sized balloons of different color are needed. Once inflated they should 
have a diameter of about 25\,cm. In one balloon we inserted a bouncy ball (super ball, with a diameter of about 5\,cm) before 
inflating. Bouncy balls are typically smooth, so that the balloon is not punctured.

\subsubsection{Presentation}
This experiment is typically performed together with a volunteer from the audience. We do not disclose to the 
audience in advance that one balloon has a bouncy ball inside. The audience member together with a member 
of the show stand approximately 5\,m apart on either side of the stage. First the empty balloon is tossed back 
and forth. It flies like a normal balloon slowing down and eventually dropping vertically, barely covering the 5\,m. 
This represents the early version of the proton which behaves as a homogeneous mass. Next the show member 
tosses the other balloon. It flies erratically and is very difficult to catch. As this behaviour is unexpected, the 
participant from the audience is usually suitably surprised. The balloon is then tossed back, again flying erratically. 
The balloon can then be punctured, revealing the bouncy ball inside. This second proton, closer to today's view, 
has substructure which is revealed through its macroscopic (compared to the size of the balloon) behaviour. This 
substructure, which takes the form of quarks inside the proton, was revealed through the use of scattering 
experiments.

\begin{figure*}[h!]	
   \centering
        \includegraphics[width=0.75\textwidth]{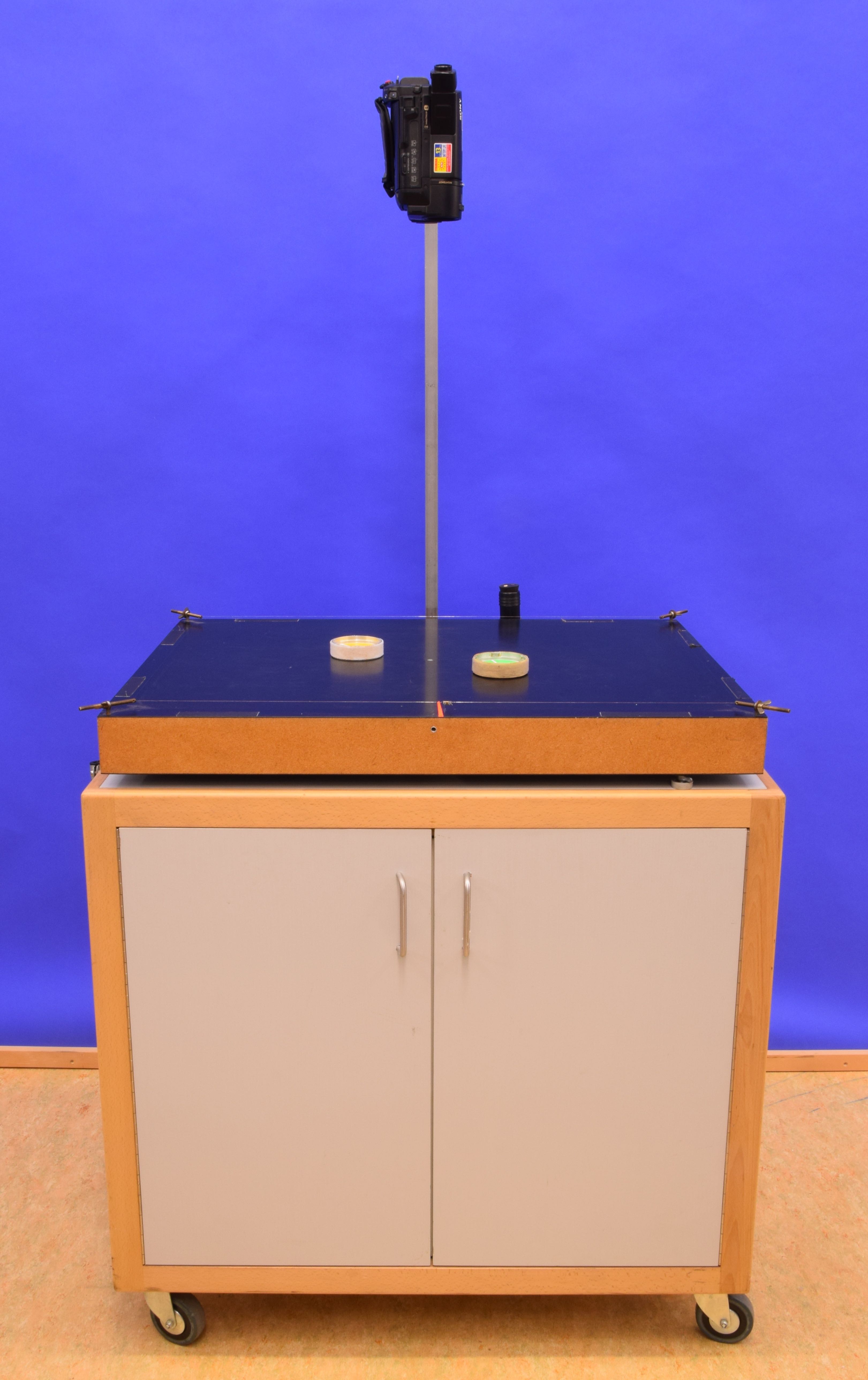}
        \caption{Overall view of the air table with two pucks on it. On top is the camera, which we use to
        project the puck interactions onto the big screen. The cabinet underneath contains the air pump.
        You can also clearly see the taut wires, which keep the pucks on the table.}
        \label{fig:air-table-big}
        \end{figure*} 

\subsection{Air Table and Strong Force}
\label{app:airtable}
        
With this experiment we wish to demonstrate how a strong, attractive short-range force can overcome
a weaker long-range repulsive force.

\subsubsection{Materials and  Technical Details}
The experiment requires an air table, as well as two or three pucks. An overall view is given in 
Fig.\,\ref{fig:air-table-big}. Our air table is 80\,cm x 60\,cm and is mounted on a cabinet on wheels. The 
cabinet is 85\,cm x 70\,cm and including the wheels, 80\,cm high. The pump for the air flow is stored 
inside the cabinet. This allows for easy transport on- and offstage. A camera which looks down onto the 
air table is connected via a metal arm to the side of the cabinet. Along the edges of the air table a taught 
wire at about half the height of the pucks guarantees that the pucks bounce back almost elastically. The 
pucks should be of identical size and shape, but with differing colors, to easily distinguish them. Each 
puck must have a magnet within it, with the same orientation, \textit{e.g.} the North pole upwards. We 
also require a Velcro ring for the pucks, which fits tightly, but does not impede their low-friction sliding 
across the table.

\begin{figure*}[h!]
	    \centering
            \includegraphics[width=0.45\textwidth]{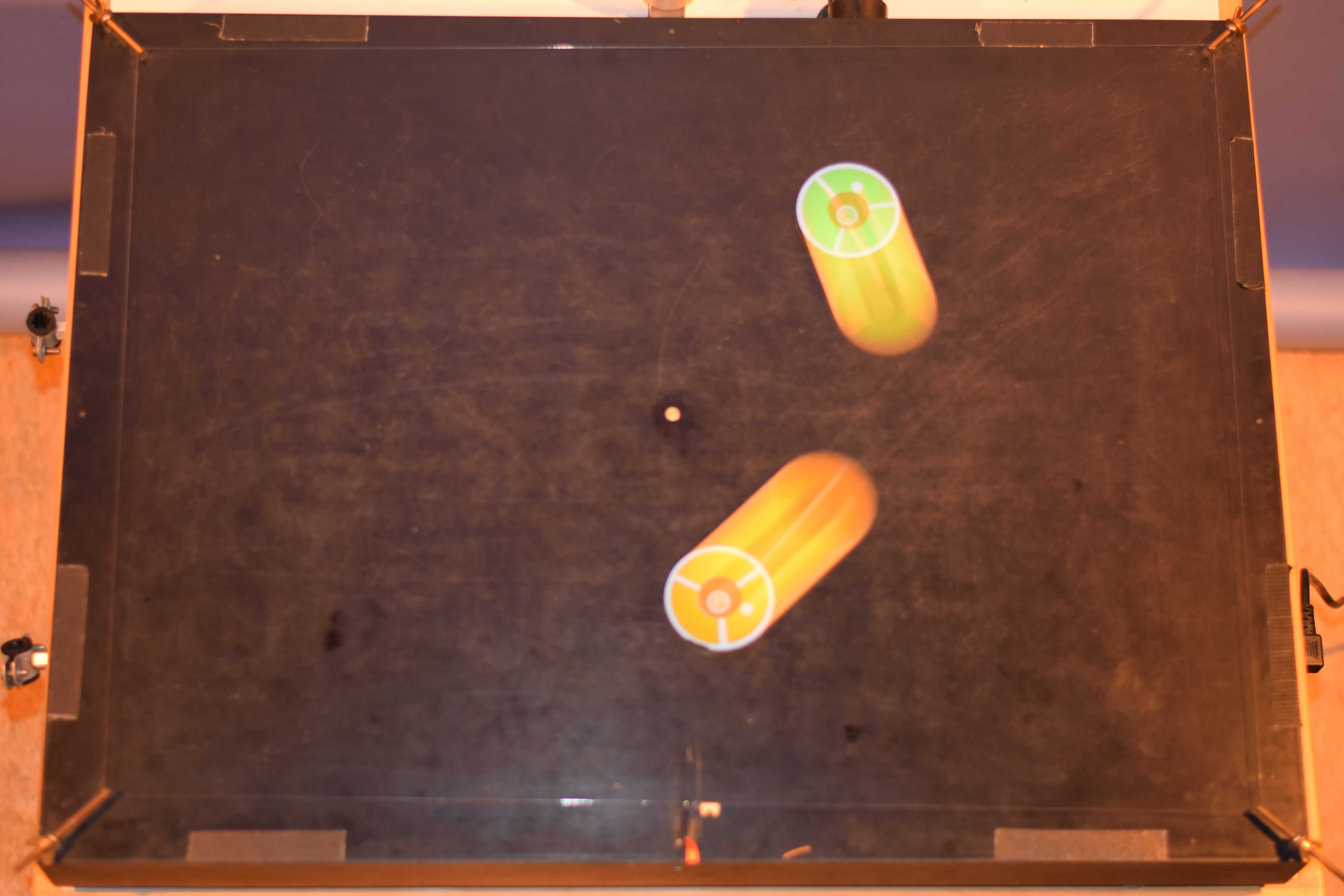} \; \includegraphics[width=0.45\textwidth]{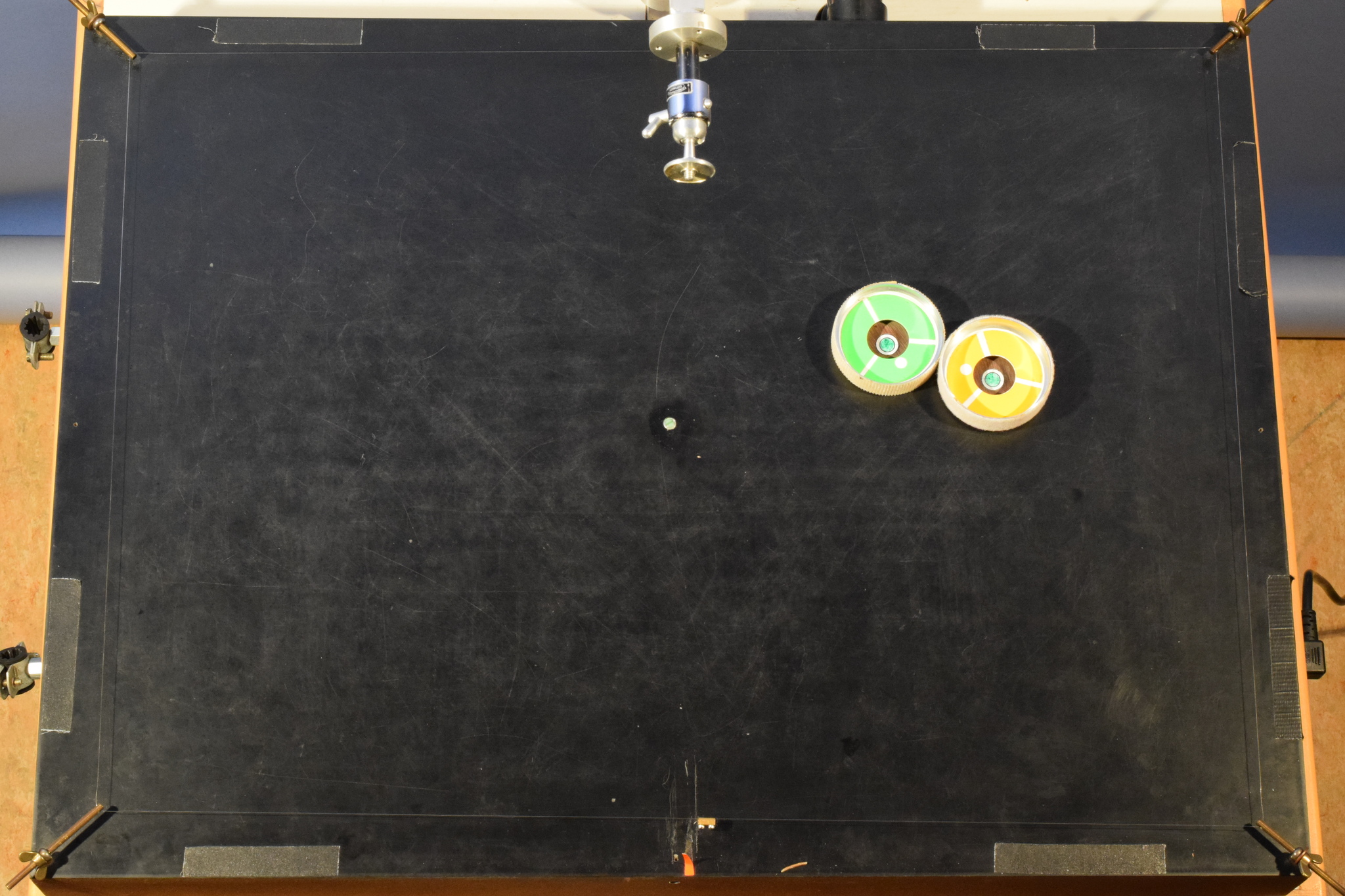}
           \\  (a) \hspace{6cm} (b)
        \caption{Top down look onto our air table, which we use as an analogy for the strong force. In (a) no Velcro 
        or strong force is present, the magnetic force causes the two protons to repulse each other.  This is shown
        here by an extended exposure in the photo. In (b) the Velcro overpowers the repulsive force of the magnets 
        at short distances and the two protons stick together. }
        \label{fig:air-table-2}
\end{figure*}

\subsubsection{Presentation}
The experiment is performed in two stages: 
First we demonstrate the repulsion of protons by the electric force for same charged particles, here however
represented by magnets, see Fig.\,\ref{fig:air-table-2}(a). Then we demonstrate the short range binding qualities 
of the strong force, see Fig.\,\ref{fig:air-table-2}(b).

We start with the two pucks on the air table and try to place them near each other. These pucks are meant to 
symbolize protons in the nucleus. The magnet within them is a representation of the repulsion created by their 
equal electric charge. When we let go of the two protons they repel each other. If we try to shoot them together, 
\textit{i.e.} use  kinetic energy, they will just deflect and fly apart because of their identical electric charge. If there 
was no other force this is what would occur in the nucleus of an atom. Then we add the Velcro to the pucks and 
collide them, with little energy. Again they repel each other. However if we collide them with higher energy, they
get close enough for the Velcro to take over. Then the two pucks stick together and start to move in unison. The 
Velcro symbolizes the strong force interaction, which has no effect over large distances but is incredibly strong 
when the two protons are close, as in the nucleus.

\subsection{Photon Clicker}
\label{app:photonclicker}

This experiment demonstrates the quantum nature of light. After a filter attenuates the light emitted by a 
laser, individual photons are detected by a photomultiplier tube which is connected to a loud speaker. In 
this way the audience can hear a clicking sound for each individual photon.

\subsubsection{History}

The photoelectric effect was discovered by Heinrich Hertz in Karlsruhe, in 1887 \cite{hertz-photon}. He noticed
that the emission of sparks in his spark-dipole, used for (producing and)  detecting electromagnetic waves, was 
enhanced by external light. Together with detailed investigations by his assistant Wilhelm Hallwachs he 
showed that the UV component of light leads to the strongest effect \cite{hallwachs}. Einstein was able
to interpret this in terms of the quantization of electromagnetic radiation and postulated the photon
\cite{einstein}.

\begin{figure}[h!]
  \includegraphics[width=0.96\textwidth]{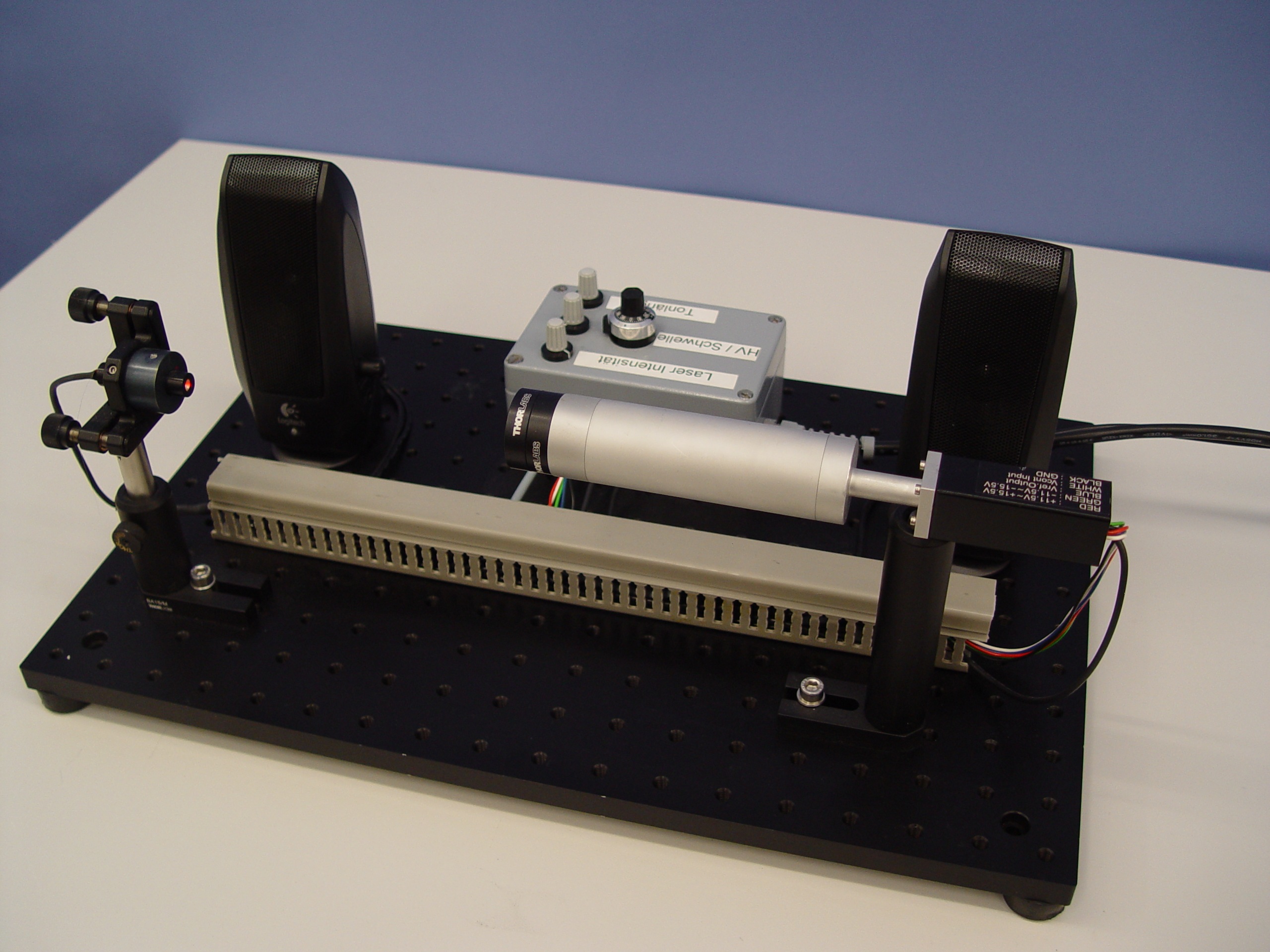}
\caption{Photon clicker experiment. The red dot on the left is the laser. The black box on the right is the photo multiplier.
The setup is described in more detail in the main text.}
\label{fig:photon-clicker}
\end{figure}

The apparatus we use in the show was originally designed in Bonn by Remmer Meyer-Fennekohl and Antoine Weis. It was part of 
the exhibition {\it `Quantenphysik heute - Quantenphysik morgen'} on quantum physics in the Deutsche Museum, 
Bonn, in the autumn of 2000 \cite{DM-BN-Ausstellung}. In Ref.~\cite{bib:photon-clicker-1} three demonstration 
experiments on the wave and particle nature of light are described. The first is exactly the setup we use here.
(The second experiment describes the diffraction of light from a double slit. In the third experiment the latter is 
refined to single photons, slowly building up the diffraction pattern.) In Ref.~\cite{bib:photon-clicker-2} these ideas 
are extended to interference. There you can hear the interference pattern. We use a rebuilt, smaller version, which
is easier to transport.

\subsubsection{Materials}
\begin{figure}[h!]
  \includegraphics[width=\textwidth]{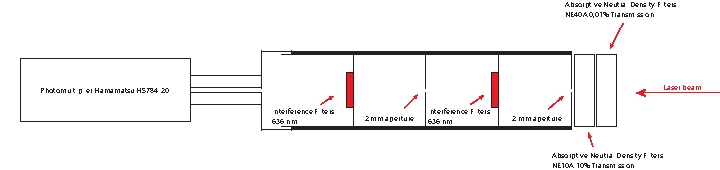}
  \caption{A schematic drawing of the attenuation and detection of the photon clicker experiment. The laser beam enters from 
  the right. The photomultiplier is on the far left. The laser beam first encounters two absorptive neutral density filters, then
  two 2\,mm apertures alternating with two 636\,nm interference filters.}
\label{zeichnung-photomultiplier}
\end{figure}

The apparatus we use in the show is shown in Fig.\,\ref{fig:photon-clicker}. On the left, we
see a red laser diode with adjustable power ($< 1\,\milli\watt$), which is used as the light source. The 
intensity is reduced to $0.001\%$ of the initial intensity by a neutral density filter with $\text{ND}=5.0$. The 
laser beam is adjusted to a pin hole mounted in an aluminum tube, which additionally shields the detector.
The detector is a Hamamatsu Photosensor Module H5784-20, which is comprised of a
photomultiplier tube, a high-voltage power supply and a low noise amplifier. See Fig.~\ref{zeichnung-photomultiplier}
for details of the attenuation in front of the photomultiplier.

The amplifier output is connected to usual active PC speakers. Everything is mounted on a plate, so 
that the alignment is stable and it is easy to carry on stage.

\subsubsection{Presentation}
Guiding the audience through the setup is especially important for this experiment, because most people 
are not used to the devices and their function and they are quite small. Depending on the size of the 
auditorium, the speaker sound should be amplified for the audience, possibly with a hand-held microphone.
The presenter should always block the beam, or turn off the speakers, while talking to avoid disturbance by 
the detector sound.

The presenter starts with the laser at high intensity. By holding
a sheet of paper between the laser and the detector she shows the red dot on the paper, indicating that the laser is on. She then introduces the photo multiplier as a device for light 
detection and switches on the speakers. The light intensity should 
be high enough to produce continuous noise. The presenter demonstrates that the sound is caused by the 
light by again blocking the laser beam with a sheet of paper several times. This eliminates the noise. Then she she reduces the 
intensity of the laser and invites the audience to listen carefully. She adjusts the light intensity 
until single clicks can be clearly distinguished and explains that you can now hear individual photons.

\subsubsection{Safety}
Standard precautions for the usage of lasers must be taken. The laser is operated at low power and
is fixed on the apparatus, greatly reducing any risk. The presenter should
not hold anything reflecting, \textit{e.g.} a wrist watch, into the beam.

\subsection{Tossing Medicine Balls on Inline Skates}
\label{app:medicineballs}

\begin{figure}[ht]
  \center{\includegraphics[width=0.85\textwidth]{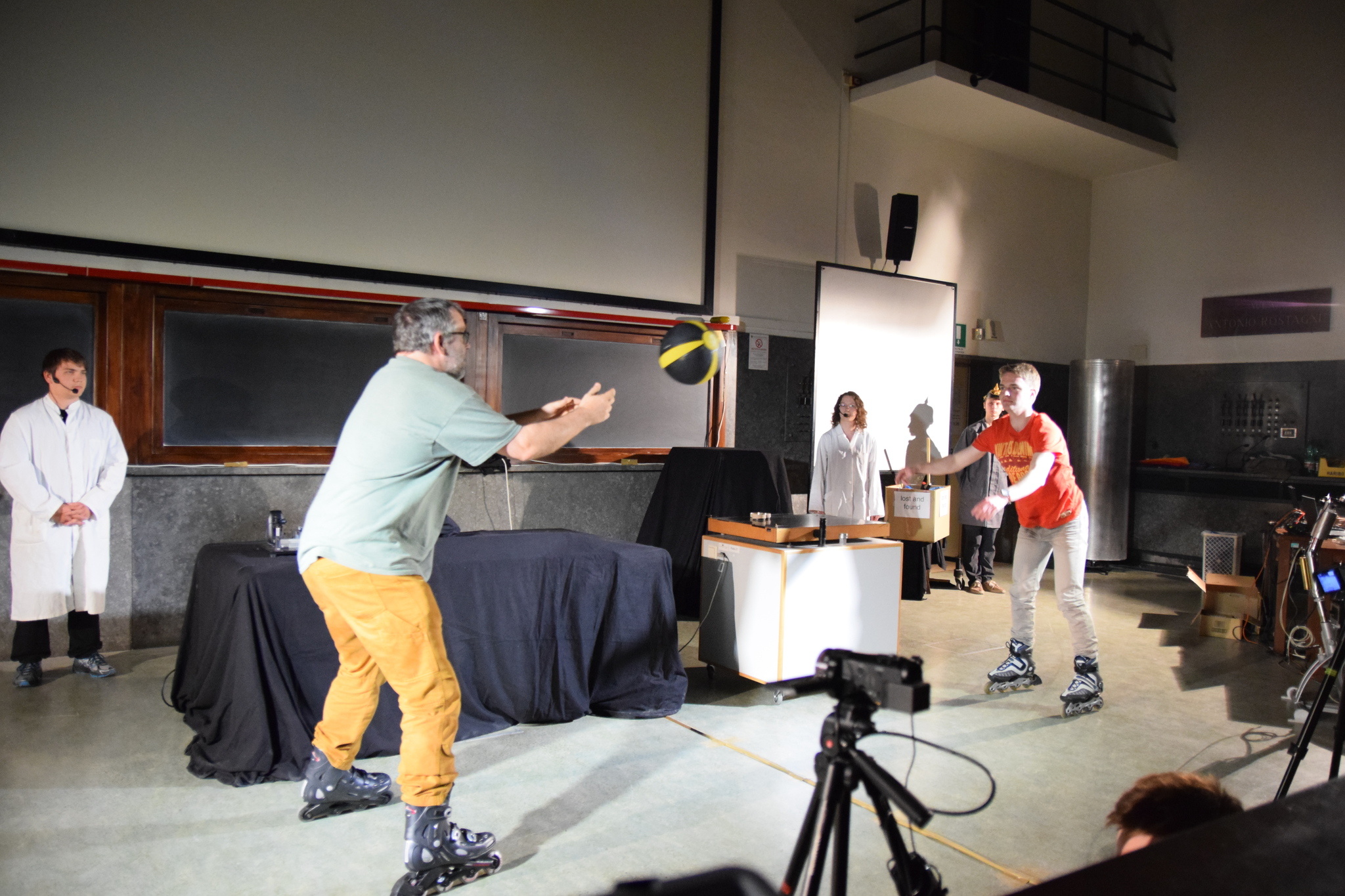}}
  \caption{AF (right) and HD (left) exchanging a medicine ball during the
  show in Padua. On the left in the back is MBr as Haimo Zobernig and on the right is KH
  as Sau Lan Wu. In front of KH  you can see the air table of experiment of App.\,\ref{app:airtable}. Partially
  covered by AF is JSchm  as the Caretaker.}
\label{stage-medicine-balls-2}
\end{figure}

In this experiment we want to visualize how two particles can exert a force on each other by exchanging
a mediating particle, as in quantum field theory. Two people wearing inline skates throw a medicine ball 
back and forth and through momentum conservation repel each other, see Figs.\,\ref{stage-medicine-balls}
and \ref{stage-medicine-balls-2}.

\subsubsection{History}
\begin{figure}[h!]
  \center{\includegraphics[width=0.9\textwidth]{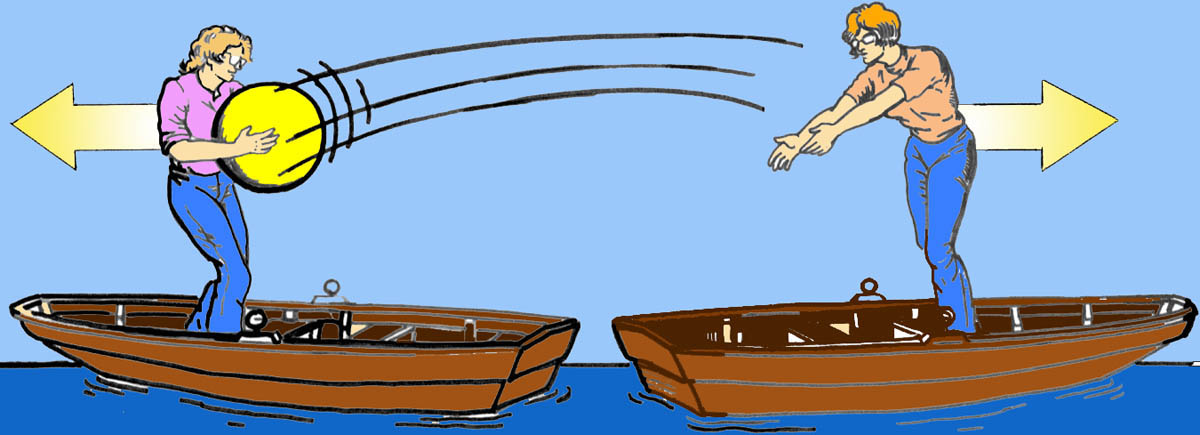}}
  \caption{Two people on rowing boats in the water exchanging a heavy object, and thus repelling each other.
  We thank Daniel Class and Don Lincoln for allowing us to use this picture.}
\label{exchange-force-boat}
\end{figure}

In the literature, a similar analogy is widely used to describe a repellent force through an exchange particle. In 
Fig.\,\ref{exchange-force-boat} we show an example where two people on separate rowing boats exchange
a large heavy object. Yellow arrows indicate their resulting motion.\footnote{It seems more likely that 
they would actually fall in the water!} However, this is highly impractical onstage. We have thus altered
this to two people on inline skates, exchanging a medicine ball.

\subsubsection{Materials and Technical Details}
This demonstration requires  a $5\,{\rm kg}$ medicine ball and two pairs of inline skates. Furthermore, for this experiment 
a lot of space is needed, and preferably a hard floor. Inline skates do not work as well on carpeted floors. Although 
we managed at ICTP Trieste.

\subsubsection{Presentation}
After announcing the two participants, in our case a professor and a student, they skate onstage, one of them 
holding the medicine ball in his hands. Both place themselves in the center of the stage about 1\,m apart. A 
moderator points out the inline skates  and the medicine ball and opens the stage for the experiment. The two 
people throw the medicine ball back and forth several times until the distance between them increased to about 
5\,m, or as far as they can throw the ball and the stage allows. Afterwards, both skate closer to each other and the 
procedure is repeated once. Both skaters stay onstage for the explanation of the experiment by the moderators
and then exit. During the explanation, as a joke, it can be pointed out that students and professors repel each
other.

\subsubsection{Extension: Collider}
In a previous show, we have extended this to include pair production and annihilation. For this we had two 
assistants walk in holding a screen, behind which they could hide, but behind which also the skaters could 
vanish. The assistants had two medicine balls with them. The two people on inline skates entered the stage 
from opposite sides and then together vanished behind the screen. Then they immediately took the medicine 
balls and threw them out. That was pair-annihilation. Here one proponent was a professor and the other an 
anti-professor, \textit{i.e.} a student. We often also displayed a Feynman graph of the process via the projector.

Next two people from the outside threw the medicine balls back to behind the screen, the two proponents on
inline skates caught them, handed them to the two holding the screen, and skated back out again: pair production.
Here we also showed the Feynman graph, by simply rotating the previous one by 180$^\degree$\!\!.

\subsubsection{Safety}
This experiment needs a lot of space onstage, which needs to be free from any cables or other things which could
trip the skaters.  Also skaters should practice throwing and catching a heavy ball while skating. If they feel unsure they 
should wear helmets.

\subsection{Relativistic Bicycle}
\label{app:rel-bicycle}

\begin{figure*}[h!]
	    \centering
            \includegraphics[width=0.47\textwidth]{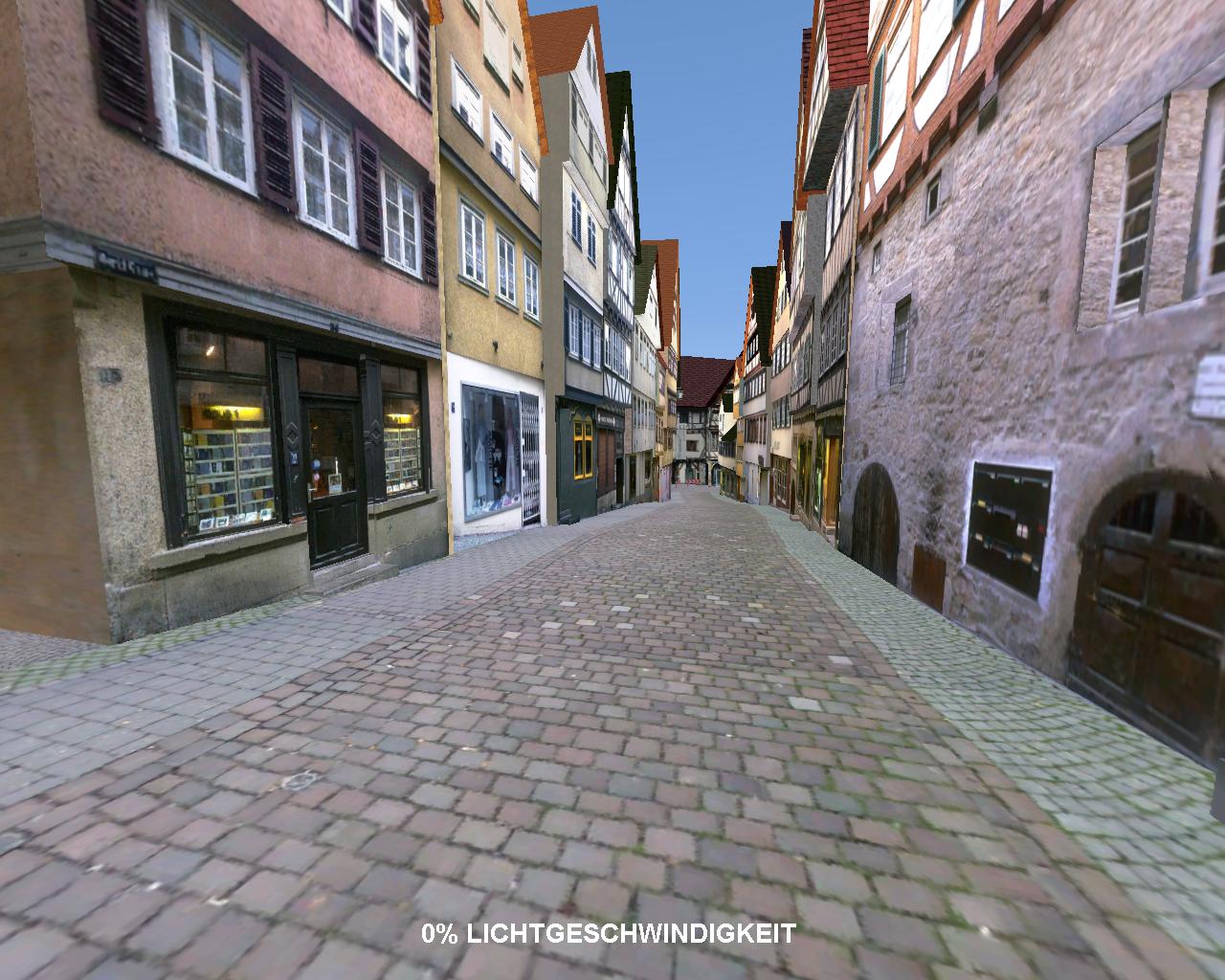} \; \includegraphics[width=0.47\textwidth]{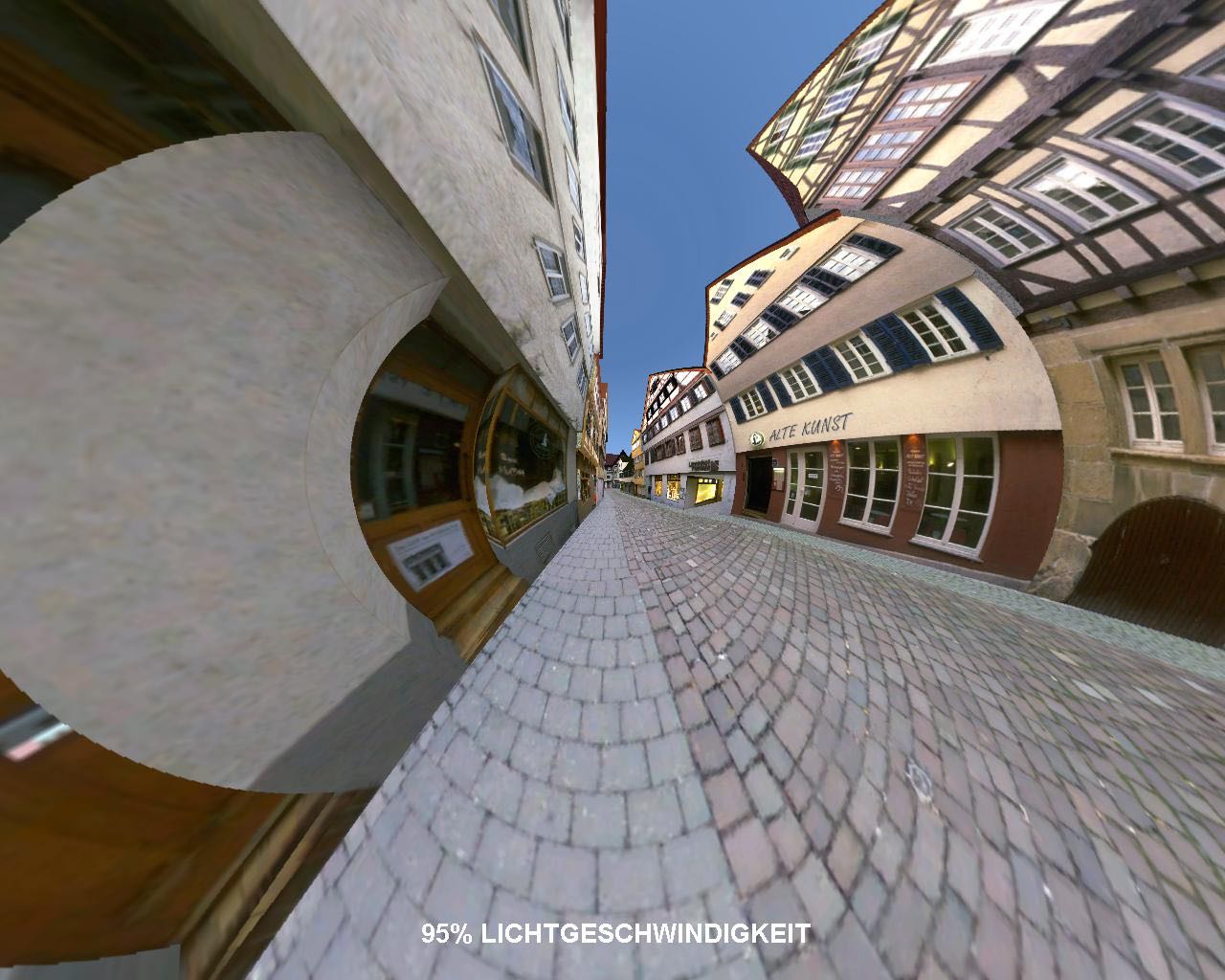}
           \\  (a) \hspace{6cm} (b)
        \caption{Screen shots of the program from T\"ubingen described in the text \cite{tuebingen-fahrrad,bicycle_software}. The
        speed of the bicycle in terms of the speed of light can be seen at the bottom of each screen. (`Lichtgeschwindigkeit' = `speed of light', in German) (a) shows the image of a 
        street in T\"ubingen at rest. (b) corresponds to a distortion at  95\,\% of the speed of light.}
                \label{fig:screen-shot}
\end{figure*} 

In this experiment a computer simulation, visualizes the world as seen from a bicycle travelling close to the 
speed of light \cite{tuebingen-fahrrad,bicycle_software}.  The simulation is hooked up to a fixed exercise 
bicycle, set onstage. ``The world'' here is a digitized 3D model of the city of T\"ubingen, see 
Fig.\,\ref{fig:screen-shot}\,(a). The speed of light has been reduced to about 30\,km/h, so that a moderately fit 
person can achieve dramatic effects. The distorted, relativistic view onscreen, see Fig.\,\ref{fig:screen-shot}\,(b),
then corresponds to the speed of the cyclist onstage. The essential feature is that objects are to first order not 
length contracted \cite{terrell} but rotated and thus distorted through the relativistic effects.

\begin{figure}[h!]
  \center{\includegraphics[width=0.94\textwidth]{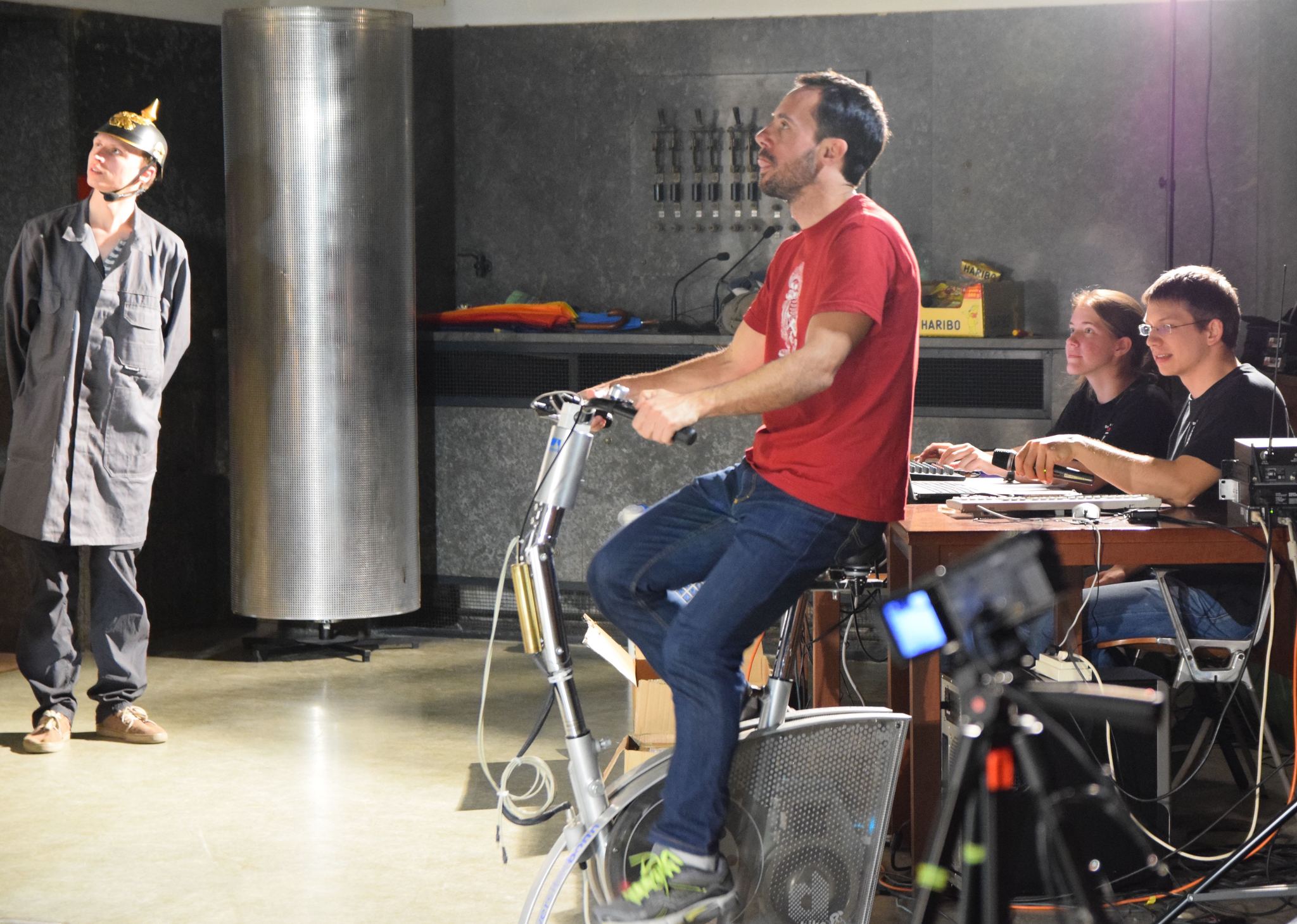}}
  \caption{LU on the relativistic bicycle during rehearsals at Padua University. On the left is JSchm as the Caretaker
  wearing his German helmet for the DESY part, described in Sect.\,\ref{sec:desy}. On the right are JH and CJ at the
  table where the lighting, sound, and projection are controlled.}
  \label{pic:relativistic-bicycle-padua}
\end{figure}

\subsubsection{History}
The relativistic effects shown here were inspired by drawings in George Gamow's book \textit{Mr. Tompkins in 
Wonderland} \cite{gamow}, in which he describes a cyclist riding through a town, where the speed of light is 
30\,km/h. The houses on the left and right are length contracted. However, A. Lampa and  J. Terrel showed that 
this does not occur \cite{lampa,terrell}. This inspired U. Kraus and M. Borchers of T\"ubingen University to 
develop a program which also includes the rotational effects due to the finite speed 
\cite{tuebingen-fahrrad,bicycle_software}. For early work on these rotational effects (aberration) see for example 
\cite{terrell,lampa,penrose}. For a more pedagogical treatment see the Physics Today article by Victor Weisskopf
\cite{weisskopf}. The essential feature is that the light recorded by the eye or by a photograph arrives 
simultaneously, but due to differing distances travelled it was not emitted simultaneously. A nice very short film 
showing the lack of contraction for a sphere is given at the website \cite{film:lampa-terrell}.

\subsubsection{Materials and Technical Details}
In the show we use  a fixed exercise bike which allows for rotation of the handle bars around a vertical axis,
as in a normal bicycle, see Figs.\,\ref{pic:relativistic-bicycle} and \ref{pic:relativistic-bicycle-padua}. Electronics 
on the bicycle register the steering and the pedal rate and transmit this to a computer (PC) via a USB connection. 
The software we use here has been developed at T\"ubingen University \cite{tuebingen-fahrrad,bicycle_software}.
It employs a digitized 3D model of the city of T\"ubingen developed by H.~B\"ulthoff (MPI Biological Cybernetics, 
T\"ubingen), see Fig.\,\ref{fig:screen-shot}. The speed of light is set to approximately 30\,km/h. The  PC is 
connected to a projector to present the simulation onstage.

\subsubsection{Presentation}
In the show the bicycle is employed as a way to imagine what it is like to zoom through an accelerator tunnel
close to the speed of light. The presenter introduces the \textit{relativistic bicycle}, shows the cables which connect 
it to the computer and thus to the screen, where the bicycle speed is indicated as a percentage of the speed of 
light. The person riding the bicycle, see Fig.\,\ref{pic:relativistic-bicycle-padua}, accelerates slowly (almost) up to the 
speed of light, brakes and repeats this a few times. On the screen, the audience can see the distortion of the buildings 
with increasing speed, see Fig.\,\ref{fig:screen-shot}. We usually play ``\textit{Bicycle Race}, by Queen" during the ride.

\subsubsection{Additional Information}
The relativistic bicycle has been onstage only in this particle physics show. It is very sturdy and also suited as a 
hands-on exhibit. In 2005 it was successfully used at the Deutsche Museum Bonn. During the intermission in our  
regular (non-particle physics) shows, we always have many experiments, which the audience can try themselves 
under supervision, including sometimes the relativistic bicycle.

\subsubsection{Safety}
The bicycle is quite heavy and stable, so a crash helmet is optional. $\ddot \smile$

%
%
%

\begin{figure}[h!]
  \center{\includegraphics[width=0.45\textwidth]{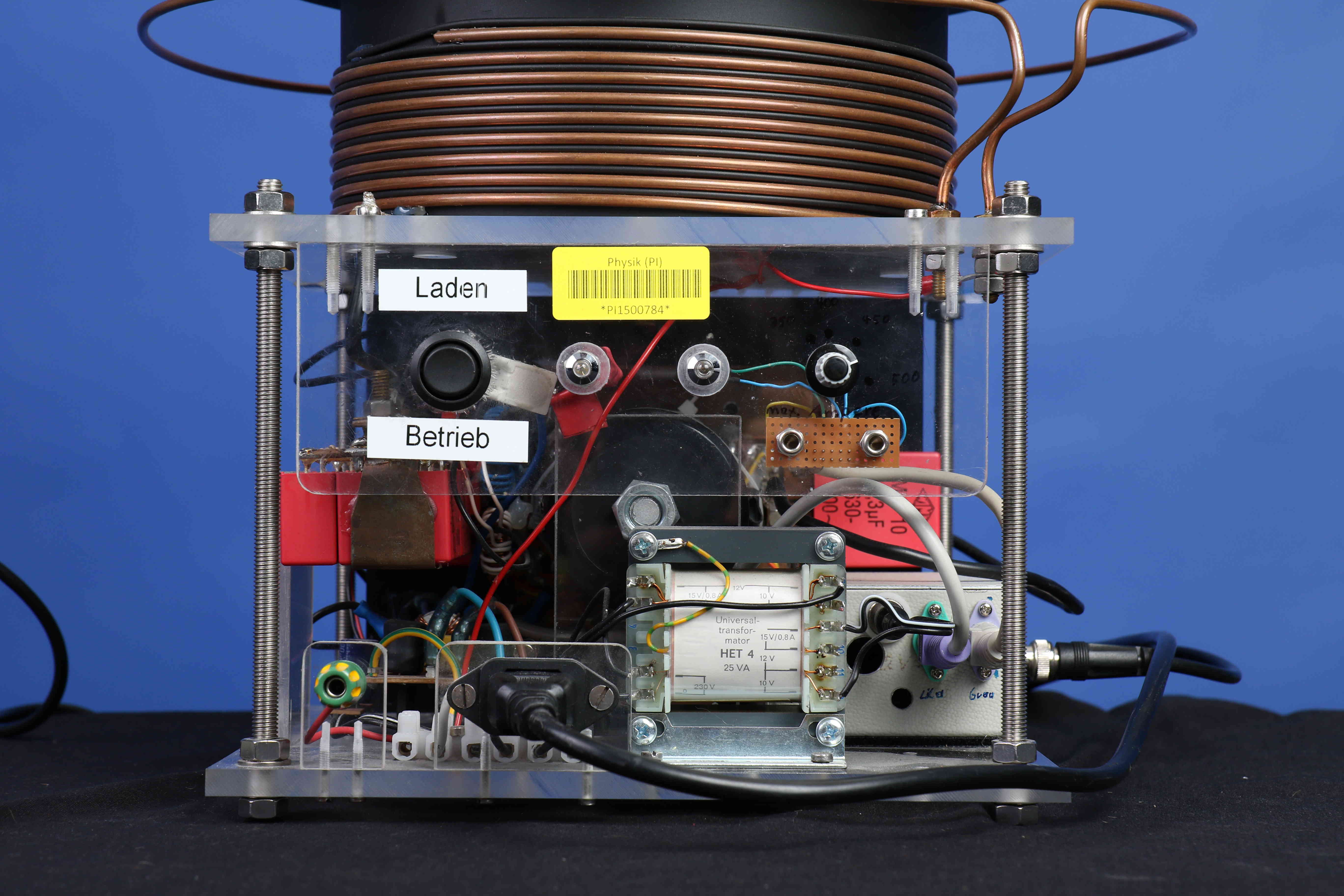} \; \includegraphics[width=0.45\textwidth]{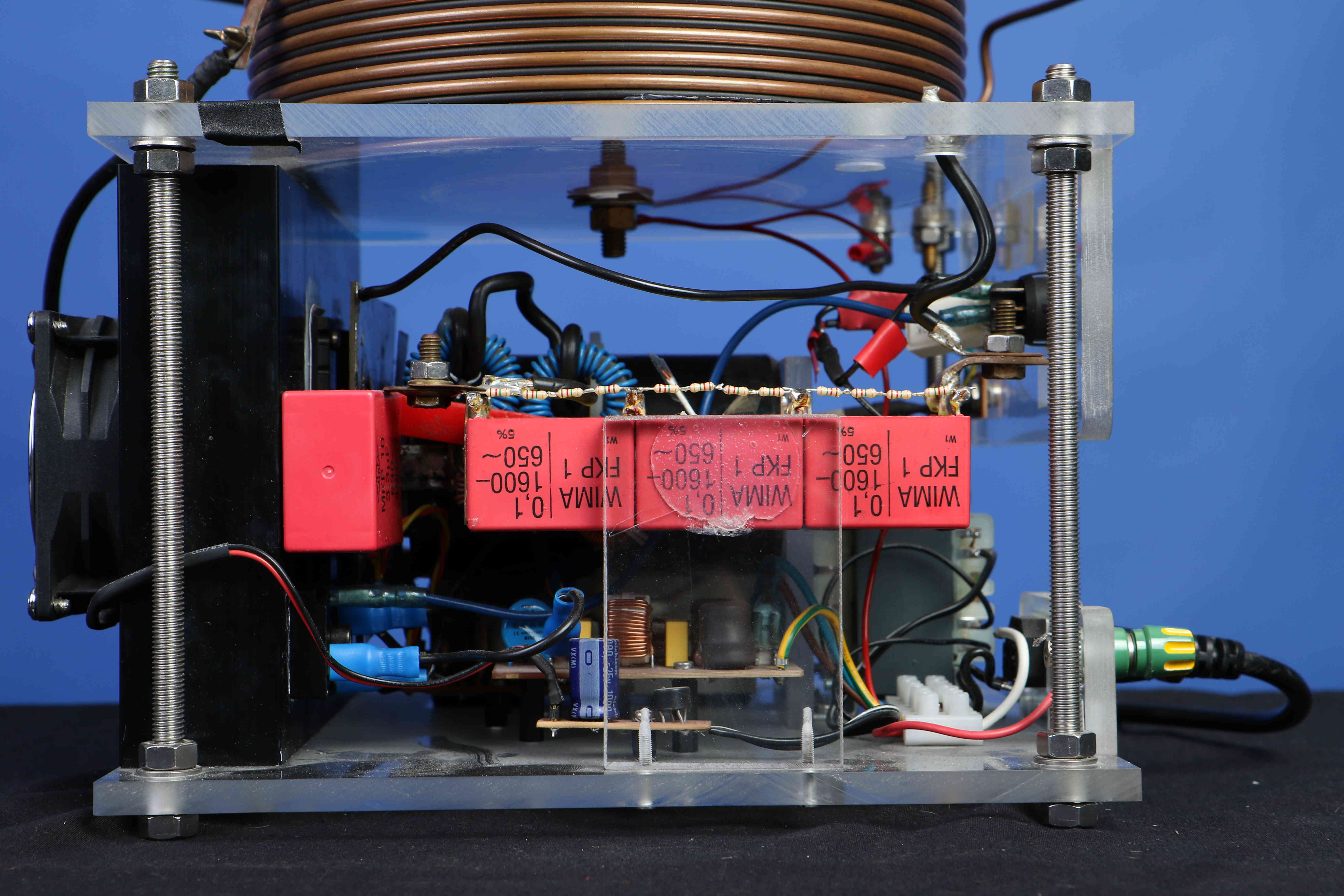} 
\\ (a) \hspace{6cm} (b)   }
  \caption{Close-up of the electronics of our Tesla coil, (a) front,  and (b) side.}
\label{img:tesla-3}
\end{figure}

\subsection{Tesla Coil}
\label{app:teslacoil}

A Tesla coil is an electrical resonant transformer circuit which is used to produce high voltage at high 
frequencies and low currents. The high voltage is discharged via extended lightning bolts. The Tesla coil we 
use creates bolts up to a length of about 1\,m.

\begin{figure}[h!]
  \center{\includegraphics[width=0.6\textwidth]{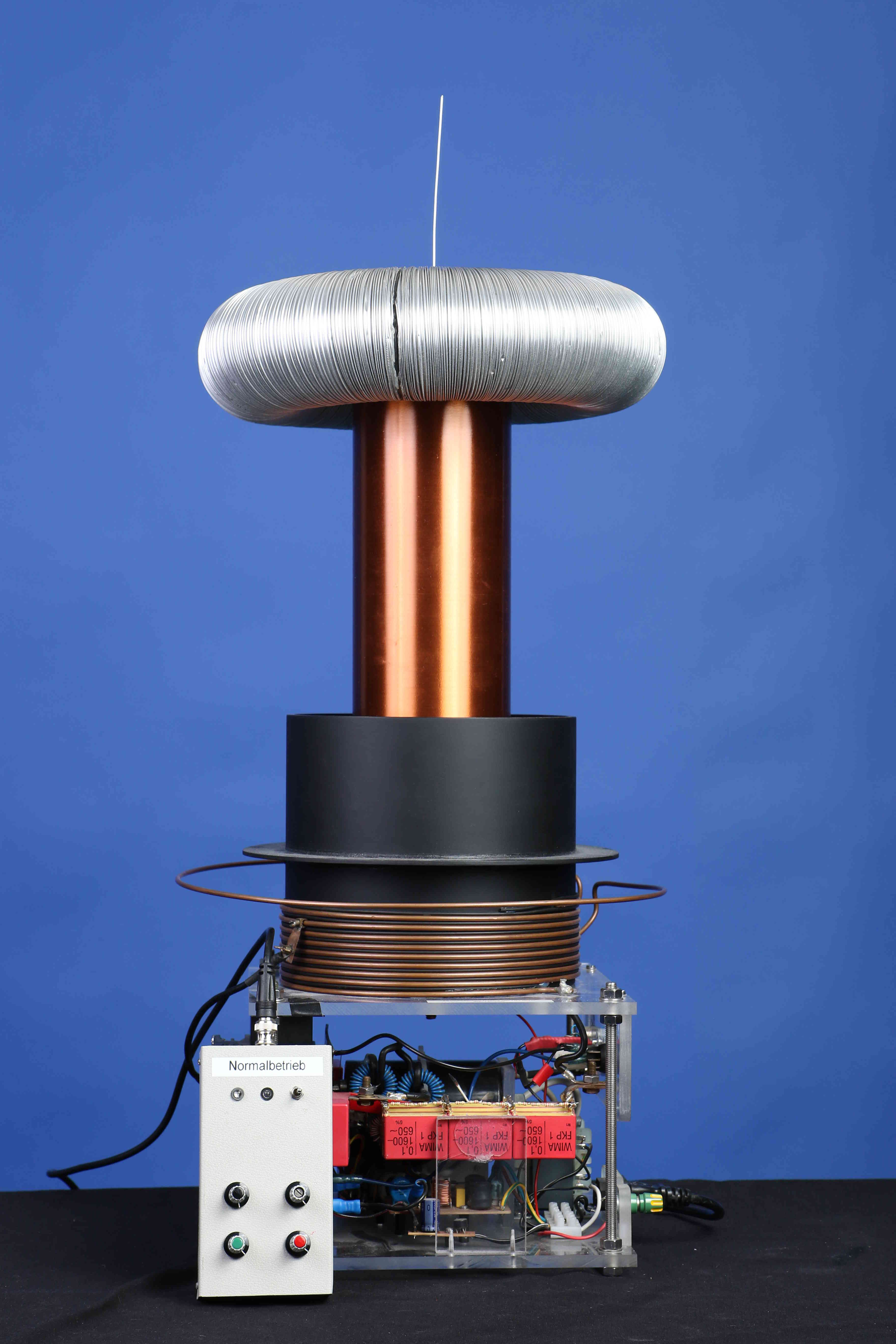}}
  \caption{Our Tesla transformer, built by Timo Poller. The gray box on the left contains the remote controls connected to the transformer via an extended cable.}
\label{img:tesla-2}
\end{figure}

\subsubsection{History}
The Tesla coil was invented by  Nikola Tesla (1856-1943) around 1891. A concise history leading up to the invention
can be found for example in \cite{tesla-book}. Today Tesla coils are used for entertainment, to show the effects of high 
voltage and frequency, but also as vacuum system leak detectors. Our Tesla coil in the show was built by Timo Poller,
a Bonn physics student, and produces about 200,000\,V output voltage, see Figs.\,\ref{img:tesla-4}, \ref{img:tesla-2},
and \ref{img:tesla-3}.

\begin{figure}[h!]
  \center{\includegraphics[width=0.65\textwidth]{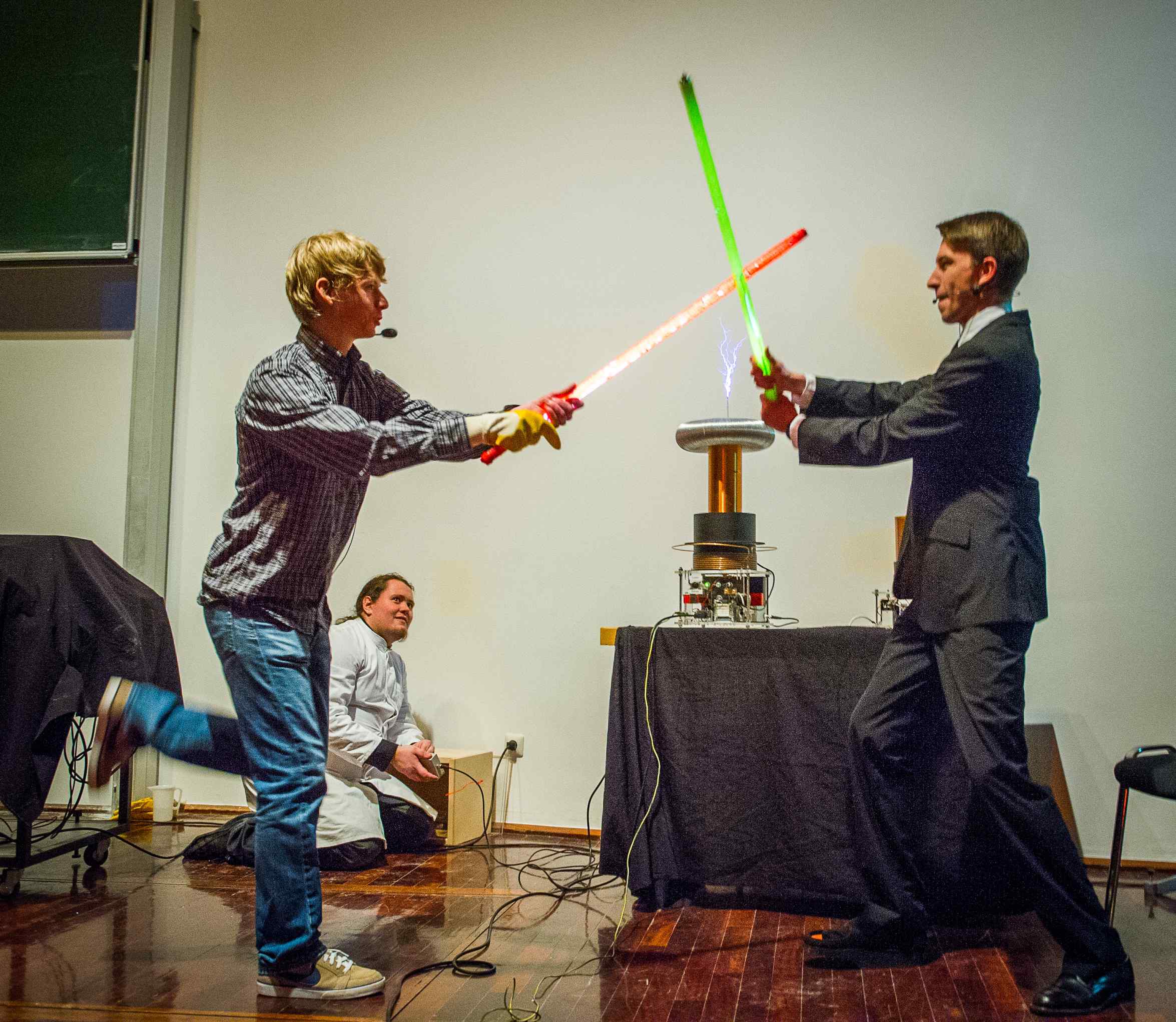} }
  \caption{Lightsaber fight using conventional fluorescent tubes wrapped in colored paper, and of course
  our Tesla transformer. This is from a previous show and was a fight between Yannick Edmonds as James Bond (on the right)
  and Erik Busley as Dr. Yes on the left. This photo was taken by Volker Lannert.}
\label{img:tesla-6}
\end{figure}

\begin{figure}[h!]
  \center{\includegraphics[width=0.85\textwidth]{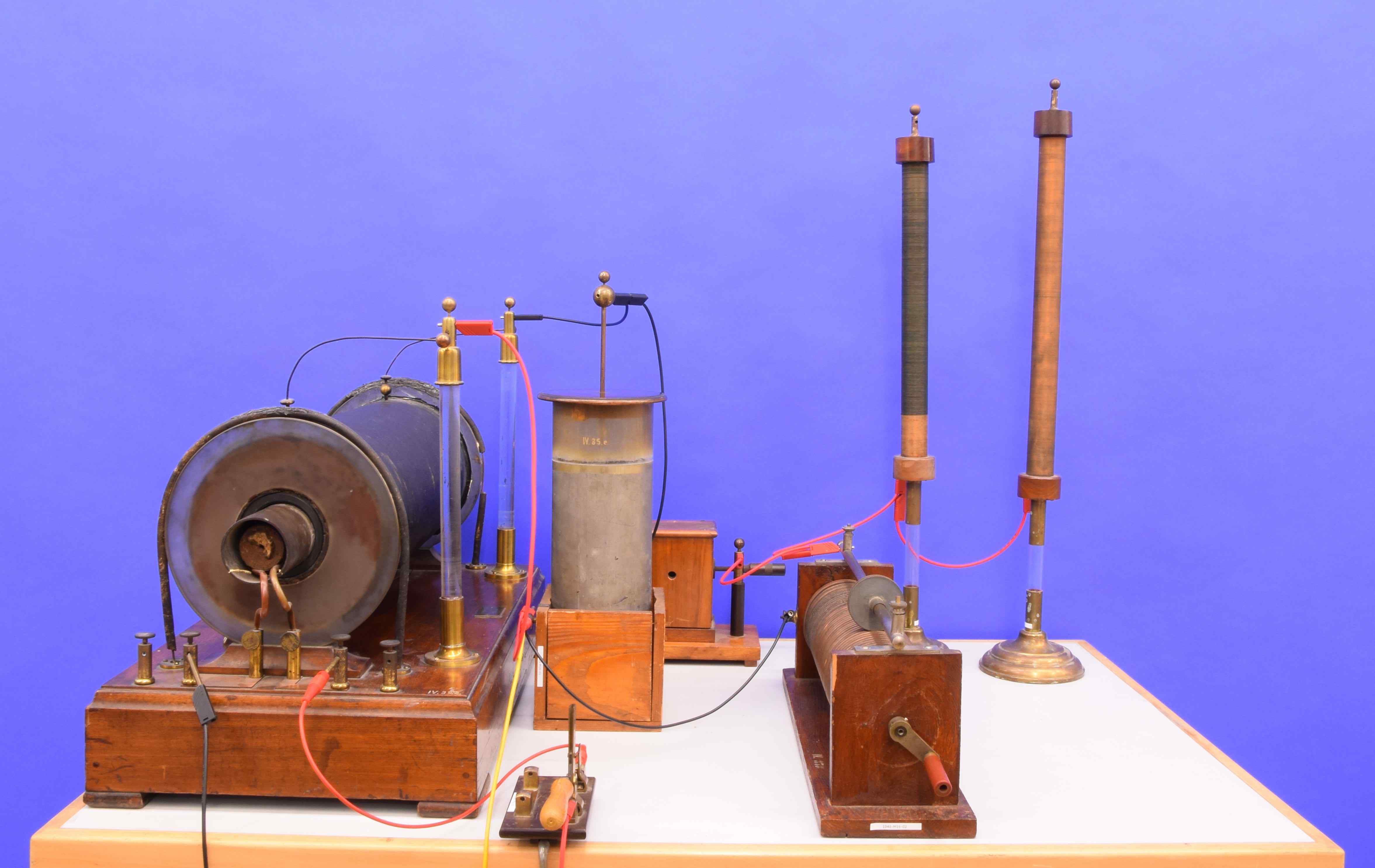}}
  \caption{Tesla transformer using an old induction coil from 1858. The large horizontal cylindrical object on the 
  left is the induction coil. The vertical cylindrical object in the middle is a Leyden jar. Just behind it the smaller wooden 
  casing contains the spark gap. Next to the right, is a variable coil. In the back, the two vertical slender cylindrical 
  objects are Seibt coils.}
  \label{img:old-tesla-1}
\end{figure}

\subsubsection{Materials and Technical Details}
A Tesla coil can be bought or built. Instructions on how to build your own can be found on YouTube at \cite{tesla-youtube}.  
Details of our apparatus are shown in Figs.\,\ref{img:tesla-2}, and \ref{img:tesla-3}. It has the dimensions 300\,mm 
x 300\,mm x 800\,mm. For easier transport, the secondary coil, the silver toroid, and the needle on top can be dismounted. 
It has a normal (230\,V) power plug and the circuitry is completely based on semiconductors. The output voltage is about 
200,000\,V. We are happy to provide construction details upon request.

\begin{figure}[h!]
  \center{\includegraphics[width=0.88\textwidth]{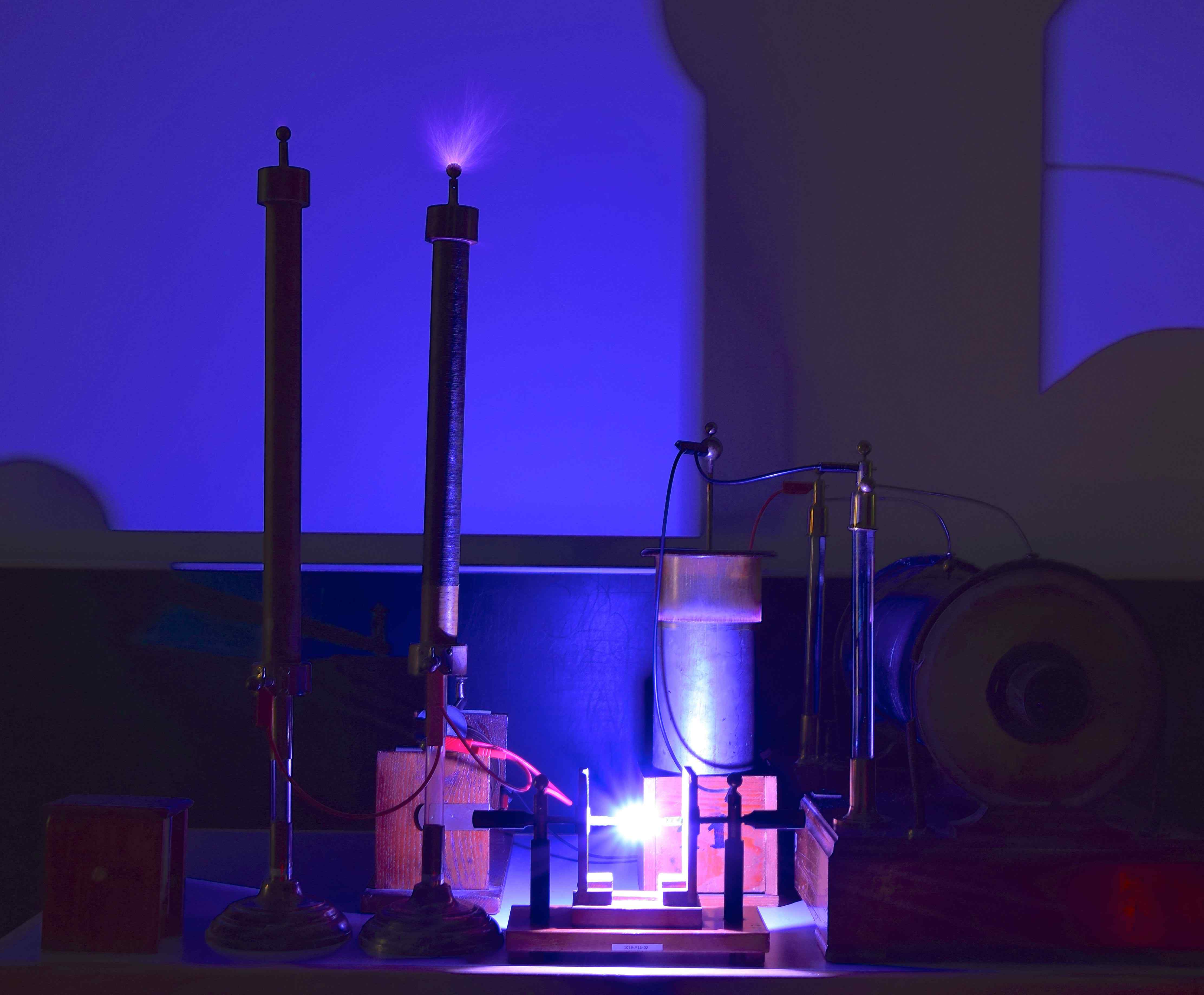}}
  \caption{Old Tesla coil in action. The box has been lifted to show the spark gap. At the top is the lightning 
  discharge of one of the Seibt'sche coils. }
  \label{img:old-tesla-2}
\end{figure}

\subsubsection{Presentation}
The Tesla coil is used in this show to demonstrate very high voltages, specifically, we used it for comparison to 
the energy of the particles in the LHC, see Figs.\,\ref{img:tesla-1}, \ref{img:tesla-4}. The lecture hall is dimmed,
and then the transformer is turned on, generating large sparks. This is also quite loud. For a brief discussion of
further applications, see the following subsubsection.

\subsubsection{Extensions}
The strong electromagnetic fields surrounding the Tesla coil can be demonstrated for example using fluorescent tubes, 
see Fig.\,\ref{img:tesla-6}. In the case of the lightsaber fight we often accompany this with the Star Wars sound track. 
This is very popular with the audience, but  would have been too much of a distraction during the introduction of the 
LHC in the particle physics show. Also music, \textit{e.g.} from a guitar or from an electronic source, can be transmitted 
via the Tesla sparks. 

\begin{figure}[h!]
  \center{\includegraphics[width=0.64\textwidth]{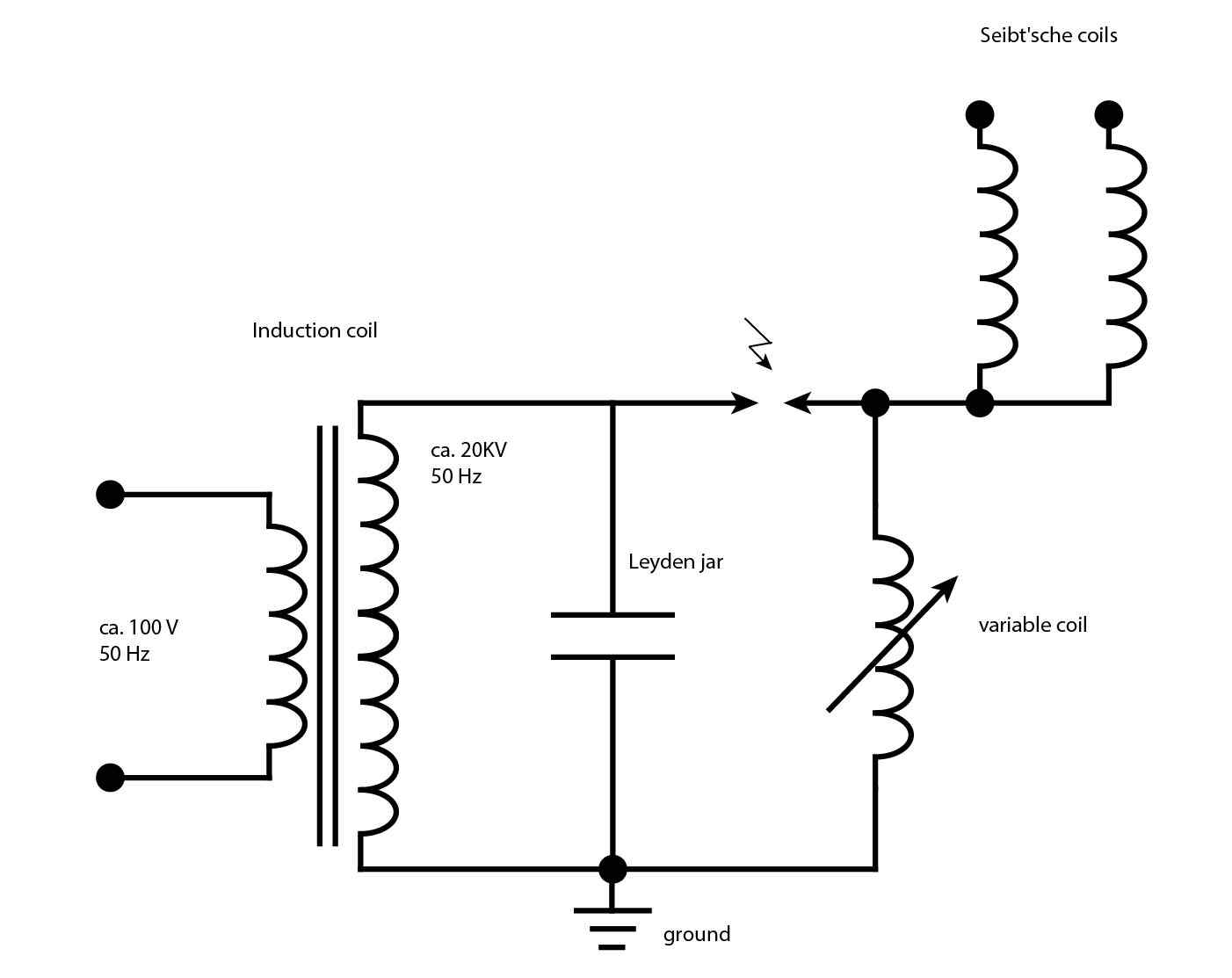}}
  \caption{Circuit drawing for the setup in Figs.\ref{img:old-tesla-1}, \ref{img:old-tesla-2}.}
  \label{img:old-tesla-3}
\end{figure}

In Bonn we also have a nice (robust!) historical one with an induction coil from 1858, see Figs.\,\ref{img:old-tesla-1}, 
\ref{img:old-tesla-2}, \ref{img:old-tesla-3}, which we have mainly used in local shows, not when travelling.

\subsubsection{Safety}
Since the Tesla coil produces high electrical fields great care must be taken. For example pacemakers (and other 
electronic equipment) can be damaged, and audience members who have one should retreat to the back of the 
auditorium. Furthermore for this specific Tesla transformer the lower part containing the circuits is not encased to 
allow for better cooling. During operation a larger distance should be maintained, and the circuits should not be 
touched. The discharges at the top should also not be touched in anyway, for this larger Tesla transformer, as the 
field strengths are too large.

\subsection{Tossing Balls: a Collider Onstage}
\label{app:ballcollider}
This experiment consists of two plastic boxes each filled with the small colored plastic balls which are usually found in children's ball pits. It is performed by two people standing a few meters apart apart. For the first trial, each one each 
picks a single ball. They throw these two balls towards each other and usually, no matter how times they try, the two balls do 
not meet. For the second trial the entire boxes of balls are thrown. Subsequently, there are almost always at least a few collisions. The principle behind the experiment is to demonstrate why at colliders we do not try to collide single particles, but rather bunches of particles.

\subsubsection{History}
We first saw this analogy used for a collider physics experiment in a science slam performed by Klaus 
Desch in Bonn on Nov. 23$^{\mathrm{rd}}$, 2011. However he used small wooden crates of mandarins.
We thought this would be too messy in Padua. We are not aware of any other usage.

\subsubsection{Materials}
Two boxes with approximately 50 plastic balls each in 2 colours. The two boxes should contain different 
colors, so that the effect is more easily seen.

\subsubsection{Technical Details}
For the balls, we ordered ball pit balls for children. The ones we used have a diameter of 54\,mm.
The size of the box should be such that they are easy to carry and lift.

\subsubsection{Presentation}
The two protagonists are standing in the LHC ring at CERN and trying to understand what happens at the LHC,
as has been explained above. Each of them takes a ball, they turn towards each other a few meters apart and 
then count down from 3. At 0 both throw the ball and, despite their best efforts, the balls pass by each other, as 
can be seen in Fig.~\ref{Fig:TwoProtons}.

They give it a second try, but quickly realize that this strategy doesn't work. Confused, they ask how this is done at the 
LHC. The answer is that the LHC applies a smarter concept. Instead of single balls, \textit{i.e.} single protons,
bunches of protons are propelled at each other.  The two time travellers want to try that as well and the
caretaker hands them each a box full of balls, symbolizing two bunches of protons.  They count down, again 
from 3, and throw the balls all at once, as can be seen in Fig.~\ref{Fig:ManyProtons}.

This time many balls collide and the result is a beautiful pattern of coloured balls on the stage floor.
Afterwards the audience is asked whether this time they saw collisions; as is to be expected, the majority answer that they did indeed see at least one collision.

\subsubsection{Safety}
People can slip and fall on the balls onstage. In our show the Caretaker sweeps them aside to avoid this eventuality.

\subsection{Visualizing the Electric Vector Field}
\label{app:paperelectric}
This is not a real experiment, it is more of a visual aid to introduce the abstract concept of a field, in this case a 
vector field. This is in contrast to the later introduced scalar field, the Higgs boson being a scalar field. The 
important point is that when a test charge is placed in a field it obtains a property. In the case of a vector field 
this property has a magnitude \textit{and} a direction. 

\subsubsection{History}
To the best of our knowledge this experiment was first used by HKD during his Science Slam in the 
Jakobshof, in Aachen, Germany, on January 16$^{\mathrm{th}}$, 2013 
\cite{herbi-higgs-slam}.\footnote{Worst restrooms, ever.}  We do not know of any previous instances 
of this experiment.

\begin{figure}[h!]
	\center{\includegraphics[width=0.7\textwidth]{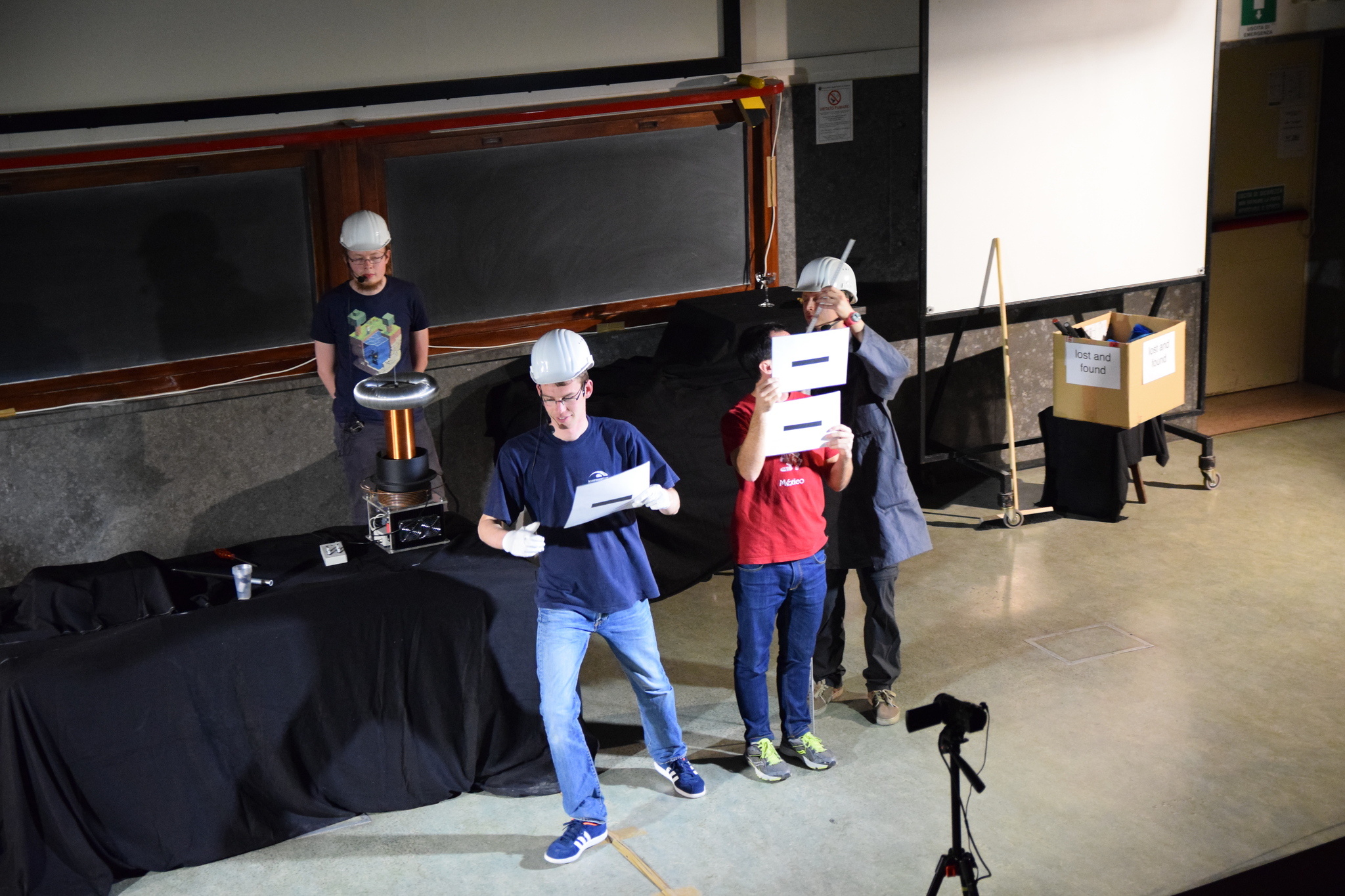}}
	\caption{TL as the test charge, repelled by LU with the double-negative charge.
		In the background,  at the blackboard, CSch. Just behind LU, the Caretaker, JSchm.}
		\label{Fig:paper-electric-field-3}
\end{figure}

\subsubsection{Materials}
To demonstrate the electric charges we used seven A4 charts in total for our presentation. These charts
have + and $-$ signs printed on them depending on which charge they represent. The charts were 
laminated to  allow for frequent usage. The charts looked as follows:
\begin{itemize}
\item One chart with one + on the front and one $-$ sign on the backside. These are the test charges.
\item Two charts, each with one + on it, and a third chart with many + signs on it.
\item Two charts, each with one $-$ on it, and a third chart with many $-$ signs on it.
\end{itemize}

\subsubsection{Presentation}
The presentation is geared up for the joke at the end, where the test charge flips sign rapidly. So this
involves some acting skills and careful timing. Patience is required.

For this demonstration we need three people. In our show these were the two main characters, MH and LU,
as well as one of the CERN scientists, here TL. MH and LU stand a few meters apart and each of 
them has a sign with a single charge on it displayed to the audience, MH a plus and LU a minus sign, 
see Fig.\,\ref{Fig:paper-electric-field-1}. Previously, via the projector a ``Faraday" slide of the electric vector 
field with field lines between two opposite point charges was shown, see Fig.\,\ref{pic:paper-electric-field-2}.
Hopefully some members of the audience have seen this in their school days. Thus this is a reenactment of that picture, 
with the field lines missing.

TL has the sign with a + on one side and a $-$ on the other, he represents the test charge, which is placed 
in the static electric field, created by the two time travellers. The scientist steps between the other two, holding 
up his sign, let's say $-$. He is then attracted to MH, who is holding the $+$ and repulsed by LU, who 
is holding a $-$. After stating this point, he strides to MH. Here it is essential to stress the conceptual point; by 
being placed \textit{in} a \textit{vector} field he receives a property (potential energy), \textit{and} a direction, 
the momentum of his motion. 

When the scientist with the probe charge reaches MH with the oppositely charged source charge, he flips 
his sign by flipping his chart. He is thus now attracted to LU, the other source charge, and repulsed by 
MH. He thus strides over to LU.

Now LU and MH double their  charges and then TL once again flips his charge. He is now attracted more strongly
to MH and thus runs over to her. After flipping his charge he runs back to LU.

Finally, LU and MH dramatically increase their charges, see the photo on the right in 
Fig.\,\ref{pic:paper-electric-field-2}. TL demands close attention of the audience, goes into a starting
position as for a race, and then quickly flips his sign back and forth. Slightly out of breath, he emphasizes
that he has moved so quickly, that the audience wasn't even able to see him move.

\subsection{Scalar Field: Analogy}
\label{app:scalar-field}
This is an extension of the previous experiment on the vector field, described in App.\,\ref{app:paperelectric}.
The main point is when you place an object in a scalar field it takes on a property, in our case temperature,
but {\it no} direction.

\subsubsection{Materials}
This requires a slide showing a weather map with temperatures, in our case Italy, see Fig.\,\ref{img:italy-map},
and one glass of water.

\subsubsection{Presentation}
The presenter takes a glass of water and asks the audience to imagine placing this at various places
on the map. It obviously acquires the local temperature, but no direction. The important point then
is that when you place an object in the background  Higgs-field, it also acquires a property, namely mass.
The stronger the interaction with the Higgs field the larger the mass.

\subsection{Visualizing the Scalar Higgs Field}
\label{app:higgscloth}

This experiment creates a visualization of the Higgs field and the mechanism through which particles interact with that field and therefore acquire mass. A large cloth is used to represent the Higgs field and the movement of objects of different weights through the swinging cloth helps create a tangible metaphor for the motion of particles of different mass.

\subsubsection{History}
The idea for this experiment originated from a Science Slam in Aachen \cite{herbi-higgs-slam} by HKD.

\subsubsection{Materials and Technical Details}
We use a large cloth of 2.50\,m$\times$1.40\,m, which represents the background Higgs field.
We use a large plush-owl as a heavy particle, a small plush-sheep as an electron, a light particle,
a big-fat green plush-frog to represent the Higgs boson, and two hammers to excite the Higgs field.

\subsubsection{Presentation}
\begin{figure}[h]
\center{\includegraphics[width=0.85\textwidth]{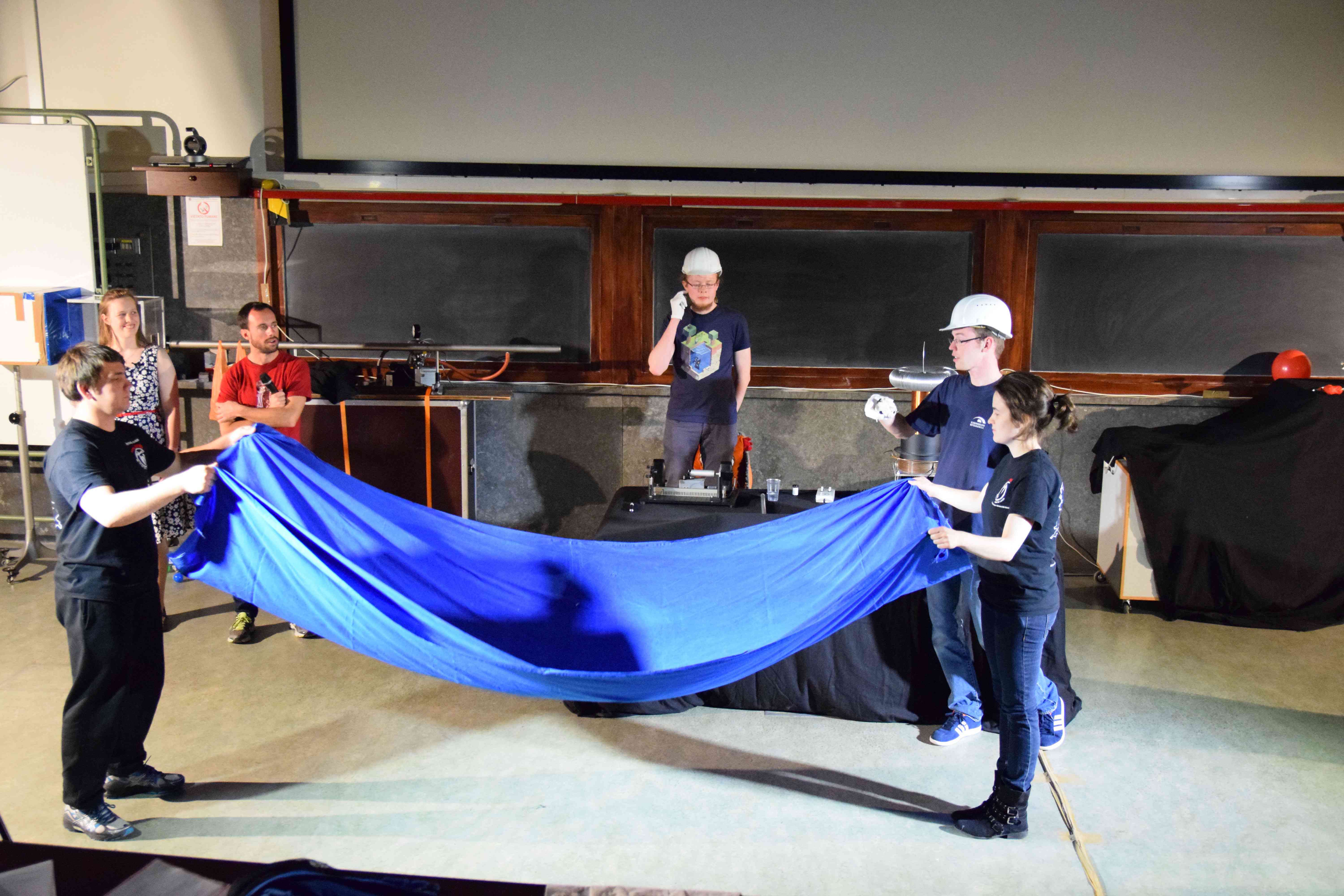}}
\caption{The swinging cloth in the image represents our fluctuating Higgs Field. One CERN scientist, TL,
is about to throw in the small plush-sheep.}
\label{fig:HiggsField}
\end{figure}

Two people grab each one end of the cloth and start swinging it slowly up and down to represent the fluctuating Higgs field,
Fig.~\ref{fig:HiggsField}.
While in principle swinging the cloth is simple, doing so with the right strength and frequency takes a bit of practice. When 
one of our two volunteers inevitably gets tired the moderator throws in a quick joke about the Higgs field's eternal nature 
and mentions that it NEVER stops.
 
To investigate the interaction of particles with the Higgs field we begin by throwing in either a ping-pong ball or, as is shown 
in Fig.~\ref{fig:LightParticle}, a tiny plush sheep.

Since the object is very light, if the two people swing the cloth properly it jumps from side to side with a very high frequency. 
In terms of our analogy, the particle is barely interacting with the field and is capable of moving through space quickly. A small 
weakness of the experiment is that the motion of the object stems from the motion of the cloth. In reality the Higgs field does 
not provide particles with their momentum. The small toy corresponds to a light particle, like an electron, whose low interaction 
with the Higgs field corresponds to a low mass.

Next, a larger toy is thrown into the cloth. While the field continues billowing, the heavy object barely jumps and moves very 
slowly or not at all through the cloth. In our case we chose to use a toy owl, which corresponds a heavy particle like the top 
quark. The particles weight comes from its heavy interactions with the Higgs field that causes it to move slowly through it.

\begin{figure}[h]
	\center{\includegraphics[width=0.9\textwidth]{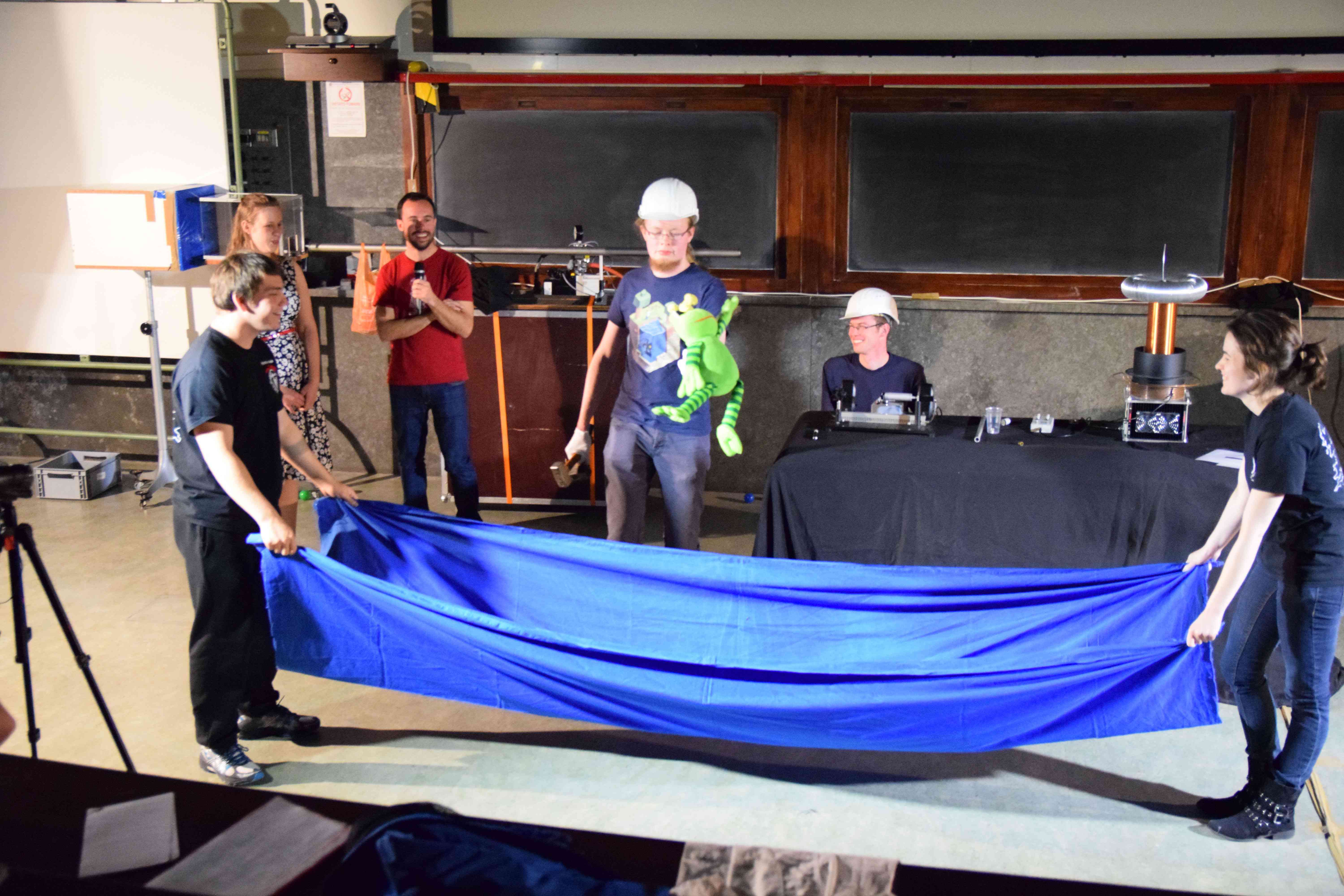}}
	\caption{Higgs boson being dropped out of the Higgs field after it was excited by the smashing of two hammers. One CERN scientist, CSch, is holding one of the hammers in his right hand. The cloth is being billowed by MBr and JSch-R.}
	\label{fig:HiggsBoson}
\end{figure}

We conclude the experiment by offering an explanation of the Higgs boson itself. It is said that the Higgs boson can be 
understood as the result of an excitation of the Higgs field. In order to ``excite" the field onstage, one of the scientists takes 
two hammers to ``hit" the Higgs field strongly. He counts down from three and finally smashes the hammers onto each other 
with the cloth in between. In this instant, someone hidden somewhere onstage throws another rather heavy plush animal or 
ball into the field. In our case we used a green frog, as can be seen in Fig.~\ref{fig:HiggsBoson}.
The two members swing the cloth so that the Higgs boson flies out and lands in the arms of the moderator. He explains that the Higgs boson is produced by the exciting the Higgs field and continues the story by asking how such an excitation is produced at the LHC.

\subsubsection{Safety}
The only real safety concern is the use of the hammers. The moderator should be careful not to hit themselves.

\subsection{The Higgs Mechanism and Eddy Currents: an Analogy}
\label{app:ewaldeddy}
\begin{figure}[h!]
    \centering
    \includegraphics[width=0.81\textwidth]{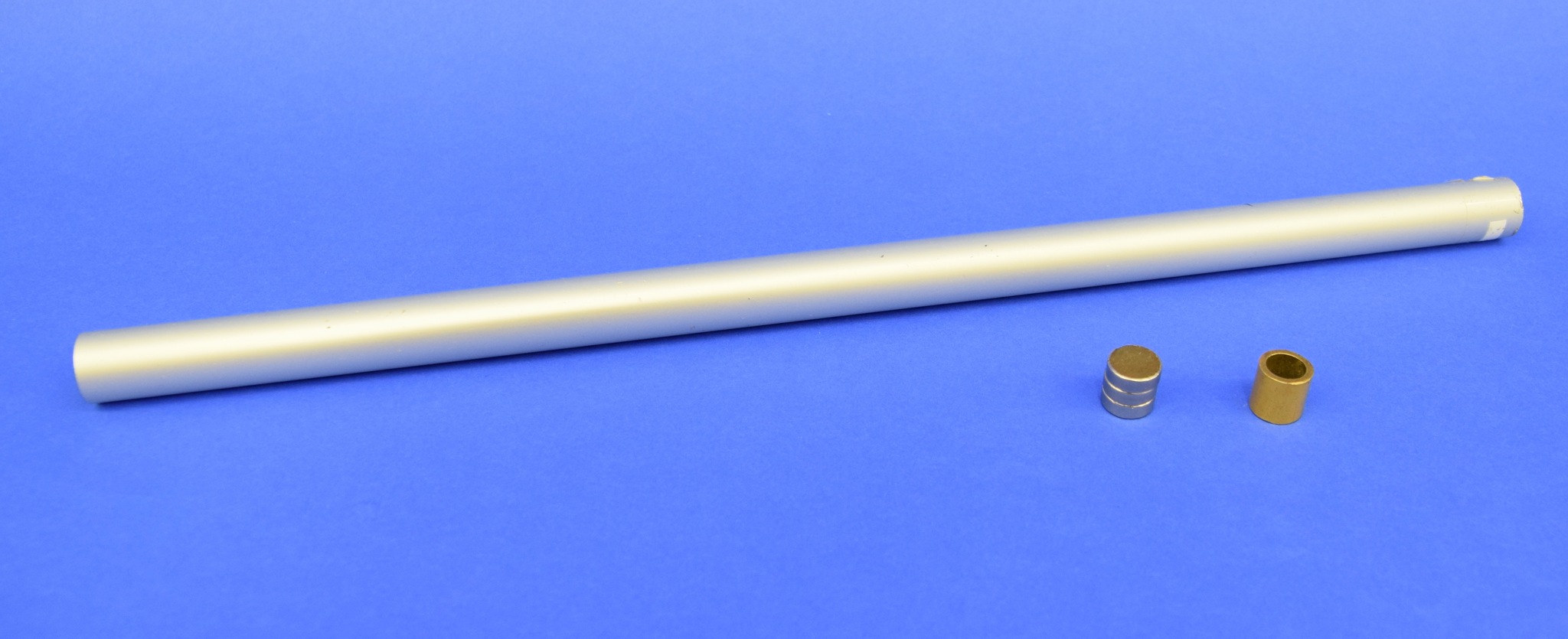}
    \caption{Materials for the eddy-current Higgs analogy experiment. The aluminum tube is 440\,mm long.}
    \label{fig:aluminumtube}
\end{figure}

For this experiment we use an aluminum tube as shown in Fig.~\ref{fig:aluminumtube}, together with a brass bolt 
as well as a strong neodymium  magnet. This is a well-known experiment, which is often used 
to show the effect of eddy currents. In can be performed also with a much larger tube, and 
larger magnets. In our group, it was EP's idea to use this experiment as an analogy for 
the Higgs mechanism. 

\subsubsection{Technical Details}
The aluminum tube we use is 440\,mm long, has an inner diameter of 16\,mm and an outer diameter of
20\,mm. We use a simple metal bolt which is made of brass (non-magnetic) and has a diameter
of 15\,mm and a height of also 15\,mm. The magnet is the same size: 15\,mm high and 15\,mm in diameter.

\subsubsection{Presentation}
The main point of this experiment within the show is the analogy with the Higgs mechanism. After
spontaneous symmetry breaking the lowest energy configuration for the Higgs field has a constant 
background value everywhere in space and time. By being placed in this background field, particles
obtain mass. The stronger they couple to the Higgs field, the larger the mass. In our experiment,
the background Higgs field is symbolized by the aluminum tube. We hold the aluminum tube vertically
in one hand. Below the lower opening we place a plastic cup. This collects the pieces
as they drop through, with  a distinctly audible ``plonk". Next we take a permanent magnet and lift
a key. Then with this permanent magnet we show that the aluminum tube is not magnetic.

For the experiment we first take the brass bolt. This represents a massless particle which does not
interact with the Higgs field, which in our physical representation means it does not interact with the 
aluminum tube. We drop this bolt
into the tube. It flies straight through and drops into the cup. In our analogy, this is a massless 
particle which flies at the speed of light, unhindered by our representation of the Higgs field, the aluminum tube.

Next we drop the neodymium magnet into the tube. Due to the eddy currents, it falls very slowly. In our
case it takes several seconds to fall into the cup. In our analogy the magnet represents  a massive particle,
for example an electron. It interacts with the aluminum tube, our representation of the Higgs field, and therefore
it can not fly at the speed of light; instead it falls very slowly.

\subsection{Light-Suits and Higgs Collision}
\label{app:lightsuits}
In this experiment five people represent different particles wearing differently coloured light-suits, where the 
lights can be turned on and off. They perform a choreography meant to mimic a Higgs-production from two colliding 
protons.

\subsubsection{History}
Light-suits are very popular in various artistic performances, with many examples on YouTube.

\subsubsection{Materials and Technical Details}

For this experiment we used five black overalls. To these we attached LED strips of various colors.
The LED strips run on 12\,V/DC. Each package is 5\,m and we used about 8\,m per overall. The
LED system we used was extendable. To supply the LED strips with electric power we provided each suit 
with eight AA rechargeable batteries, a battery holder for parallel connection, thin wire, and a switch. This gives
the person in the suit control of the light. In the show, as an additional visual effect, we use two yellow flashing 
warning lights which indicate the activated accelerator.

\begin{figure*}[h!]
	    \centering
    \includegraphics[width=0.475\textwidth]{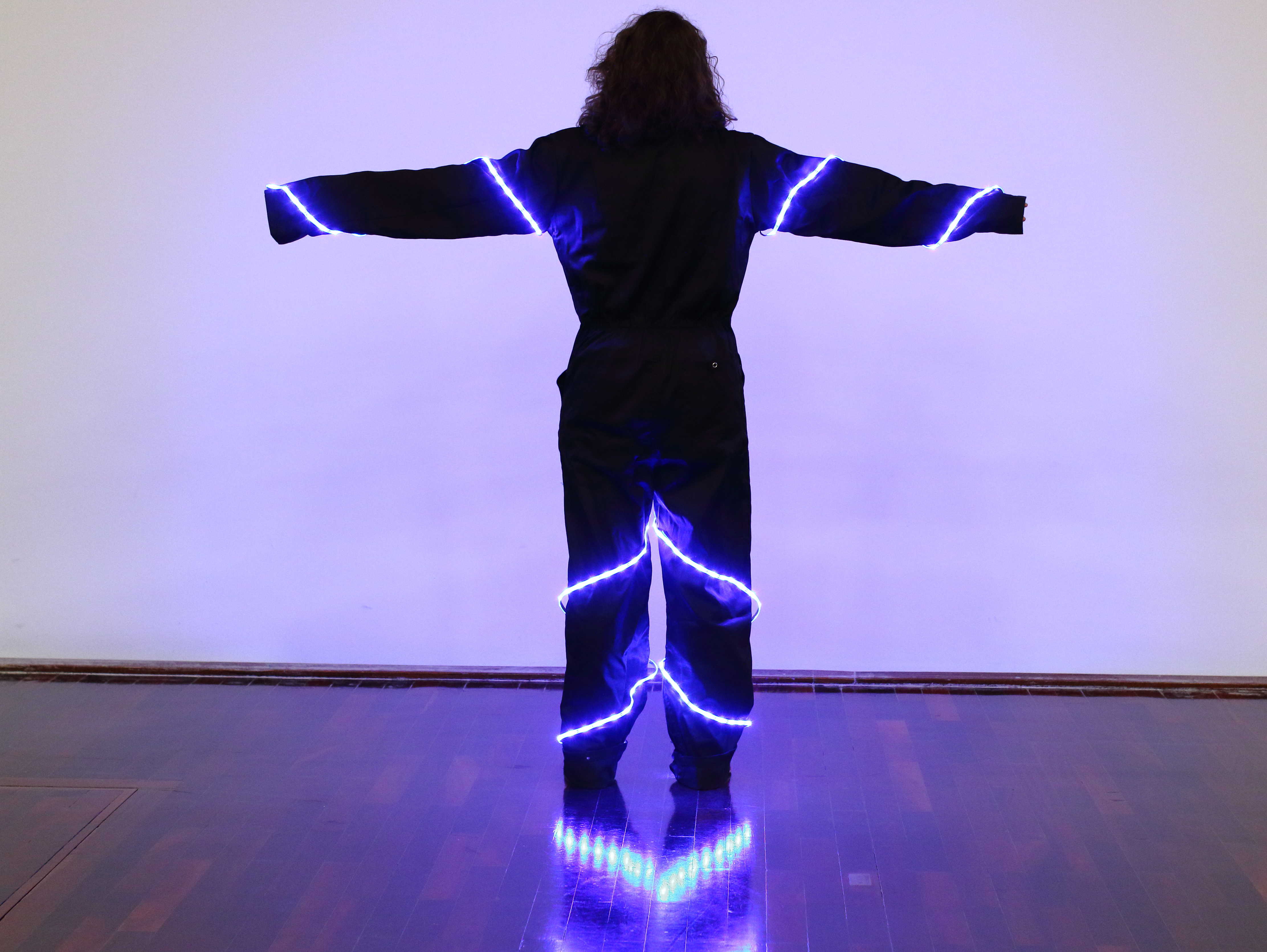}\;\includegraphics[width=0.475\textwidth,height=5.6cm]{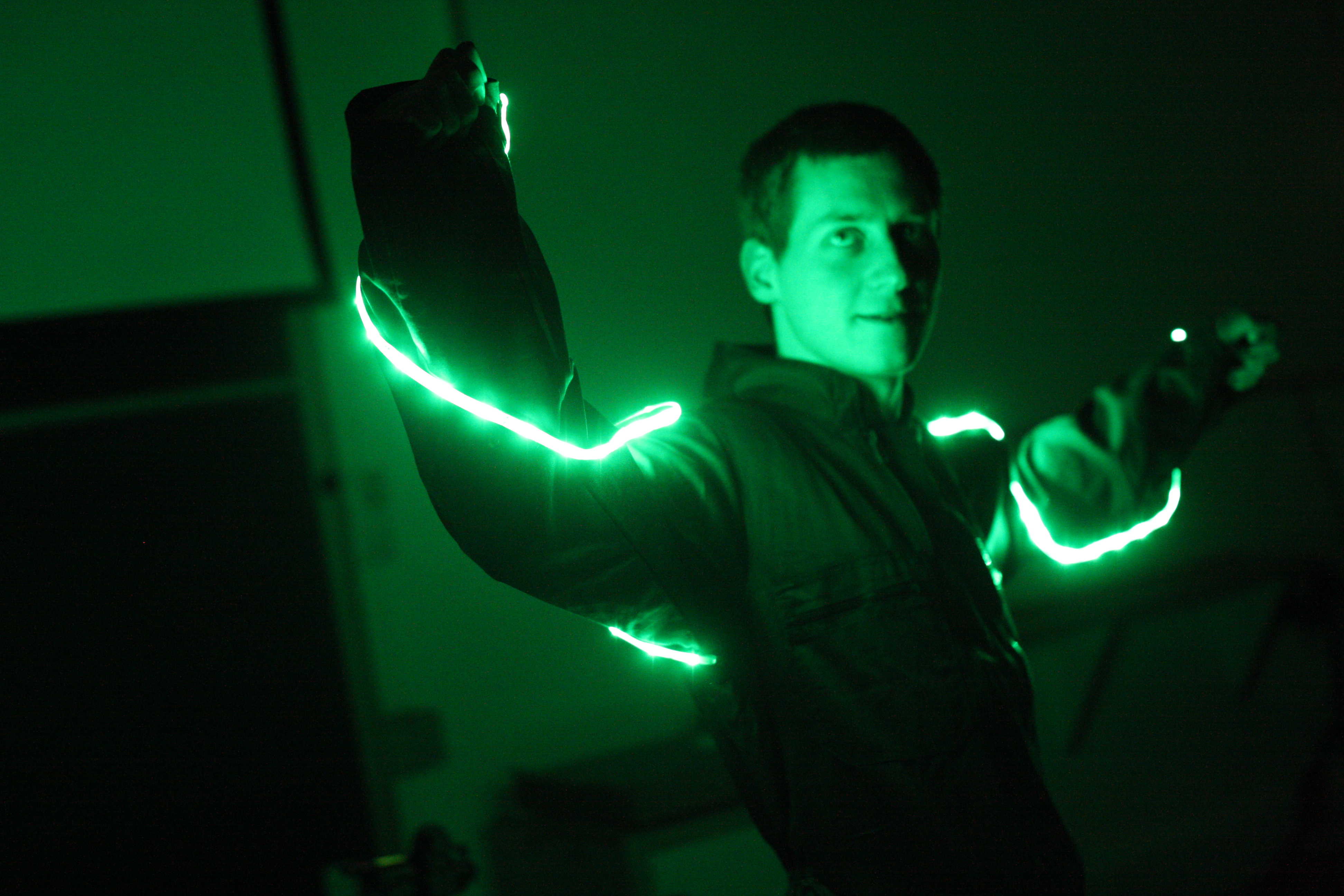}
           \\  (a) \hspace{6cm} (b)
        \caption{(a) One of the blue light suits, representing a photon. (b) A close-up of the green light suit for the 
        Higgs boson, with MKr.}
        \label{fig:light-suits}
\end{figure*} 

We decided to show the decay of the Higgs boson into two photons, therefore we have two white light-suits representing 
protons, two blue light-suits for photons, and a green light-suit for the Higgs boson, see Fig.~\ref{fig:light-suits}. For each 
suit the LED strip was cut into four pieces, and wound up around the arms and legs of the overall. It is best to fix the LED 
strip to the overall by using small pieces of  hook-and-loop tape. For the power supply, each LED strip is connected with 
the battery holder in one of the pockets of the overall. The circuit has a switch that controls all lights on a given suit. The 
switch is connected with a long wire such that it can be placed at the end of the sleeve. The five suits available to us were 
mainly sewed and assembled by Dagmar Fa{\ss}bender and MKo.

\subsubsection{Presentation}
Onstage the flashing yellow lights are switched on and the moderator warns that the accelerator has been turned on. 
The people onstage escape, while pointing out that the audience has to stay.

When the stage gets dark the five people in the light-suits, which are switched off, enter the stage. The Higgs boson and 
the two photons place themselves kneeling in the center of the stage, where the Higgs is about 1.5\,m away from the 
photons. The protons place themselves at the extreme left and right ends of the stage. Some form of dance music starts,
the protons light up and move towards each other. In the first approach they miss  and go to the opposite edge of the stage 
where they turn off their lights again for a short time. They re-light, approach again and this time ``collide", right next to the 
Higgs boson. While the protons kneel down and turn off their lights, the Higgs boson stands and lights up. It moves slowly 
towards the position of the photons and is flashing using the switch, indicating instability. When the Higgs boson approaches 
the photons it decays and the person kneels down and switches of the light. Now the photons light up and move quickly along 
two corridors through the audience to the back of the hall.

The reaction is explained after the first performance. Everything is rewound by the Caretaker, who can control time. All the 
particles perform the reaction backward but a bit faster. When the Higgs is backing up we play the warning beeping sound that
trucks and vans use. We play the sound of a music tape being rewound for the other particles. Then the performance is repeated 
more slowly and explained by the Caretaker. This last experiment marks the end of the show.

\subsubsection{Safety}
The performance requires a lot of space and a dark stage, which should be free of anything people can trip over.
Special attention is needed since, due to a previous experiment, small plastic balls can still be scattered onstage.

\end{appendix}


\begin{thebibliography}{99}
\bibitem{Oxford-Accelerate}
A full length video of the Oxford show `Accelerate' can be viewed at
\url{www2.physics.ox.ac.uk/accelerate}. We are not aware if this show has been performed since.
\end{thebibliography}
\end{document}